\documentclass[twocolumn]{aastex63} 

\usepackage{amsmath}
\usepackage{amssymb}
\usepackage{empheq}
\usepackage{mathrsfs}
\usepackage{booktabs}
				
\usepackage{amssymb}

\shorttitle{Uniform Keplerian Fits to Warm Jupiters}
\shortauthors{Morgan et al.}
\graphicspath{{./}{figures/}}
\begin{document}


\title{Exploring Warm Jupiter Migration Pathways with Eccentricities. I. \\ Catalog of Uniform Keplerian Fits to Radial Velocities of 200 Warm Jupiters}

\correspondingauthor{Marvin Morgan}
\email{marv08@utexas.edu }

\author[0000-0003-4022-6234]{Marvin Morgan}
\affiliation{Department of Astronomy, The University of Texas at Austin, Austin, TX 78712, USA}
\affiliation{Department of Physics, University of California, Santa Barbara, Santa Barbara, CA 93106, USA}

\author[0000-0003-2649-2288]{Brendan P. Bowler}
\affiliation{Department of Physics, University of California, Santa Barbara, Santa Barbara, CA 93106, USA}
\affiliation{Department of Astronomy, The University of Texas at Austin, Austin, TX 78712, USA}

\author[0000-0001-6532-6755]{Quang H. Tran}
\affiliation{Department of Astronomy, Yale University, New Haven, CT 06511, USA}
\affiliation{Department of Astronomy, The University of Texas at Austin, Austin, TX 78712, USA}

\author[0000-0001-9957-9304]{Robert A. Wittenmyer}
\affiliation{University of Southern Queensland, Centre for Astrophysics, West Street, Toowoomba, QLD 4350 Australia}

\author[0000-0001-7294-5386]{Duncan J. Wright}
\affiliation{University of Southern Queensland, Centre for Astrophysics, West Street, Toowoomba, QLD 4350 Australia}

\author[0000-0002-4891-3517]{George Zhou}
\affiliation{University of Southern Queensland, Centre for Astrophysics, West Street, Toowoomba, QLD 4350 Australia}

\author[0000-0002-0692-7822]{Tyler R. Fairnington}
\affiliation{University of Southern Queensland, Centre for Astrophysics, West Street, Toowoomba, QLD 4350 Australia}

\begin{abstract}
 Giant planets are expected to predominantly form beyond the water ice line and occasionally undergo inward migration. Unlike hot Jupiters, which can result from high-eccentricity tidal migration, warm Jupiters between 0.1-1 AU ($\approx$10--365 d) are in many ways more challenging to explain because they reside outside the tidal influence of their host stars. Warm Jupiters should therefore preserve traces of their origins as their eccentricities are directly related to their past interactions. We analyze the eccentricities of 200 warm Jupiters orbiting 194 Sun-like host stars (with FGKM spectral types) using 18,587 RV measurements across 40 high-resolution spectrographs. RVs are compiled from the literature and are supplemented with 540 new observations from MINERVA-Australis at Mt. Kent Observatory and the Habitable-zone Planet Finder spectrograph at McDonald Observatory's Hobby-Eberly Telescope, which are timed to improve eccentricity constraints by sampling orbits near periastron passage. The overarching goal of this program is to establish the relative importance of giant planet migration channels through the largest homogeneous analysis of warm Jupiter orbital properties to date. In particular, we evaluate and compare the impact of different system architectures and host star characteristics on the population-level eccentricity distributions of warm Jupiters. Here, we present the target sample, observations, orbit fitting procedure, and parameter summary statistics of our survey. All orbit fit solutions, parameter posterior chains, and merged RV tables for each system are made publicly available. 
\end{abstract}

\section{Introduction} \label{sec:intro}

Early radial velocity observations of old Sun-like stars revealed a population of Jovian planets with orbital separations that deviate from the giant planets found in our own solar system (\citealt{CampbellWalker1988}; \citealt{MarcyBenitz1989}; \citealt{Walker1992}; \citealt{Mayor1995}; \citealt{Santos2003}; \citealt{Udry2003}). Since then, the past 30 years of planet discoveries has shown that the giant planet occurrence rate peaks between about 1--10 AU and decreases interior and exterior to this range (\citealt{Cumming2008}; \citealt{Bowler2016}; \citealt{Fernandes2019}; \citealt{Nielsen2019}; \citealt{Wittenmyer2020}; \citealt{Fulton2021}). Jovian-mass planets are expected to form in this region where disk conditions are conducive to giant planet core assembly and gas accretion before the dispersal of the protoplanetary gas disk around 5--10 Myr (\citealt{Pollack1996}; \citealt{Lecar2006}; \citealt{KennedyKenyon2008}). 

This poses a challenge to explain the origin of giant planets found on close-in orbits. As opposed to hot Jupiters (\emph{P} $\lesssim$10~d), which may arrive on short periods in part through high-eccentricity migration followed by tidal circularization (\citealt{DawsonJohnson2018}; \citealt{FortneyDawson2021}), warm Jupiters with orbital periods ranging from $\approx$10--365 d are a population of giant planets that are situated within the water ice line yet beyond the region of tidal influence from their host star (\citealt{Marcy1996}; \citealt{Wittenmyer2010}; \citealt{Santerne2016}).

The formation and evolution of  warm Jupiters remains elusive (see \citealt{DawsonJohnson2018} and references therein). They may have migrated to their present-day locations through long-range disk migration (\citealt{GoldreichTremaine1980}; \citealt{Lin1996}; \citealt{KleyNelson2012}), gravitational interactions such as planet-planet scattering (\citealt{RasioFord1996}; \citealt{IdaLin2004}; \citealt{Chatterjee2008};  \citealt{BeaugNesvorn2012}; \citealt{Petrovich2014}), or secular interactions (\citealt{Kozai1962}; \citealt{Lidov1962}; \citealt{WuMurray2003}; \citealt{WuLithwick2011}; \citealt{DawsonChiang2014}; \citealt{Petrovich2015}; \citealt{Naoz2016}; \citealt{Petrovich2016}). Under the right conditions, in situ formation is also possible (\citealt{Batygin2016}; \citealt{Boley2016}).

These processes, in principal, should imprint distinct features on the obliquities (\citealt{Albrecht2012}; \citealt{LiWinn2016}; \citealt{Albrecht2022}; \citealt{Rice2022}; \citealt{Morgan2024}) and eccentricities (\citealt{DongHuangTESS2021}; \citealt{KaneWittenmyer2024}) of warm Jupiters. Eccentricities should therefore play a pivotal role in disentangling the relative significance of these migration mechanisms. A comprehensive census of warm Jupiter orbital properties may help unravel the intricacies of giant planet formation, migration, and evolution. Substantial effort has been devoted to discovering and characterizing this population of giant planets (e.g. \citealt{FischerValenti2005}; \citealt{Johnson2010}; \citealt{Dawson2013ApJ}), however, a comprehensive analysis of their eccentricities as a population has not been carried out. In particular, because eccentricity distributions can be asymmetric, multi-modal, and pile up against 0, compiling heterogeneously reported constraints from previously published orbit fits will not correctly capture the true individual constraints and population-level behavior of warm Jupiters; instead, a systematic reanalysis of orbit constraints is needed. In this work we aim to establish the relative importance of warm Jupiter migration channels through the largest homogeneous analysis of warm Jupiter orbital properties to date. 

The paper is structured as follows. In Section \ref{sec:Target_Sample} we discuss our warm Jupiter target selection and properties of their host stars. In Section \ref{sec:observations} we present new spectroscopic observations from our multi-hemisphere campaign to improve eccentricity measurements for systems with the poorest constraints in our sample. Next, we describe our orbit refitting and model selection framework in Section \ref{sec:orbit_refitting}. In Section \ref{sec:results} we contextualize our results and evaluate the impact of the new RV measurements from our observational campaign. Finally, we discuss implications and outline forthcoming future studies that will make use of these results in Section \ref{sec:Conclusion}.

\section{Target Sample}\label{sec:Target_Sample}

\begin{figure}[t]  
 \centering  
    \includegraphics[width=\linewidth]{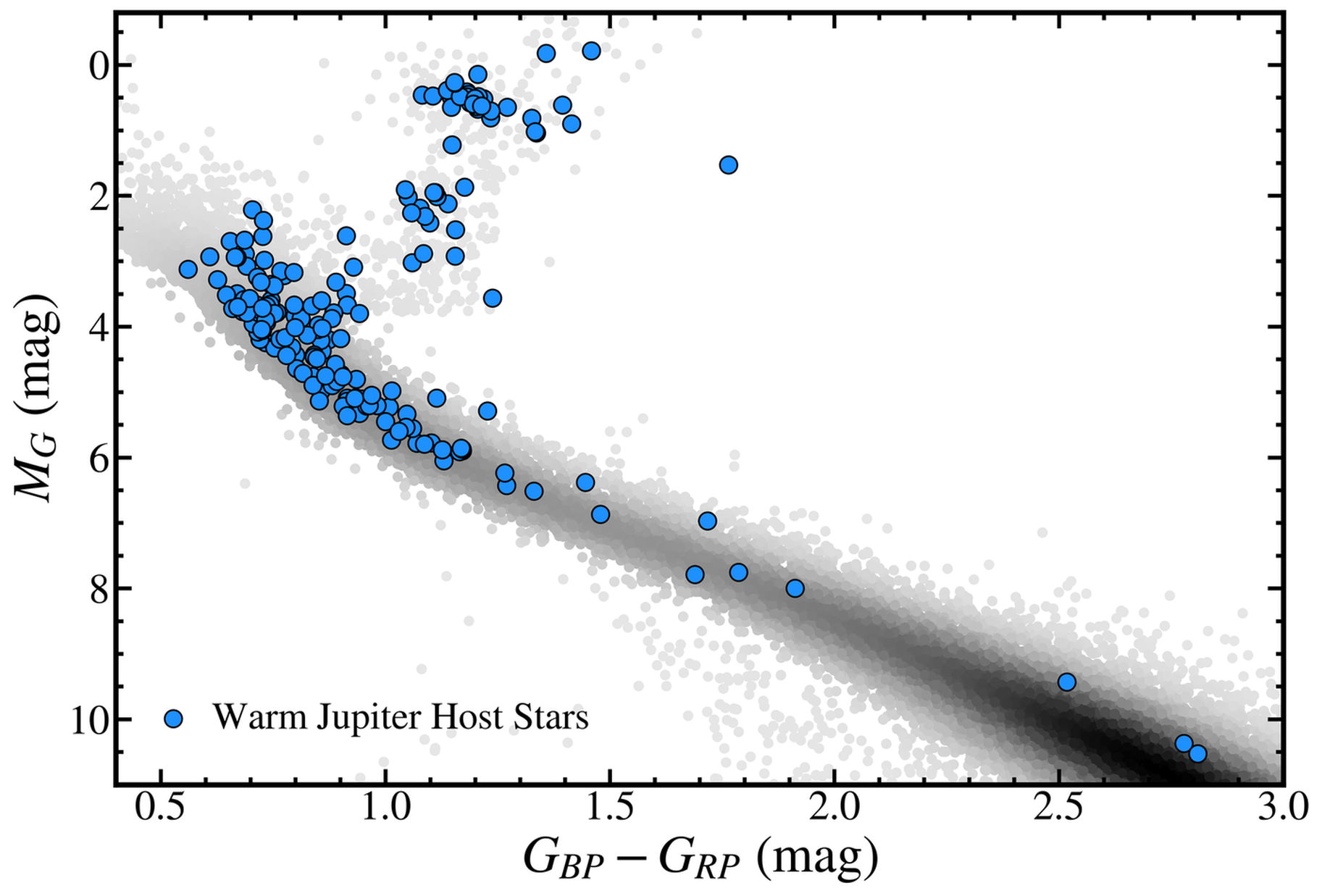}
    \caption{Warm Jupiter host stars in our sample plotted on a Gaia $M_{G}$ vs $G_{BP}$ - $G_{RP}$ color-magnitude diagram. Our analysis is comprised of both main-sequence and post-main-sequence FGKM stars.}
    \label{fig:Gaia_CMD}
\end{figure}

\begin{figure*}
\begin{center}
{\includegraphics[width=\linewidth]{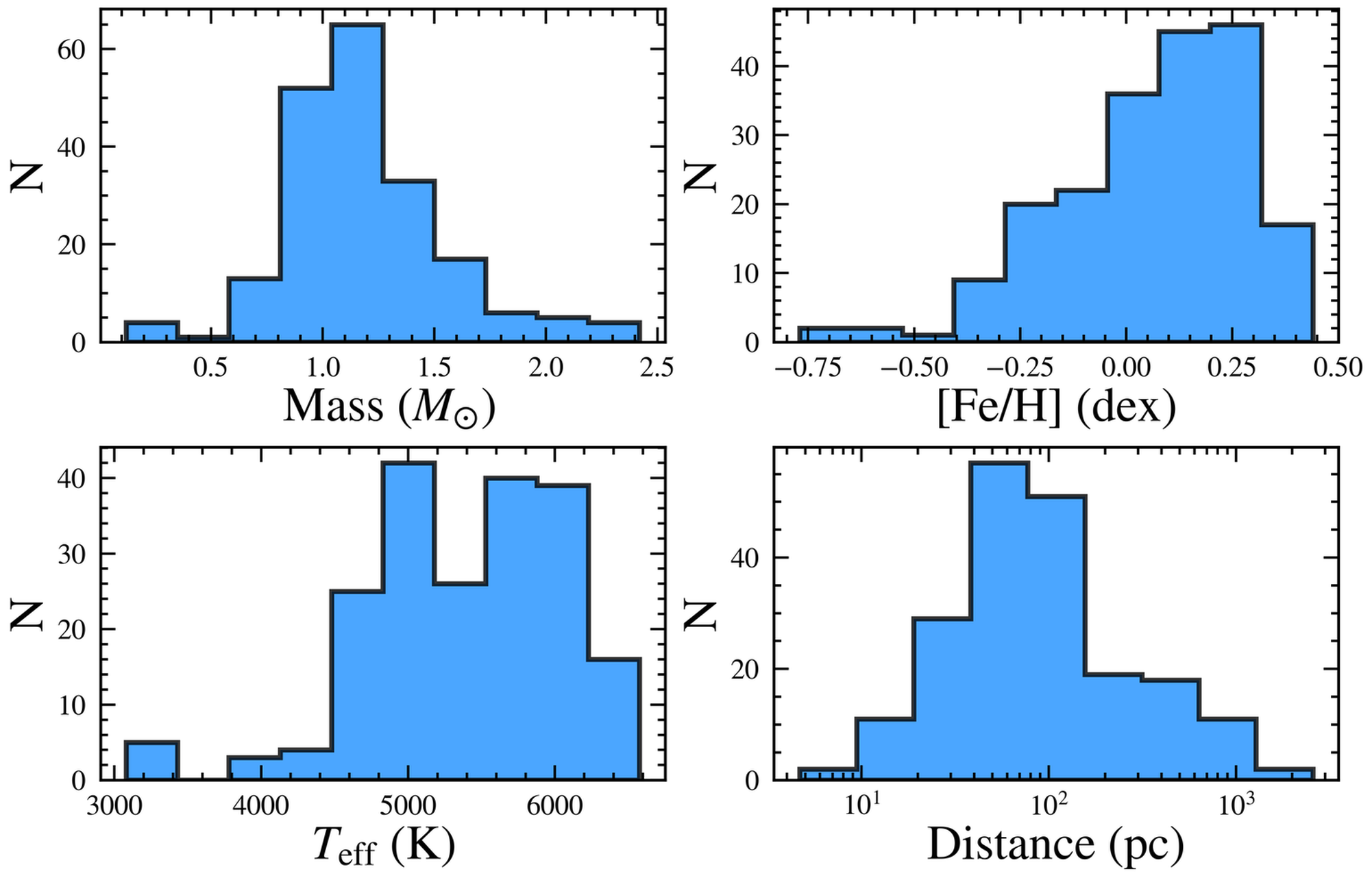}}
\caption{Distribution of stellar host masses, metallicities, effective temperatures, and distances for 194 host stars of warm Jupiters in our sample.} 
\label{fig:stellar_histogram}
\end{center}
\end{figure*}

\begin{figure*}
\begin{center}
{\includegraphics[width=\linewidth]{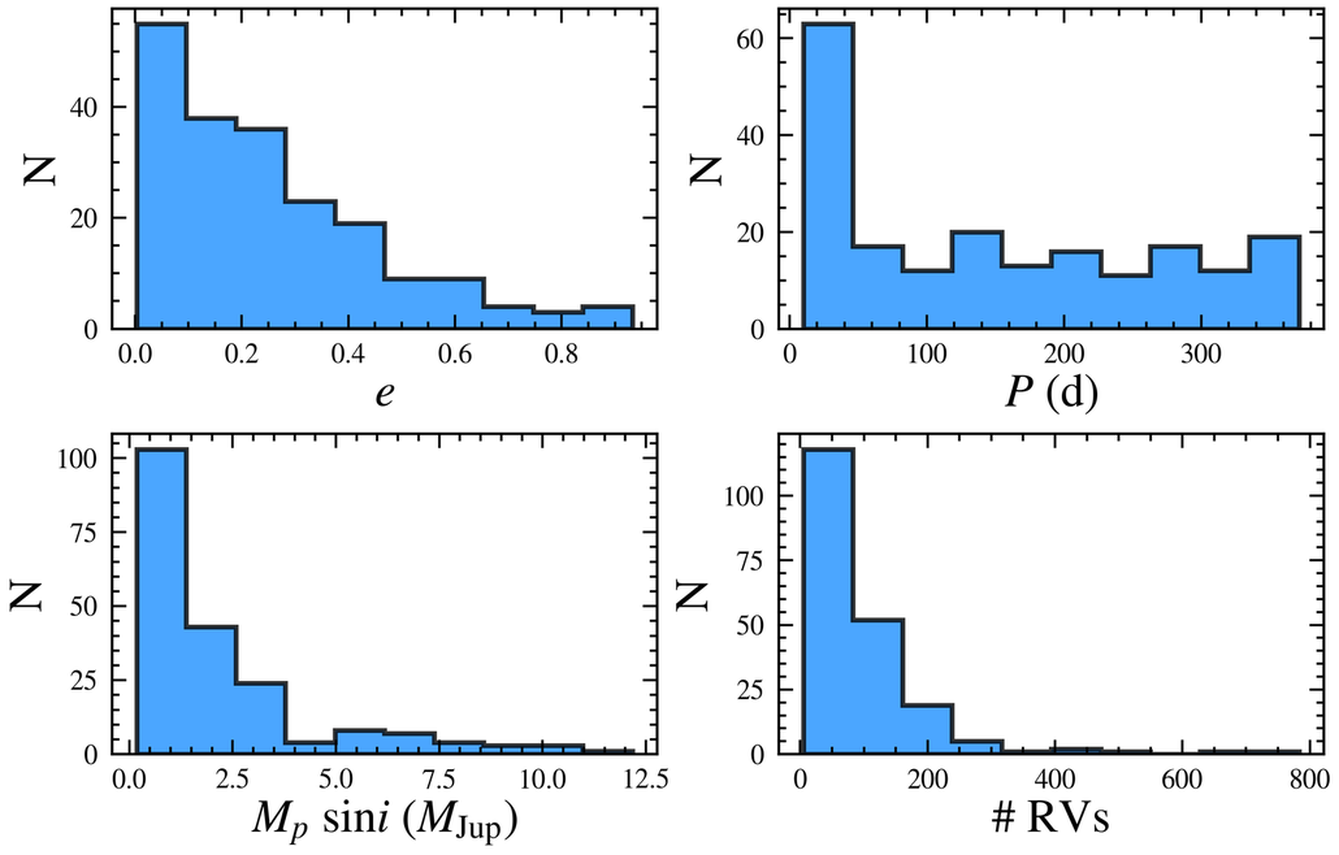}}
\caption{Distribution of eccentricities, orbital periods, minimum masses, and number of radial velocities per target for all planets in our analysis. Eccentricities, orbital periods, and minimum masses are derived from the final fits using \texttt{RadVel}.} 
\label{fig:planet_histogram}
\end{center}
\end{figure*}

Our sample of warm Jupiters originates from the NASA Exoplanet Archive (\citealt{Akeson2013}) as of May 2022. We first selected all planets confirmed by radial velocity measurements in order to homogeneously refit the orbits. We then selected planets with orbital periods between 10--365 days ($\approx$ 0.1--1 AU), and a measured minimum mass of $m_p \sin i$ = 0.3--13 $M_\mathrm{Jup}$, or a radius $>$ 8 ${\emph{R}}_{\oplus}$, ensuring that low mass brown dwarfs and sub-Jovian size planets are excluded. Although transit ingress/egress durations can provide information about orbital eccentricities (\citealt{MandelAgol2002}; \citealt{HolmanMurray2005}), these constraints are generally quite broad so we do not consider transiting planets that have not also been recovered with RVs. 

Hot Jupiters are also removed because their orbits have been shaped by tidal circularization. Warm Jupiters, on the other hand, are expected to be largely undisturbed by tidal forces. For a main-sequence Sun-like star, a separation $\gtrsim$ 0.1 AU corresponds to a circularization timescale of $>$ $1$ Gyr (\citealt{RasioTides1996}; \citealt{Lubow1997}). P-type circumbinary planets (planets orbiting around more than one host star) are removed to avoid the potential impact of a second central body on the planetary system's formation and evolution.

Two planets, HD 102272 b (\citealt{Niedzielski2009}) and HD 17092 b (\citealt{NiedzielskiKonacki2007}), were excluded from the analysis due to the absence of reported radial velocity measurements, resulting in 206 planets. HD 114762, originally thought to be a Jovian planet (\citealt{1989Natur.339...38L}; \citealt{1991ApJ...380L..35C}), has since been identified as a binary star (\citealt{Kiefer2019}; \citealt{Winn2022}; \citealt{Holl2023}) and was subsequently removed from our analysis. Furthermore, five planets (Kepler-418 b, KOI-12 b, Kepler-9 b, Kepler-9 c, and Kepler-539 b) were ultimately removed due to unconstrained fits in this study. Kepler-418 b (\citealt{Tingley2014}) has an insufficient number of RV measurements needed to independently converge on a unique orbit fit without the transit information which was used in the original fit. Similiarly, the orbits of Kepler-9 b (\citealt{Holman2010S}; \citealt{Borsato2019}), Kepler-9 c, and Kepler-539 b (\citealt{Mancini2016}) could not be reproduced without transit information as well. In particular, the period, $P_{0}$, and time of inferior conjunction, $T_{C}$, for these planets are tightly constrained from transit light curves but are not used in this study in order to maintain uniformity of the refitting methodology. Most of KOI-12 b's (\citealt{Bourrier2015koi12}) radial velocity data set are dominated by Rossiter-McLaughlin measurements, which are used to measure the sky-projected spin-orbit angle between a planet's orbital plane and its host stars equatorial plane. Altogether, this yields a sample of 200 warm Jupiters orbiting 194 host stars. 132 of these warm Jovian planets are in apparently single-planet systems based on the available observations, and 68 reside in known multi-planet systems. 

\subsection{Host Stars}\label{sec:Host_Stars}
One of the overarching goals of this effort is to examine how planet eccentricities track with properties of the host star.  We have therefore assembled information about stellar metallicity, mass, evolutionary stage, age, and multiplicity. Host star properties in this study are compiled from a multitude of sources and are usually from detailed characterization as part of radial velocity and transit surveys (See Figure \ref{fig:Gaia_CMD}). Each campaign has its own specific focus, depending on the nature of the science goals (see \citealt{Gaudi2005}; \citealt{Fischer2016}; \citealt{Perryman2018}). For instance, while some programs focus on identifying planets around Sun-like main sequence stars, others are designed to study the impact of planet formation and migration around stars of varying metallicities, or as a function of evolutionary stage, or are dedicated to the discovery of Jupiter analogs. 

 Figure \ref{fig:stellar_histogram} shows the distribution of stellar masses, metallicities, effective temperatures, and distances compiled from the Strasbourg astronomical Data Center's VizieR catalog access tool (\citealt{Ochsenbein2000}), the NASA Exoplanet Archive (\citealt{Akeson2013}), and Gaia (\citealt{GaiaCollaboration2023}; \citealt{BailerJones2021}).  Most masses range from about 0.5--2 $M_{\odot}$, metallicities from $-$0.50 to $+$0.5 [Fe/H], effective temperatures from $\approx$ 4000--6500 K, and distances out to $\sim$ 500 parsecs (most being within 100 pc). 

Stars in this sample were observed with various spectral resolving power  and wavelength coverage in the optical and near infrared. These are summarized in Table \ref{tab:spectrographs_instruments}. Each individual instrument in the uniform Keplerian fit has its own jitter values to reduce inhomogeneity in the analysis. Treating jitter as a free parameter for each instrument is intended to mitigate potential differences in both measurement and astrophysical uncertainties. For instance, a more active star may exhibit larger uncertainties in the optical compared to the near-infrared, and the jitter term absorbs this wavelength-dependent variability. Additionally, differences in instrumental precision or stability stemming from varying spectral resolutions ($\approx$40,000 $<$ \emph{R} $<$ 100,00) or environmental control are also accounted for with this prescription.

\section{Observations}\label{sec:observations}

Here we describe new RV observations obtained as part of this program from December 2021 to July 2023 with the Habitable-zone Planet Finder spectrograph (Section \ref{sec:Habitable Zone Planet Finder}) and MINERVA-Australis (Section \ref{MINERVA-Australis}).

Among the full sample of warm Jupiters available for this experiment, the precision of existing eccentricity constraints varies with reported uncertainties extending to as high as $\sigma_{e}$ $\sim$ 0.2. All targets have existing RV detections of warm Jupiters, but the lack of sufficient phase coverage, modest number of RVs, or moderate stellar jitter in some cases resulted in a poor constraint on the eccentricity. To make optimal use of all systems for our population-level eccentricity analysis, we launched a multi-hemisphere survey partitioned between the Hobby-Eberly Telescope's (HET) Habitable-zone Planet Finder spectrograph (HPF) in the north and MINERVA-Australis for southern targets to improve the eccentricities of the least-constrained $\sim$$10\%$ of the sample. This resulted in 20 targets altogether, 10 observed with HPF and 10 with MINERVA-Australis. We focused increased cadence near periastron, where the acceleration is greatest and the maximum information content about the orbital eccentricity is located (e.g. \citealt{Endl2006}; \citealt{Blunt2019}; \citealt{Bergmann2021}). We add a total of 540 new radial velocity measurements which are presented below.

\subsection{Habitable-zone Planet Finder Spectrograph}\label{sec:Habitable Zone Planet Finder}
Northern targets in our sample were observed with HPF on the 10-m Hobby Eberly Telescope at McDonald Observatory (\citealt{Ramsey1998}). HPF is a near-infrared ($\lambda$ = 810--1270 nm) high-resolution spectrograph (\emph{R} $\equiv$ $\lambda$/$\Delta$$\lambda$ $\approx$ 50,000) calibrated with a laser frequency comb, with on-sky RV precisions at the $\sim$1 m/s level (\citealt{Mahadevan2012}; \citealt{Metcalf2019}; \citealt{Tran2021}). The brightest ($J$ $<$ 8 mag) 10 northern targets with reasonable $v \sin i$ values (generally $\lesssim$ 5–10 km/s) were prioritized. Observations were carried out using the Hobby-Eberly Telescope's queue scheduling mode (\citealt{Shetrone2007}). 180 radial velocity measurements were collected from December 2021 to July 2023. 

Relative HPF RVs are measured with a least-squares matching algorithm based on the \texttt{SERVAL} \citep{Zechmeister2018} package following the procedures described in \citet{Tran2021}. All 1D spectra are first corrected for barycentric motion with the \texttt{barycorrpy} \citep{Kanodia2018}. Then, high frequency pixel-to-pixel variations are removed with a median master flat spectrum. A master science template is created by scaling all spectra to the highest-S/N spectrum with multiplicative third-order polynomials and applying a uniform cubic basic (\textit{B}-spline) regression. $\kappa$-sigma clipping is applied to the template spectrum to remove large outliers and data points near the spectrum edge. Telluric and sky emission features are masked during this process.

Using this master science template, we simultaneously fit for the RV shifts and polynomial coefficients. The template is Doppler shifted in steps of 50 m s$^{-1}$ from $-5$ km s$^{-1}$ to 5 km s$^{-1}$ and the $\chi^2$ residuals between the shifted template and science spectrum are calculated. The RV shift and variance for each spectral order are estimated from this $\chi^2$ parabola. The final relative RV value for each spectrum is the weighted mean and standard error among the eight \'echelle orders detailed in \citet{Tran2021}. For the RV of a single epoch, three contiguous exposures are obtained and binned together using a weighted average.

\subsection{MINERVA-Australis}\label{MINERVA-Australis}
We obtained new observations from MINERVA-Australis to improve the eccentricities of warm Jupiter targets in the southern hemisphere. MINERVA-Australis is comprised of four 0.7m CDK700 telescopes that feed light into an optical high-resolution spectrograph ($\lambda$ = 500--630nm) with $\sim$ 1 m/s precision radial velocity capabilities (\citealt{Wittenmyer2018}; \citealt{Addison2019}). The brightest ($G$ $<$ 10 mag) 10 southern targets with reasonable $v \sin i$ values (generally $\lesssim$ 5–10 km/s) were selected for further observations. 360 radial velocity measurements were taken from January 2022 to July 2023. 

RVs are derived through a cross correlation function, where the matched template is the mean spectrum of each telescope. Each fiber is treated as an independent instrument and the spectra are analyzed individually. Radial velocity measurements with deviations greater than 3$\sigma$ from the mean are clipped from each individual fiber for all systems. 

\begin{deluxetable*}{lcccccl}
\renewcommand\arraystretch{0.9}
\tabletypesize{\footnotesize}
\setlength{\tabcolsep} {.1cm}
\tablewidth{0pt}
\tablecolumns{5}
\tablecaption{Spectrographs used to collect RV data for warm Jupiters \label{tab:spectrographs_instruments}}
\tablehead{ \colhead{Telescope} & \colhead{Instrument} & \colhead{$\#$ RVs} & \colhead{$\lambda$} &\colhead{Reference} \\
& & &\colhead{(nm)}&
}
\startdata
Whippie 1.5-m & AFOE & 78 & 450--700 & 1 \\
Lick/Levy 2.4-m & APF & 444 & 374--970 & 2 \\
Fairborn AST 2-m & AST & 136 & 492--710 & 3 \\
BOAO 1.8-m & BOES & 165 & 350--1050 & 4,5 \\
Calar Alto 2.2-m & CAFE & 28& 365--980 & 6 \\
Calar Alto/CAHA 3.5-m & CARMENES & 368& 550--1700 & 7,8 \\
ESO CAT 1.4-M & CES &95 & 360--1100 & 9,10 \\
CTIO 1.5-m & CHIRON & 262& 410--870 & 11,12 \\
ESO Euler 1.2-m & CORALIE & 1817 & 381--681 & 13 \\
Tautenburg 2-m & COUDE & 212& 463--737 & 14 \\
ESO Paranal VLT 8-m & CRIRES & 4 & 1000--5000 & 15,16 \\
OHP 1.9-m & ELODIE & 396 & 390--681 & 17,18 \\
CFHT 3.6-m & ESPaDOnS &29 & 370--1050 & 19 \\
MPG 2.2-m & FEROS & 438& 360--920 & 20,21 \\
ESO LA Silla 1-m & FIDEOS &34 & 400--680 &22,23 \\
NOT 2.5-m & FIES &52 & 370--730 & 24 \\
TNG 3.6-m & GIANO &7 & 950--2500 & 25,26 \\
Lick/Shane 3-m & HAMILTON & 1620& 340--1100 &27,28 \\
ESO LA Silla 3.6-m & HARPS &1261 & 380--690 & 29,30  \\
TNG 3.6-m & HARPS-N &301 & 378--691 & 31 \\
Subaru 8.2-m & HDS & 35&  300--1000 & 32 \\
OAO 1.9-m & HIDES &1690 & 500--620 & 33 \\
Keck-I 10-m & HIRES & 4547 & 300--1100 & 34,35 \\
McDonald HET 9.2-m & HPF & 195 & 810--1270 & 36,37 \\
McDonald HET 9.2-m & HRS & 1134 & 390--1100 & 38 \\
Sloan 2.5-m & KeckET &16 &  500--590 & 39 \\
KPNO 2.1-m & KPNO ET & 40& 500--564 & 40,41,42 \\
Sloan 2.5-m & MARVELS & 25& 500--570 & 43,44,45 \\
Magellan II 6.5-m & MIKE &48 & 320--1000 & 46,47,48 \\
Mount Kent Obs./ 4$\times$0.7-m & MINERVA-Australis & 539 & 500--630 & 49,50 \\
LCOGT/ 6 $\times$ 1-m & NRES & 33& 390--860 &51 \\
Magellan II 6.5-m & PFS & 48& 390--620 & 52,53 \\
TNG 3.6-m & SARG & 29& 370--1000 & 54,55 \\
SONG 1-m & SONG & 40 & 440--690 & 56 \\
OHP 1.9-m & SOPHIE/SOPHIE+ & 1460 & 387--694 & 57,58 \\
Tillinghast 1.5-m & TRES & 106&  385--910 & 59 \\
McDonald 2.7-m & Tull &313 & 387--1045 & 60,61 \\
AAT 3.9-m & UCLES & 451& 300--1100 & 62,63,64 \\
Xinglong 2.2-m & XSCES & 31 & 330--1100 & 65,66 \\
Xinglong 2.2-m & XSHRS & 60 & 330--1100 & 66 \\
\enddata
\tablecomments{Telescopes and instruments used to discover and characterize warm Jupiters in our sample. Listings are ordered by instrument name.}
\tablerefs{(1) \citet{BrownNoyes1994}; (2) \citet{Vogt2014}; (3) \citet{EatonWilliamson2004}; (4) \citet{Izumiura2005}; (5) \citet{Kim2007}; (6) \citet{Aceituno2013}; (7) \citet{Quirrenbach2012}; (8) \citet{Reiners2018}; (9) \citet{Hatzes1996}; (10) \citet{Kurster1999}; (11) \citet{Schwab2010}; (12) \citet{Tokovinin2013PASP..125.1336T};
(13) \citet{Queloz2000}; (14) \citet{Hatzes2003}; (15) \citet{Kaeufl2004}; (16) \citet{Bean2010}; (17) \citet{Mayor1995}; (18) \citet{Baranne1996}; (19) \citet{Donati2006}; (20) \citet{Kaufer1999}; (21) \citet{Setiawan2008}; (22) \citet{Tala2014}; (23) \citet{Vanzi2018}; (24) \citet{Telting2014}; (25) \citet{Oliva2012}; (26) \citet{Oliva2012b}; (27) \citet{Vogt1987}; (28) \citet{Cumming1999}; (29) \citet{Mayor2003}; (30) \citet{Mayor2009}; (31) \citet{Cosentino2012}; (32) \citet{Noguchi2002}; (33) \citet{Sato2005}; (34) \citet{Vogt1994}; (35) \citet{Vogt2000}; (36) \citet{Mahadevan2012}; (37) \citet{Metcalf2019}; (38) \citet{Tull1998}; (39) \citet{Ge2006B}; (40) \citet{Ge2003}; (41) \citet{Ge2006a}; (42) \citet{Mahadevan2008}; (43) \citet{Ge2008}; (44) \citet{Ge2009a}; (45) \citet{Ge2009b};  (46) \citet{Bernstein2003}; (47) \citet{LopesMorales2008}; (48) \citet{Minniti2009};
(49) \citet{Wittenmyer2018}; (50) \citet{Addison2019}; (51) \citet{Eastman2014}; (52) \citet{Crane2008}; (53) \citet{Crane2010}; (54) \citet{Gratton2001}; (55) \citet{Gratton2004}; (56) \citet{Andersen2014}; (57) \citet{Perruchot2008}; (58) \citet{Bouchy2013}; (59) \citet{Furesz2008}; (60) \citet{Tull1995}; (61) \citet{CochranHatzes1994}; (62) \citet{Diego1990}; (63) \citet{Tinney2001}; (64) \citet{Horton2012}; (65) \citet{Zhao200}; (66) \citet{Fan2016}. } 
\end{deluxetable*}

\begin{deluxetable}{ccc}
\tabletypesize{\scriptsize}
\tablecaption{Parameter priors for MCMC analysis. \label{tab:priors}}
\tablehead{
    \colhead{Parameter} & \colhead{Prior} & \colhead{Description}
}
\startdata
\emph{$P$} & $\mathcal{U}$[1, 400] & Period (d) \\
\emph{e} & $\mathcal{U}$[0, 1] & Eccentricity \\
$\omega$ & $\mathcal{U}$[0, 2$\pi$] & Argument of Periastron (rad) \\
\emph{$T_{C}$} & $\mathcal{U}$[$\tau$-$\frac{1}{2}$$P_{0}$, $\tau$+$\frac{1}{2}$$P_{0}$] & Time of Conjunction (BJD) \\
\emph{K} & $\mathcal{J}$[1, 1000] & RV Semi-Amplitude 
(m $\text{s}^{-1}$) \\
\emph{$dv/dt$} & $\mathcal{U}$[--2.75, 2.75] & Linear Trend (m $\text{s}^{-1} \text{d}^{-1}$)\\
\emph{\emph{$d^{2}v/dt^{2}$}} & $\mathcal{U}$[--0.0075, 0.0075] & Curvature (m $\text{s}^{-1} \text{d}^{-2}$)\\
\emph{$\sigma_\mathrm{jit}$} & $\mathcal{U}$[0, 100] & Jitter (m $\text{s}^{-1}$) \\ 
\hline
\hline
\enddata
\tablenotetext{a}{$\tau$ refers to an arbitrary reference time.}
\tablenotetext{b}{$\mathcal{U}$ and $\mathcal{J}$ represent uniform and Jeffreys priors, respectively. $\mathcal{U}(a,b) = 1/(b-a)$ between $a$ and $b$, and 0 elsewhere. $\mathcal{J}(a,b) \propto 1/x$ between $a$ and $b$.}
\end{deluxetable}

\subsection{Instrument Offsets}\label{sec:spectrograph_data}
Several instruments listed in Table \ref{tab:spectrographs_instruments} have undergone upgrades to improve precision and efficiency. In cases where a significant offset (as documented in the literature) is mentioned, we divide the RVs for those spectrographs into two distinct datasets with an unknown velocity break to accurately model the radial velocity measurements both before and after the upgrade. 

For instance, HPF underwent a vacuum warm-up and cool-down cycle for maintenance in May 2022, resulting in a division of the radial velocity measurements during our observational campaign into pre- and post-warmup periods. In addition, CORALIE underwent an upgrade in June 2007 to increase the efficiency of the instrument on bright targets and to improve RV accuracy on fainter targets (\citealt{Segransan2010}). This upgrade led to RV offsets that could reach up to $\simeq$ 20 m $\text{s}^{-1}$ prompting the designation of RVs before and after the upgrade as CORALIE-98 and CORALIE-07, respectively. During the summer and winter of 2020, CHIRON underwent a multi-month shutdown and a statistically significant offset was observed (\citealt{Rodriguez2021}). Therefore, for TOI-1478 b we treat the CHIRON observations as two different instruments, CHIRON1 and CHIRON2. 

 HIRES was upgraded in August 2004 which helped increase the precision of typical observations from $\sim$3 m $\text{s}^{-1}$ to $\sim$1 m $\text{s}^{-1}$ (\citealt{Butler2006}). We opted not to include this offset in our analysis as the HIRES break has been shown to be comparable to the RV precision (see further details outlined in \citealt{Rosenthal2021}). In the case of HD 130322 (\citealt{2015ApJ...803....8H}), the HIRES data is modeled separately for the pre- and post-upgrade periods, which prompted us to follow a similar approach. To ensure convergence with two unique systems, HD 224693 and HIP 14810, we treated the HIRES instrument as two independent instruments. For HD 224693, RV measurements collected from \citet{Ment2018} are referred to as HIRES1 and those collected from \citet{Johnson2006} are referred to as HIRES. Similarly, for HIP 14810,  RV measurements collected from \citet{Ment2018} are referred to as HIRES1 and those collected from \citet{Wright2009} are referred to as HIRES. 

\section{Orbit Re-fitting}\label{sec:orbit_refitting}
We compile literature measurements from discovery papers and follow-up surveys to generate the most complete and up-to-date compilation of radial velocity measurements for warm Jupiters in our sample, as of May 2022. Including our new observations with HPF and MINERVA-Australis, this amounts to 18,587 RV measurements from 40 spectrographs acquired over the past 35 years. The median number of RV measurements per star is 70 with a range of 5--783 and the median baseline is 7 years with a range of 30 days to 30 years. 

We use the \texttt{RadVel} package to model the radial velocity time series data and re-fit Keplerian orbits (\citealt{Fulton2018}). \texttt{RadVel} implements an affine-invariant Markov chain Monte Carlo (MCMC) posterior sampling routine, \texttt{emcee} (\citealt{Foreman-Mackey2013}), which generates the parameter posterior distributions for individual systems. The eccentricity distribution functions are then used for our broader population-level statistical analysis. The derived eccentricity posterior distributions for each system will be used in forthcoming papers in this series to analyze the underlying eccentricity distribution of warm Jupiters as a population. 

We utilize a standard framework for orbit fitting and posterior sampling in our analysis by fitting for orbital period ($P$), time of inferior conjunction ($T_{C}$), velocity semi-amplitude ($K$), and parameterized forms of eccentricity ($e$) and argument of periastron ($\omega$), $\sqrt{e}$ $\cos \omega$ and $\sqrt{e}$ $\sin \omega$. Through binary star observations, \citet{LucySweeney1971AJ.....76..544L} found that the eccentricity posterior is not well sampled for orbits with small eccentricities. To avoid this bias, $\sqrt{e}$ $\cos \omega$ and $\sqrt{e}$ $\sin \omega$ are used to aid with convergence when $e$ is small and $\omega$ is uncertain (\citealt{Ford2006}; \citealt{Anderson2011}). Uniform priors in $\sqrt{e}$ $\cos \omega$ and $\sqrt{e}$ $\sin \omega$ impose a uniform prior for $e$ and a non-uniform (but weakly constrained) prior in $\omega$. Initial parameter guesses use results of orbit fits from individual discovery and follow-up papers. 

Equation 2 from \citet{Rosenthal2021} summarizes the Keplerian model used to fit the RV data, 
\begin{equation}
m(t) = \sum_{n=1}^N K(t|P_{n}, T_{C,n}, e_{n}, \omega_{n}, K_{n}) + \gamma + \dot{\gamma}(t - t_0) + \ddot{\gamma}(t - t_0)^2
\label{keplerian_motion}
\end{equation}

\noindent where \emph{n} is the number of planets, $P_{n}$ is the orbital period for planet $n$, $T_{C,n}$ is the time of inferior conjunction, $e_{n}$ is the eccentricity, $\omega_{n}$ is the argument of periastron, $K_{n}$ is the radial velocity semi-amplitude, $\gamma$ is the instrumental offset, $\dot{\gamma}$ is a linear acceleration term, $\ddot{\gamma}$ is a quadratic acceleration term, and $t_0$ is an arbitrary reference epoch, chosen here to be the first radial velocity measurement in each system. $t_0$ is important because it defines the reference time for calculating the inferior conjunction and any radial acceleration terms. \texttt{RadVel} also fits for ``jitter" (unaccounted for astrophysical and instrumental noise) which is added in quadrature with the measurement uncertainties (\citealt{Fulton2018}). Indeed, some systems do not have reported time of inferior conjunction ($T_{C,n}$) values, but rather, they have reported time of periastron ($T_{P}$) values. In such cases, we transform the reported $T_{P}$ to $T_{C}$ and allow for the prior to freely fit for $T_{C}$. We note that a small number of RV measurements relative to the number of free parameters could lead to overfitting, but all but two systems have at least 10 RV measurements, with a median of 70 RVs per target. Due to the limited RV coverage for K2-114 and K2-115, both of which have fewer than 10 measurements, we adopt single Keplerian fits to minimize the risk of overfitting. Additionally, for the sake of model simplicity, orbit fits from the literature for some systems are reported with fixed eccentricities of 0. In our analysis, we incorporate greater flexibility by allowing eccentricity  to vary in each system.  

We use a HardBounds prior---a linearly uniform distribution bounded between two values---for the period, time of conjunction, jitter, linear acceleration, and quadratic acceleration to constrain the parameter values within a physically meaningful range while maintaining equal probability density across the specified bounds. A Jeffreys prior with upper and lower bounds is used for the RV semi-amplitude to ensure positive values and prevent bias toward small values of $K$. To aid with convergence, we adopt a uniform prior on the eccentricity from 0--1, following the standard basis parameterization using $\sqrt{e}$ $\cos \omega$ and $\sqrt{e}$ $\sin \omega$. These weakly informative priors are listed in Table \ref{tab:priors}.

\subsection{Model Selection}\label{sec:model_selection}
Combining radial velocity datasets allows for a longer baseline of observations, ultimately facilitating the discovery of long-term RV trends that are caused by distant giant planets, brown dwarfs, and low-mass stars (e.g., \citealt{Crepp2014}; \citealt{Lagrange2019}; \citealt{Bowler2021A}; \citealt{Bowler2021B}; \citealt{Dalal2021}; \citealt{Barbato2023}; \citealt{Matthews2024}). To generate a consistent comparison among each of the orbit fits in systems with signals of only a single giant planet, we re-fit each dataset three times: with a single-planet Keplerian orbit and no RV trend in the model, a single planet plus a linear trend (a radial acceleration), and a single planet with both linear and curvature terms (changes in radial acceleration). Each MCMC run contains at least 30 walkers and 2 $\times$ 10$^4$ steps. We utilize the built-in Gelman-Rubin (\citealt{GelmanRubin1992}; GR) statistic in \texttt{RadVel} (which imposes a minimum GR value of 1.03) coupled with trace and corner plots to assess convergence (\citealt{Foreman-Mackey2016}; \citealt{HoggForeman-Mackey2018}). 

We then test the goodness of the fit among the three models using the Bayesian Information Criterion (BIC):
\begin{equation}
    BIC = -2 \cdot \ln(\mathcal{L}) + k \cdot \ln(n)
\end{equation}
where $\ln(\mathcal{L})$ is the log-likelihood, \emph{k} is the number of free parameters in the model, and \emph{n} is the number of data points. The BIC value takes into account the complexity of the model and the goodness of fit through the $\chi^2$ value. We compare the Keplerian model against the linear trend and curvature (quadratic) models. Following the interpretation of $\Delta$BIC values from \citet{Trotta2007}, we then select the statistically favored fit if the $\Delta$BIC is $>$ 10 from the next-best fit and more complex model. For instance, a Keplerian fit is preferred over a Keplerian-plus-linear trend model if the $\Delta$BIC value between these two is less than 10, even if the model with the trend produces a slightly better $\chi^2$ value.

Orbits of all warm Jupiters that reside in previously known multi-planet systems are fit by summing the Keplerian orbit of each planet. For the specific cases of HD 11506 (\citealt{2015ApJ...799...89G}), HD 163607 (\citealt{Giguere2012}; \citealt{Luhn2019}), HD 168443 (\citealt{2007ApJ...654..625W}; \citealt{2011ApJ...743..162P}), HD 191939 (\citealt{Lubin2022}; \citealt{Orell-Miquel2023}), and HD 38529 (\citealt{2001ApJ...551.1107F}; \citealt{2009ApJS..182...97W}; \citealt{2009ApJ...693.1084W}; \citealt{Luhn2019}) we include linear trends to best fit the data following the setup outlined in previous studies.

The distribution of eccentricities, orbital periods, and minimum masses from the final \texttt{RadVel} fits can be found in Figure \ref{fig:planet_histogram}. We find that warm Jupiters exhibit a wide range of eccentricities spanning 0--0.93, with a preference for low- to moderate values below 0.4. The behavior of warm Jupiter eccentricities as a function of their orbital properties and host star characteristics is explored in more detail in companion papers in this series (Morgan et al., in press). The average $m_p \sin i$ is $\sim$1 $M_\mathrm{Jup}$, with minimum masses extending to 12.17 $M_\mathrm{Jup}$. The orbital period distribution peaks between 10--40 days and remains nearly uniform at longer periods. This is likely due to detection biases favoring transiting planets (which make up 22$\%$ of the sample) and the relative ease of discovering and following up warm Jupiters at smaller separations from their host stars.

\begin{figure*}
  \centering

  \begin{minipage}[b]{0.31\textwidth}
    \centering
    \includegraphics[width=\linewidth]{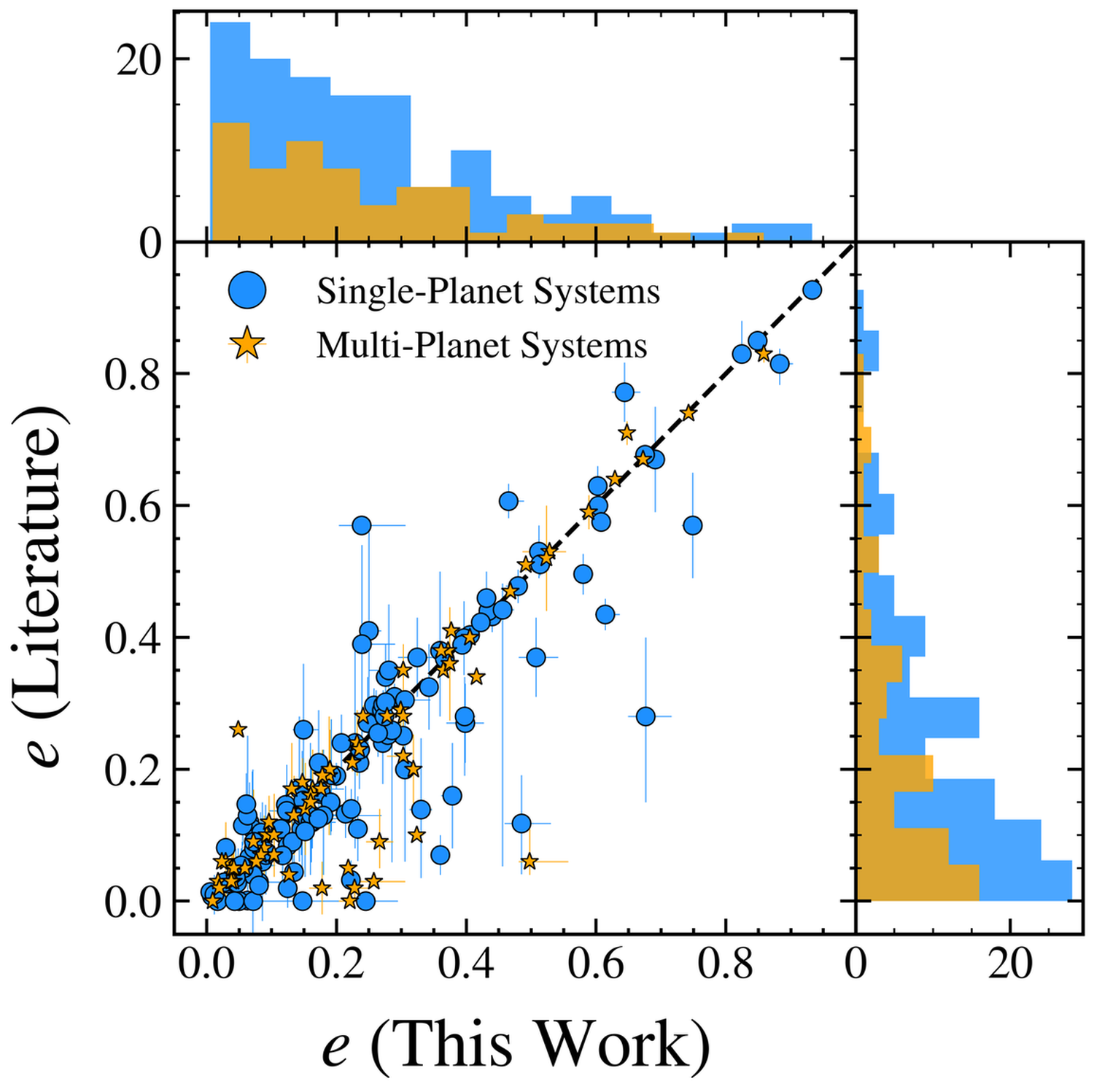}
  \end{minipage}
  \hfill
  \begin{minipage}[b]{0.31\textwidth}
    \centering
    \includegraphics[width=\linewidth]{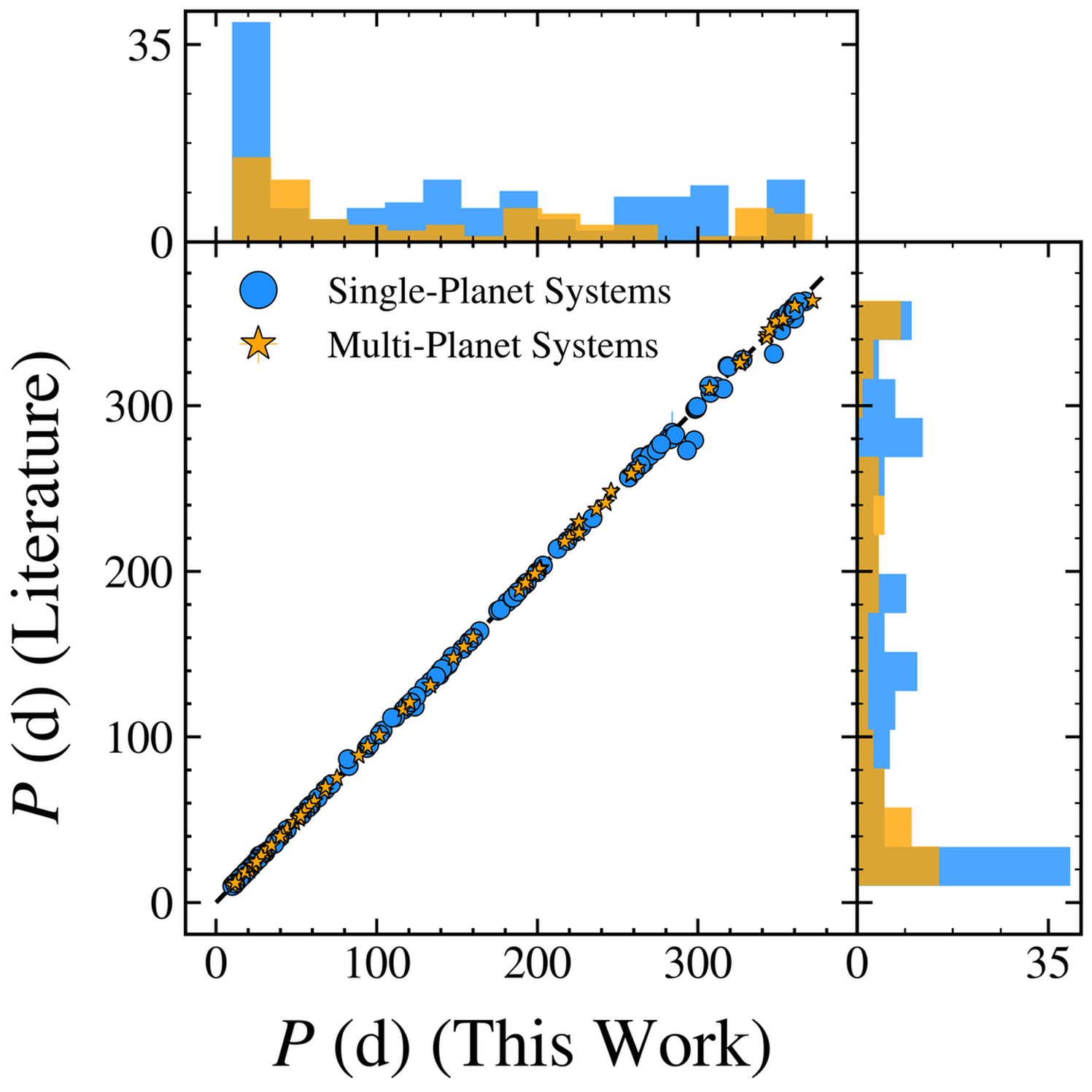}
  \end{minipage}
  \hfill
  \begin{minipage}[b]{0.31\textwidth}
    \centering
    \includegraphics[width=\linewidth]{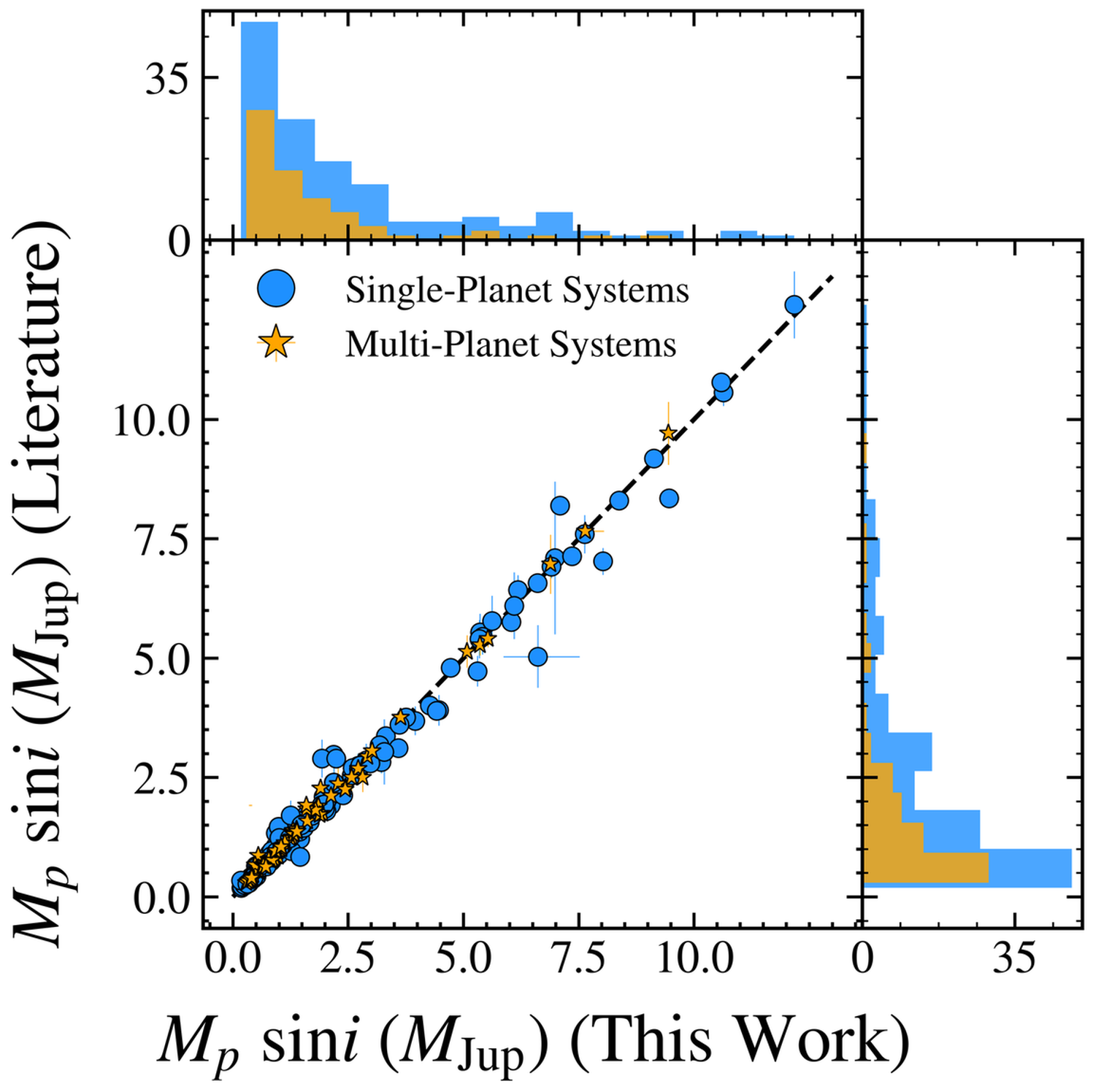}
  \end{minipage}

  \caption{Comparison of eccentricity, orbital period, and minimum mass measurements from the literature and those from our orbit refits. Warm Jupiters in single planet systems are denoted by blue circles and those in multi-planet systems are represented by orange stars.  Literature solutions were selected from the original fits in the discovery papers, with updated fits from follow-up papers utilized when available. References to earlier work used in this figure are reported in Table \ref{tab:planet_params}. }
  \label{fig:literature_compariosns}
\end{figure*}

\section{Results}\label{sec:results}

\subsection{General properties of Warm Jupiters and Host Stars}\label{sec:warm_jupiters_host_stars}

Here, we present results from the best-fit orbital solutions obtained for all 200 planets listed in Table \ref{tab:planet_params}. The Maximum a Posteriori (MAP) value\footnote{Note that we report the global MAP value, which differs from the marginalized mean, median, and mode for for each parameter. (\citealt{HoggForeman-Mackey2018}).}, and 68$\%$ highest density interval about the median for each fitted parameters posterior probability distribution are reported in Table \ref{tab:planet_params}.  A comparison of literature values and those obtained in our orbit refits can be found in Figure \ref{fig:literature_compariosns}. Stellar properties for all 194 hosts are reported in Table \ref{tab:host_stars}. Mass ($M_{*}$), metallicity ([Fe/H]), and distances (pc) are collected from discovery papers, Vizier (\citealt{Ochsenbein2000}), the NASA Exoplanet Archive (\citealt{Akeson2013}), and Gaia DR3 (\citealt{GaiaCollaboration2023}; \citealt{BailerJones2021}). 

Linear accelerations ($\dot{\gamma}$) and, when relevant, second-order quadratic terms ($\ddot{\gamma}$) are reported in Table \ref{tab:additional_planet_params}. There are 132 single-planet warm Jupiters orbiting 132 individual host stars. Additionally, 68 warm Jupiters in multi-planet systems orbit a total of 62 host stars, with 6 of these stars hosting two warm Jupiters. Of the 62 host stars with warm Jupiters in multi-planet systems, 51 host at least two giant planets. The number of planets, the total number of radial velocity measurements, the mean error on those measurements, the observational baseline, and the BIC values are provided in Table \ref{tab:additional_planet_params}.

\subsection{RV Trends and Distant Companions}\label{sec:acceleration_trends}
Many systems in this sample are expected to have wide stellar, substellar, and planetary companions. Multiplicity surveys of solar-type stars have found that almost half are in binary or higher-order systems with a period distribution peaking at $\sim$ 50 AU (\citealt{Raghavan2010}; \citealt{Kraus2013}). The distant companion rate for systems hosting warm Jupiters is unconstrained. However, for the similar population of hot Jupiter systems, the occurrence rate of distant planetary companions to stars harboring hot Jupiters from long-term RV surveys is $\approx$70$\%$ $\pm$ 8$\%$ with masses between 1--13 $M_\mathrm{Jup}$ and orbital semi-major axes between 1--20 AU  (\citealt{Knutson2014}; \citealt{Bryan2016}). AO Imaging surveys have found that 47$\%$ $\pm$ 7$\%$ of hot Jupiters have stellar companions with semimajor axes between 50 and 2000 AU (\citealt{Ngo2016}). 

Among our best-fit models, 5 multi-planet and 54 single-planet systems show either a long-term linear $\dot{\gamma}$ (25 systems) or quadratic $\ddot{\gamma}$ (34 systems) acceleration, suggesting the presence of a distant companion. Close-in orbiting giant planets and wide stellar companions can produce degeneracies in mass and orbital separation that produce the measured reflex motion. These radial accelerations will be explored in greater detail in a future study in this series.

\subsection{Systems with New RVs}\label{sec:New_RV_Systems}
Here, we discuss updated orbital parameters from our uniform fits with a focus on planetary mass, orbital separation, and eccentricity results for individual planets in our observational campaign with HPF and MINERVA-Australis (see Table \ref{tab:planet_params} for additional details).

\subsubsection{Apparently Single Planet Systems}
\begin{enumerate}

\item BD+15 2940 is a K0 giant hosting a warm Jupiter with an eccentricity of 0.26$^{+0.10}_{-0.10}$ and a minimum mass of 1.1 $M_\mathrm{Jup}$ on a $137.48 \pm 0.34$ day orbit (\citealt{Nowak2013}). It is one of the closest-orbiting planets around an evolved giant and could potentially be an engulfment candidate. The initial RV measurements were acquired using HET/HRS by the Penn State-Toruń Planet Search, extending until 2011. We revisited this system over a decade later using HPF with new measurements spanning from January 2022 to May 2022. We obtain an improved constraint on the eccentricity, 0.15$^{+0.02}_{-0.02}$, and derive a minimum mass of $1.05 \pm 0.02$ $M_\mathrm{Jup}$ and period of 138.87$^{+0.02}_{-0.02}$ days.

\item BD+55 362, a K3 star, harbors a $0.72 \pm 0.08$ $M_\mathrm{Jup}$ companion on a $265 \pm 1$ day orbit with an eccentricity of 0.27$^{+0.06}_{-0.06}$. \citet{Dalal2021} adopted a model that incorporates a linear drift and quadratic trend for their SOPHIE+ measurements with reported values of $1.76 \pm 0.04$ m s$^{-1}$ yr$^{-1}$ and $1.4365 \pm 0.0005$ m s$^{-1}$ yr$^{-2}$, respectively. We obtain new RV measurements from August 2022 to November 2022 using HPF. The new orbit fit yields an eccentricity of 0.40$^{+0.03}_{-0.03}$, minimum mass of $0.63 \pm 0.03$ $M_\mathrm{Jup}$, and orbital period of 266.18$^{+0.70}_{-0.70}$ days. The BIC analysis favors the model with a linear trend and curvature, with values of --15.25$^{+0.99}_{-0.95}$ m s$^{-1}$ yr$^{-1}$ and 1.36$^{+0.11}_{-0.11}$ m s$^{-1}$ yr$^{-2}$, respectively .  

\item HD 128356 b is a $0.89 \pm 0.07$ $M_\mathrm{Jup}$ giant planet around a K3V star orbiting every $298.2 \pm 1.6$ days with an eccentricity of 0.57$^{+0.08}_{-0.08}$ (\citealt{Jenkins2017}). CORALIE and HARPS measurements were acquired up until 2014 to confirm this warm Jupiter. With our additional added RVs collected by MINERVA-Australis from February 2022 to May 2023, we find a somewhat higher  eccentricity of 0.75$^{+0.01}_{-0.01}$ along with a consistent period of 298.19$^{+0.08}_{-0.07}$ days and minimum mass of 0.99$^{+0.04}_{-0.03}$ $M_\mathrm{Jup}$.

\item The K0 subgiant HD 148427 hosts a $0.96 \pm 0.10$ $M_\mathrm{Jup}$ warm Jupiter with a period of $331 \pm 3$ days and eccentricity of 0.16$^{+0.08}_{-0.08}$ (\citealt{Fischer2009}). HAMILTON RV measurements were obtained until 2009 to confirm the presence of this long-period planetary companion. Together with these published RVs, our additional observations collected over a decade later with MINERVA-Australis spanning February 2022 to May 2023 reveal an eccentricity of 0.38$^{+0.01}_{-0.01}$, orbital period of 347.14$^{+0.23}_{-0.22}$ days, and minimum mass of $1.27 \pm 0.02$ $M_\mathrm{Jup}$. Our BIC analysis favors the model with a linear trend of $-5.07 \pm 0.09$ m s$^{-1}$ yr$^{-1}$.  

\item  The G2 star HD 164509 harbors a $0.48 \pm 0.09$ $M_\mathrm{Jup}$ companion with an eccentricity of 0.26$^{+0.14}_{-0.14}$ and orbits its host every $282.4 \pm 3.8$ days (\citealt{Giguere2012}). Initial and follow-up RV measurements from \citet{Ment2018} were taken with HIRES and show a linear trend of $-5.1 \pm 0.7$ m s$^{-1}$ yr$^{-1}$. More recently, we obtained 15 new RVs with HPF from March 2022 to May 2022 and the combined orbit fit yields an eccentricity of 0.27$^{+0.01}_{-0.01}$, a period of 281.29$^{+0.12}_{-0.11}$ days, and minimum mass of 0.46$^{+0.004}_{-0.004}$ $M_\mathrm{Jup}$. In addition we recover a linear trend of --3.52$^{+0.03}_{-0.04}$ m s$^{-1}$ yr$^{-1}$, a result similar to the trend identified by \citet{Giguere2012}.  

\item \citet{LopesButler2008} discovered HD 205739 b, a 1.37$^{+0.07}_{-0.09}$ $M_\mathrm{Jup}$ companion on a $279.8 \pm 0.1$ day orbit with an eccentricity of 0.27$^{+0.07}_{-0.07}$ around an F-type star. MIKE RV measurements taken from 2004 to 2007 show a linear trend of $23.70 \pm 0.07$ m s$^{-1}$ yr$^{-1}$ indicating the presence of another outer body in the system. Our additional RV measurements taken with MINERVA-Australis fifteen years later, from May 2022 to May 2023, recover a linear trend of 21.92$^{+0.36}_{-0.34}$ m s$^{-1}$ yr$^{-1}$, broadly similar to results from \citet{LopesButler2008}. We find an eccentricity of 0.25$^{+0.01}_{-0.01}$, an orbital period of 282.67$^{+0.27}_{-0.29}$ days, and a minimum mass of $1.44 \pm 0.03$ $M_\mathrm{Jup}$ for the warm Jupiter.

\item HD 208487 b is a $0.45 \pm 0.05$ $M_\mathrm{Jup}$ warm Jupiter orbiting a G-type star every $130 \pm 1$ days with an eccentricity of 0.32$^{+0.10}_{-0.10}$ (\citealt{Tinney2005}). We re-observed this system from April 2022 to April 2023 with MINERVA-Australis, generating a 25 year baseline when combined with the initial UCLES RVs that were obtained from 1998--2004. We derive an eccentricity of 0.23$^{+0.02}_{-0.02}$, along with a minimum mass and orbital period of $0.48 \pm 0.01$ $M_\mathrm{Jup}$ and $129 \pm 0.05$ days, respectively.

\item HD 211403 is an F-type star hosting a 5.54$^{+0.39}_{-0.38}$ $M_\mathrm{Jup}$ companion with a period and eccentricity of $223.80 \pm 0.41$ days and 0.08$^{+0.06}_{-0.04}$, respectively (\citealt{Demangeon2021}). ELODIE, SOPHIE, and SOPHIE+ RV measurements were obtained until 2018. We acquired new HPF measurements from December 2021 to September 2022. Our combined fit resulted in an eccentricity of 0.13$^{+0.01}_{-0.01}$, an orbital period of 223.62$^{+0.10}_{-0.09}$ days, and minimum mass of $5.36 \pm 0.05$ $M_\mathrm{Jup}$. Our BIC analysis favors the accelerating model with linear and curvature terms of -81.46$^{+8.97}_{-10.08}$ m s$^{-1}$ yr$^{-1}$, and 3.96$^{+0.49}_{-0.43}$ m s$^{-1}$ yr$^{-2}$, respectively.

\item \citet{Niedzielski2015} discovered a warm Jupiter with a mass of 1.47$^{+0.20}_{-0.12}$ $M_\mathrm{Jup}$, period of $148.6 \pm 0.7$, and eccentricity of 0.38$^{+0.12}_{-0.10}$ around the K0 giant HD 216536. Our recently acquired HPF RVs (December 2021 to November 2022) combined with the original HRS RV measurements collected between 2005 and 2013 establish a seventeen-year baseline. We calculate that the companion has an eccentricity of 0.36$^{+0.02}_{-0.02}$, orbital period of 147.01$^{+0.08}_{-0.07}$ days, and a minimum mass of $0.99 \pm 0.02$ $M_\mathrm{Jup}$.

\item HD 216770 harbors a 0.65 $M_\mathrm{Jup}$ companion orbiting every $118.45 \pm 0.55$ days with an eccentricity of 0.37$^{+0.06}_{-0.06}$ (\citealt{Mayor2004A}). We combine CORALIE RVs taken from 2000 to 2002 with MINERVA-Australis RVs obtained two decades later from June 2022 to September 2022. The combined orbit fit yields a solution with an eccentricity of 0.51$^{+0.03}_{-0.03}$, period of $123.58 \pm 0.03$ days, and minimum mass of $0.49 \pm 0.02$ $M_\mathrm{Jup}$.

\item HD 219139 b is a 0.78$^{+0.05}_{-0.20}$ $M_\mathrm{Jup}$ companion with a period of 275.5$^{+2.3}_{-1.0}$ days and eccentricity of 0.11$^{+0.09}_{-0.08}$ (\citealt{Teng2022}). Our orbit fit combines the original RV measurements taken with HIDES  with new HPF RVs, yielding an eccentricity of 0.07$^{+0.01}_{-0.01}$, orbital period of $275.60 \pm 0.13$ days, and a minimum mass of $0.76 \pm 0.01$ $M_\mathrm{Jup}$. 

\item \citet{Johnson2006} detected a warm Jupiter with a mass of 0.33 $M_\mathrm{Jup}$ and eccentricity of 0.48$^{+0.05}_{-0.05}$ orbiting HD 33283 once every $18.179 \pm 0.007$ days. We combine the original HIRES RVs with updated HIRES RVs (\citealt{Ment2018}) and our recently acquired MINERVA-Australis RVs. Our orbit fit finds an eccentricity of 0.40$^{+0.01}_{-0.01}$, period of $18.1993 \pm 0.0002$ days, and a minimum mass of $0.325 \pm 0.003$ $M_\mathrm{Jup}$.

\item HD 38283, a F-type star, hosts a $0.34 \pm 0.02$ $M_\mathrm{Jup}$ warm Jupiter on a $363.2 \pm 1.6$ day orbit with an eccentricity = 0.41$^{+0.16}_{-0.16}$ (\citealt{Tinney2011}). We combined the inital UCLES RVs, which were obtained from 1998 to 2010, with our new MINERVA-Australis RVs acquired from February 2022 to September 2022. The combined data resulted in an eccentricity of 0.25$^{+0.01}_{-0.01}$, period of $365.70 \pm 0.15$, and minimum mass = $0.386 \pm 0.006$ $M_\mathrm{Jup}$.

\item HR 810, also known as $\iota$ Horologii and HD 17051, harbors a $2.26 \pm 0.18$ $M_\mathrm{Jup}$ companion with a period of $320.1 \pm 2.1$ days and eccentricity = 0.161$^{+0.069}_{-0.069}$ (\citealt{Kurster2000}; \citealt{Butler2001}; \citealt{Naef2001A375205N}). Our RV analysis amounts to a 30 year baseline by combining measurements taken with the UCLES, COUDE, and CORALIE instruments from 1992--2001 together with new MINERVA-Australis RVs (February 2022 to September 2022). We calculate an eccentricity of 0.252$^{+0.006}_{-0.006}$, orbital period of $311.95 \pm 0.11$ days, and minimum mass = $2.62 \pm 0.02$ $M_\mathrm{Jup}$.

\item NGC 2682 Sand 364, also known as BD+12 1917, is a K-type star hosting a $1.54 \pm 0.24$ $M_\mathrm{Jup}$ warm Jupiter orbiting once every $121.71 \pm 0.31$ days with an eccentricity of 0.35$^{+0.08}_{-0.08}$ (\citealt{Brucalassi2014}; \citealt{Brucalassi2017}). We combine RV measurements acquired between 2003 and 2014 by the CORALIE, HARPS, SOPHIE, AND HRS instruments with new HPF measurements collected from December 2021 to March 2023. This establishes a baseline of measurements spanning approximately twenty years. We derive an eccentricity of 0.28$^{+0.02}_{-0.02}$, period of 121.31$^{+0.07}_{-0.07}$ days, and minimum mass = $1.66 \pm 0.04$ $M_\mathrm{Jup}$. The BIC analysis favors the model with a linear trend of 2.75$^{+0.67}_{-0.65}$ m s$^{-1}$ yr$^{-1}$.

\subsubsection{Two Planet Systems}

\item HD 207832 hosts two giant planets, one of which, HD 207832 b, is a warm Jupiter with period of 161.97$^{+0.97}_{-0.78}$ days, minimum mass = 0.56$^{+0.06}_{-0.03}$ $M_\mathrm{Jup}$, and eccentricity = 0.13$^{+0.18}_{-0.05}$ (\citealt{Haghighipour2012}). The initial HIRES RVs in combination with updated RVs from HIRES (2004--2016; \citealt{Ment2018}) and MINERVA-Australis RVs (February 2022 to September 2022) creates a baseline of eighteen years with 176 measurements. The combined orbit fit yields an eccentricity of 0.32$^{+0.01}_{-0.01}$, period of $160.06 \pm 0.02$ days, and minimum mass = $0.546 \pm 0.004$ $M_\mathrm{Jup}$.

\item HD 67087, an F-type star, harbors two giant planets, where HD 67087 b is classified as a warm Jupiter with eccentricity = 0.17$^{+0.07}_{-0.07}$, period of 352.2$^{+1.7}_{-1.6}$ days, and a minimum mass = 3.06$^{+0.22}_{-0.20}$ $M_\mathrm{Jup}$. The outer giant planet, HD 67087 c, is an eccentric cold Jupiter with $e$ = 0.76$^{+0.17}_{-0.24}$, a period of 2374$^{+193}_{-156}$ days, and a minimum mass = 4.85$^{+10.0}_{-3.61}$ $M_\mathrm{Jup}$  (\citealt{Harakawa2015}). The warm Jovian was detected using the HDS and HIDES instruments from 2004 to 2014. We combined these measurements with new HPF RVs from December 2021 to March 2023 and derive an eccentricity of 0.13$^{+0.01}_{-0.01}$, orbital period of 352.96$^{+0.07}_{-0.08}$ days, and minimum mass = $3.02 \pm 0.02$ $M_\mathrm{Jup}$. The outer giant planet, HD 67087 c, yields an $e$ = 0.711$^{+0.010}_{-0.010}$, period of 2375.52$^{+18.76}_{-15.09}$ days, and minimum mass of 5.85$^{+0.14}_{-0.15}$ $M_\mathrm{Jup}$.


\item HD 75784 b is a warm Jupiter in a system hosting two giant planets, with an eccentricity of 0.09$^{+0.06}_{-0.06}$, period of $341.2 \pm 1.1$ days, and a minimum mass of $1.00 \pm 0.14$ $M_\mathrm{Jup}$. The outer giant planet, HD 75784 c, is a cold Jupiter with $e$ = $0.49 \pm 0.09$, a period of $7900 \pm 2000$ days, and a minimum mass of $5.64 \pm 0.72$ $M_\mathrm{Jup}$   (\citealt{2015ApJ...799...89G}; \citealt{Luhn2019}). We combined HIRES RVs from 2004 to 2014 with new HPF RVS acquired from December 2021 to March 2023. For the inner warm Jupiter we derive an eccentricity of 0.10$^{+0.01}_{-0.01}$, period of 342.66$^{+0.03}_{-0.04}$ days, and a minimum mass of 1.05$^{+0.01}_{-0.01}$ $M_\mathrm{Jup}$. For HD 75784 c we derive an $e$ = 0.489$^{+0.004}_{-0.004}$, orbital period of 8205.22$^{+19.30}_{-26.30}$ days, and minimum mass of 6.02$^{+0.05}_{-0.07}$ $M_\mathrm{Jup}$.

\item \citet{LoCurto2010} discovered HIP 5158 b, a warm Jupiter orbiting its K-type host star once every 345.72$^{+5.37}_{-5.37}$ days with a minimum mass of 1.42 $M_\mathrm{Jup}$ and eccentricity of 0.52$^{+0.08}_{-0.08}$. A second giant planet was later discovered in the system with an eccentricity of $0.14 \pm 0.10$, a period of $9017 \pm 3180$ days, and a minimum mass = $15.04 \pm 10.55$ $M_\mathrm{Jup}$ (\citealt{Feroz2011}). HARPS (2004 -- 2009) and new MINERVA-Australis RVs obtained from August 2022 to September 2022 are combined and yield an eccentricity of 0.5231$^{+0.0044}_{-0.0046}$, period of 344.57$^{+0.23}_{-0.22}$ days, and a minimum mass of $1.37 \pm 0.01$ $M_\mathrm{Jup}$. To aid with convergence we modeled HIP 5158 as a single-planet Keplerian with a quadratic trend, following the setup outlined in the discovery paper.

\item The K-type star HIP 65407 is the host star to two warm Jupiters that orbit just outside the 12:5 resonance, making it one of the six known systems with two confirmed warm Jupiters through RV observations as of May 2022 (\citealt{Hebrard2016}). The inner companion, HIP 65407 b, orbits its host every $28.13 \pm 0.02$ days with a minimum mass of $0.43 \pm 0.03$ $M_\mathrm{Jup}$ and eccentricity of 0.14$^{+0.07}_{-0.07}$. HIP 65407 c has a minimum mass of $0.78 \pm 0.05$ $M_\mathrm{Jup}$, an eccentricity of 0.12$^{+0.04}_{-0.04}$, and orbital period = $67.30 \pm 0.08$ days (\citealt{Hebrard2016}). We combine HPF RVs obtained in May 2022 with SOPHIE and SOPHIE+ RVs acquired from 2011 to 2015 for a total of 71 measurements. We find an eccentricity of 0.15$^{+0.01}_{-0.01}$, period of $28.13 \pm 0.003$ days, and a minimum mass of $0.42 \pm 0.003$ $M_\mathrm{Jup}$ for HIP 65407 b. For HIP 65407 c, we infer an eccentricity of 0.10$^{+0.01}_{-0.01}$, period of $67.30 \pm 0.01$ days, and a minimum mass of $0.78 \pm 0.005$ $M_\mathrm{Jup}$. 

\subsection{Systems with Poor Convergence}\label{sec:Challenges_with_Convergence}
Here, we discuss systems that required additional attention for convergence (see solutions in Table \ref{tab:planet_params}).

\item HD 191939 is a G9V star hosting six planets. Notably for this analysis, HD 191939 e is identified with a mass of $0.340 \pm 0.009$ $M_\mathrm{Jup}$, an eccentricity fixed at 0, and orbits at a distance of $0.397 \pm 0.005$ AU (\citealt{Lubin2022}). While there are four other planets within the system, they are not giant planets and, consequently, have relatively low semi-amplitude signals ($K$ $\approx$1--3 m s$^{-1}$) compared to HD 191939 e ($K$ =17.73 m s$^{-1}$). Therefore, in our modeling of the RVs, we exclusively focus on the warm Jupiter. The RV measurements from the APF and HIRES instruments indicate a preference for a model that incorporates a linear trend and curvature. 
The long-term trend showed by the RVs is associated with HD 191939 f, a high-mass planet with an unconstrained period (\citealt{Orell-Miquel2023}). While fitting the orbit of the warm Jupiter, we allowed the model to also fit for changing acceleration terms. \citet{Lubin2022} find a linear trend of 41.64$^{+2.19}_{-2.19}$ m s$^{-1}$ yr$^{-1}$, accompanied by a curvature value of --7.99$^{+2.66}_{-2.66}$ m s$^{-1}$ yr$^{-2}$. \citet{Orell-Miquel2023} find a linear trend of 69.47$^{+0.77}_{-0.77}$ m s$^{-1}$ yr$^{-1}$, accompanied by a curvature value of --3.03$^{+0.09}_{-0.09}$ m s$^{-1}$ yr$^{-2}$. In our analysis, we derive an eccentricity of 0.04$^{+0.003}_{-0.003}$, semi-major axis of 0.40$^{+0.0001}_{-0.0001}$ AU and a mass of $0.360 \pm 0.001$ $M_\mathrm{Jup}$. In addition we find a linear trend of 44.96$^{+1.14}_{-0.70}$ m s$^{-1}$ yr$^{-1}$, and a quadratic drift of --7.77$^{+0.04}_{-0.63}$ m s$^{-1}$ yr$^{-2}$, respectively. 

\item HD 59686 A is a K III giant hosting the warm Jupiter HD 59686 Ab ($P$ = 299.36$^{+0.26}_{-0.31}$ days; $e$ = 0.05$^{+0.03}_{-0.02}$; $M_\mathrm{p} \sin i$ = 6.92$^{+0.18}_{-0.24}$ $M_\mathrm{Jup}$) in a tight eccentric binary system (\citealt{Ortiz2016}). The orbit fit of the HD 59686 system is dominated by the stellar companion HD 59686 B ($a_{B}$ = 13.56$^{+0.18}_{-0.14}$ AU; $e$ = 0.73; $M_\mathrm{p} \sin i$ = 554.9$^{+1.2}_{-0.9}$ $M_\mathrm{Jup}$). Consequently, we model this complex system as a two-planet system treating HD 59686 Ab as the inner companion and HD 59686 B as the outer companion. We derive an eccentricity of $e$ = 0.048$^{+0.002}_{-0.002}$, orbital period of 299.36$^{+0.028}_{-0.029}$ days, and $M_\mathrm{p} \sin i$ = 6.92$^{+0.01}_{-0.02}$ $M_\mathrm{Jup}$ for HD 59686 Ab.

\item The warm Jupiter host star HD 87646 A is a part of a close binary system, with the two stars separated by $\sim$22 AU (\citealt{Ma2016}). The primary star, HD 87646 A, harbors both a warm Jupiter HD 87646 b ($P$ = $13.481 \pm 0.001$ d; $e$ = $0.05 \pm 0.02$; $M_\mathrm{p} \sin i$ = $12.4 \pm 0.7$$M_\mathrm{Jup}$) and brown dwarf HD 87646 c ($P$ =  $674 \pm 4$ d; $e$ = 0.50; $M_\mathrm{p} \sin i$ = $57.0 \pm 3.7$ $M_\mathrm{Jup}$). We utilize a two-planet Keplerian model for the orbit fit of this unique warm Jupiter-brown dwarf-binary star system. Our analysis of HD 87646 b finds an eccentricity of $e$ = 0.057$^{+0.002}_{-0.002}$, orbital period of $13.48167 \pm 0.00007$ d, and $M_\mathrm{p} \sin i$ = 12.17$^{+0.02}_{-0.03}$ $M_\mathrm{Jup}$.

\item 55 Cancri (55 Cnc) is a five-planet system with over twenty years of combined spectroscopic monitoring from the Tull (2.7m) and HRS spectrographs at McDonald Observatory, HIRES, HAMILTON, HARPS, HARPS-N, and SOPHIE (\citealt{Bourrier2018}). 55 Cnc b is a 0.8 $M_\mathrm{Jup}$ warm Jupiter with an eccentricity typically fixed at 0, and orbits its host star every $14.6516 \pm 0.0001$ days. We modeled this complex system using the two largest planets, b and d, as the other three planets induce small semi-amplitudes compared to these two giant planets. Our two-planet Keplerian model yields an eccentricity of 0.009$^{+0.001}_{-0.001}$, orbital period of $14.65218 \pm 0.00001$ days, and $M_\mathrm{p} \sin i$ = 0.8015$^{+0.0005}_{-0.0005}$ $M_\mathrm{Jup}$ for 55 Cnc b.

\item Kepler-65 hosts three compact transiting sub-Neptunes (Kepler-65 b, c, and d with respective periods of $\approx$ 2, 6 and 8 days), along with a long-period warm Jupiter ($\approx$ 259 days), Kepler-65 e (\citealt{Mills2019}). The three sub-Neptunes in the Kepler-65 system induce insignificant semi-amplitudes compared to planet e, leading us to concentrate our modeling efforts primarily on the warm Jupiter, which dominates the orbit fit. Our one-planet Keplerian model with Kepler-65 e produces an eccentricity of $e$ = 0.303$^{+0.006}_{-0.008}$ and $M_\mathrm{p} \sin i$ = 0.662$^{+0.006}_{-0.008}$ $M_\mathrm{Jup}$.

\item Gliese 876 (GJ 876), an M dwarf with $\approx$ 20 years of spectroscopic monitoring from the HIRES, CARMENES, ELODIE, HAMILTON, HARPS, and CORALIE spectrographs hosts four planets (\citealt{Marcy2001GJ876}; \citealt{Rivera2005GJ876};  \citealt{Correia2010GJ876}; \citealt{Rivera2010GJ876}; \citealt{Nelson2016GJ876}; \citealt{Trifonov2018GJ876}). Two of the four planets, GJ 876 c and b, are coplanar warm Jupiters in a 2:1 mean-motion resonance. They have masses of 0.86 and 2.64 $M_\mathrm{Jup}$, eccentricities of $0.265 \pm 0.002$ and $0.031 \pm 0.001$, and periods of approximately $30.259 \pm 0.010$ and $61.065 \pm 0.012$ days, respectively. The inner and outermost planets in the system, GJ 876 d and e, induce small semi-amplitudes on the host star, leading us to concentrate our modeling efforts on the two warm Jupiters that dominate the orbit fit. Our two-planet Keplerian model with GJ 876 c and b yields eccentricities of 0.049$^{+0.002}_{-0.001}$ and 0.018$^{+0.001}_{-0.001}$, and minimum masses = 0.610$^{+0.001}_{-0.001}$ and 1.902$^{+0.001}_{-0.002}$ $M_\mathrm{Jup}$, respectively.

\item Kepler-419 hosts two giant planets, one of which, Kepler-419 b, is a warm Jupiter with a period of 69.7546$^{+0.0007}_{-0.0009}$ days, mass of 2.5$^{+0.3}_{-0.3}$ $M_\mathrm{Jup}$, and eccentricity of 0.833$^{+0.013}_{-0.013}$ (\citealt{2014ApJ...791...89D}). The outer giant planet, Kepler-419 c, is a cold Jupiter with $e$ = $0.184 \pm 0.002$, a period of $675.47 \pm 0.11$ days, and a minimum mass of $7.3 \pm 0.4$ $M_\mathrm{Jup}$. To aid with convergence, we modeled Kepler-419 as a single-planet Keplerian with a quadratic trend, following the setup outlined in the discovery paper (\citealt{2014ApJ...791...89D}). Our analysis of Kepler-419 b finds an eccentricity of $e$ = 0.858$^{+0.007}_{-0.006}$, orbital period of $68.14 \pm 0.03$ d, and $M_\mathrm{p} \sin i$ = 2.81$^{+0.07}_{-0.07}$ $M_\mathrm{Jup}$.

\item For the warm Jupiter-hosting systems BD-08 2823, K2-290, Kepler-117, Kepler-1514, KOI-94, and TOI-1670 we opted to model only the giant planets to aid with convergence. Specifically, we excluded the following companions: the inner Uranus-mass planet orbiting BD-08 2823 (\citealt{Hebrard2010}), the inner mini-Neptune in the K2-290 system (\citealt{2019MNRAS.484.3522H}), the inner Neptune-mass planet orbiting Kepler-117 (\citealt{Bruno2015}), the inner Earth-sized planet in the Kepler-1514 system (\citealt{2021AJ....161..103D}), the two super-Earth's and Neptune-size planets orbiting KOI-94 (\citealt{2013ApJ...768...14W}), and the inner sub-Neptune in the TOI-1670 system (\citealt{Tran2022}). The smaller planets in these systems are either tightly constrained from transit light curves or induce small semi-amplitudes on the host star, leading us to concentrate our modeling efforts on the warm Jupiter in the system.

\item The G-type star KOI-142 (Kepler-88) is the host to three planets including the warm Jupiter KOI-142 c which has an orbital period of 22.26492$^{+0.00067}_{-0.00067}$ days, minimum mass of 0.674$^{+0.016}_{-0.016}$$M_\mathrm{Jup}$, and eccentricity of 0.05724$^{+0.00045}_{-0.00045}$ (\citealt{2013ApJ...777....3N}; \citealt{2020AJ....159..242W}). The orbit of the inner-most planet, the transiting sub-Neptune sized planet KOI-142 b, proved to be challenging to constrain with only RVs. Our two-planet Keplerian model with KOI-142 c produces an eccentricity of $e$ = 0.023$^{+0.005}_{-0.005}$, period = 22.2693$^{+0.0006}_{-0.0006}$ days, and $M_\mathrm{p} \sin i$ = 0.642$^{+0.003}_{-0.003}$ $M_\mathrm{Jup}$.

\item  HD 160691, also known as $\mu$ Arae, hosts four planets, one of which is the warm Jupiter HD 160691 e with an orbital period of 310.55$^{+0.83}_{-0.83}$ days, mass of 0.52 $M_\mathrm{Jup}$, and eccentricity of 0.0666$^{+0.0122}_{-0.0122}$ (\citealt{Pepe2007}). The inner Neptune-mass planet ($\approx$ 10 days), HD 160691 d, induces a negligible semi-amplitude compared to the three outer giant planets in the system causing us to concentrate our modeling efforts on the three larger Jupiter-mass planets. The three planet Keplerian model with HD 160691 e yields an eccentricity of 0.124$^{+0.004}_{-0.004}$, $M_\mathrm{p} \sin i$ = 0.480$^{+0.006}_{-0.007}$ $M_\mathrm{Jup}$, and orbital period of 308.904$^{+0.158}_{-0.157}$ days.

\item We opted to only use the CORALIE RV data when modeling the orbit of HD 72892 b (\citealt{Jenkins2017}) due to challenges in achieving convergence with the complete dataset. The additional 8 HARPS RVs, collected over a $\approx$21 day period, were excluded from the orbital fit. The orbit fit yields an eccentricity of 0.422$^{+0.003}_{-0.004}$, minimum mass of 5.41$^{+0.04}_{-0.05}$ $M_\mathrm{Jup}$, and orbital period of 39.469$^{+0.005}_{-0.005}$ days, consistent with the values reported in \citet{Jenkins2017}.

\item In the case of KOI-1257 b and HD 113337 b, the Keplerian model without any acceleration terms is statistically favored. Although the curvature models are disfavored, they represent a more realistic scenario consistent with the findings in the discovery papers. The Keplerian model without acceleration terms may be favored due to the penalization of model complexity with the BIC and the additional parameters required by a more complex model.
\citet{Santerne2014} and \citet{Borgniet2014} find significant linear and quadratic accelerations in both systems. \citealt{Santerne2014} report a trend of $135 \pm 80$ m s$^{-1}$ yr$^{-1}$, and $-283 \pm 75$ m s$^{-1}$ yr$^{-2}$ revealing the presence of an outer companion. Our solution yields a trend of $493.09^{+1.05}_{-0.71}$ m s$^{-1}$ yr$^{-1}$, and $-109.85^{+3.72}_{-4.09}$  m s$^{-1}$ yr$^{-2}$. \citealt{Borgniet2014} find a trend of $-41.2 \pm 3.6$ m s$^{-1}$ yr$^{-1}$ and $-15.9 \pm 0.6$ m s$^{-1}$ yr$^{-2}$. We infer a trend of $-61.25^{+0.27}_{-0.17}$ m s$^{-1}$ yr$^{-1}$, and $-16.58^{+0.09}_{-0.08}$  m s$^{-1}$ yr$^{-2}$.

\end{enumerate}

\section{Conclusion}\label{sec:Conclusion}
In this uniform analysis of warm Jupiter orbits, we have presented Keplerian orbital fits of 200 planets orbiting 194 stars based on 18,587 radial velocity (RV) measurements acquired from 40 spectrographs over the course of the past 35 years. Eccentricity posterior distributions for individual systems will be used for population-level statistical analyses in future studies. As part of this, we have also presented our targeted spectroscopic observations from a multi-hemisphere campaign with HPF and MINERVA-Australis to refine the eccentricities of a subset of giant planets in our analysis.

 Future papers in the Exploring Warm Jupiter Migration Pathways With Eccentricities series will examine how eccentricities depend on properties of their host stars and environmental context, such as planet multiplicity and stellar multiplicity. Together these will rely on warm Jupiter eccentricities as a tracer of how and when they form, migrate, and dynamically interact on Gyr timescales.

\begin{longrotatetable}
\begin{deluxetable*} {@{}cccccccccc@{}}
\centerwidetable
\renewcommand\arraystretch{0.87}
\setlength{\tabcolsep}{.05cm}
\tabletypesize{\tiny}
\tablecaption{Best-fitted solutions for warm Jupiters\label{tab:planet_params}}
\tablehead{
  \colhead{Planet Name}  & \colhead{${e}$} & \colhead{$a$} & \colhead{$T_\mathrm{c}$} & \colhead{$P$} & \colhead{$\omega$} & \colhead{$K$} & \colhead{$M_p$ sin$i$} & \colhead{Ref} \\
  &   & \colhead{(AU)} & \colhead{(BJD -- 2450000)} & \colhead{(days)} & \colhead{($^{\circ}$)} & \colhead{(m s$^{-1}$)} & \colhead{($M_\mathrm{Jup}$)}  
}
\startdata
\hline
14 And b & 0.023 (0.010, 0.010) & 0.82987 (0.00032, 0.00034) & 2904.943 (0.894, 0.871) & 186.168 (0.108, 0.113) & 29.51 (16.07, 31.63) & 99.94 (0.83, 0.79) & 4.748 (0.037, 0.040) & 1 \\
24 Boo b & 0.0545 (0.0072, 0.0068) & 0.190865 (0.000004, 0.000004) & -6.387 (0.214, 0.222) & 30.61056 (0.00100, 0.00094) & 307.094 (8.744, 8.769) & 51.834 (0.414, 0.402) & 0.7913 (0.0063, 0.0062) & 2 \\
4 UMa b & 0.440 (0.003, 0.003) & 0.87189 (0.00059, 0.00062) & 3009.49 (0.57, 0.59) & 268.123 (0.273, 0.287) & 22.99 (0.64, 0.65) & 213.36 (0.90, 0.96) & 6.980 (0.031, 0.030) & 3 \\
55 Cnc b & 0.009 (0.001, 0.001) & 0.1135577 (0.0000001, 0.0000001) & 5803.337 (0.003, 0.003) & 14.65218 (0.00001, 0.00001) & 57.82 (3.89, 3.59) & 70.89 (0.04, 0.04) & 0.8015 (0.0005, 0.0005) & 4, 5, 6  \\
70 Vir b & 0.3978 (0.0002, 0.0002) & 0.4809564 (0.0000004, 0.0000003) & 5306.62 (0.01, 0.01) & 116.6927 (0.0001, 0.0001) & 358.55 (0.03, 0.03) & 314.70 (0.07, 0.07) & 7.352 (0.002, 0.002) & 7, 8, 9, 10  \\
8 UMi b & 0.049 (0.024, 0.025) & 0.45564 (0.00020, 0.00019) & 4148.00 (1.69, 1.79) & 93.617 (0.063, 0.060) & 10.20 (264.84, 49.62) & 50.12 (1.07, 1.15) & 1.413 (0.032, 0.033) & 11 \\
91 Aqr b & 0.032 (0.002, 0.002) & 0.70093 (0.00003, 0.00003) & 3429.03 (0.15, 0.11) & 181.152 (0.013, 0.013) & 151.35 (3.28, 3.18) & 92.86 (0.18, 0.19) & 3.234 (0.006, 0.006) & 12 \\
BD+14 4559 b & 0.222 (0.007, 0.007) & 0.64321 (0.00038, 0.00035) & 3294.70 (0.86, 0.88) & 269.172 (0.239, 0.222) & 91.14 (1.65, 1.68) & 55.07 (0.39, 0.41) & 1.060 (0.008, 0.008) & 13 \\
\enddata
\end{deluxetable*}
\end{longrotatetable}

\startlongtable
\begin{deluxetable*} {lccccccccl}
\centering 
\renewcommand\arraystretch{0.95}
\setlength{\tabcolsep}{.2cm}
\tabletypesize{\small}
\tablecaption{Host star properties.\label{tab:host_stars}}
\tablehead{
     \colhead{Hoststar} & \colhead{SpT} & \colhead{$T_\mathrm{eff}$} & \colhead{$M_\star$} & \colhead{[Fe/H]}  & \colhead{Distance} & \colhead{Ref.} \\
        &  & \colhead{(K)} & \colhead{($M_\odot$)} & \colhead{(dex)} & \colhead{(pc)} & & 
        }
\startdata
\hline
14 And & K0 III & $4813^{+20}_{-20}$ & $2.20^{+0.10}_{-0.20}$ & $-0.24^{+0.03}_{-0.03}$ & $75.85^{+0.42}_{-0.44}$ & 1  \\
24 Boo & G3 IV & $4893^{+15}_{-15}$ & $0.99^{+0.19}_{-0.13}$ & $-0.77^{+0.03}_{-0.03}$ & $94.18^{+0.34}_{-0.40}$ & 2  \\
4 UMa & K1 III & $4415^{+70}_{-70}$ & $1.23^{+0.15}_{-0.15}$ & $-0.25^{+0.04}_{-0.04}$ & $75.63^{+0.62}_{-0.60}$ & 3  \\
55 Cnc & K0 IV & $5172^{+18}_{-18}$ & $0.91^{+0.01}_{-0.01}$ & $0.35^{+0.10}_{-0.10}$ & $12.58^{+0.01}_{-0.01}$ & 4  \\
70 Vir & G4 V & $5481^{+100}_{-100}$ & $1.06^{+0.03}_{-0.03}$ & $-0.06^{+0.01}_{-0.01}$ & $18.09^{+0.03}_{-0.03}$ & 5,6,7   \\
8 UMi & K0 & $4847^{+7}_{-7}$ & $1.44^{+0.19}_{-0.19}$ & $-0.03^{+0.02}_{-0.02}$ & $162.68^{+0.42}_{-0.30}$ & 8,9   \\
91 Aqr & K0 III & $4665^{+18}_{-18}$ & $1.40^{+0.10}_{-0.10}$ & $-0.03^{+0.10}_{-0.10}$ & $45.28^{+0.28}_{-0.24}$ & 10  \\
BD-08 2823 & K3 V & $4746^{+63}_{-63}$ & $0.74^{+0.07}_{-0.07}$ & $-0.07^{+0.03}_{-0.03}$ & $41.30^{+0.02}_{-0.02}$ & 11  \\
\enddata
\end{deluxetable*}

\startlongtable
\begin{deluxetable*} {cccccccccccccl}
\centering 
\renewcommand\arraystretch{0.95}
\setlength{\tabcolsep}{.05cm}
\tabletypesize{\scriptsize}
\tablecaption{Additional details of each planetary system including signals from long-term companions.\label{tab:additional_planet_params}}
\tablehead{
    \vspace{2pt} \\
     \colhead{Planet} & \colhead{$\#$ Planets} & \colhead{$N_\mathrm{data}$} & \colhead{$\bar{\text{\emph{$\sigma$}}}_{RV}$} & \colhead{$\Delta$$t$} & \colhead{$N_\mathrm{inst}$} & \colhead{$\chi^{2}_{\nu}$} & \colhead{$\Delta$$\mathrm{BIC_{K}}$} & \colhead{$\Delta$$\mathrm{BIC_{L-K}}$} & \colhead{$\Delta$$\mathrm{BIC_{Q-K}}$} & \colhead{$\dot{\gamma}$} & \colhead{$\ddot{\gamma}$} \\
  &   &  &  \colhead{(m s$^{-1}$)} & \colhead{(years)} & & & & & & \colhead{(m s$^{-1}$ yr$^{-1}$)} & \colhead{(m s$^{-1}$ yr$^{-2}$) } }
\startdata
\hline
14 And b & 1 & 34 & 5.86 & 4.07 & 1 & 0.96 & 0.00 & -8.41 & -8.33 & $\cdots$ & $\cdots$ \\
24 Boo b & 1 & 149 & 6.70 & 13.15 & 1 & 0.98 & 0.00 & 8329.69 & 8150.76 & $\cdots$ & $\cdots$ \\
4 UMa b & 1 & 60 & 6.80 & 2.43 & 1 & 0.98 & 0.00 & -205.76 & -197.83 & -11.89 (0.83, 0.75) & $\cdots$ \\
55 Cnc b & 5 & 783 & 1.67 & 25.33 & 7 & 1.00 & $\cdots$ & $\cdots$ & $\cdots$ & $\cdots$ & $\cdots$ \\
70 Vir b & 1 & 345 & 4.36 & 29.05 & 3 & 0.95 & 0.00 & -83.94 & -36.84 & -0.42 (0.01, 0.01) & $\cdots$ \\
8 UMi b & 1 & 21 & 13.76 & 5.17 & 1 & 1.04 & 0.00 & 4.97 & -48.32 & 12.05 (1.69, 1.68) & -2.51 (0.31, 0.33) \\
91 Aqr b & 1 & 160 & 6.05 & 12.31 & 2 & 0.99 & 0.00 & 61.50 & 249.36 & $\cdots$ & $\cdots$ \\
BD+14 4559 b & 1 & 43 & 8.33 & 5.04 & 1 & 1.02 & 0.00 & -1304.77 & -1397.08 & -24.25 (1.31, 1.26) & 3.49 (0.35, 0.33) \\
\enddata
\tablecomments{\scriptsize The number of planets in column 2 details the number of planets in the system not the number of planets fit. For single planet systems we present BIC values for all 3 models (Keplerian ($\Delta$$\mathrm{BIC_{K}}$), Keplerian+Linear ($\Delta$$\mathrm{BIC_{L-K}}$), or Keplerian+Quadratic ($\Delta$$\mathrm{BIC_{Q-K}}$)). Reduced $\chi^{2}$ values are reported for every system. The number of parameters for each system can be computed as $(5 \times N_\mathrm{P}) + (2 \times N_\mathrm{i})$ + acceleration terms, where $N_\mathrm{P}$ is the number of planets and $N_\mathrm{i}$ is the number of instruments in the model. The mean error on the RV's ($\bar{\text{\emph{$\sigma$}}}_{RV}$), time baseline of the RVs ($\Delta$t), and number of RVs ($N_\mathrm{data}$) are also presented. For systems where an acceleration is favored, the MAP value is reported along with the upper and lower bounds of the 68.3$\%$ highest density interval regions of the posterior probability distribution.}
\end{deluxetable*}

\section{Acknowledgments}
We thank William D Cochran, Kyle Franson, and Lillian Jiang for insightful conversations. The authors would also like to thank Chad Bender, Steven Janowiecki, Greg Zeimann, Suvrath Mahadevan, the HPF team, and all the resident astronomers and telescope operators at the HET for supporting these observations and data processing. The authors would also like to thank the MINERVA-Australis team and all the resident astronomers and telescope operators at M-A for supporting these observations and data processing. We are grateful to Sylvain Korzennik for RV data of HD 89744 b, Michaela Dollinger and Artie Hatzes for RV data on 4 UMA b, and Damien Segransan and Michel Mayor for RV data on HD 63765 b. We thank the many observers who contributed to the measurements reported in this work.

B.P.B. acknowledges support from the National Science Foundation grant AST-1909209, NASA Exoplanet Research Program grant 20-XRP20$\_$2-0119, and the Alfred P. Sloan Foundation.

This work made use of observations obtained with the Hobby-Eberly Telescope (HET), which is a joint project of the University of Texas at Austin, the Pennsylvania State University, Ludwig-Maximillians-Universitaet Muenchen, and Georg-August Universitaet Goettingen. The HET is named in honor of its principal benefactors, William P. Hobby and Robert E. Eberly. The HPF team acknowledges support from NSF grants AST-1006676, AST-1126413, AST-1310885, AST-1517592, AST-1310875, ATI 2009889, ATI-2009982, AST-2108512, and the NASA Astrobiology Institute (NNA09DA76A) in the pursuit of precision radial velocities in the NIR. The HPF team also acknowledges support from the Heising-Simons Foundation via grant 2017-0494.  We would like to acknowledge that the HET is built on Indigenous land. Moreover, we would like to acknowledge and pay our respects to the Carrizo $\&$ Comecrudo, Coahuiltecan, Caddo, Tonkawa, Comanche, Lipan Apache, Alabama-Coushatta, Kickapoo, Tigua Pueblo, and all the American Indian and Indigenous Peoples and communities who have been or have become a part of these lands and territories in Texas, here on Turtle Island.

We acknowledge the Texas Advanced Computing Center (TACC) at The University of Texas at Austin for providing high performance computing, visualization, and storage resources that have contributed to the results reported within this paper. Data presented herein were obtained with the MINERVA-Australis facility at the Mt. Kent Observatory from telescope time allocated through the NN-EXPLORE program. NN-EXPLORE is a scientific partnership of the National Aeronautics and Space Administration and the National Science Foundation. This publication makes use of The Data $\&$ Analysis Center for Exoplanets (DACE), which is a facility based at the University of Geneva (CH) dedicated to extrasolar planets data visualisation, exchange and analysis. DACE is a platform of the Swiss National Centre of Competence in Research (NCCR) Planets, federating the Swiss expertise in Exoplanet research. The DACE platform is available at https://dace.unige.ch.

This research has made use of the VizieR catalog access tool,
CDS, Strasbourg, France (doi:10.26093/cds/vizier). The
original description of the VizieR service was published in
2000, A$\&$AS 143, 23. This work has made use of data from the European Space Agency (ESA) mission {\it Gaia} (\url{https://www.cosmos.esa.int/gaia}), processed by the {\it Gaia} Data Processing and Analysis Consortium (DPAC,
\url{https://www.cosmos.esa.int/web/gaia/dpac/consortium}). Funding for the DPAC has been provided by national institutions, in particular the institutions participating in the {\it Gaia} Multilateral Agreement.

\facility{HET (HPF) and MINERVA-Australis}

\software{\texttt{radvel} \citep{Fulton2018}, \texttt{numpy} \citep{Harris2020}, \texttt{matplotlib}  \citep{Hunter2007}, \texttt{corner} \citep{Foreman-Mackey2016}, \texttt{emcee} \citep{Foreman-Mackey2013}, \texttt{ArviZ} \citep{arviz2019}, and \texttt{pandas} \citep{reback2020pandas}} 

\appendix
\section{New Radial Velocity Observations of Warm Jupiter Hosts}\label{sec:Observational_Campaign}

Here we report the relative radial velocity observations of the warm Jupiter host stars observed with HPF and MINERVA-Australis. All RV tables for targets observed with HPF and MINERVA-Australis can be found in machine-readable format (see Table \ref{tab:BD_15_2940}).

\begin{deluxetable}{lcccc}[h] 
\renewcommand\arraystretch{0.9}
\tabletypesize{\large}
\setlength{\tabcolsep}{0.5cm}
\tablewidth{0pt}
\tablecolumns{5}
\tablecaption{New Relative Radial Velocities of BD+15 2940\label{tab:BD_15_2940}}
\tablehead{
    \colhead{System} &
    \colhead{BJD} &
    \colhead{RV (m s$^{-1}$)} &
    \colhead{$\sigma_{RV}$ (m s$^{-1}$)} &
    \colhead{Instrument}
}
\startdata
BD+15 2940&  2459604.01778&    18.74&    6.06&    HPF1 \\
BD+15 2940&  2459625.96012&    -22.48&    9.67&    HPF1 \\
BD+15 2940&  2459626.94975&    -16.96&    14.77&    HPF1 \\
BD+15 2940&  2459629.94852&    -45.17 &  7.51&    HPF1 \\
BD+15 2940&  2459631.94374&    -27.85&    6.35&    HPF1 \\
BD+15 2940&  2459632.94328&    -33.31&    10.62&    HPF1 \\
BD+15 2940&  2459635.93873&    -22.14&    17.21&    HPF1 \\
BD+15 2940&  2459680.81257&    -2.99&    8.96&    HPF1 \\
BD+15 2940&  2459683.98918&    9.76&    3.61&    HPF1 \\
BD+15 2940&  2459685.79253&    -24.85&    3.38&    HPF1 \\
BD+15 2940&  2459686.98267&    0.76&    4.87&    HPF1 \\
BD+15 2940&  2459687.98372&    -42.39&    1.72&    HPF1 \\
BD+15 2940&  2459690.78599&    -45.37&    5.09&    HPF1 \\
BD+15 2940&  2459691.96513&    -50.92&    2.79&    HPF1 \\
BD+15 2940&  2459693.96830&    -13.20&    7.92&    HPF1 \\
BD+15 2940&  2459697.96413&    -15.55&    4.66&    HPF1 \\
BD+15 2940&  2459705.73549&    29.58&    2.95&    HPF1 \\
BD+15 2940&  2459706.74279&    32.60&    7.93&    HPF1 \\
BD+15 2940&  2459709.73420&    51.63&    15.31&    HPF1 \\
BD+15 2940&  2459710.73096&    34.36&    8.43&    HPF1 \\
BD+15 2940&  2459712.72906&    35.92&    5.47&    HPF1 \\
BD+15 2940&  2459712.93233&    29.67&    10.07&    HPF1 \\
BD+15 2940&  2459713.72432&    30.22&    7.38&    HPF1 \\
BD+15 2940&  2459713.93013&    26.62&    2.99&    HPF1 \\
BD+15 2940&  2459714.72178&    36.20&    8.06&    HPF1 \\
BD+15 2940&  2459714.91427&    17.74&    4.27&    HPF1 \\
BD+15 2940&  2459715.91303&    0.76&    0.52&    HPF1 \\
BD+15 2940&  2459717.90645&    9.43&    1.93&    HPF1 \\
\enddata
\tablecomments{Relative radial velocities for BD+15 2940 are displayed here.}
\end{deluxetable}

\clearpage
\section{Orbit Fits}\label{sec:Orbit_Fits}

\begin{figure}
\hskip -0.8 in
 \centering
 \begin{minipage}{\textwidth}
   \centering
   \includegraphics[width=\linewidth]{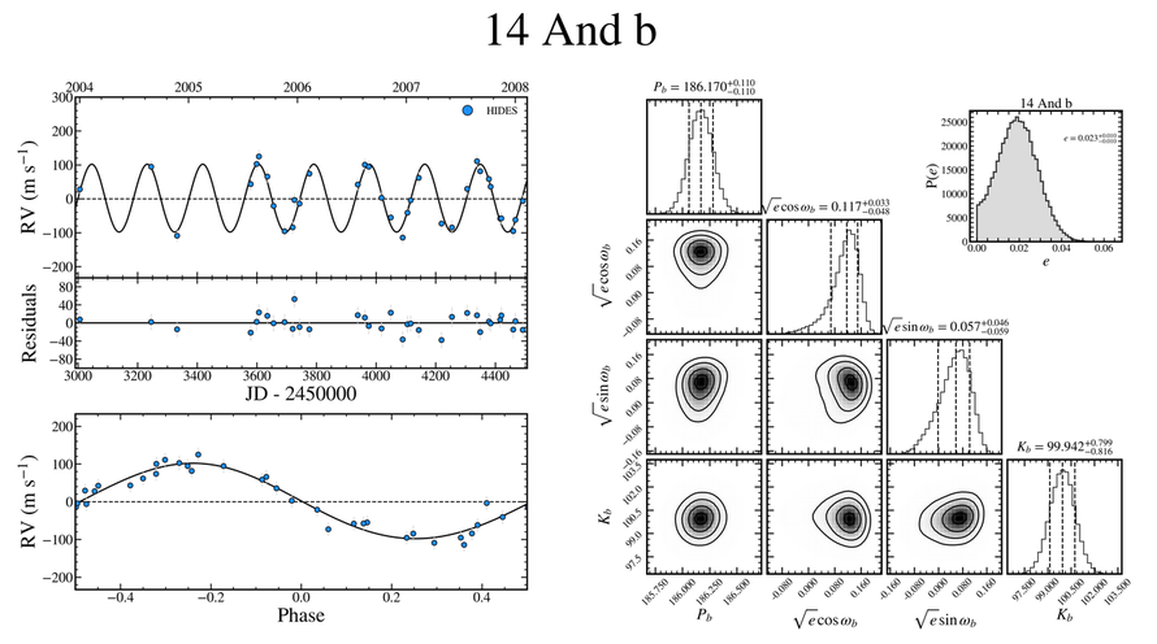}\\
   \vskip .3 in
   \includegraphics[width=\linewidth]{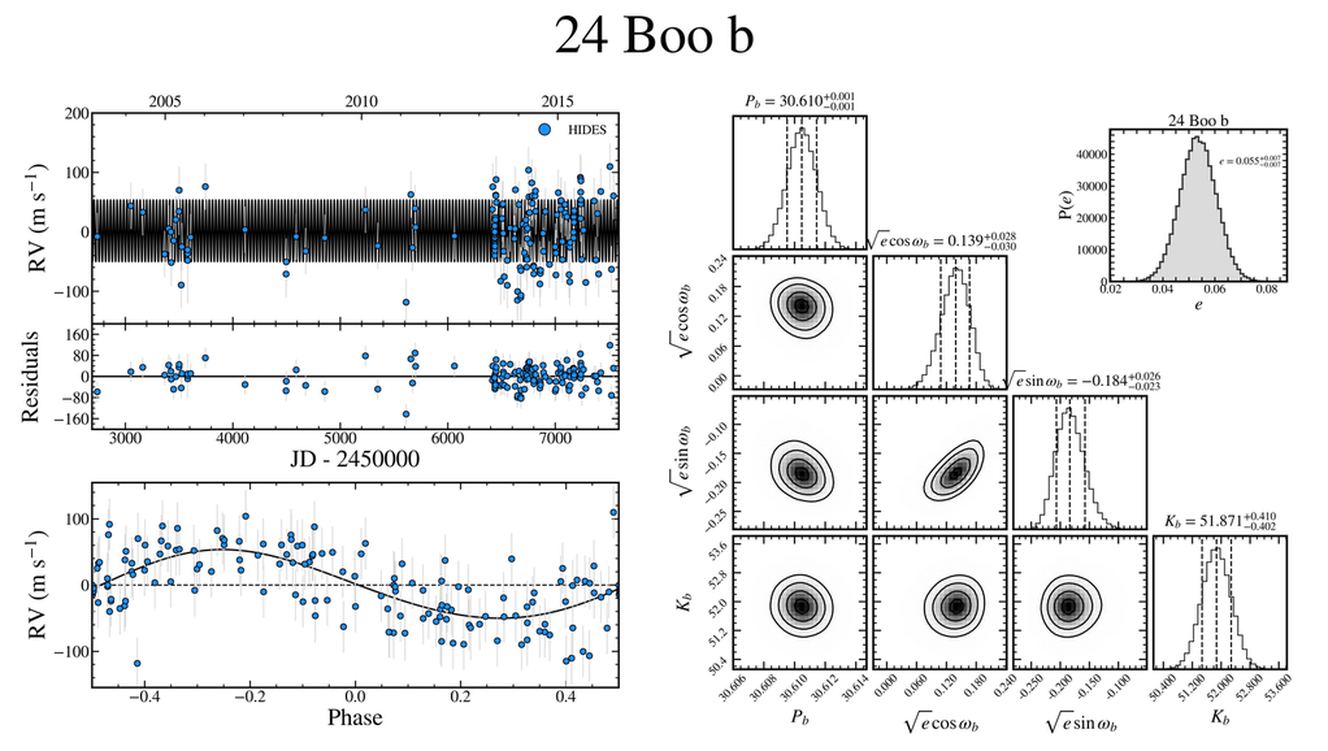}
 \end{minipage}
 \caption{Example summary plot of the known warm Jupiters 14 And b and 24 Boo b using \texttt{RadVel}. Left: Panel showing the RV timeseries plot combined with the residuals to the model and the phase folded orbit of the planet. Right: Corner plot with the joint posterior distributions and a histogram of the eccentricity posterior distribution of the planet.}
 \label{fig:Combined_Plots1}
\end{figure}

\begin{figure}
\hskip -0.8 in
 \centering
 \begin{minipage}{\textwidth}
   \centering
   \includegraphics[width=\linewidth]{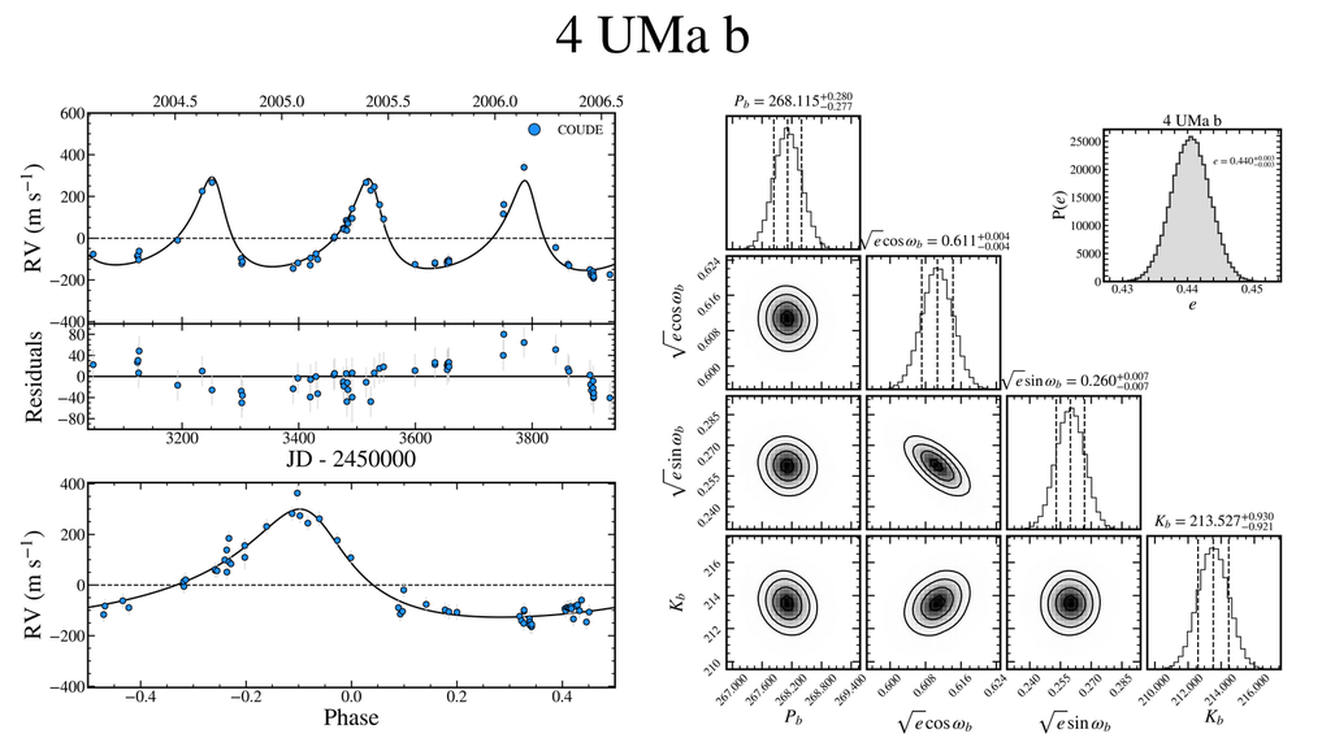}\\
   \vskip .3 in
   \includegraphics[width=\linewidth]{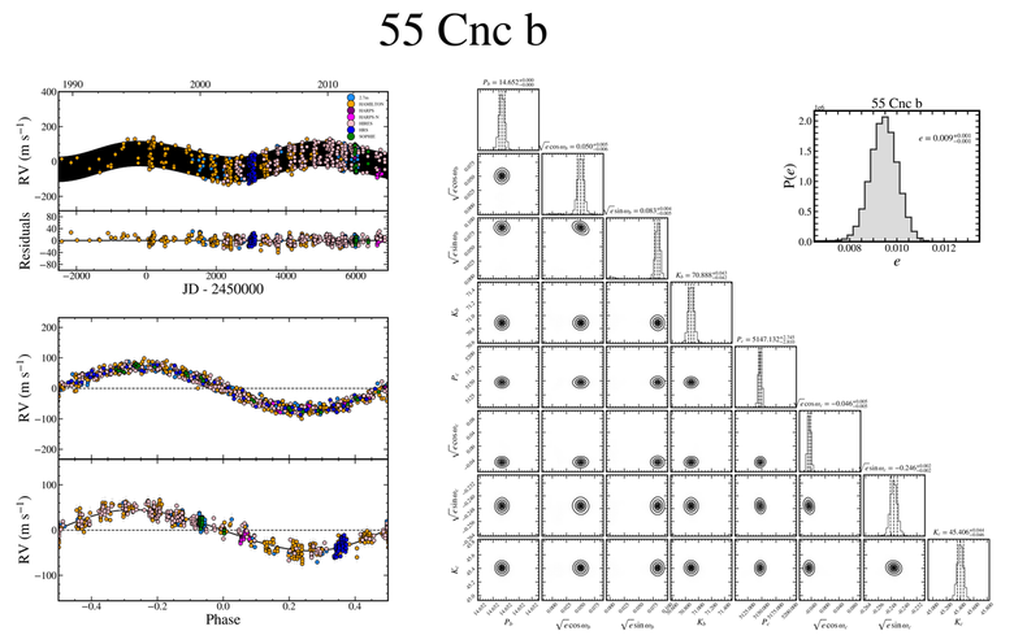}
 \end{minipage}
 \caption{Summary of results for the warm Jupiters 4 UMa b and 55 Cnc b.}
 \label{fig:Combined_Plots2}
\end{figure}

\begin{figure}
\hskip -0.8 in
 \centering
 \begin{minipage}{\textwidth}
   \centering
   \includegraphics[width=\linewidth]{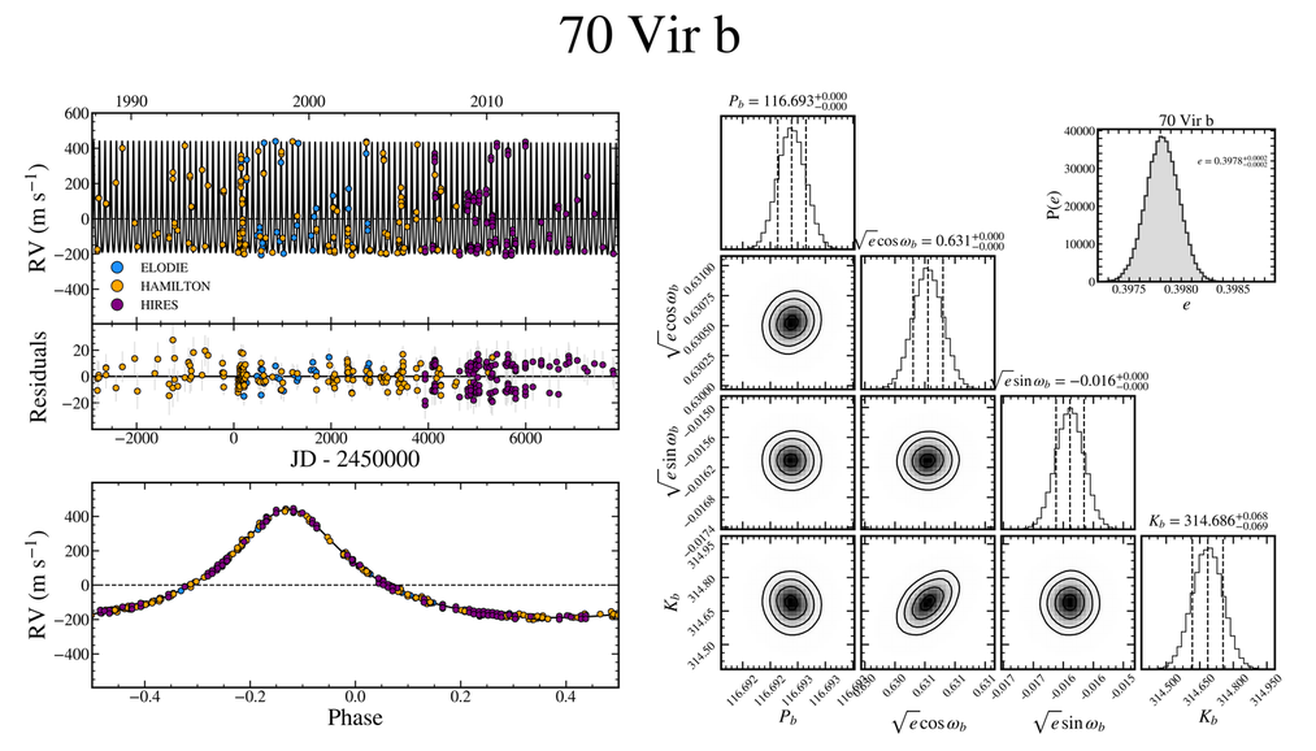}\\
   \vskip .3 in
   \includegraphics[width=\linewidth]{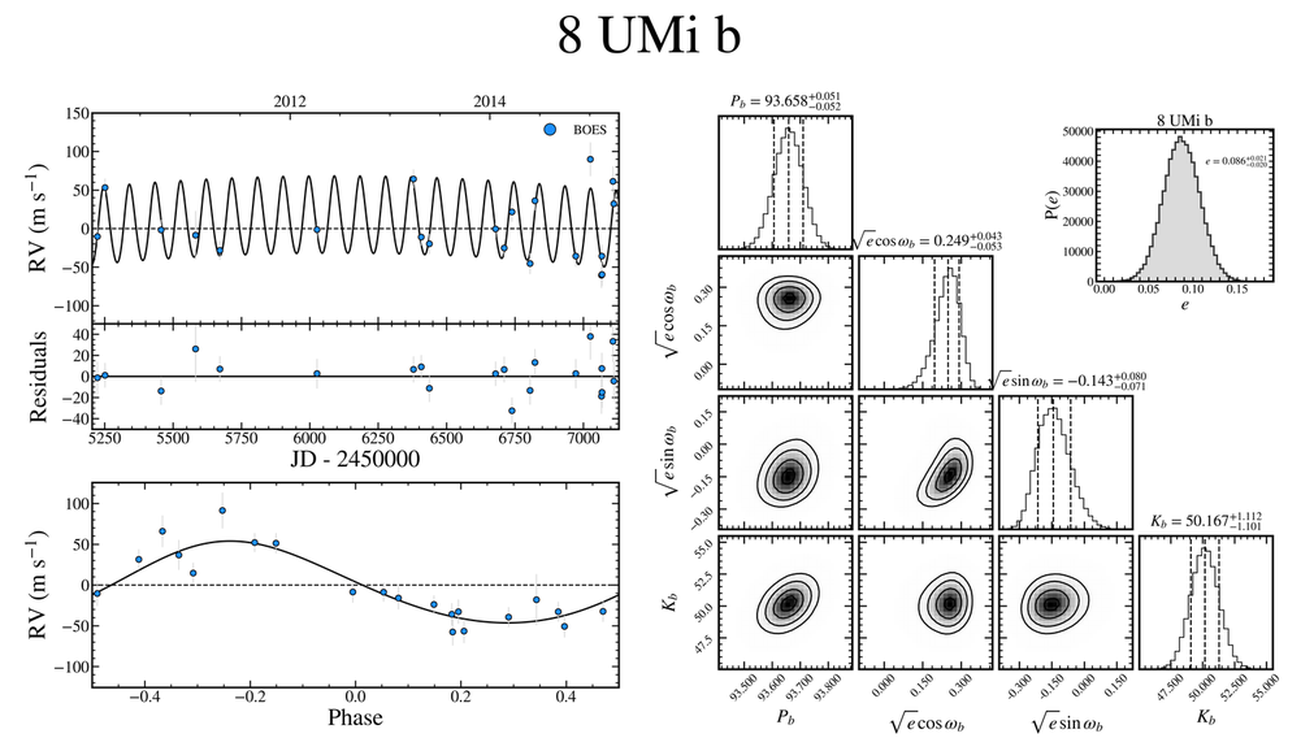}
 \end{minipage}
 \caption{Summary of results for the warm Jupiters 70 Vir b and 8 UMi b.}
 \label{fig:Combined_Plots3}
\end{figure}

\begin{figure}
\hskip -0.8 in
 \centering
 \begin{minipage}{\textwidth}
   \centering
   \includegraphics[width=\linewidth]{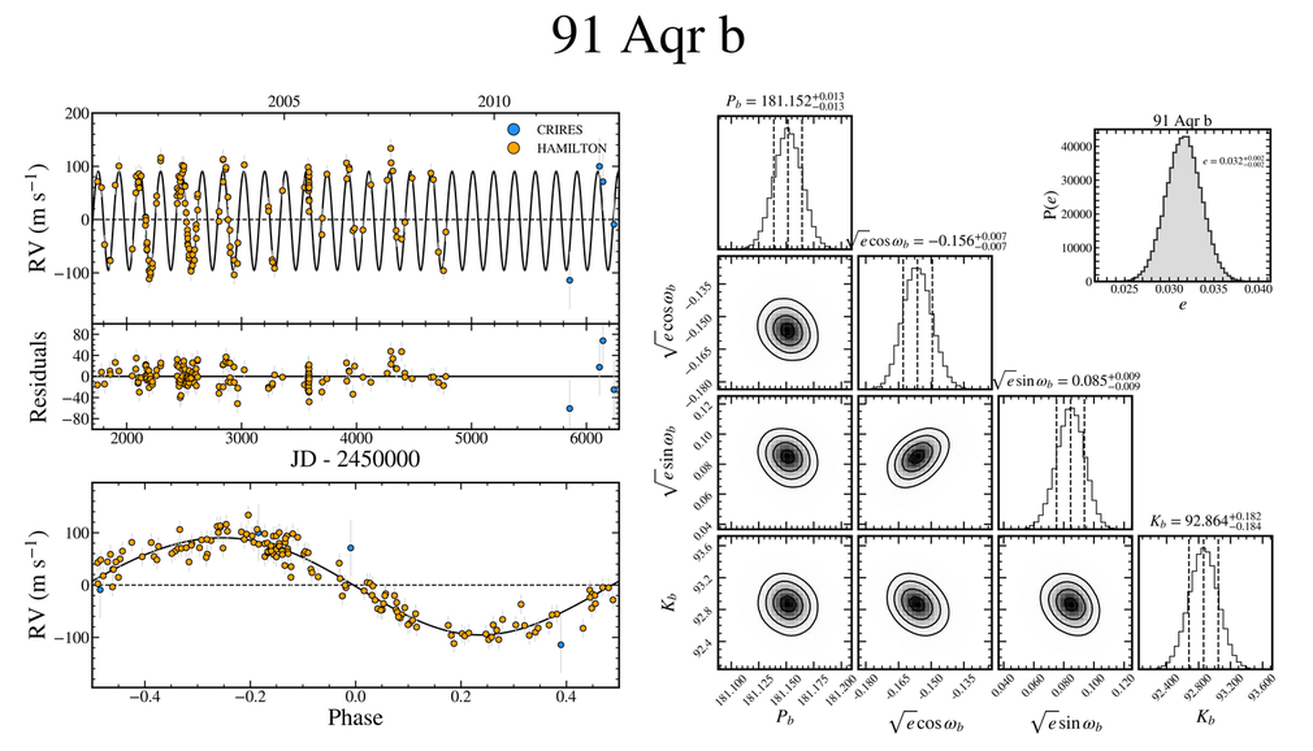}\\
   \vskip .3 in
   \includegraphics[width=\linewidth]{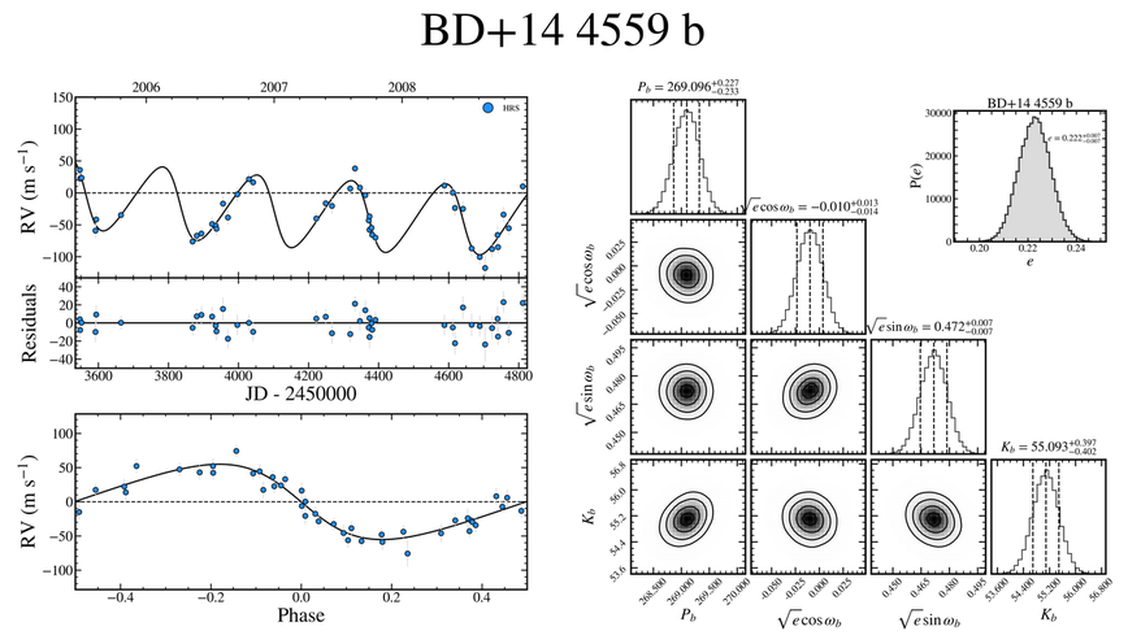}
 \end{minipage}
 \caption{Summary of results for the warm Jupiters 91 Aqr b and BD+14 4559 b.}
 \label{fig:Combined_Plots4}
\end{figure}

\begin{figure}
\hskip -0.8 in
 \centering
 \begin{minipage}{\textwidth}
   \centering
   \includegraphics[width=\linewidth]{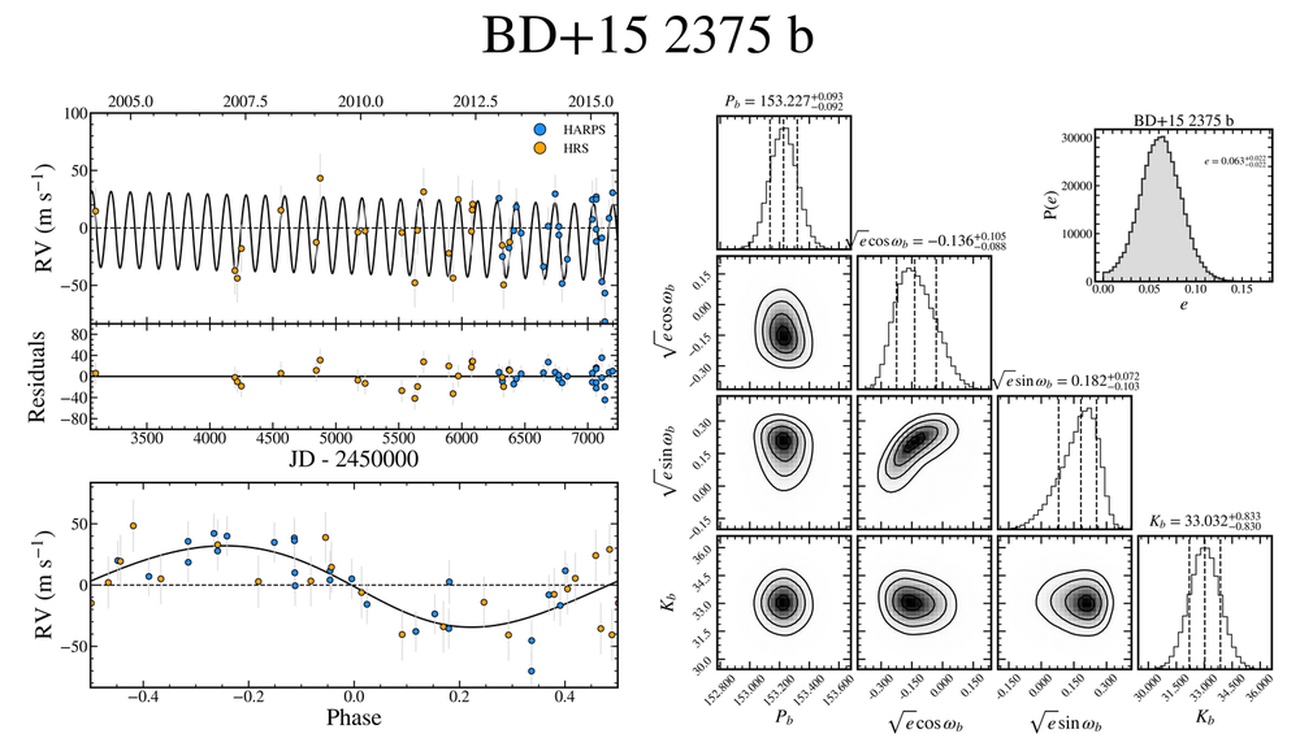}\\
   \vskip .3 in
   \includegraphics[width=\linewidth]{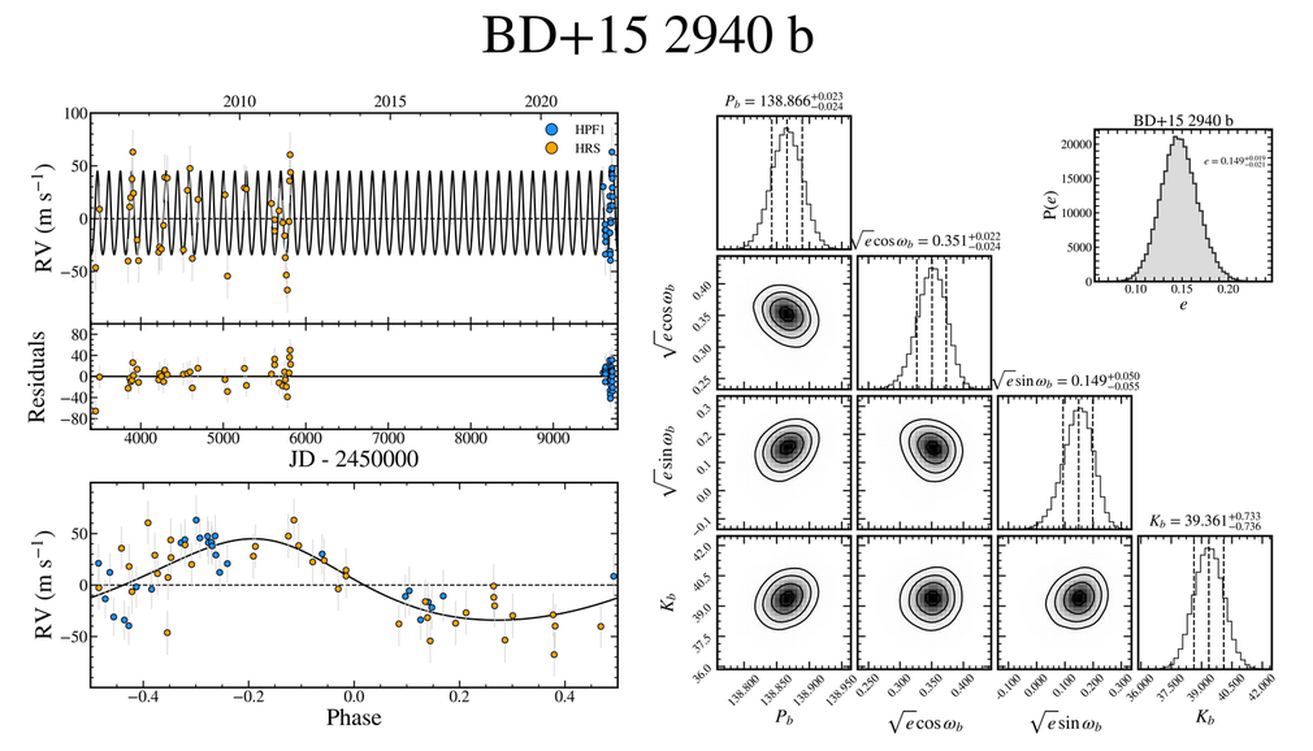}
 \end{minipage}
 \caption{Summary of results for the warm Jupiters BD+15 2375 b and BD+15 2940 b.}
 \label{fig:Combined_Plots5}
\end{figure}

\begin{figure}
\hskip -0.8 in
 \centering
 \begin{minipage}{\textwidth}
   \centering
   \includegraphics[width=\linewidth]{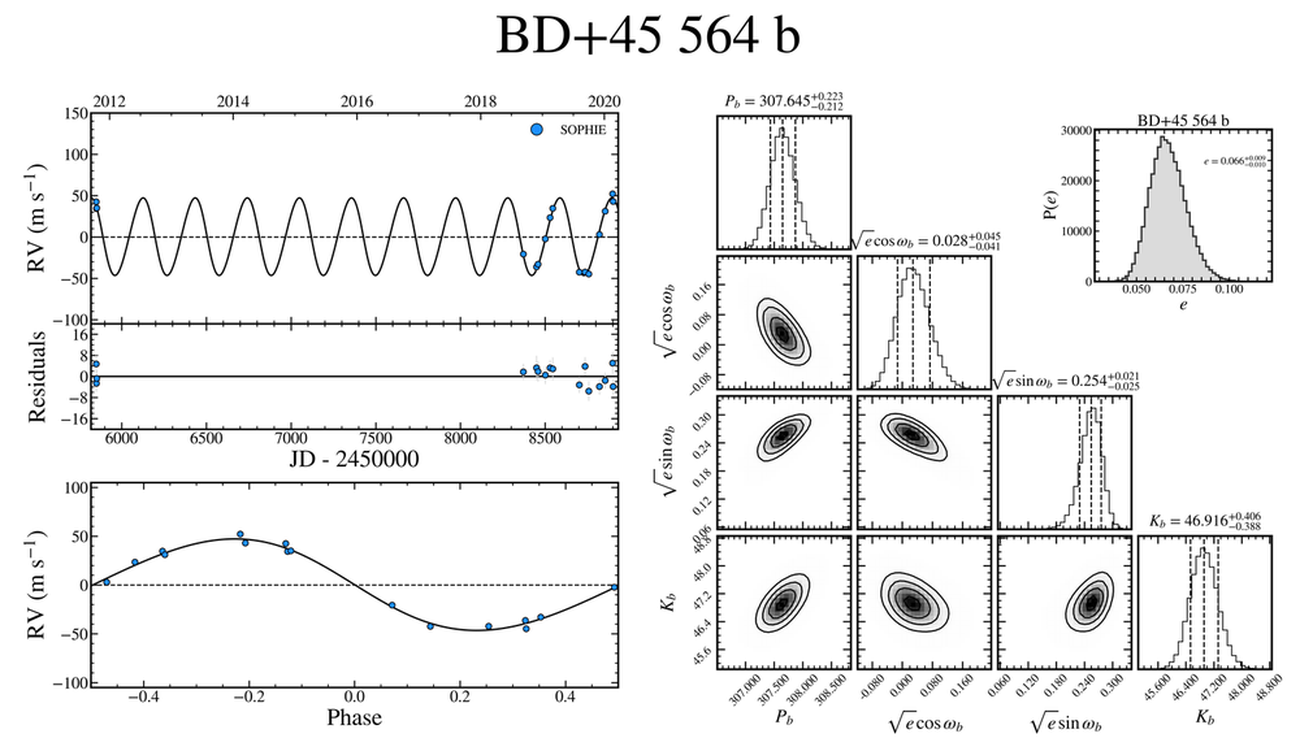}\\
   \vskip .3 in
   \includegraphics[width=\linewidth]{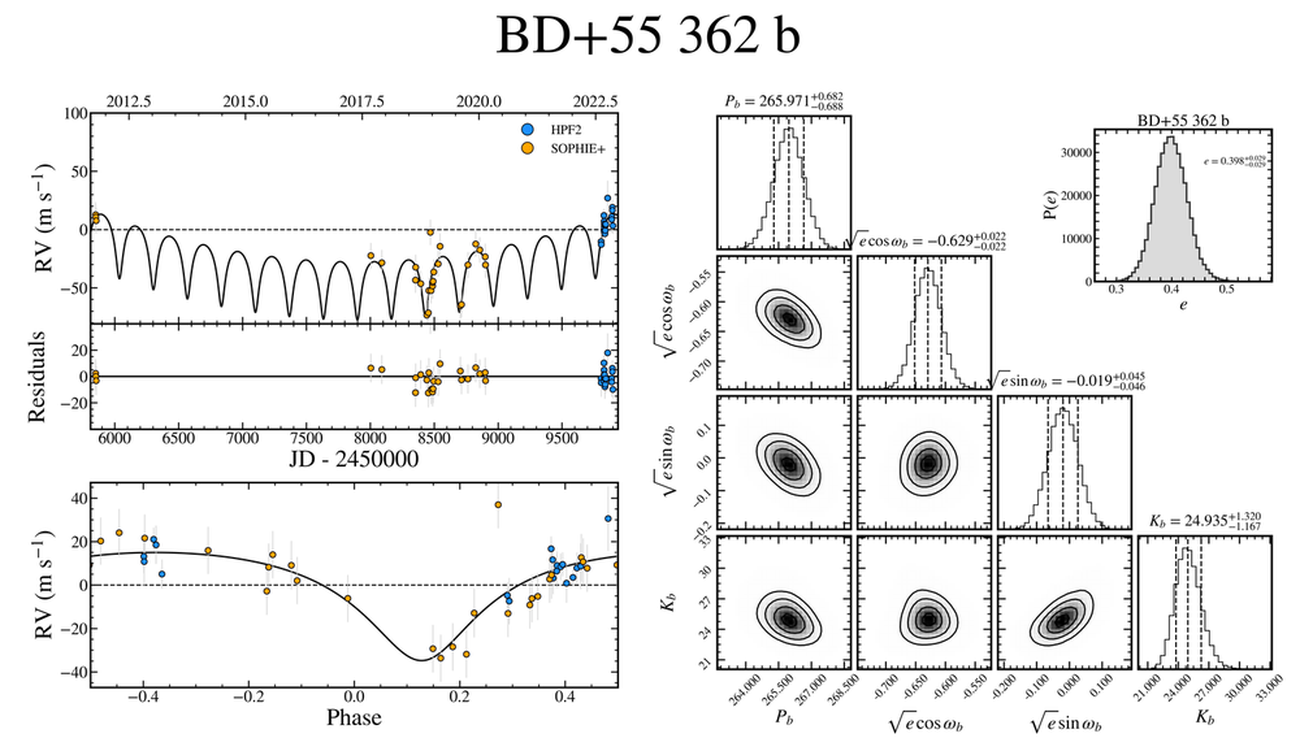}
 \end{minipage}
 \caption{Summary of results for the warm Jupiters BD+45 564 b and BD+55 362 b.}
 \label{fig:Combined_Plots6}
\end{figure}

\begin{figure}
\hskip -0.8 in
 \centering
 \begin{minipage}{\textwidth}
   \centering
   \includegraphics[width=\linewidth]{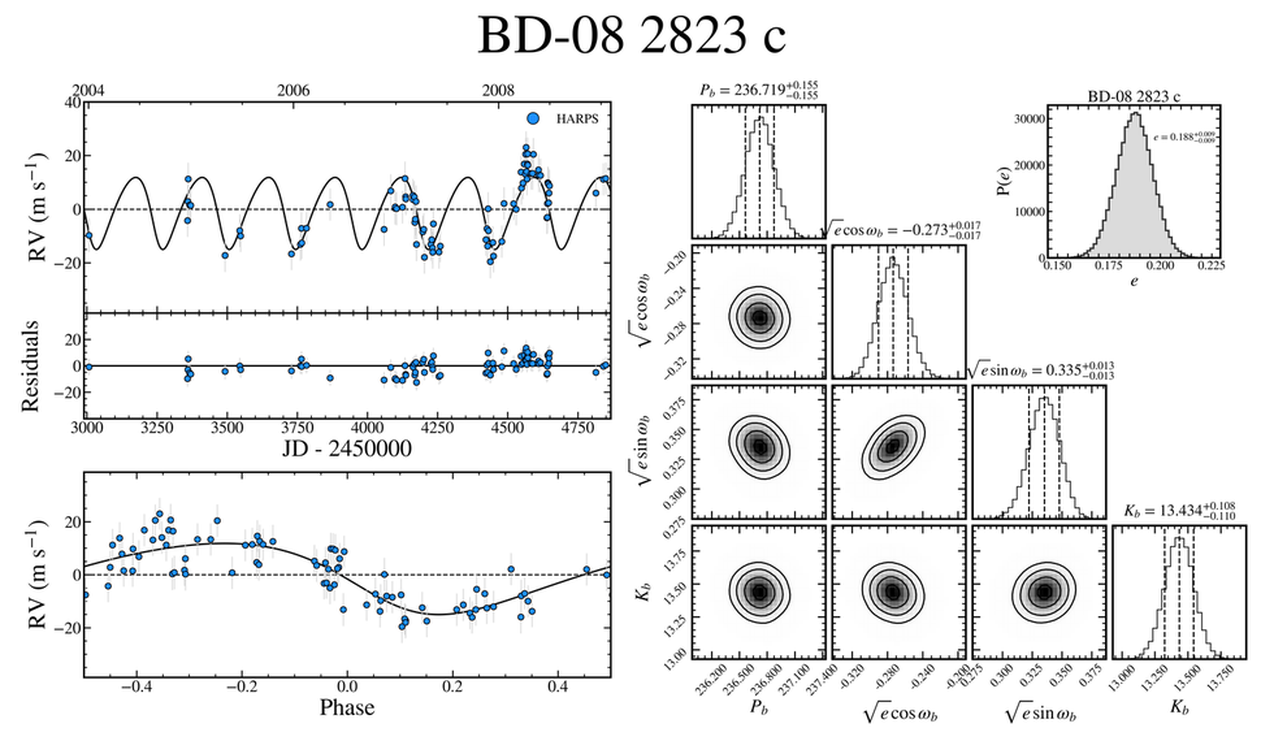}\\
   \vskip .3 in
   \includegraphics[width=\linewidth]{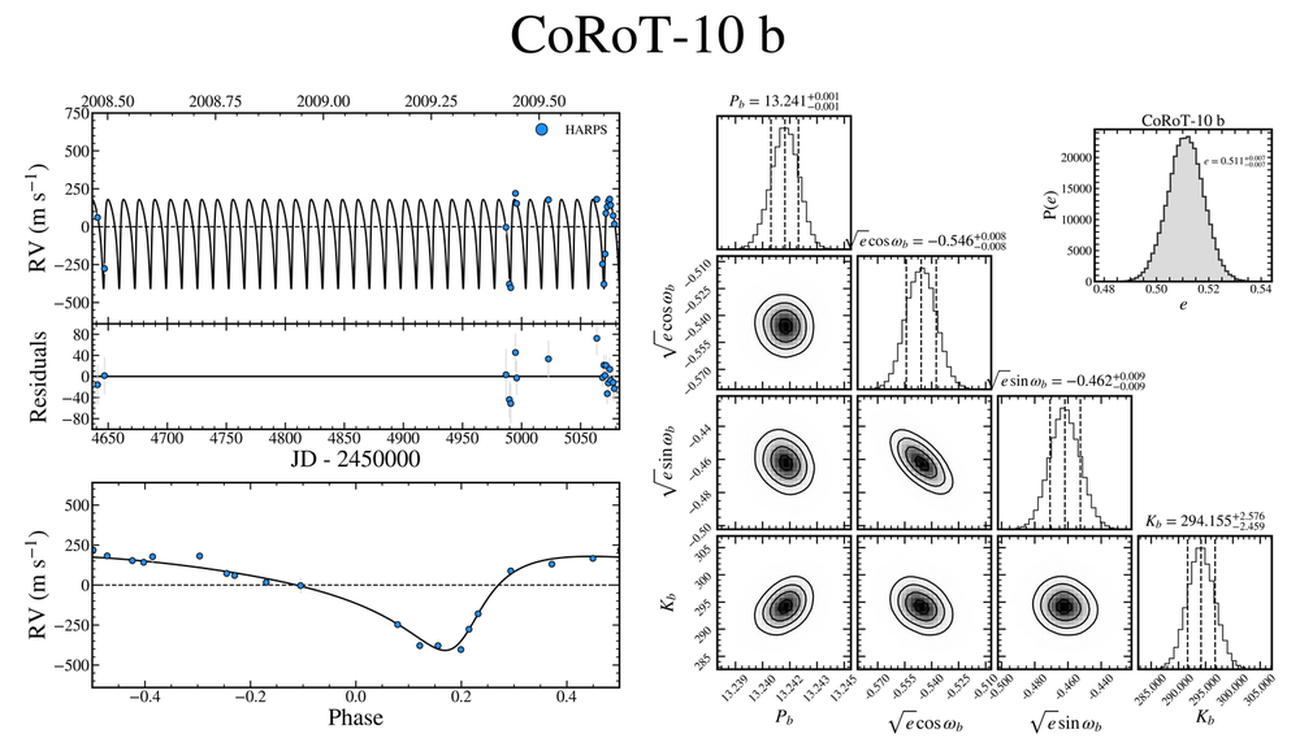}
 \end{minipage}
 \caption{Summary of results for the warm Jupiters BD-08 2823 c and CoRoT-10 b.}
 \label{fig:Combined_Plots7}
\end{figure}

\begin{figure}
\hskip -0.8 in
 \centering
 \begin{minipage}{\textwidth}
   \centering
   \includegraphics[width=\linewidth]{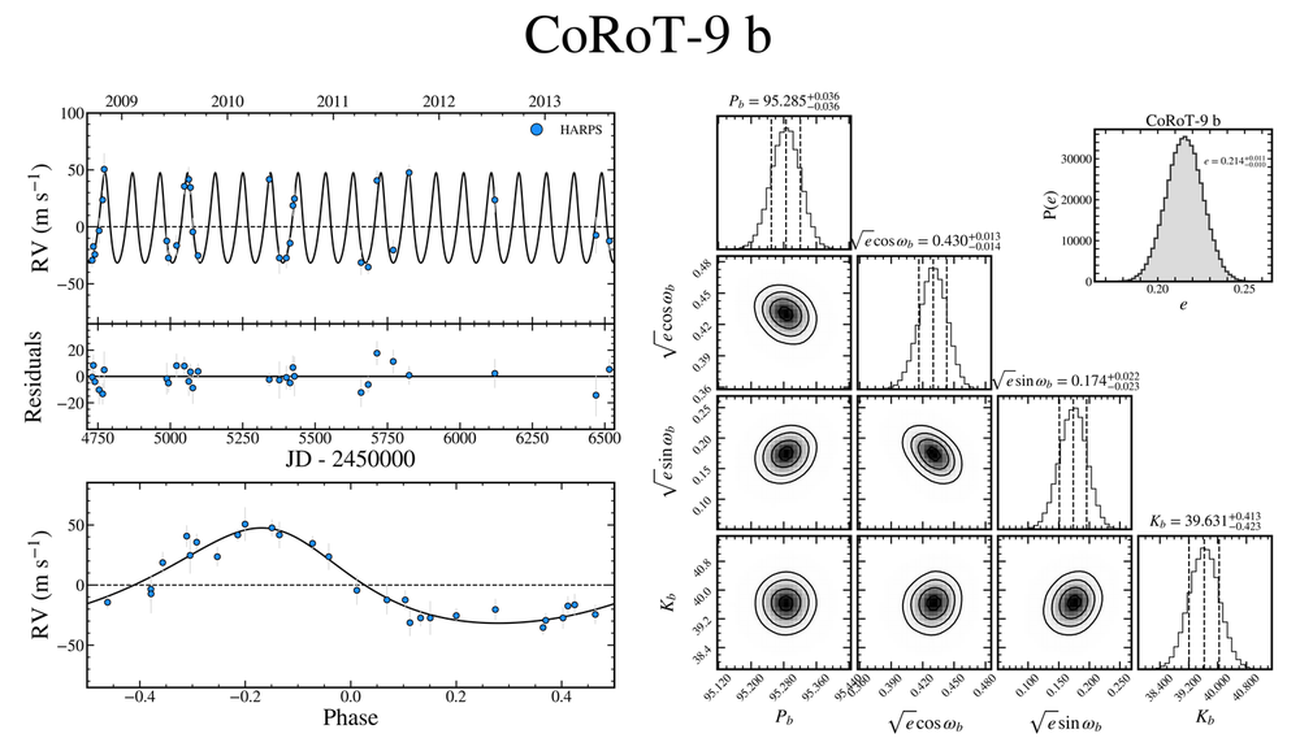}\\
   \vskip .3 in
   \includegraphics[width=\linewidth]{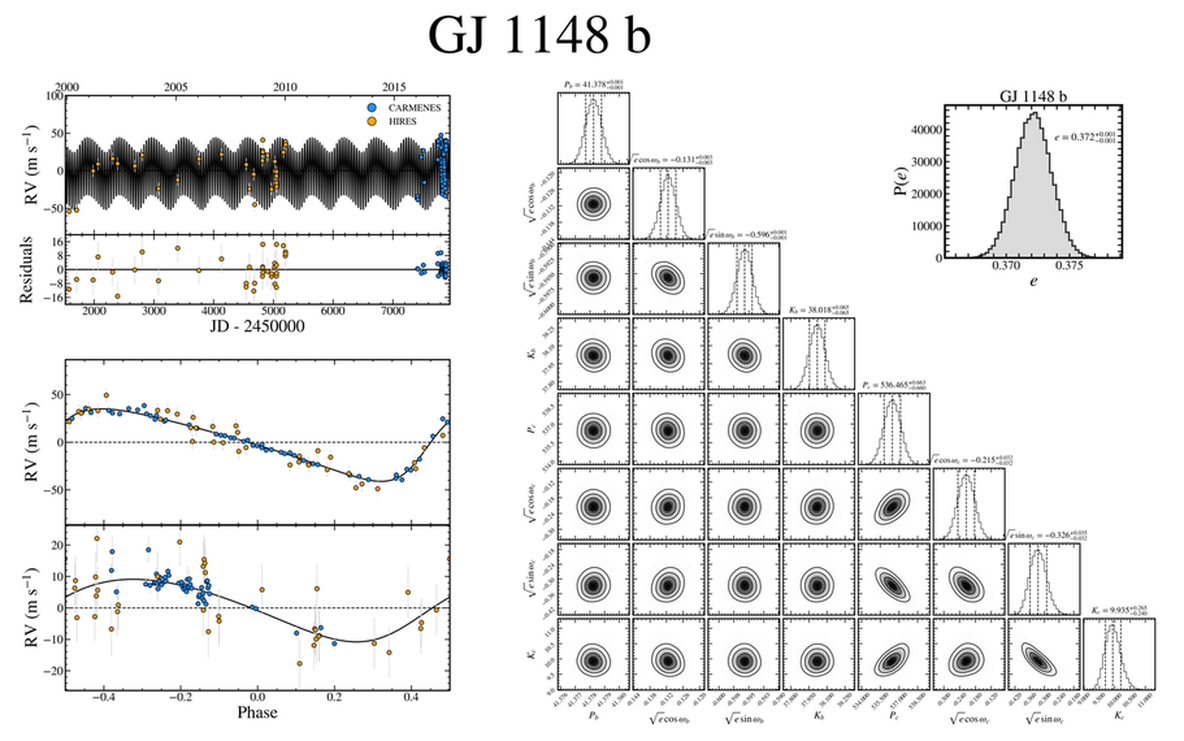}
 \end{minipage}
 \caption{Summary of results for the warm Jupiters CoRoT-9 b and GJ 1148 b.}
 \label{fig:Combined_Plots8}
\end{figure}

\begin{figure}
\hskip -0.8 in
 \centering
 \begin{minipage}{\textwidth}
   \centering
   \includegraphics[width=\linewidth]{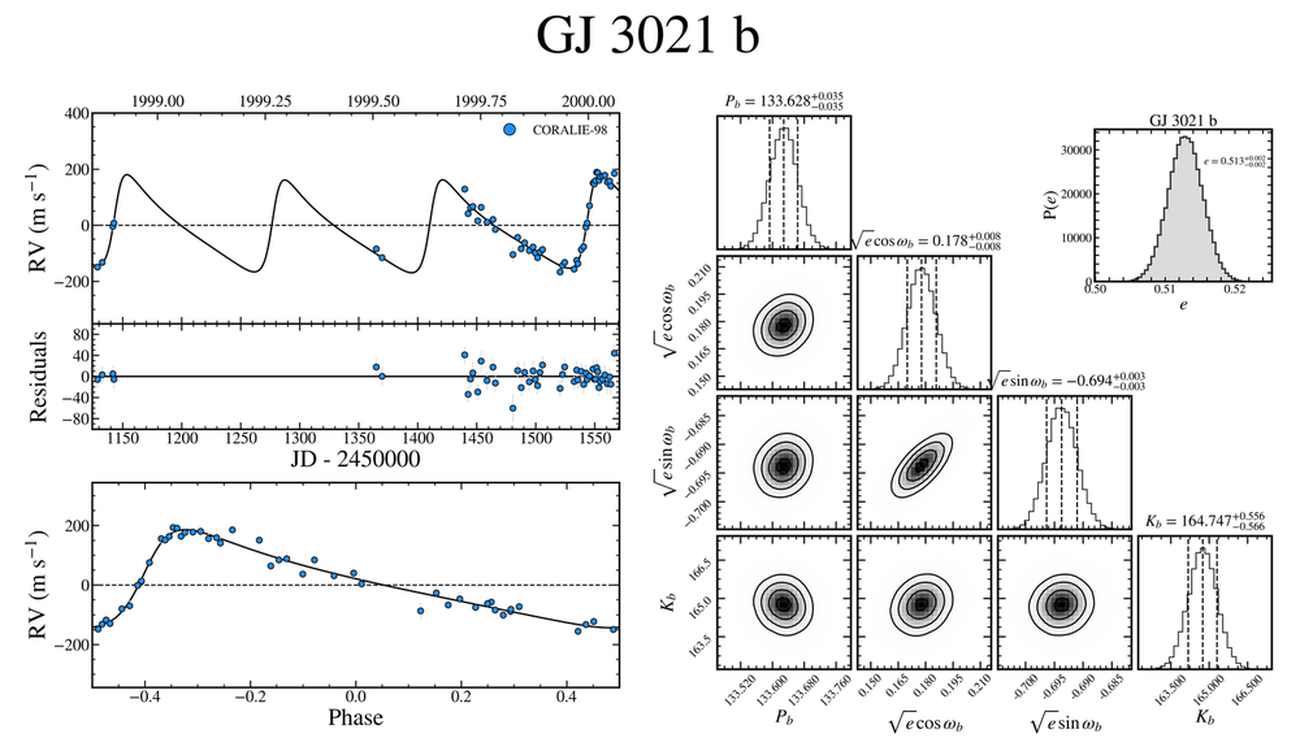}\\
   \vskip .3 in
   \includegraphics[width=\linewidth]{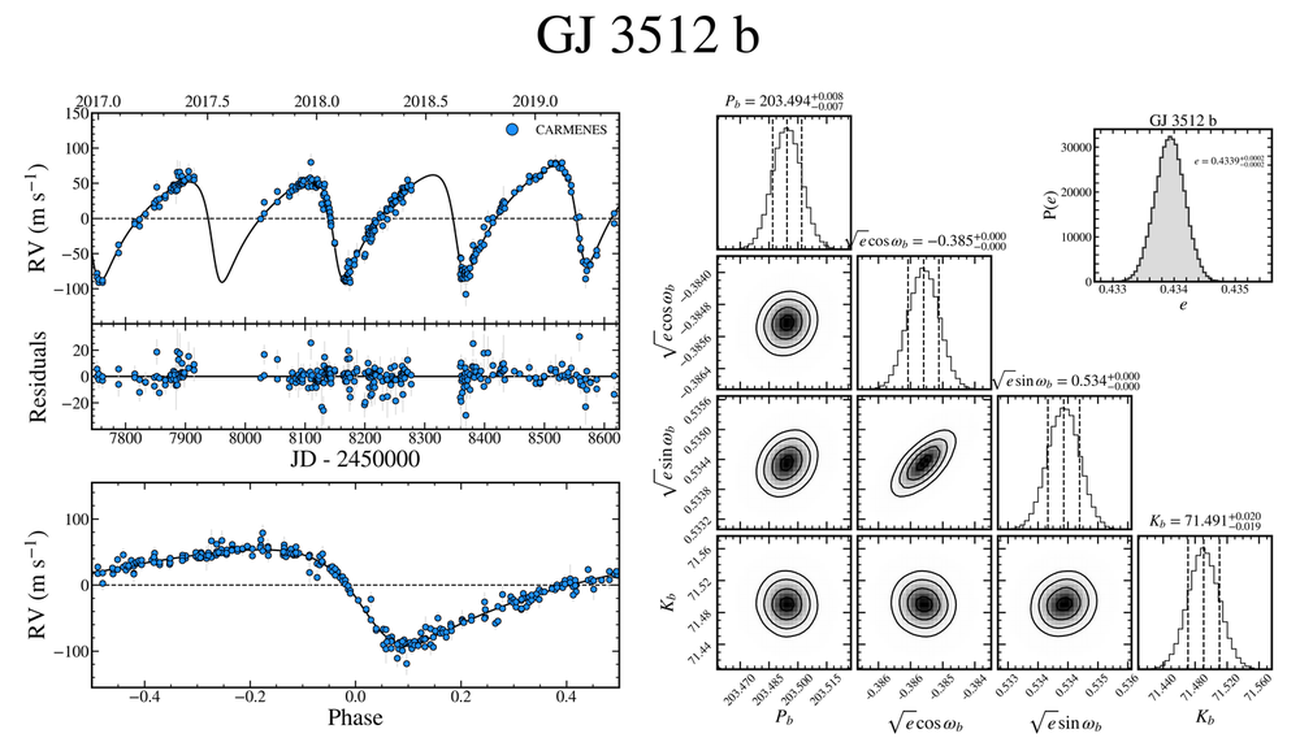}
 \end{minipage}
 \caption{Summary of results for the warm Jupiters GJ 3021 b and GJ 3512 b.}
 \label{fig:Combined_Plots9}
\end{figure}

\begin{figure}
\hskip -0.8 in
 \centering
 \begin{minipage}{\textwidth}
   \centering
   \includegraphics[width=\linewidth]{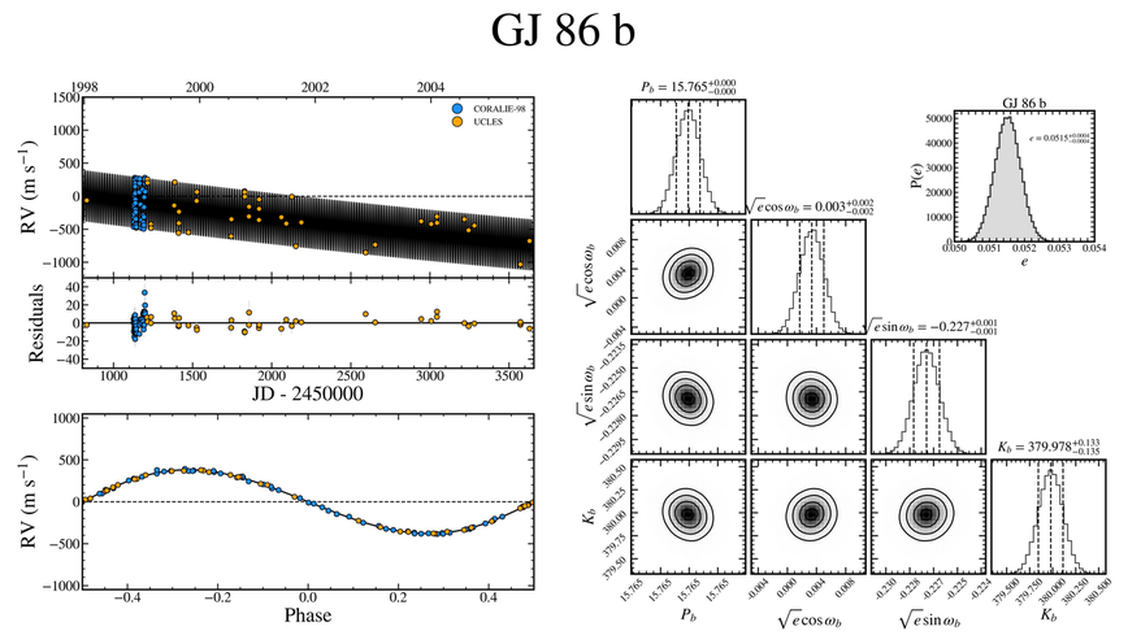}\\
   \vskip .3 in
   \includegraphics[width=\linewidth]{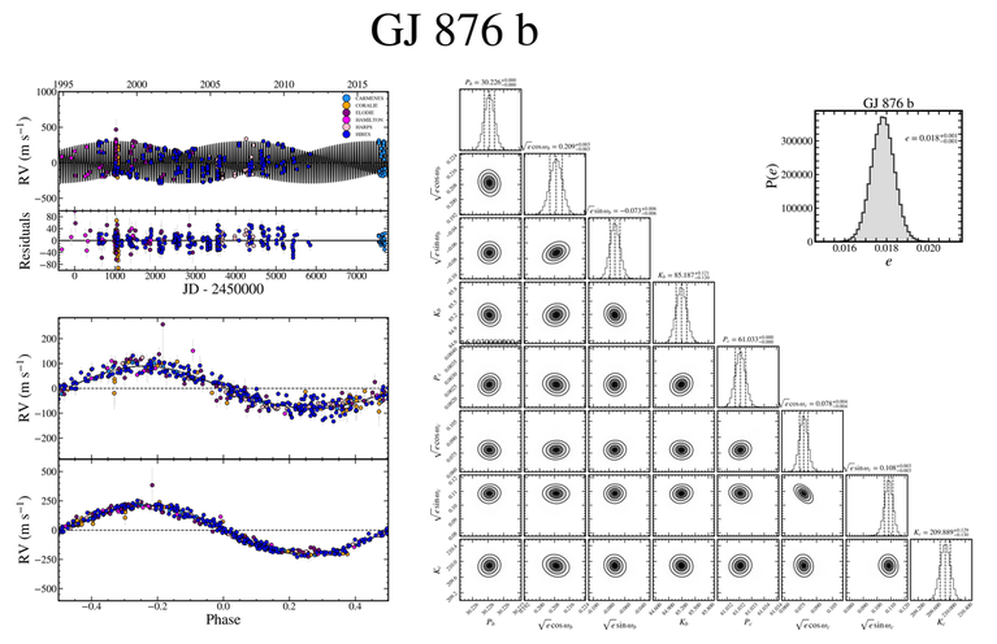}
 \end{minipage}
 \caption{Summary of results for the warm Jupiters GJ 86 b and GJ 876 b.}
 \label{fig:Combined_Plots10}
\end{figure}

\begin{figure}
\hskip -0.8 in
 \centering
 \begin{minipage}{\textwidth}
   \centering
   \includegraphics[width=\linewidth]{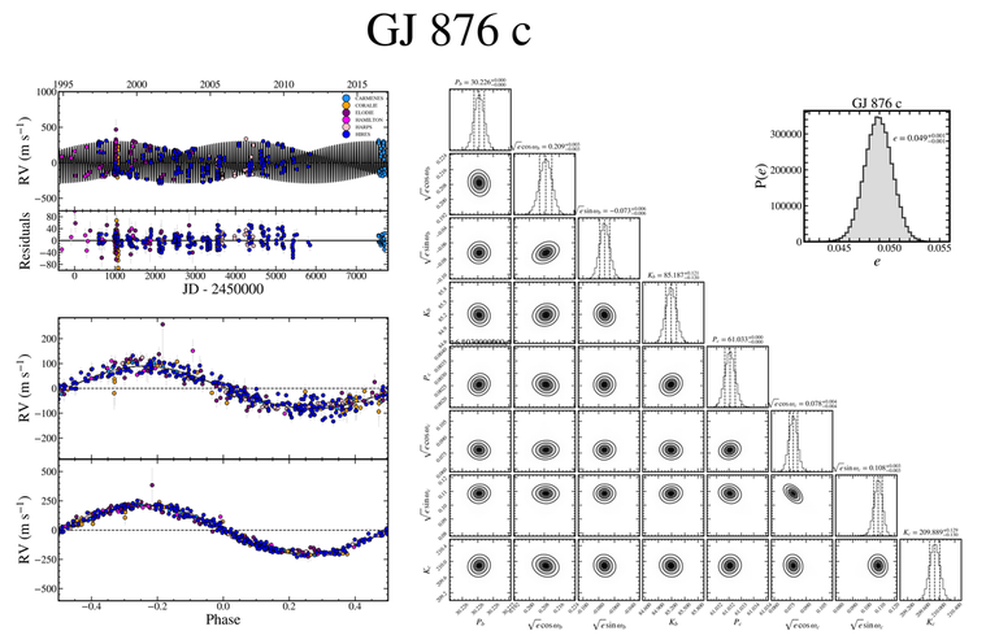}\\
   \vskip .3 in
   \includegraphics[width=\linewidth]{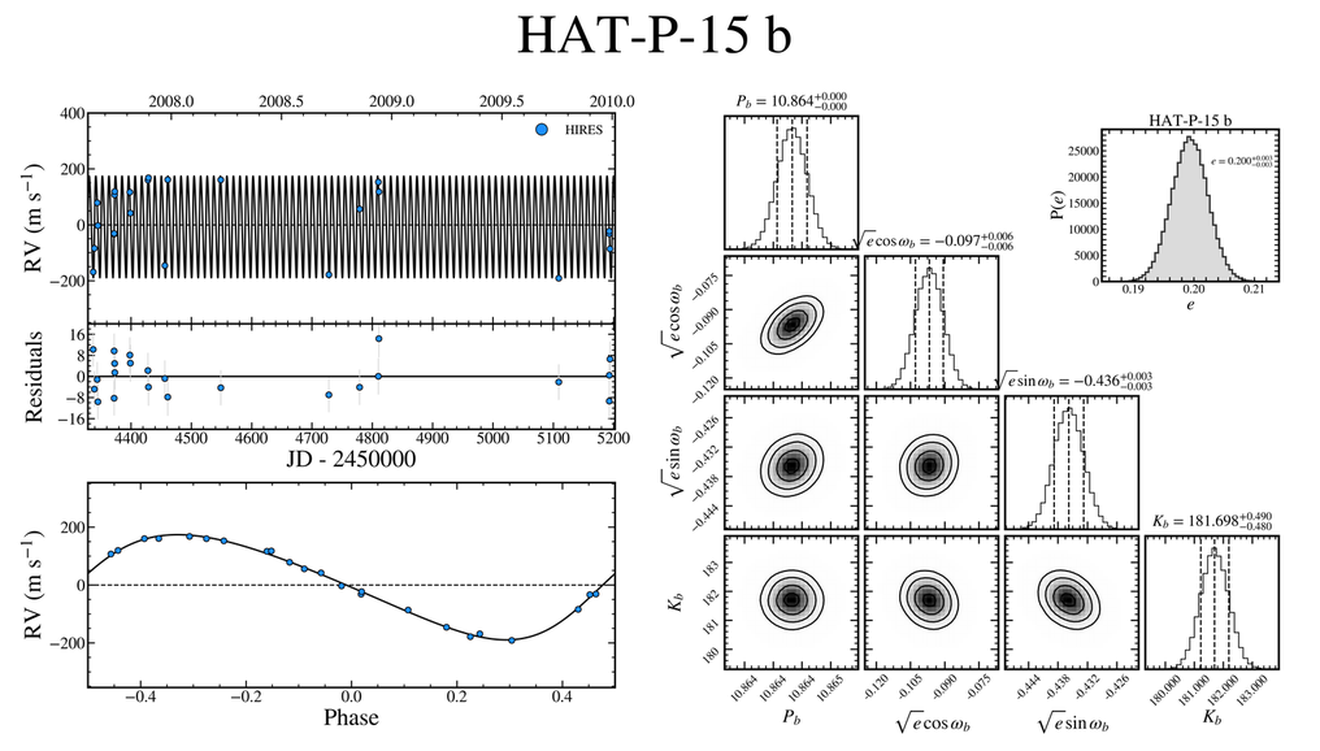}
 \end{minipage}
 \caption{Summary of results for the warm Jupiters GJ 876 c and HAT-P-15 b.}
 \label{fig:Combined_Plots11}
\end{figure}

\begin{figure}
\hskip -0.8 in
 \centering
 \begin{minipage}{\textwidth}
   \centering
   \includegraphics[width=\linewidth]{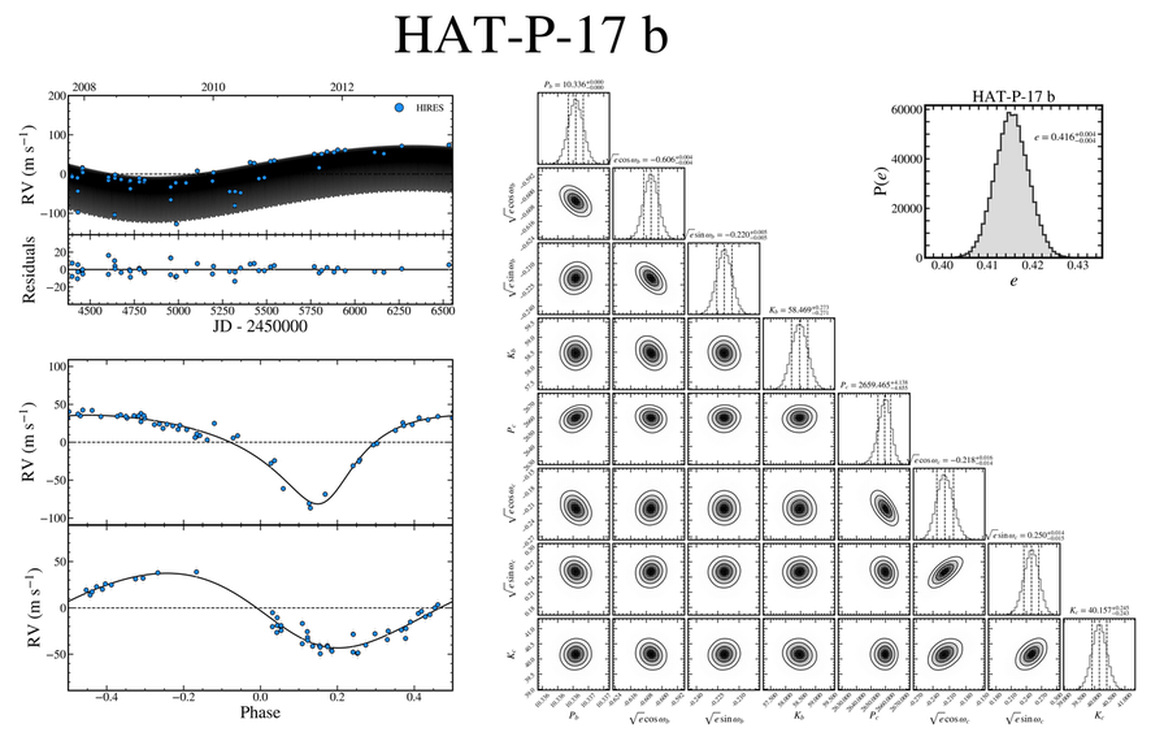}\\
   \vskip .3 in
   \includegraphics[width=\linewidth]{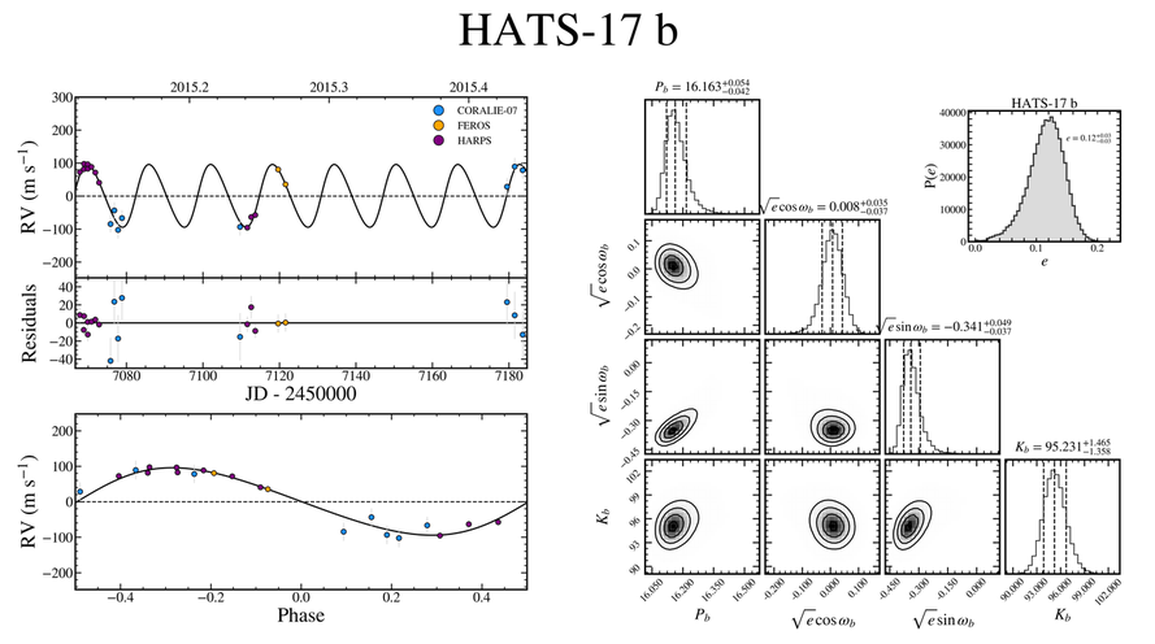}
 \end{minipage}
 \caption{Summary of results for the warm Jupiters HAT-P-17 b and HATS-17 b.}
 \label{fig:Combined_Plots12}
\end{figure}

\begin{figure}
\hskip -0.8 in
 \centering
 \begin{minipage}{\textwidth}
   \centering
   \includegraphics[width=\linewidth]{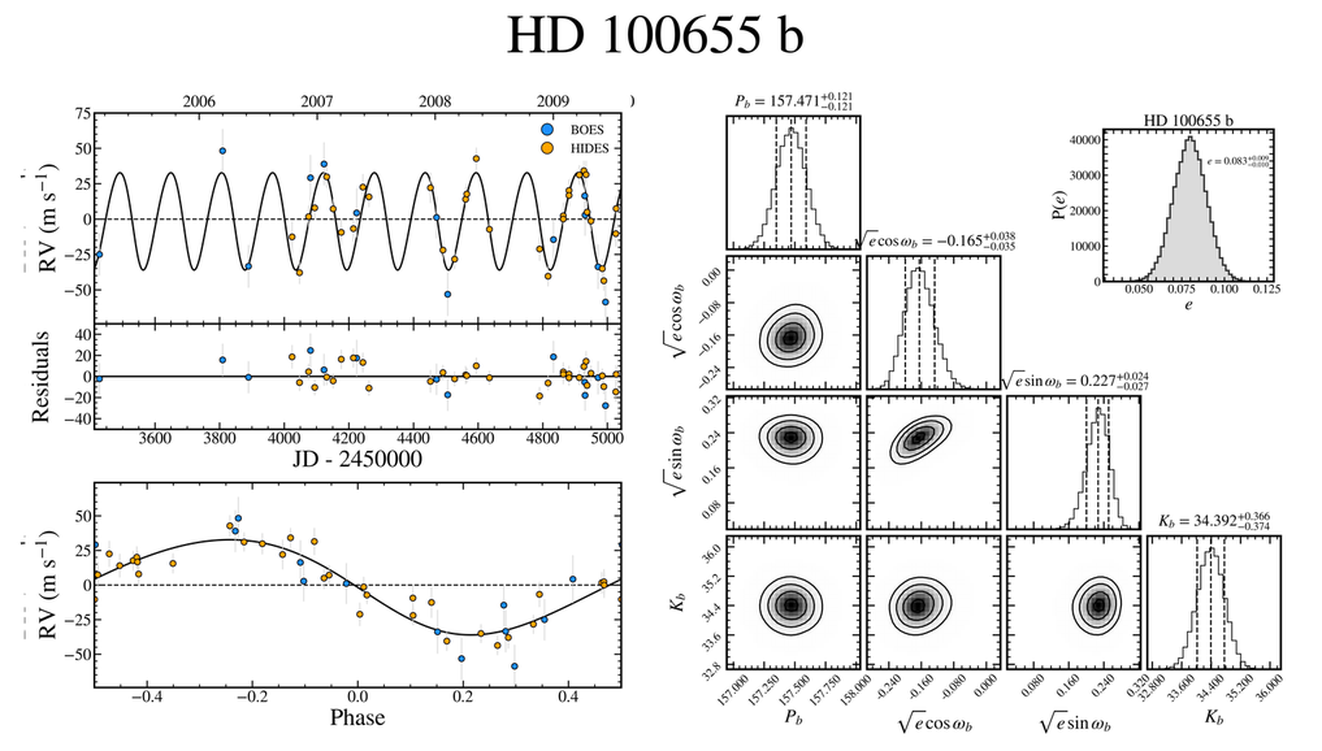}\\
   \vskip .3 in
   \includegraphics[width=\linewidth]{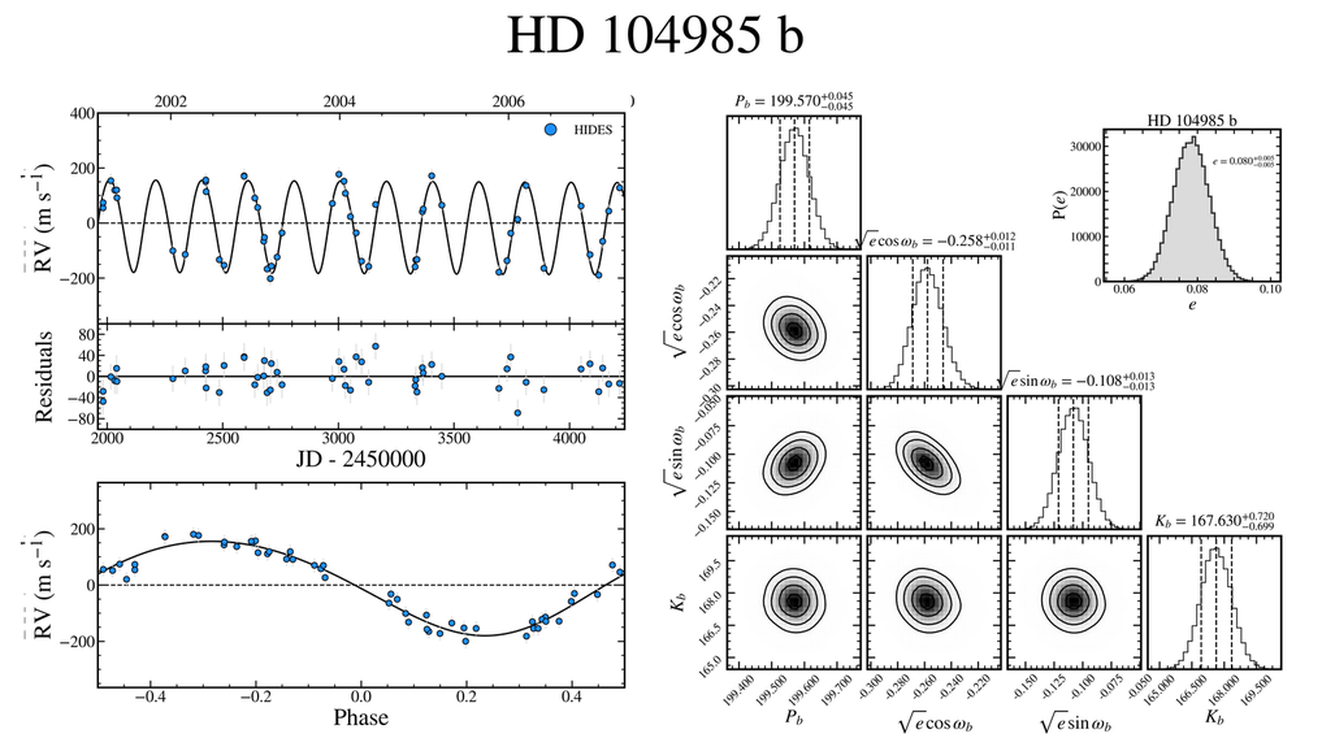}
 \end{minipage}
 \caption{Summary of results for the warm Jupiters HD 100655 b and HD 104985 b.}
 \label{fig:Combined_Plots13}
\end{figure}

\begin{figure}
\hskip -0.8 in
 \centering
 \begin{minipage}{\textwidth}
   \centering
   \includegraphics[width=\linewidth]{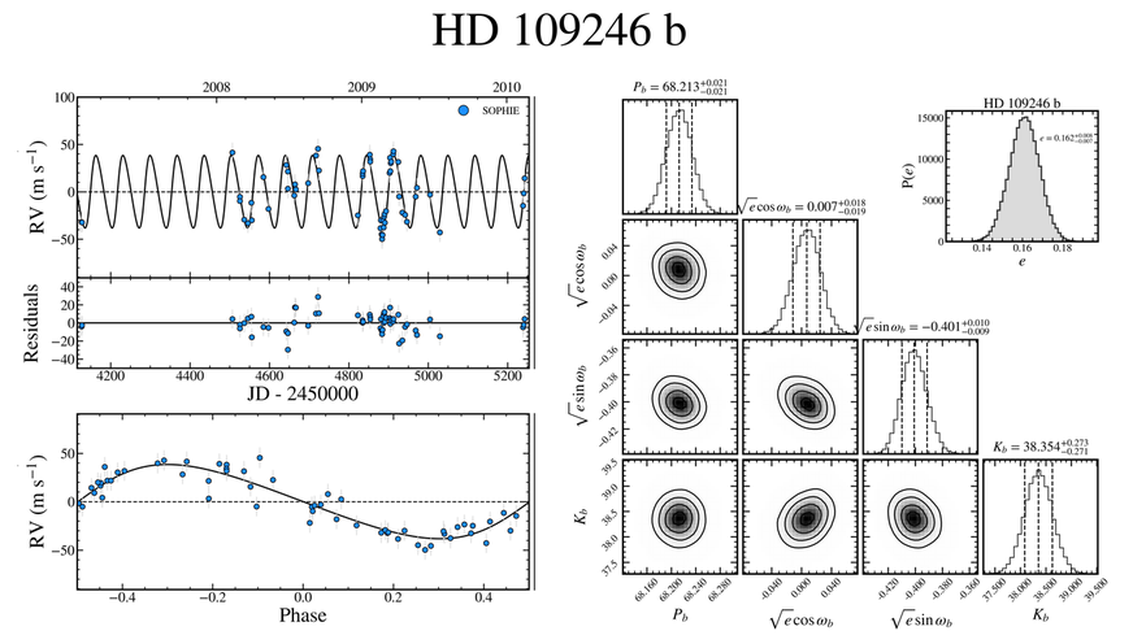}\\
   \vskip .3 in
   \includegraphics[width=\linewidth]{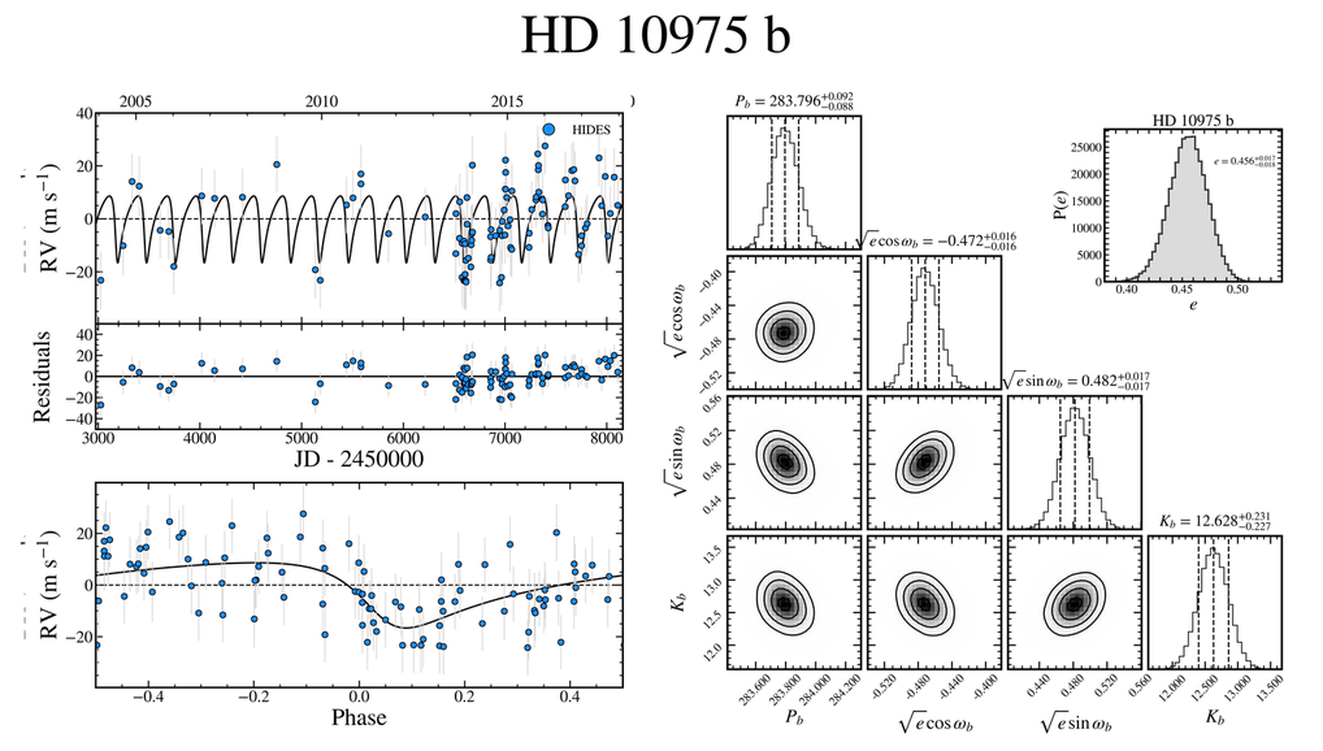}
 \end{minipage}
 \caption{Summary of results for the warm Jupiters HD 109246 b and HD 10975 b.}
 \label{fig:Combined_Plots14}
\end{figure}

\begin{figure}
\hskip -0.8 in
 \centering
 \begin{minipage}{\textwidth}
   \centering
   \includegraphics[width=\linewidth]{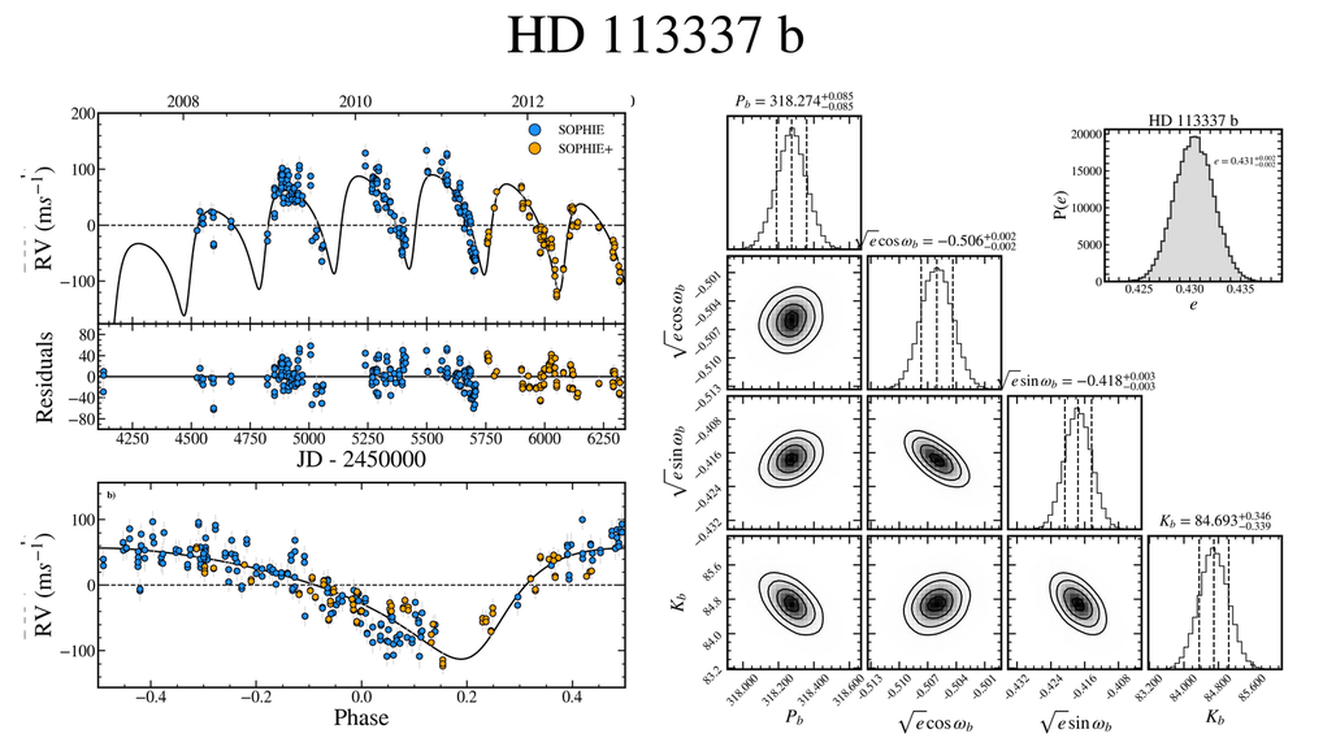}\\
   \vskip .3 in
   \includegraphics[width=\linewidth]{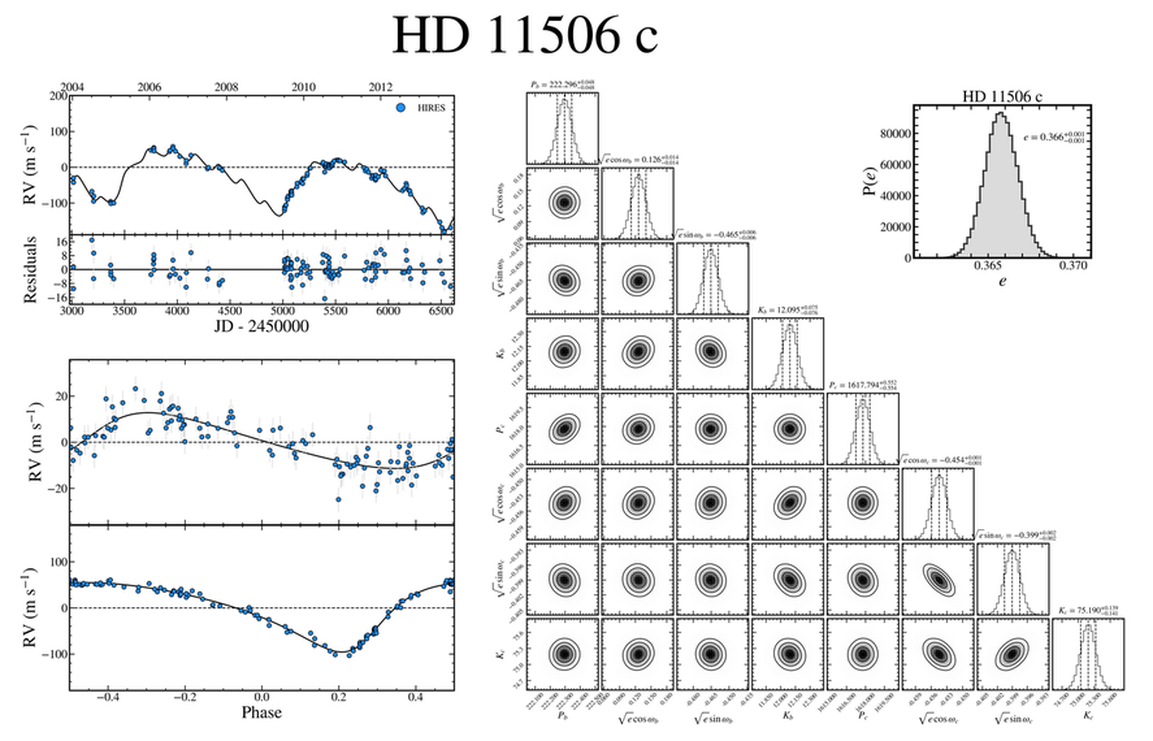}
 \end{minipage}
 \caption{Summary of results for the warm Jupiters HD 113337 b and HD 11506 c.}
 \label{fig:Combined_Plots15}
\end{figure}

\begin{figure}
\hskip -0.8 in
 \centering
 \begin{minipage}{\textwidth}
   \centering
   \includegraphics[width=\linewidth]{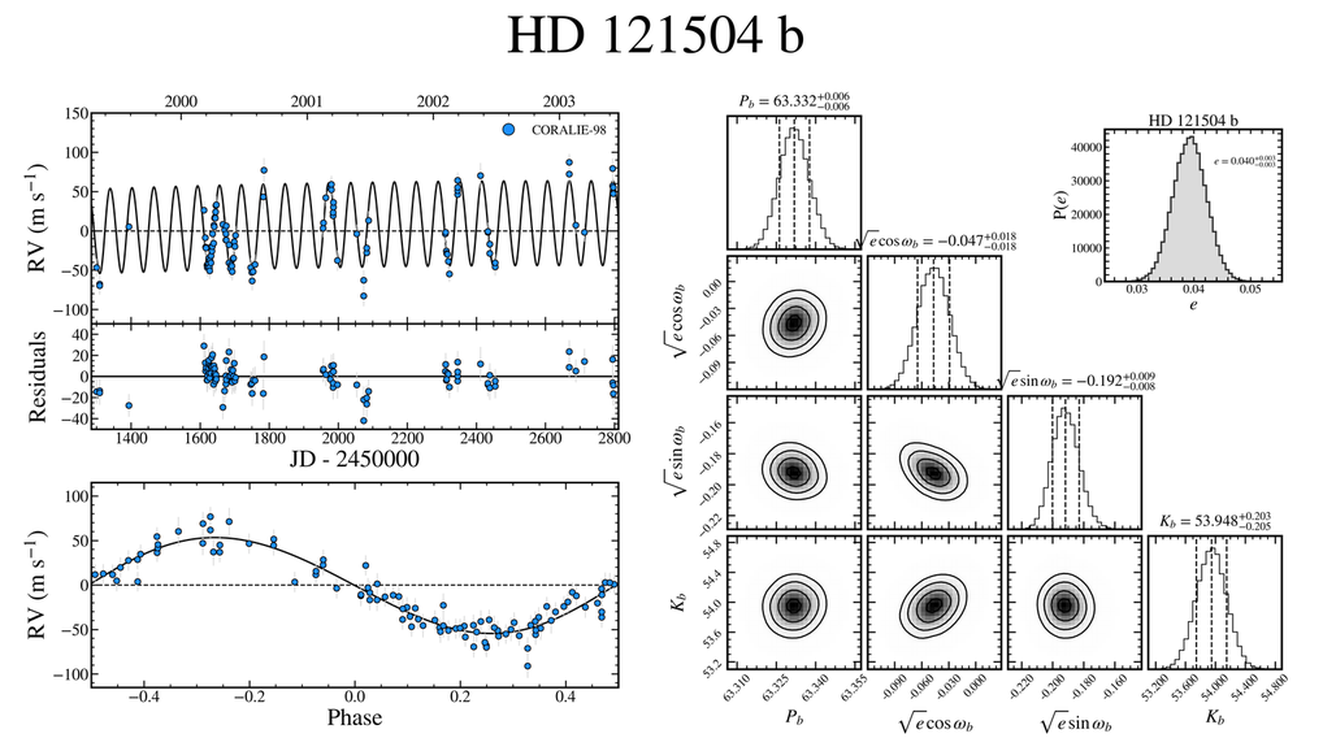}\\
   \vskip .3 in
   \includegraphics[width=\linewidth]{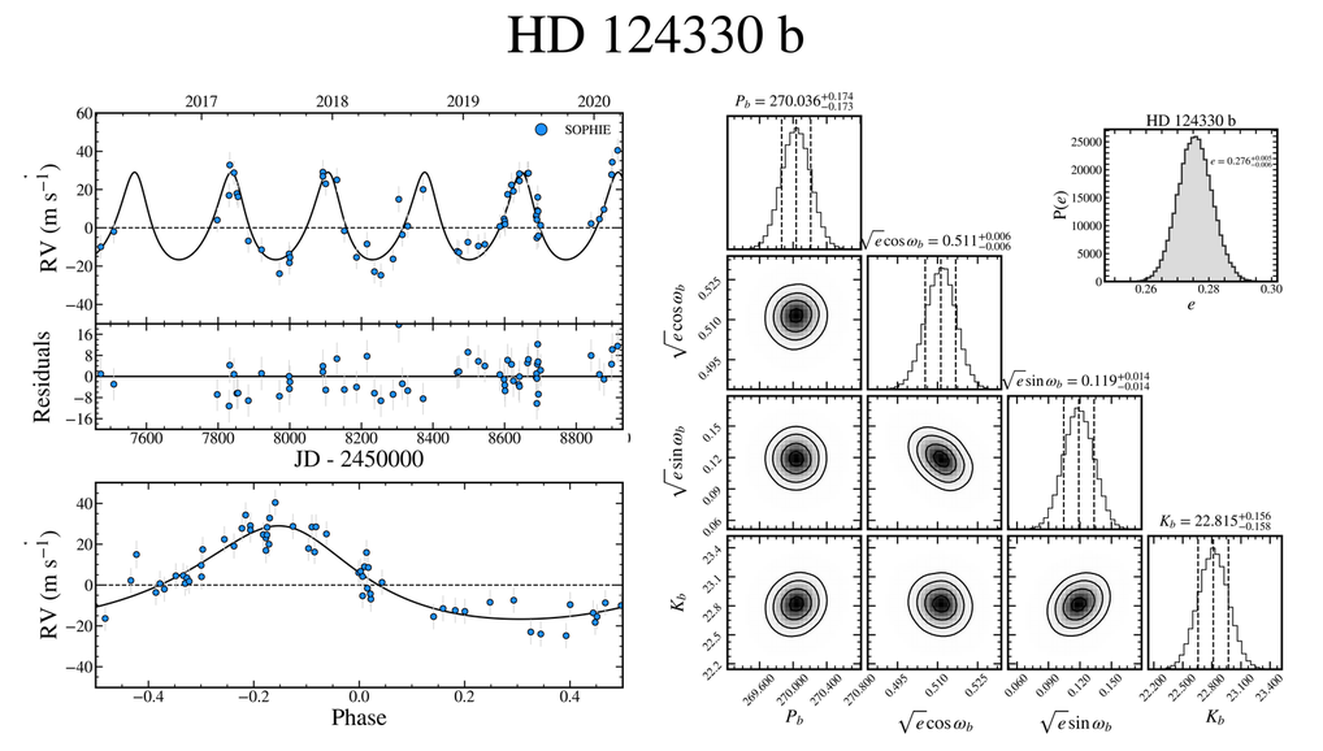}
 \end{minipage}
 \caption{Summary of results for the warm Jupiters HD 121504 b and HD 124330 b.}
 \label{fig:Combined_Plots16}
\end{figure}

\begin{figure}
\hskip -0.8 in
 \centering
 \begin{minipage}{\textwidth}
   \centering
   \includegraphics[width=\linewidth]{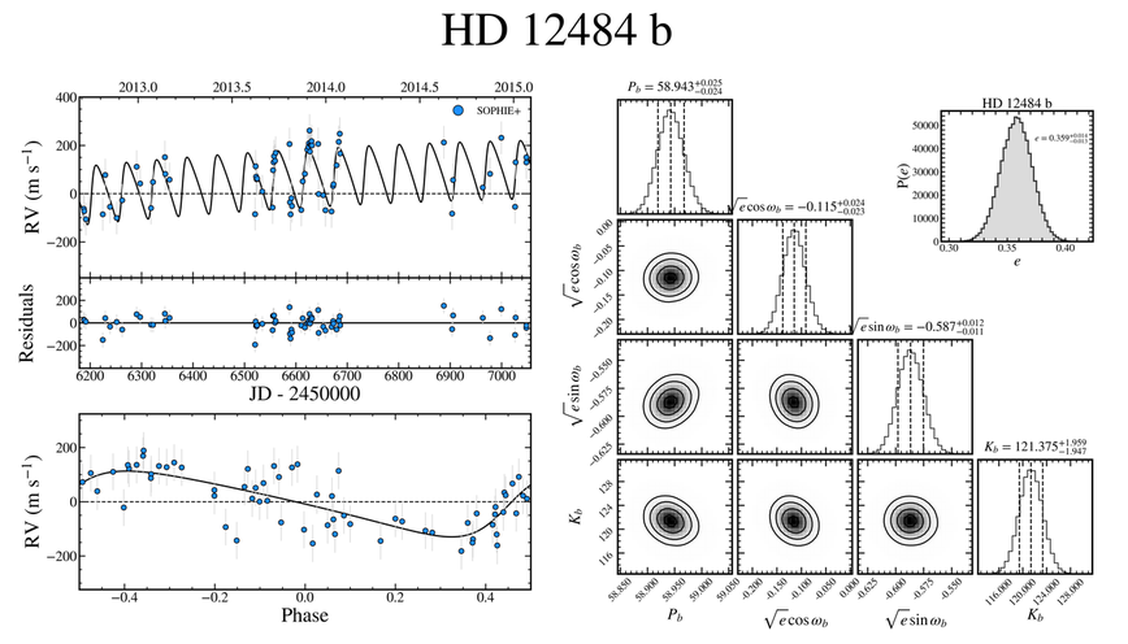}\\
   \vskip .3 in
   \includegraphics[width=\linewidth]{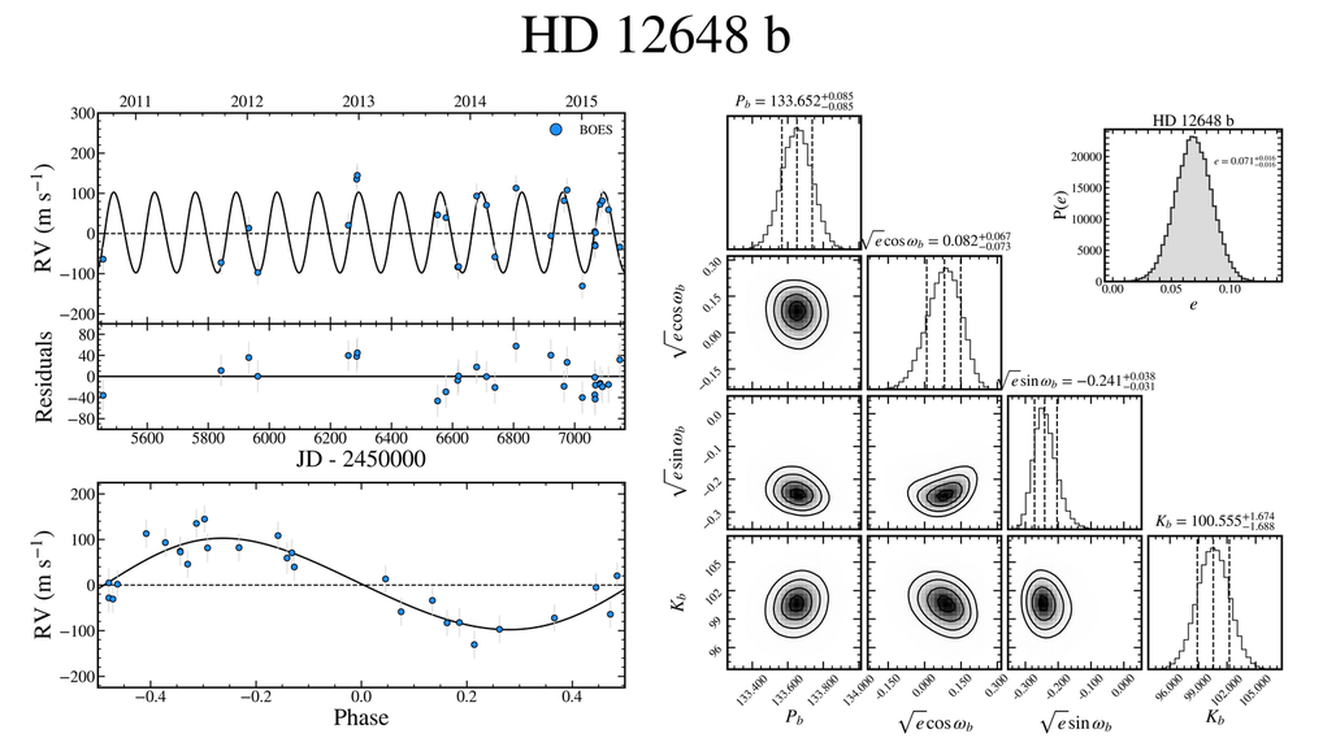}
 \end{minipage}
 \caption{Summary of results for the warm Jupiters HD 12484 b and HD 12648 b.}
 \label{fig:Combined_Plots17}
\end{figure}

\begin{figure}
\hskip -0.8 in
 \centering
 \begin{minipage}{\textwidth}
   \centering
   \includegraphics[width=\linewidth]{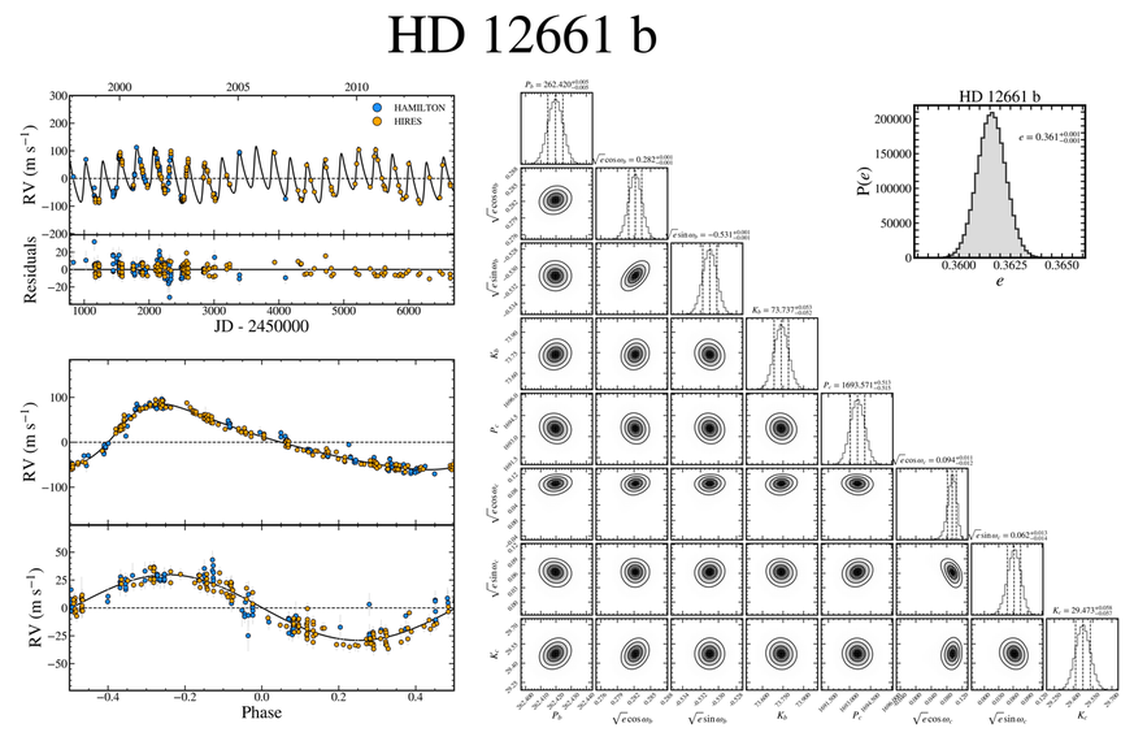}\\
   \vskip .3 in
   \includegraphics[width=\linewidth]{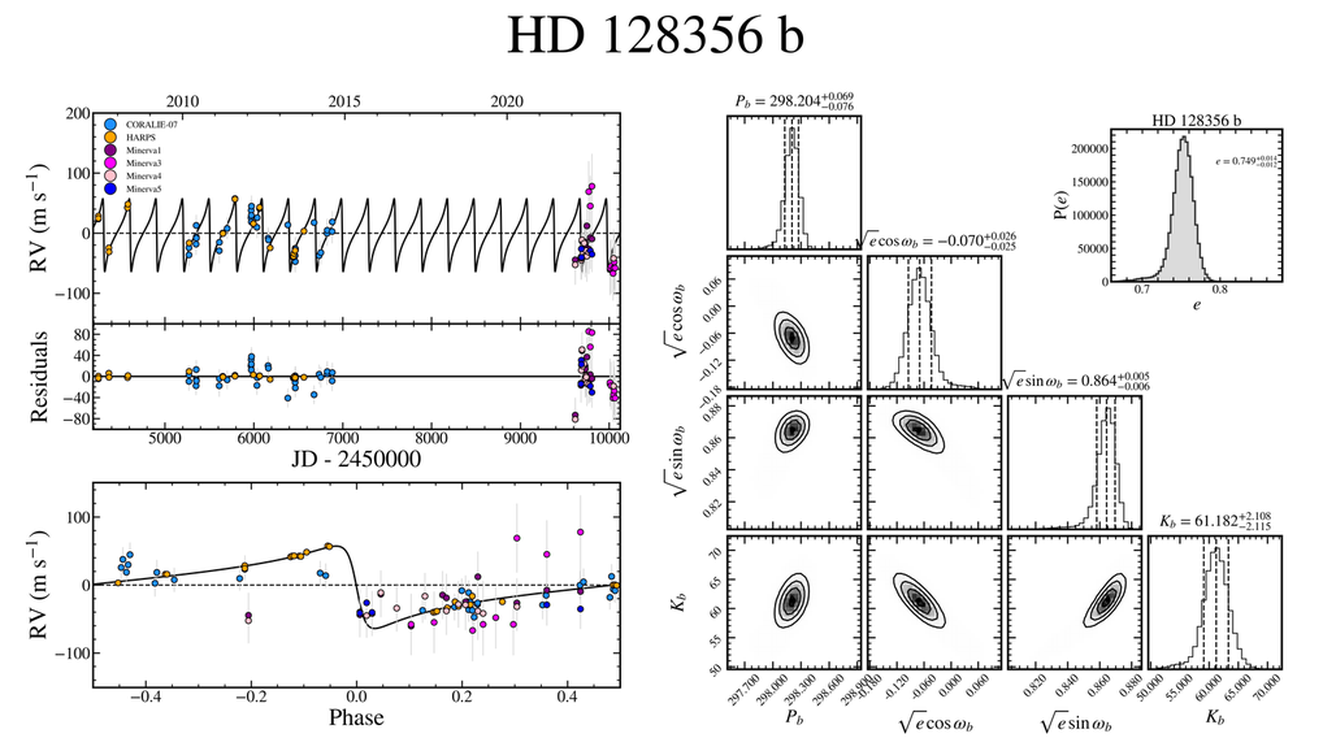}
 \end{minipage}
 \caption{Summary of results for the warm Jupiters HD 12661 b and HD 128356 b.}
 \label{fig:Combined_Plots18}
\end{figure}

\begin{figure}
\hskip -0.8 in
 \centering
 \begin{minipage}{\textwidth}
   \centering
   \includegraphics[width=\linewidth]{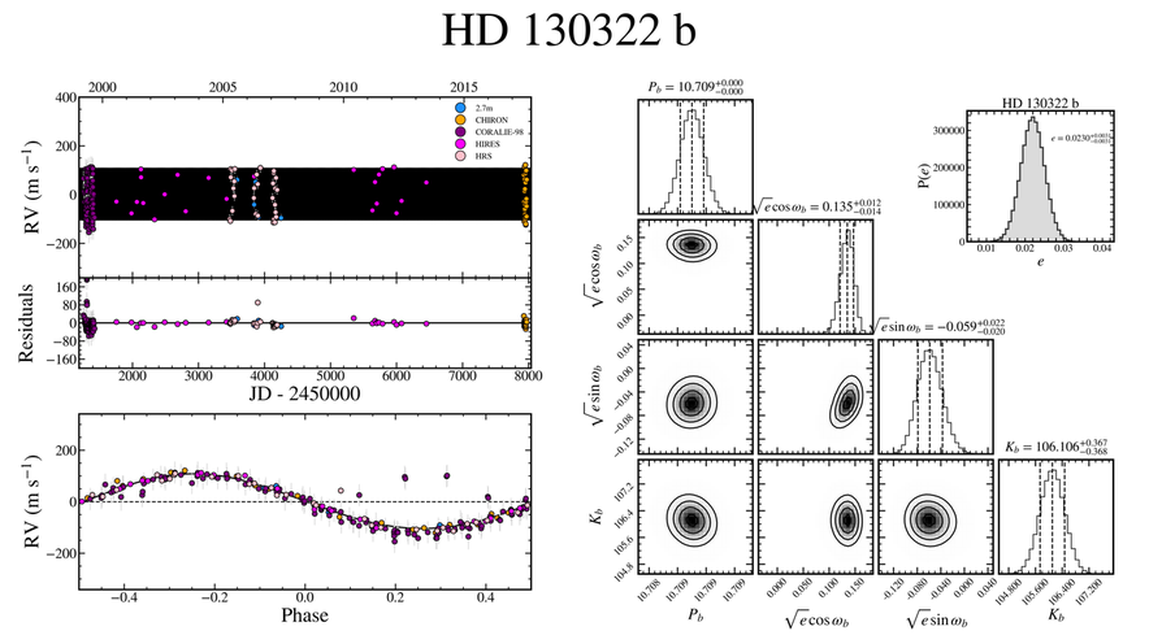}\\
   \vskip .3 in
   \includegraphics[width=\linewidth]{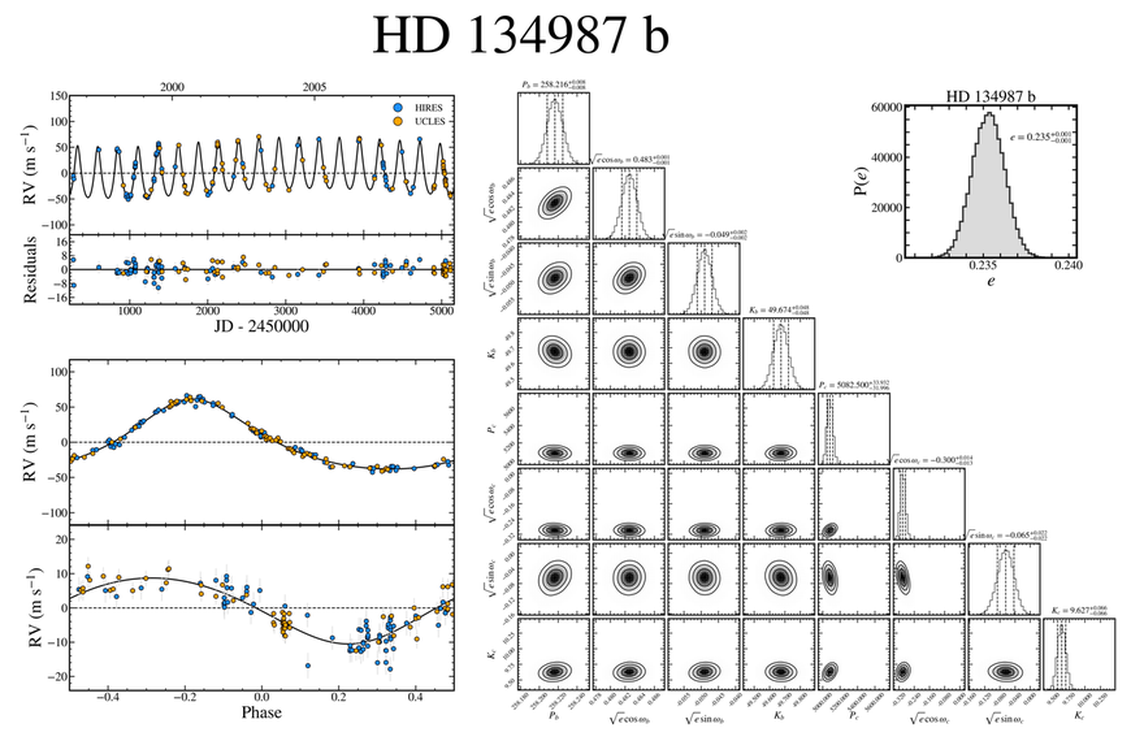}
 \end{minipage}
 \caption{Summary of results for the warm Jupiters HD 130322 b and HD 134987 b.}
 \label{fig:Combined_Plots19}
\end{figure}

\begin{figure}
\hskip -0.8 in
 \centering
 \begin{minipage}{\textwidth}
   \centering
   \includegraphics[width=\linewidth]{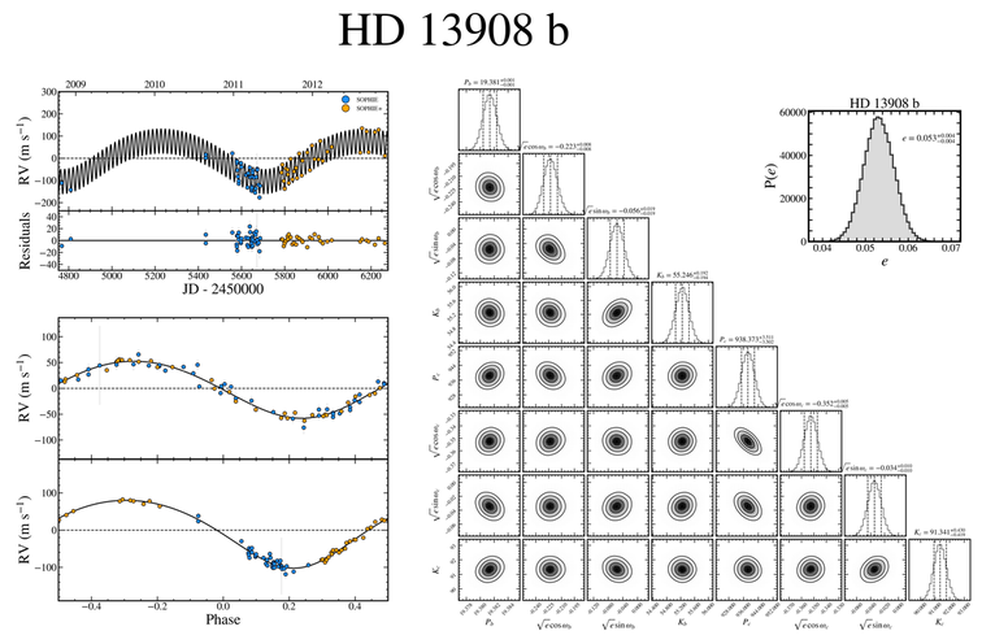}\\
   \vskip .3 in
   \includegraphics[width=\linewidth]{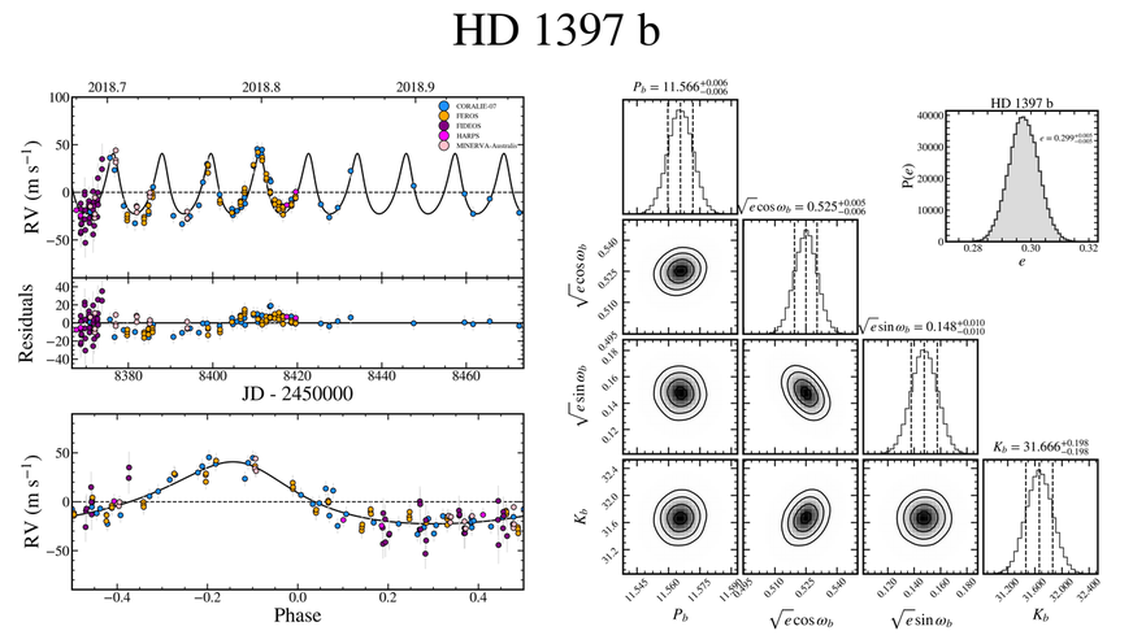}
 \end{minipage}
 \caption{Summary of results for the warm Jupiters HD 13908 b and HD 1397 b.}
 \label{fig:Combined_Plots20}
\end{figure}

\begin{figure}
\hskip -0.8 in
 \centering
 \begin{minipage}{\textwidth}
   \centering
   \includegraphics[width=0.9\linewidth]{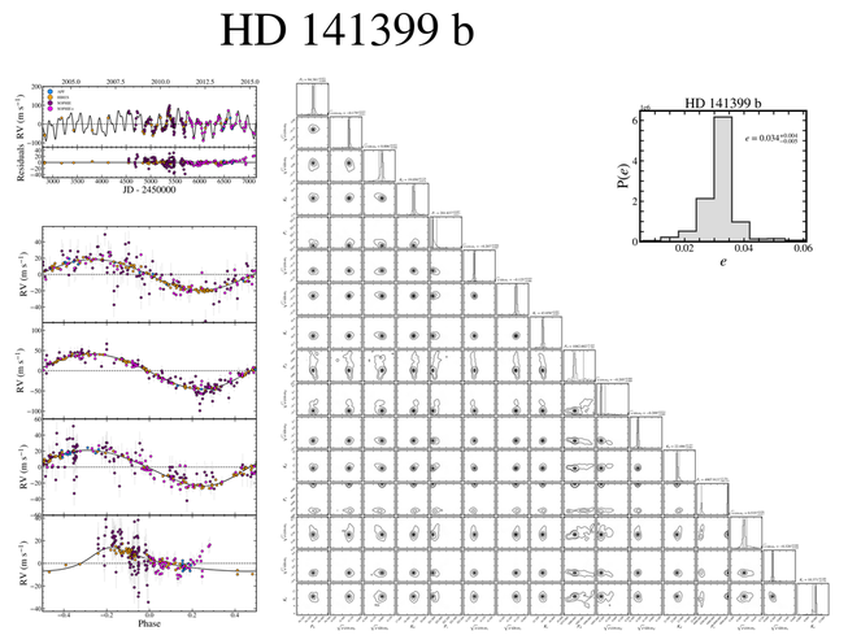}\\
   \vskip .1 in
   \includegraphics[width=0.9\linewidth]{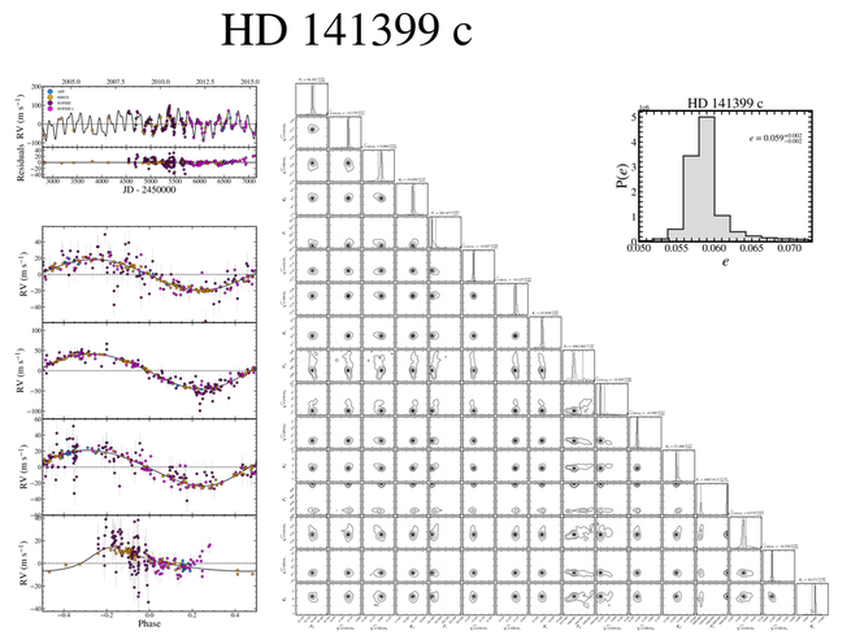}
 \end{minipage}
 \caption{Summary of results for the warm Jupiters HD 141399 b and HD 141399 c.}
 \label{fig:Combined_Plots21}
\end{figure}

\begin{figure}
\hskip -0.8 in
 \centering
 \begin{minipage}{\textwidth}
   \centering
   \includegraphics[width=\linewidth]{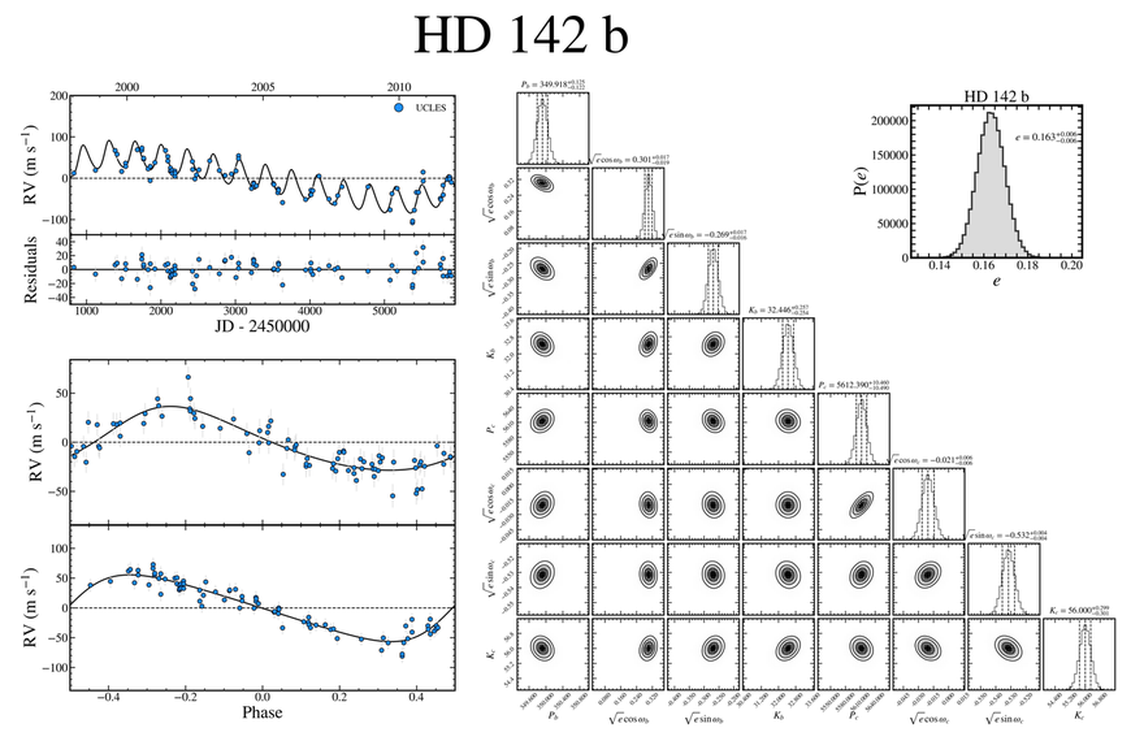}\\
   \vskip .3 in
   \includegraphics[width=\linewidth]{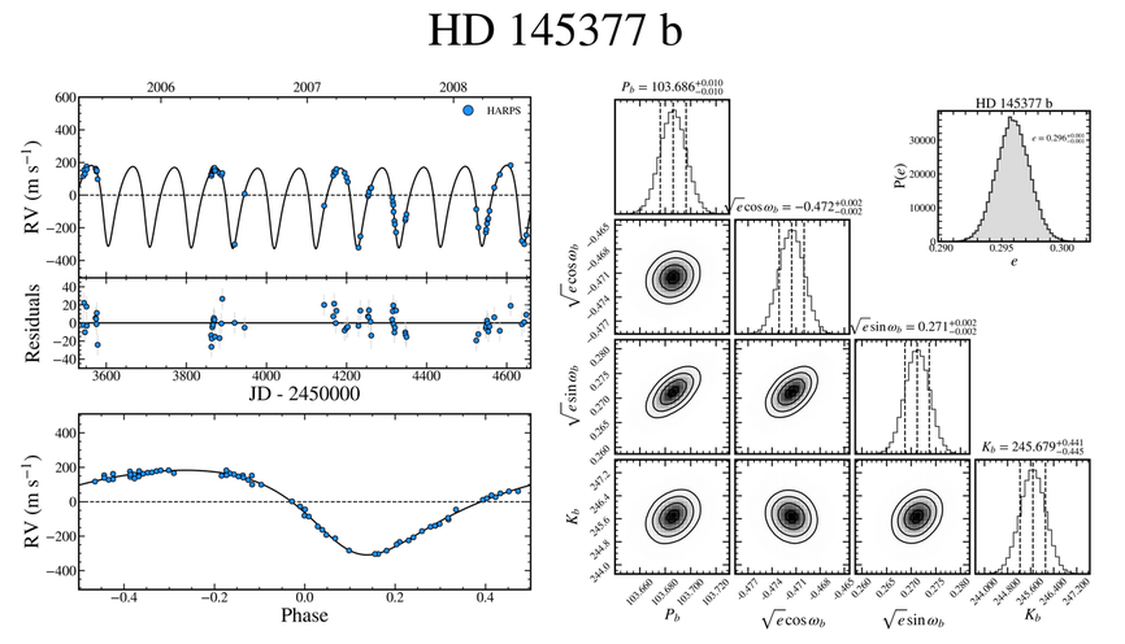}
 \end{minipage}
 \caption{Summary of results for the warm Jupiters HD 142 b and HD 145377 b.}
 \label{fig:Combined_Plots22}
\end{figure}

\begin{figure}
\hskip -0.8 in
 \centering
 \begin{minipage}{\textwidth}
   \centering
   \includegraphics[width=\linewidth]{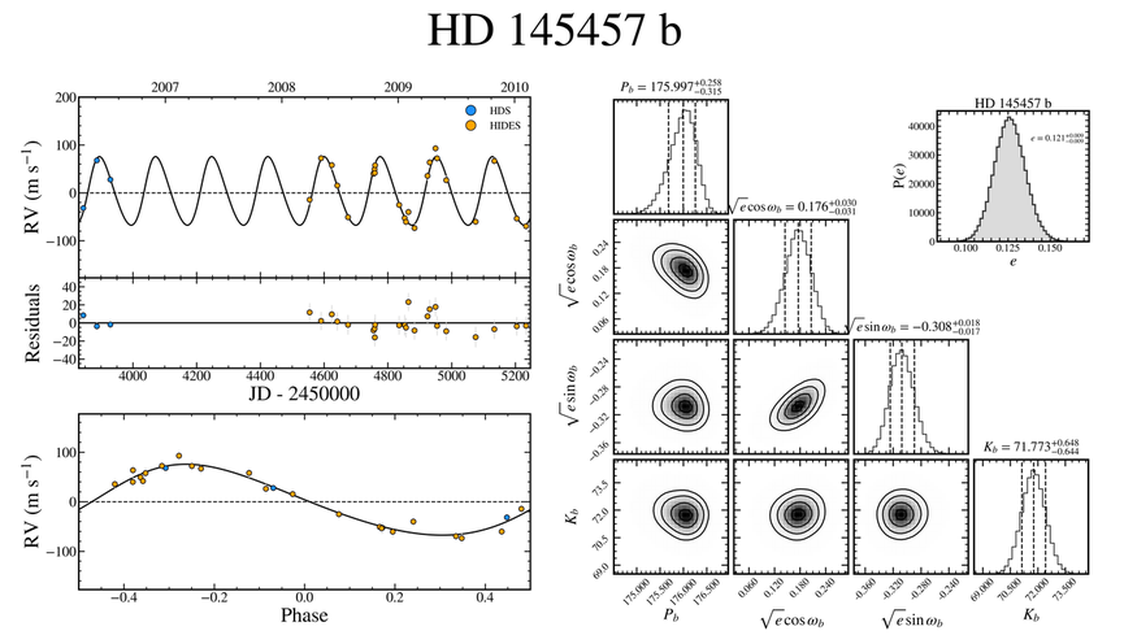}\\
   \vskip .3 in
   \includegraphics[width=\linewidth]{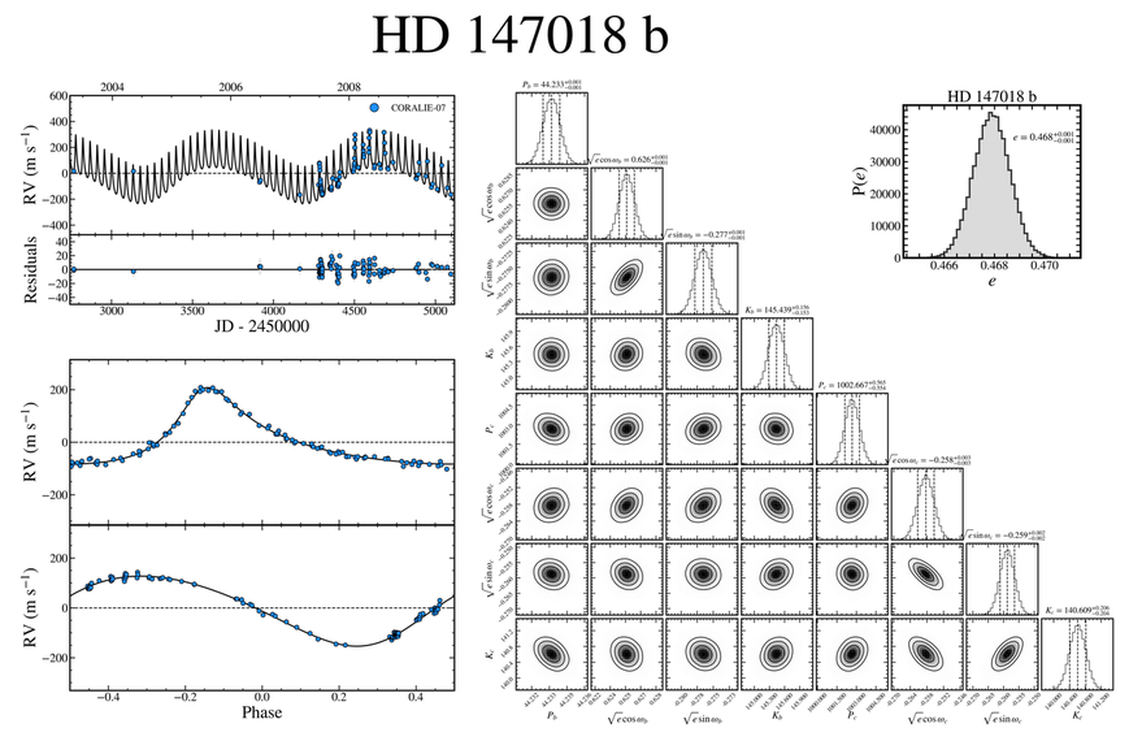}
 \end{minipage}
 \caption{Summary of results for the warm Jupiters HD 145457 b and HD 147018 b.}
 \label{fig:Combined_Plots23}
\end{figure}

\begin{figure}
\hskip -0.8 in
 \centering
 \begin{minipage}{\textwidth}
   \centering
   \includegraphics[width=\linewidth]{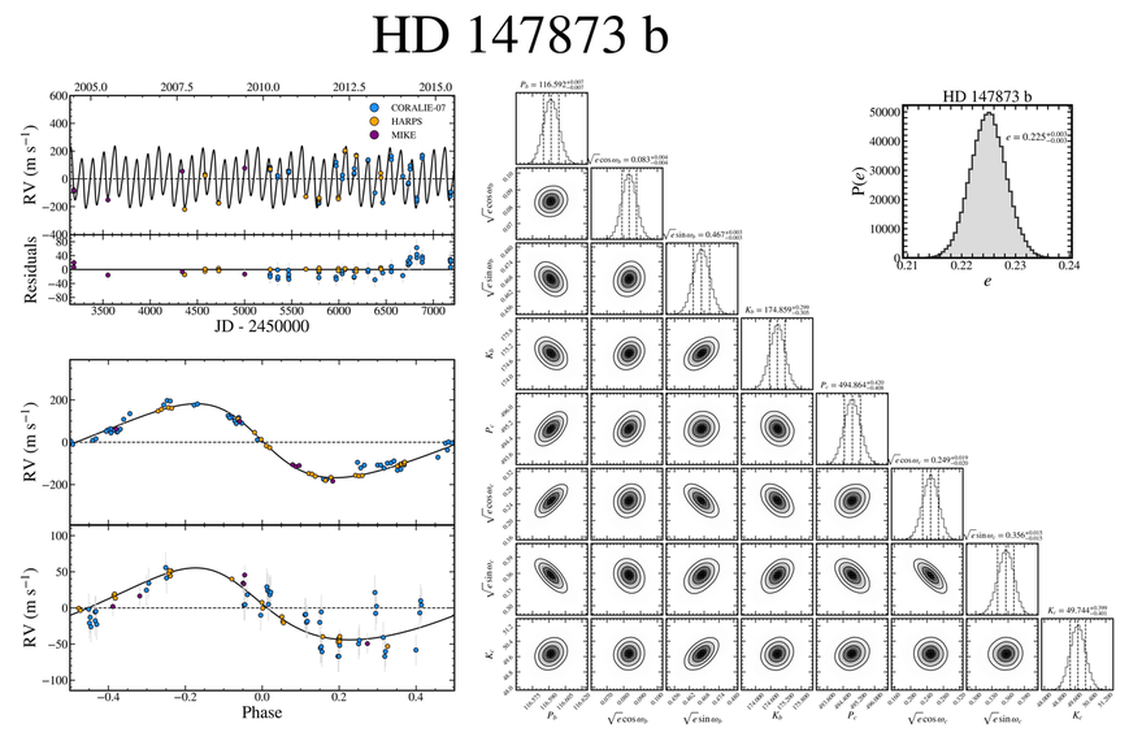}\\
   \vskip .3 in
   \includegraphics[width=\linewidth]{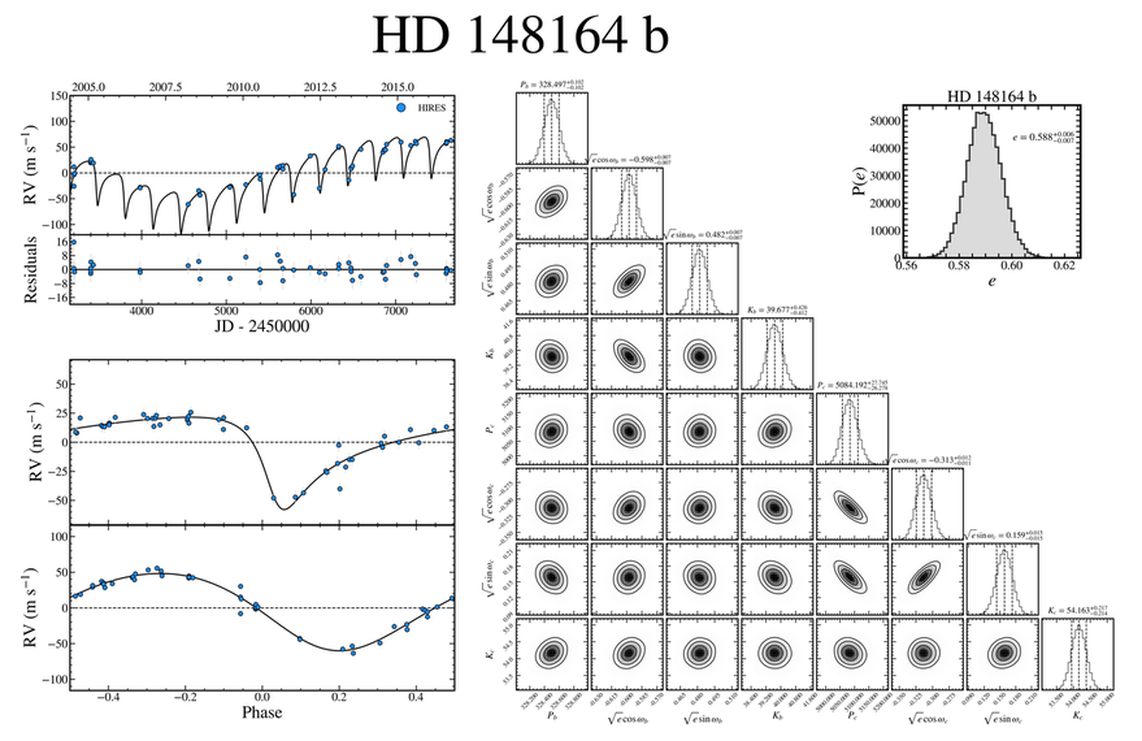}
 \end{minipage}
 \caption{Summary of results for the warm Jupiters HD 147873 b and HD 148164 b.}
 \label{fig:Combined_Plots24}
\end{figure}

\begin{figure}
\hskip -0.8 in
 \centering
 \begin{minipage}{\textwidth}
   \centering
   \includegraphics[width=\linewidth]{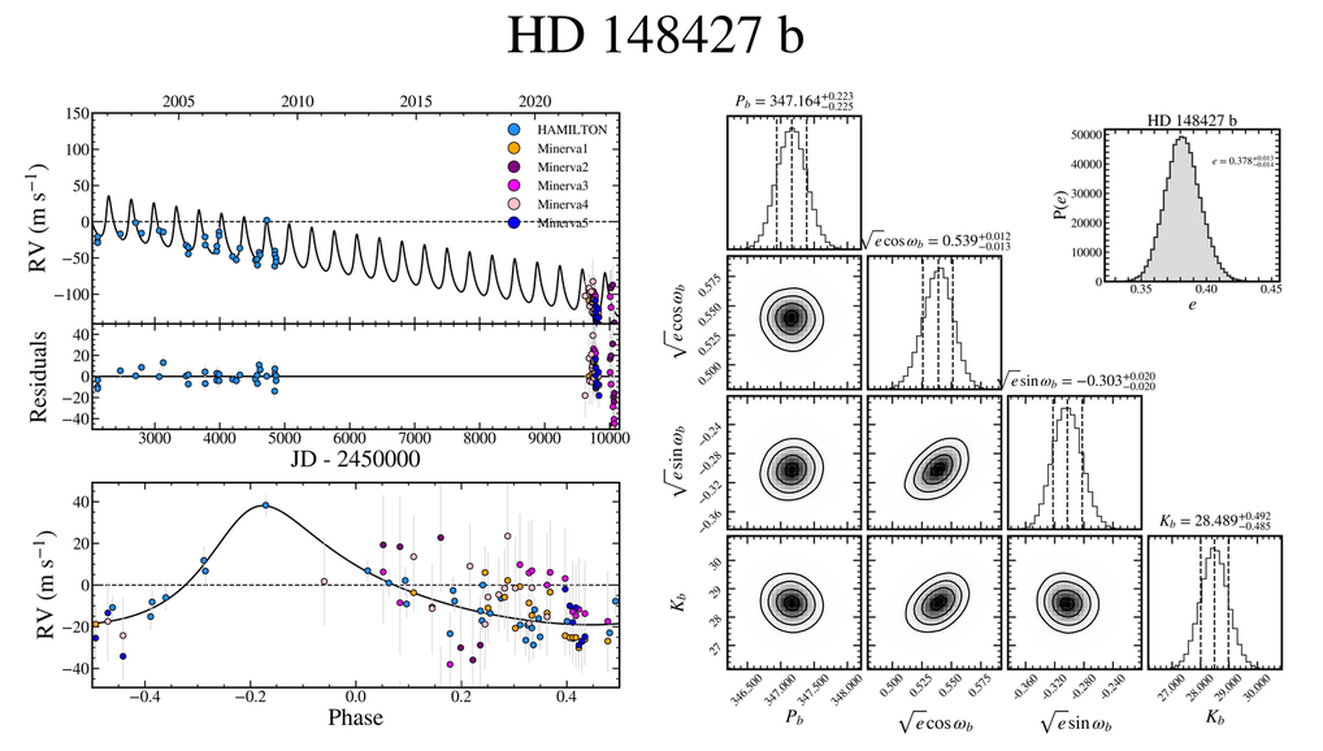}\\
   \vskip .3 in
   \includegraphics[width=\linewidth]{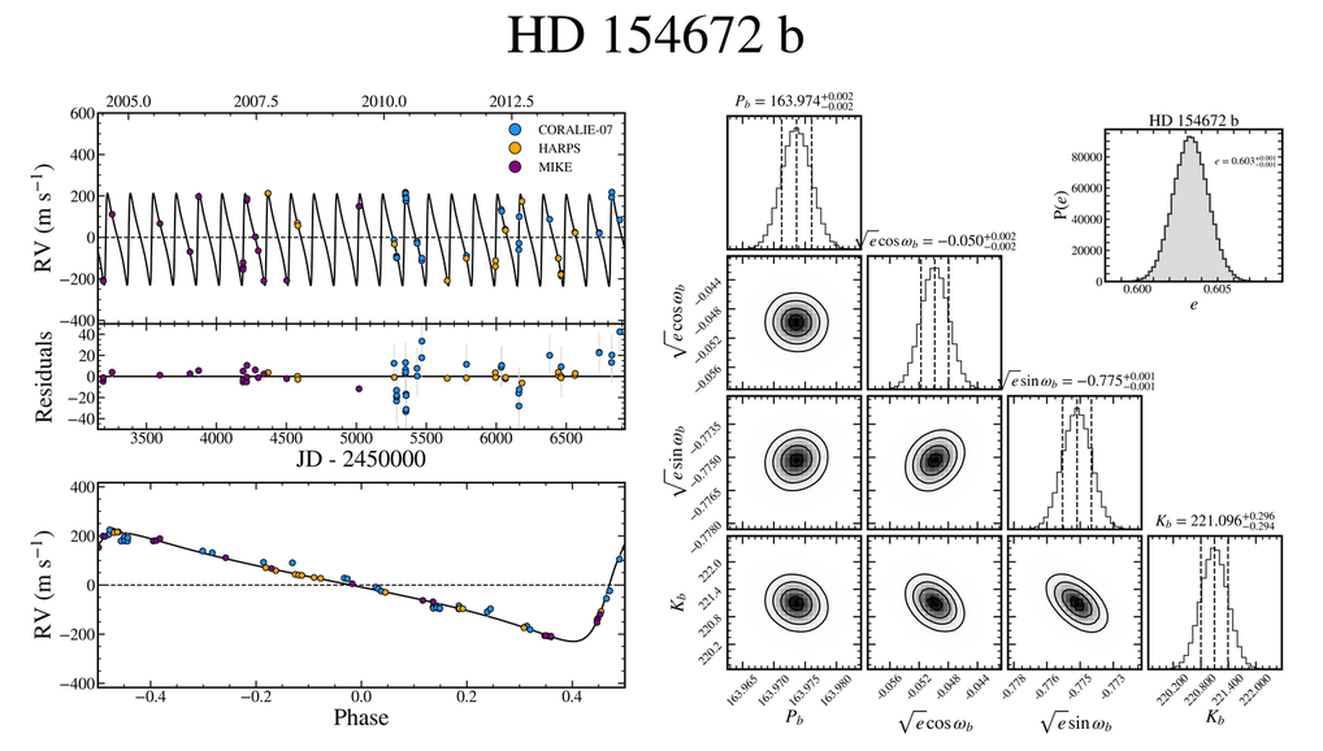}
 \end{minipage}
 \caption{Summary of results for the warm Jupiters HD 148427 b and HD 154672 b.}
 \label{fig:Combined_Plots25}
\end{figure}

\begin{figure}
\hskip -0.8 in
 \centering
 \begin{minipage}{\textwidth}
   \centering
   \includegraphics[width=\linewidth]{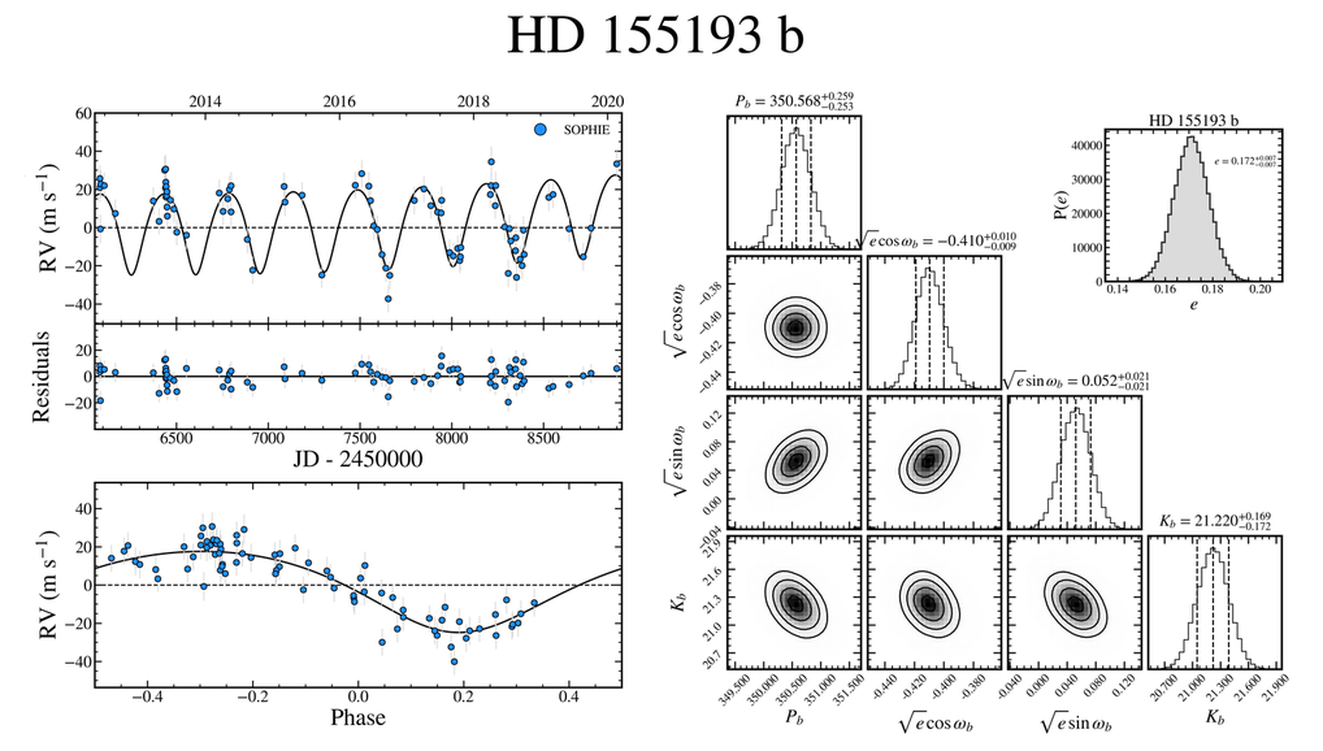}\\
   \vskip .3 in
   \includegraphics[width=\linewidth]{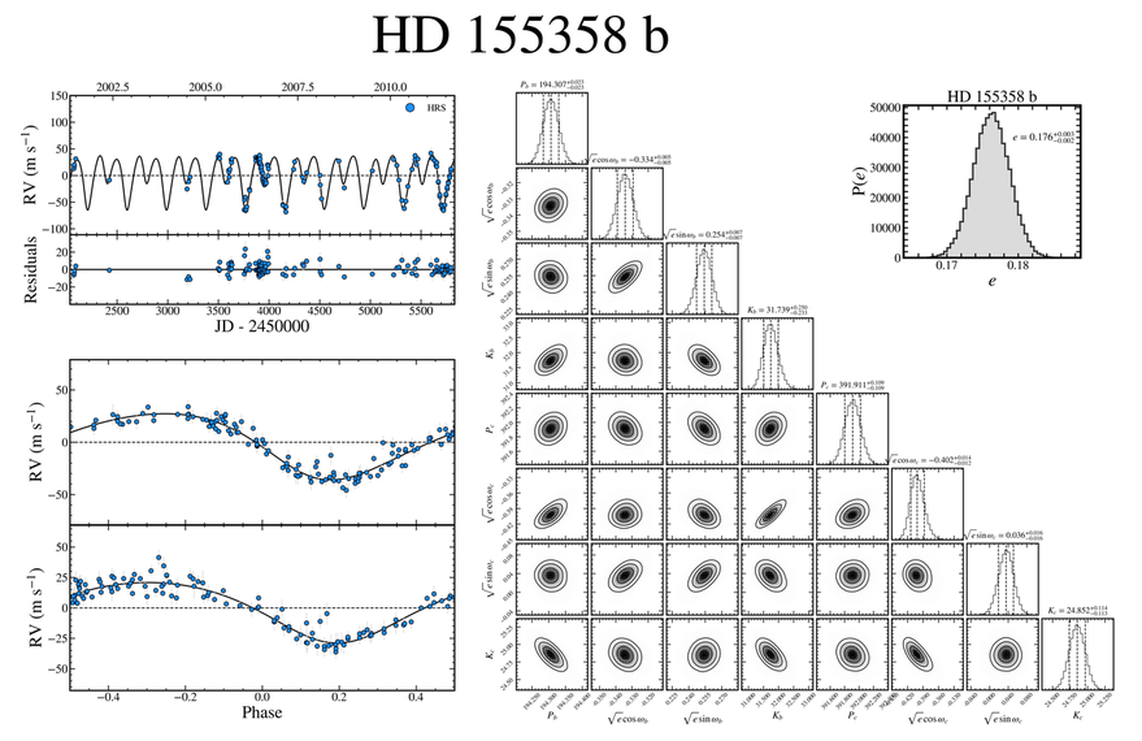}
 \end{minipage}
 \caption{Summary of results for the warm Jupiters HD 155193 b and HD 155358 b.}
 \label{fig:Combined_Plots26}
\end{figure}

\begin{figure}
\hskip -0.8 in
 \centering
 \begin{minipage}{\textwidth}
   \centering
   \includegraphics[width=\linewidth]{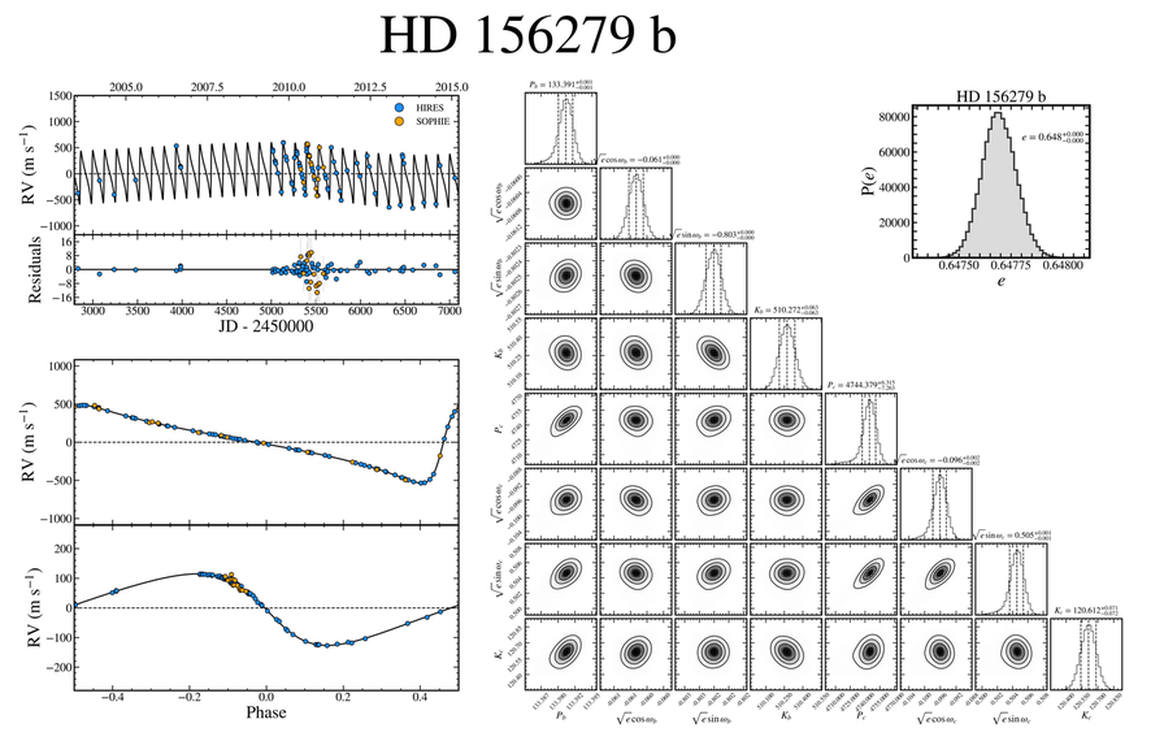}\\
   \vskip .3 in
   \includegraphics[width=\linewidth]{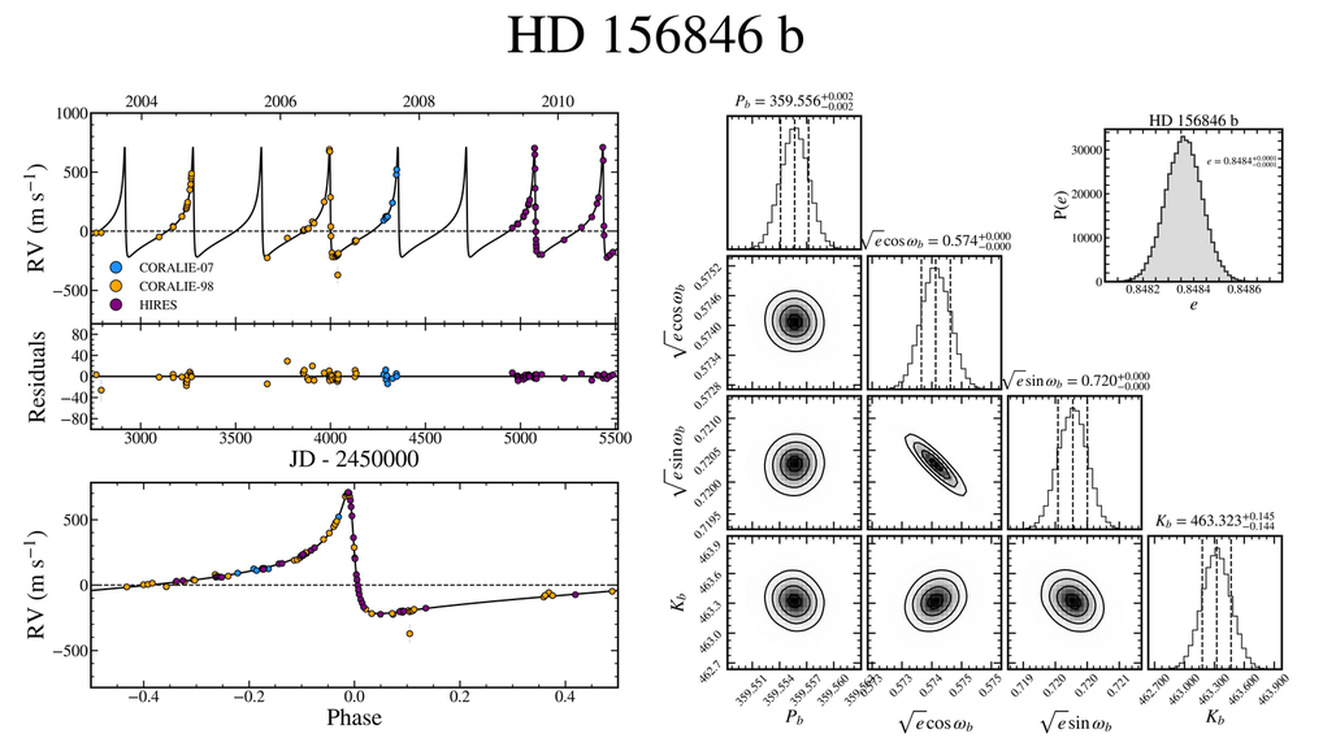}
 \end{minipage}
 \caption{Summary of results for the warm Jupiters HD 156279 b and HD 156846 b.}
 \label{fig:Combined_Plots27}
\end{figure}

\begin{figure}
\hskip -0.8 in
 \centering
 \begin{minipage}{\textwidth}
   \centering
   \includegraphics[width=\linewidth]{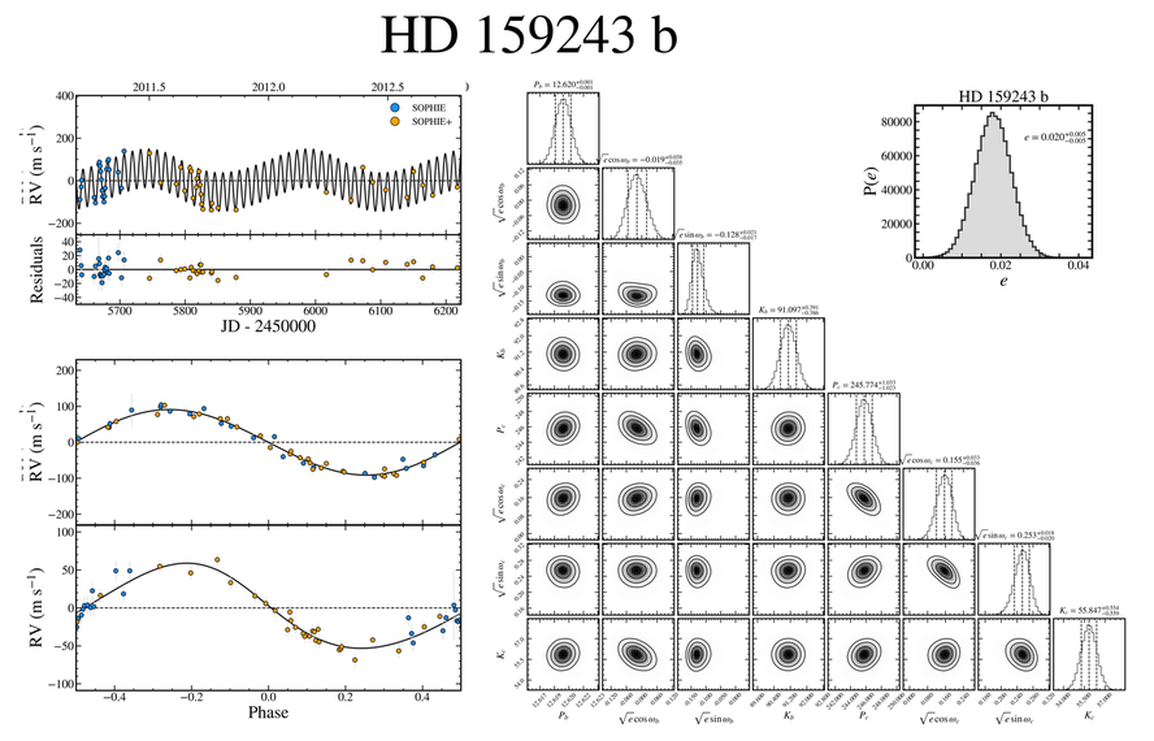}\\
   \vskip .3 in
   \includegraphics[width=\linewidth]{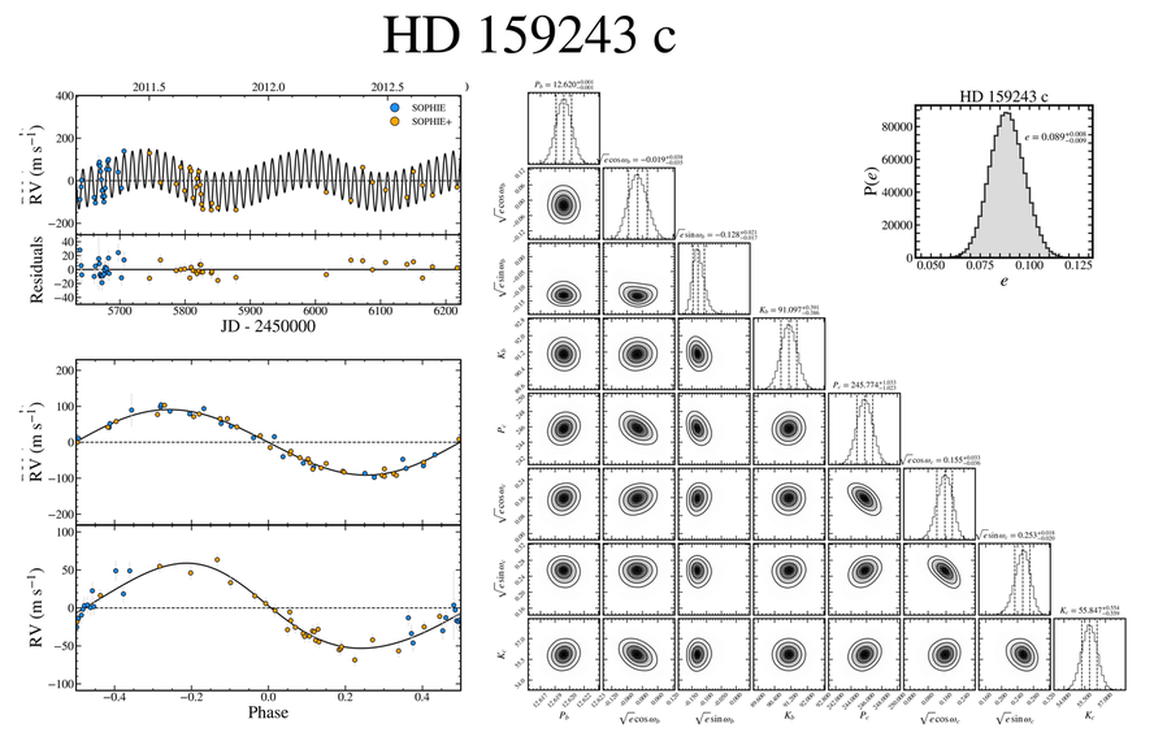}
 \end{minipage}
 \caption{Summary of results for the warm Jupiters HD 159243 b and HD 159243 c.}
 \label{fig:Combined_Plots28}
\end{figure}

\begin{figure}
\hskip -0.8 in
 \centering
 \begin{minipage}{\textwidth}
   \centering
   \includegraphics[width=\linewidth]{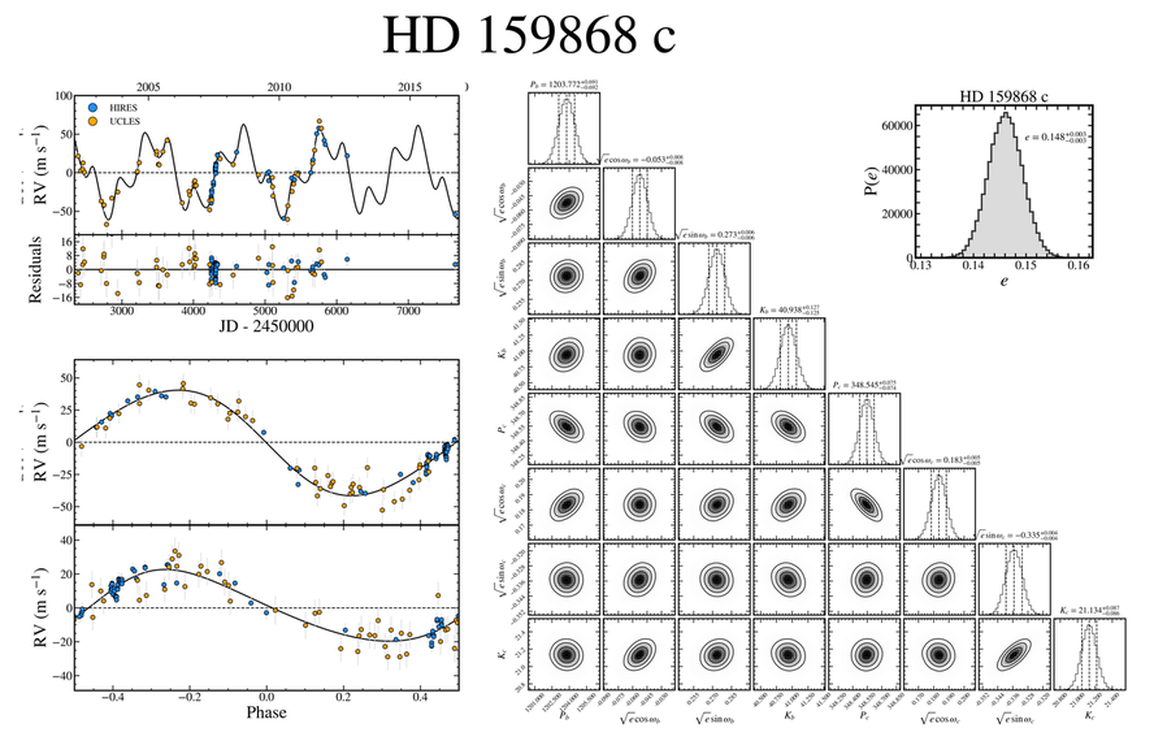}\\
   \vskip .2 in
   \includegraphics[width=0.9\linewidth]{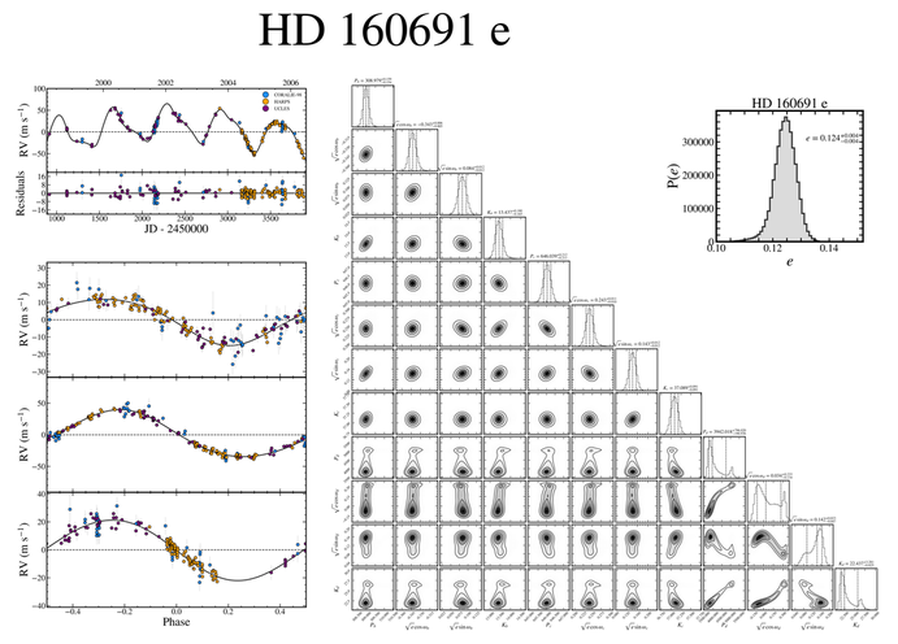}
 \end{minipage}
 \caption{Summary of results for the warm Jupiters HD 159868 c and HD 160691 e.}
 \label{fig:Combined_Plots29}
\end{figure}

\begin{figure}
\hskip -0.8 in
 \centering
 \begin{minipage}{\textwidth}
   \centering
   \includegraphics[width=\linewidth]{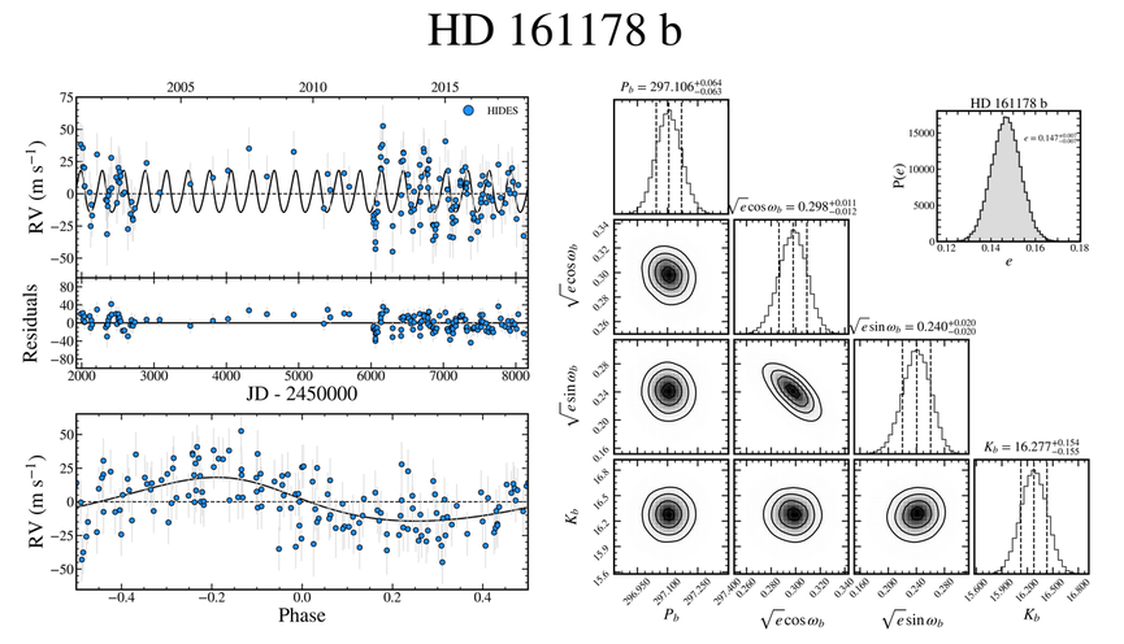}\\
   \vskip .3 in
   \includegraphics[width=\linewidth]{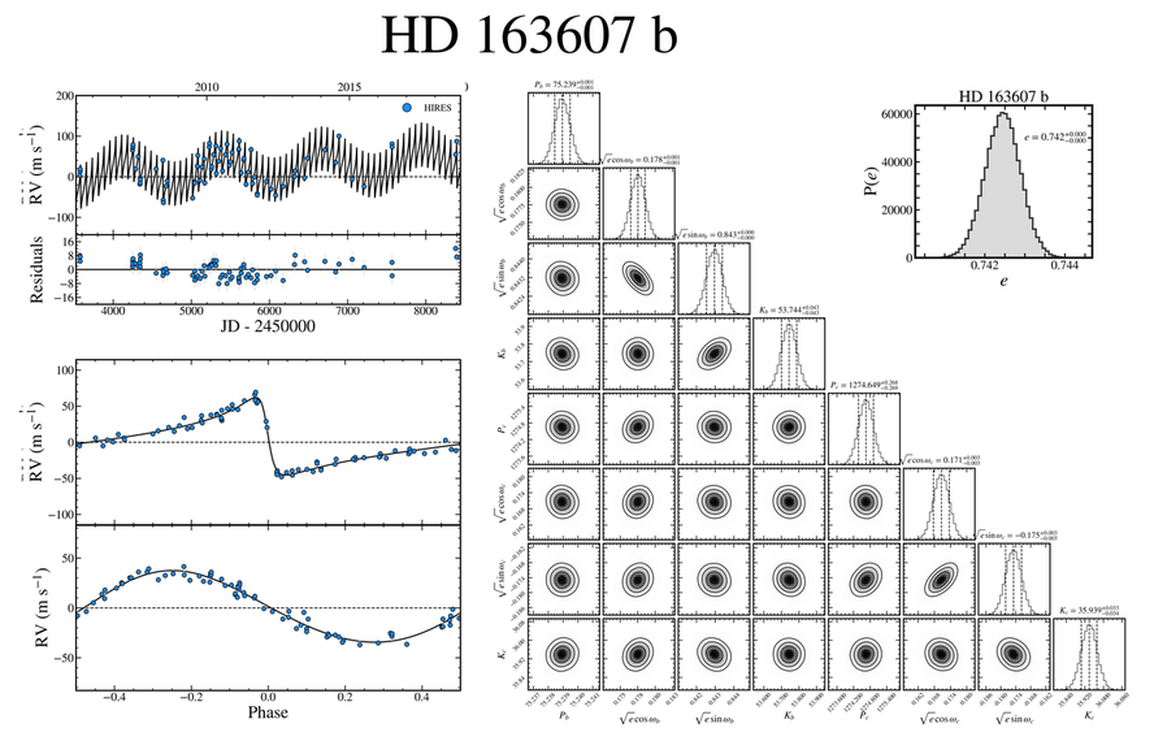}
 \end{minipage}
 \caption{Summary of results for the warm Jupiters HD 161178 b and HD 163607 b.}
 \label{fig:Combined_Plots30}
\end{figure}

\begin{figure}
\hskip -0.8 in
 \centering
 \begin{minipage}{\textwidth}
   \centering
   \includegraphics[width=\linewidth]{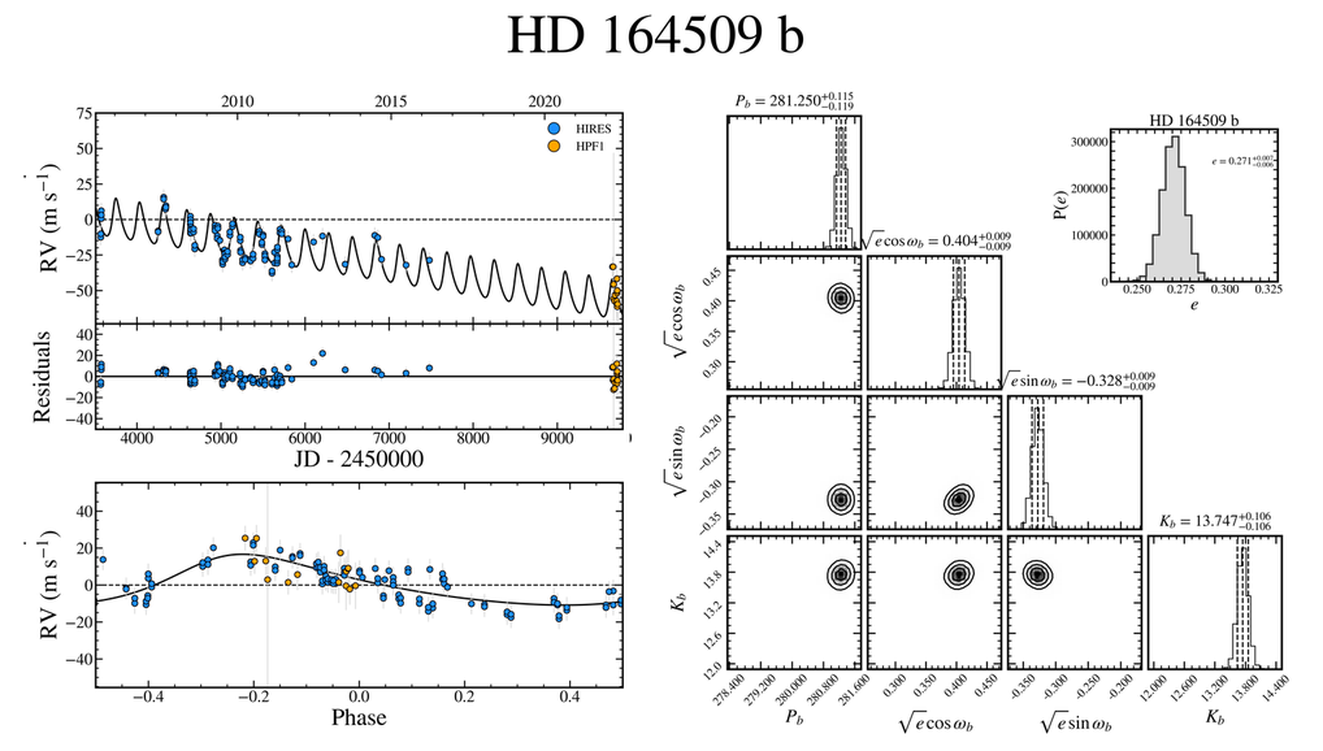}\\
   \vskip .3 in
   \includegraphics[width=\linewidth]{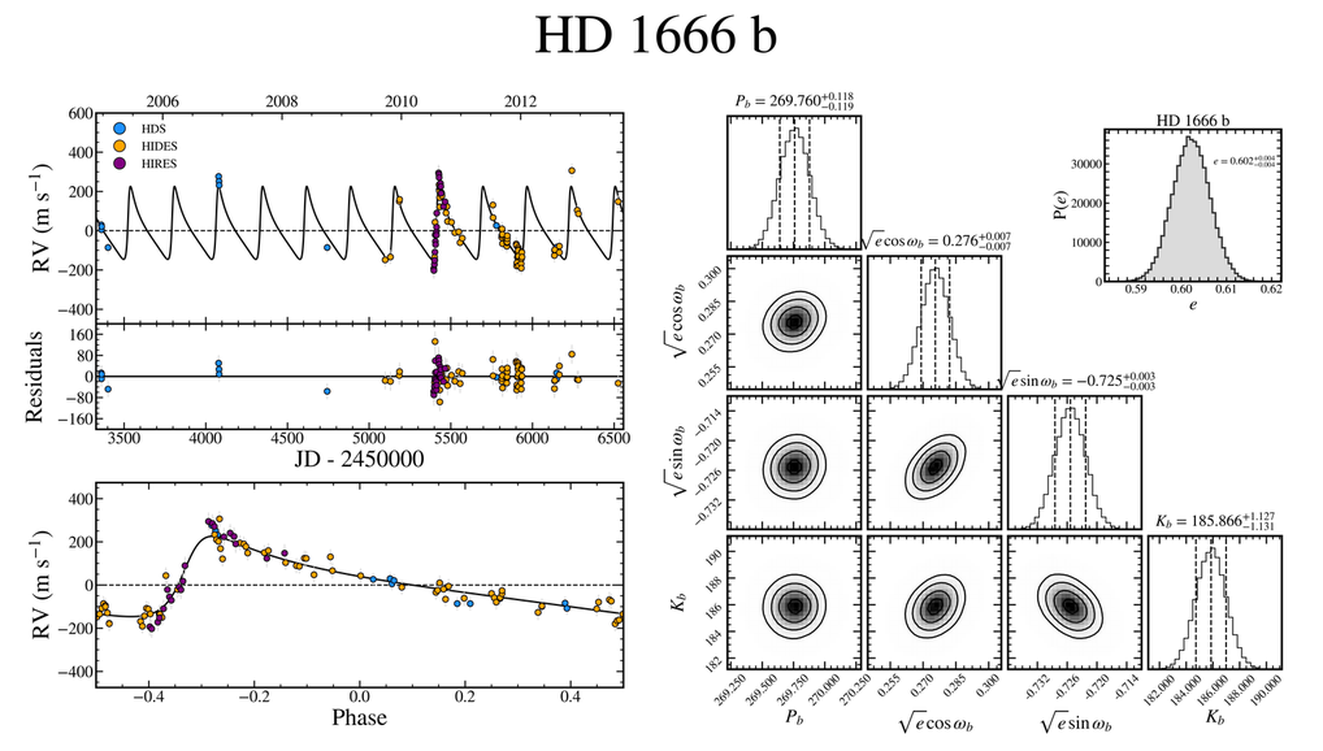}
 \end{minipage}
 \caption{Summary of results for the warm Jupiters HD 164509 b and HD 1666 b.}
 \label{fig:Combined_Plots31}
\end{figure}

\begin{figure}
\hskip -0.8 in
 \centering
 \begin{minipage}{\textwidth}
   \centering
   \includegraphics[width=\linewidth]{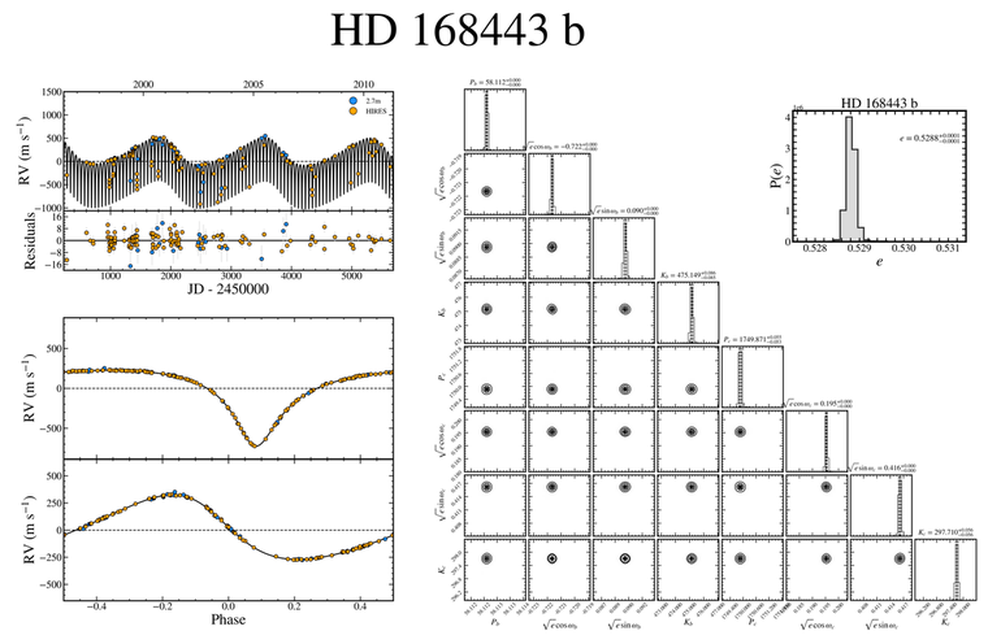}\\
   \vskip .3 in
   \includegraphics[width=\linewidth]{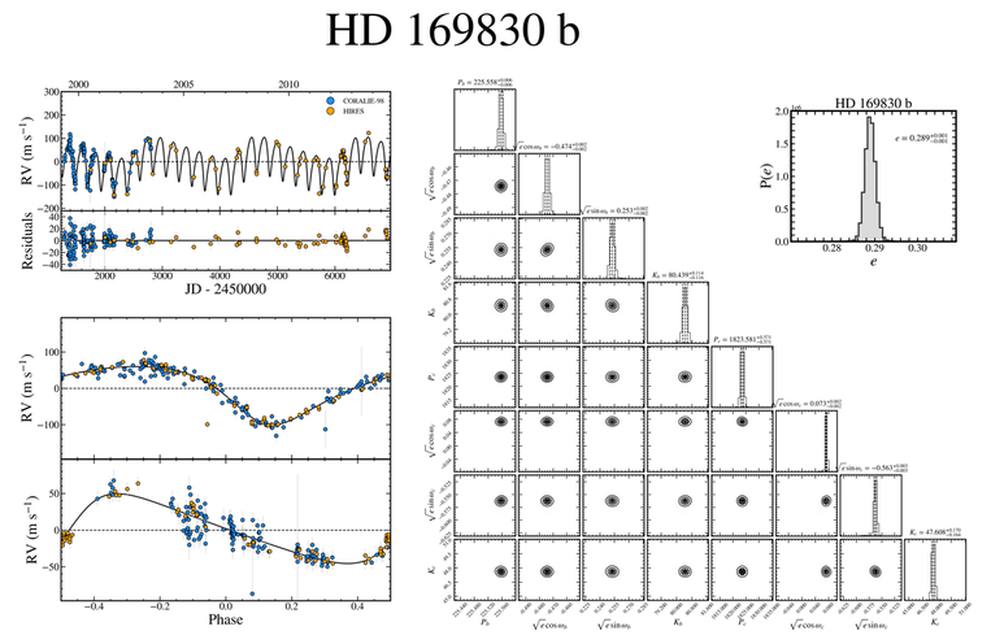}
 \end{minipage}
 \caption{Summary of results for the warm Jupiters HD 168443 b and HD 169830 b.}
 \label{fig:Combined_Plots32}
\end{figure}

\begin{figure}
\hskip -0.8 in
 \centering
 \begin{minipage}{\textwidth}
   \centering
   \includegraphics[width=\linewidth]{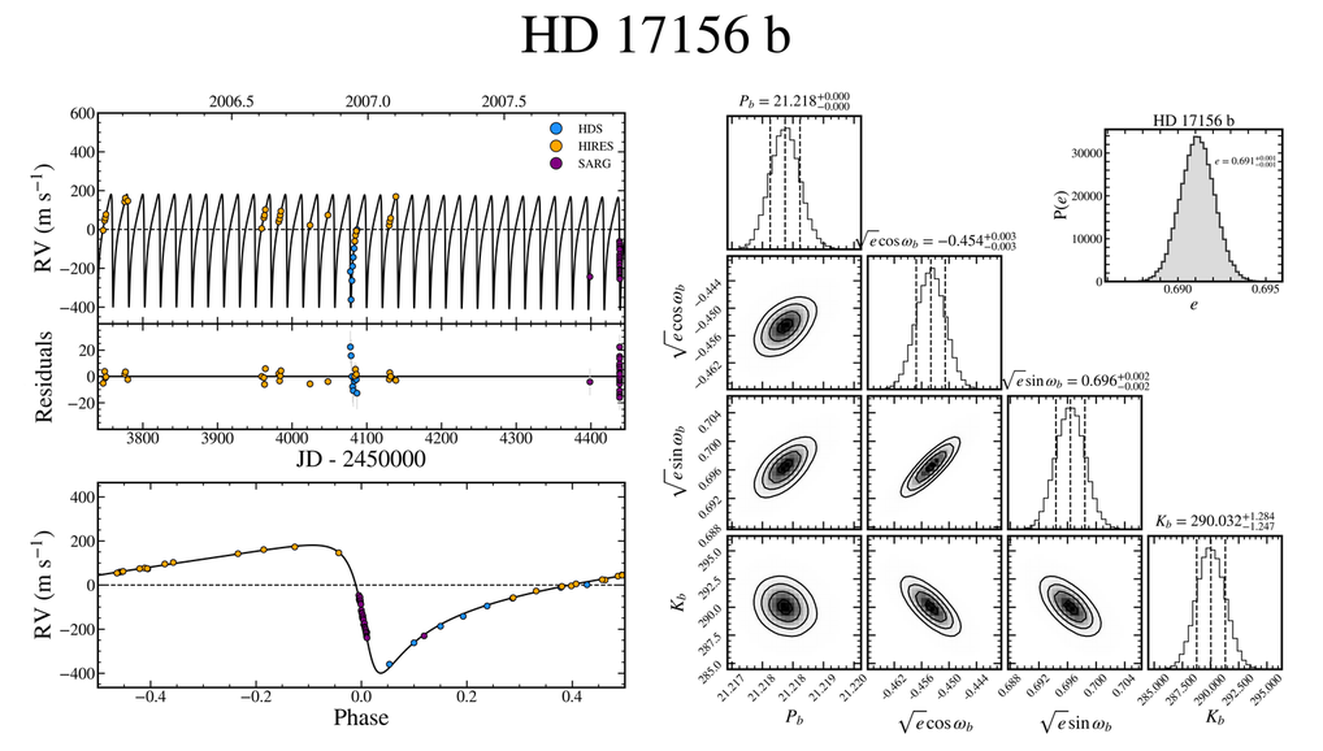}\\
   \vskip .3 in
   \includegraphics[width=\linewidth]{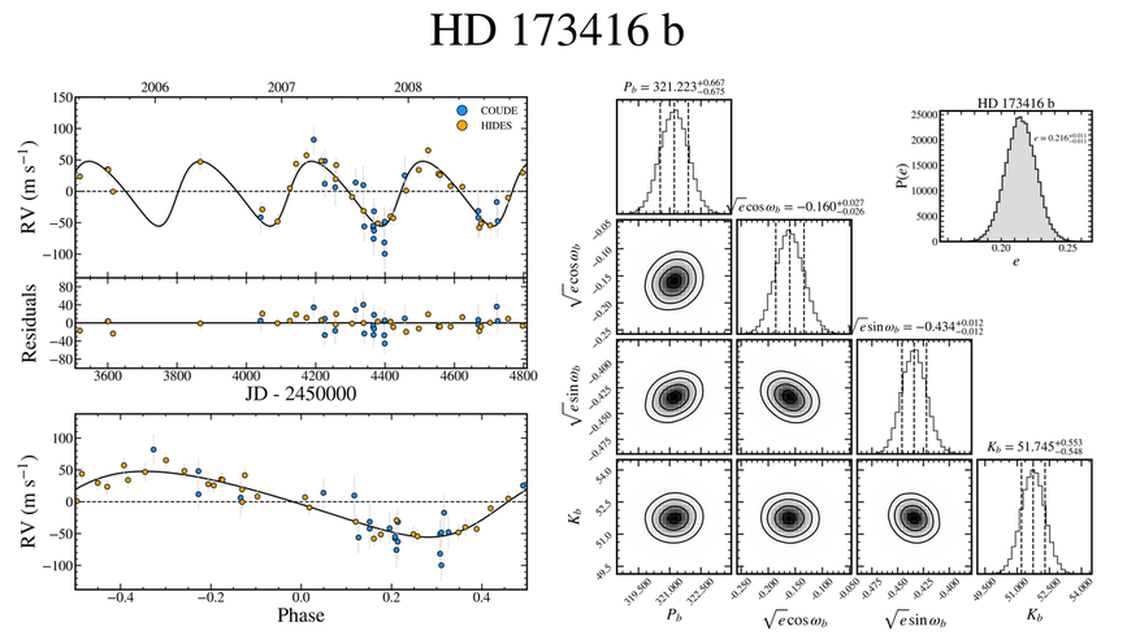}
 \end{minipage}
 \caption{Summary of results for the warm Jupiters HD 17156 b and HD 173416 b.}
 \label{fig:Combined_Plots33}
\end{figure}

\begin{figure}
\hskip -0.8 in
 \centering
 \begin{minipage}{\textwidth}
   \centering
   \includegraphics[width=\linewidth]{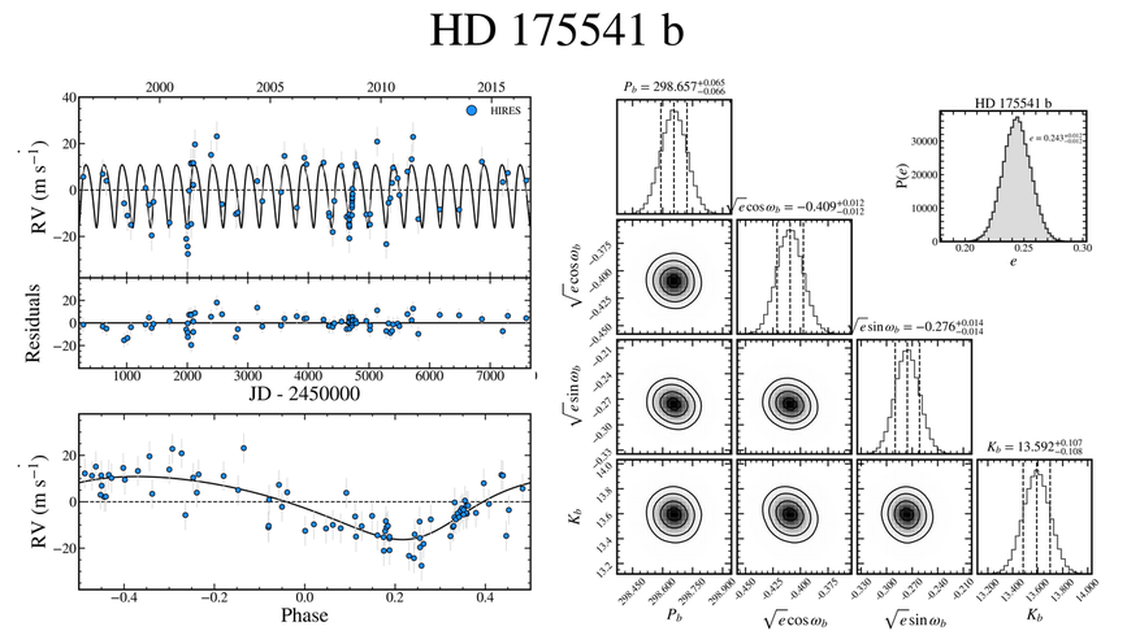}\\
   \vskip .3 in
   \includegraphics[width=\linewidth]{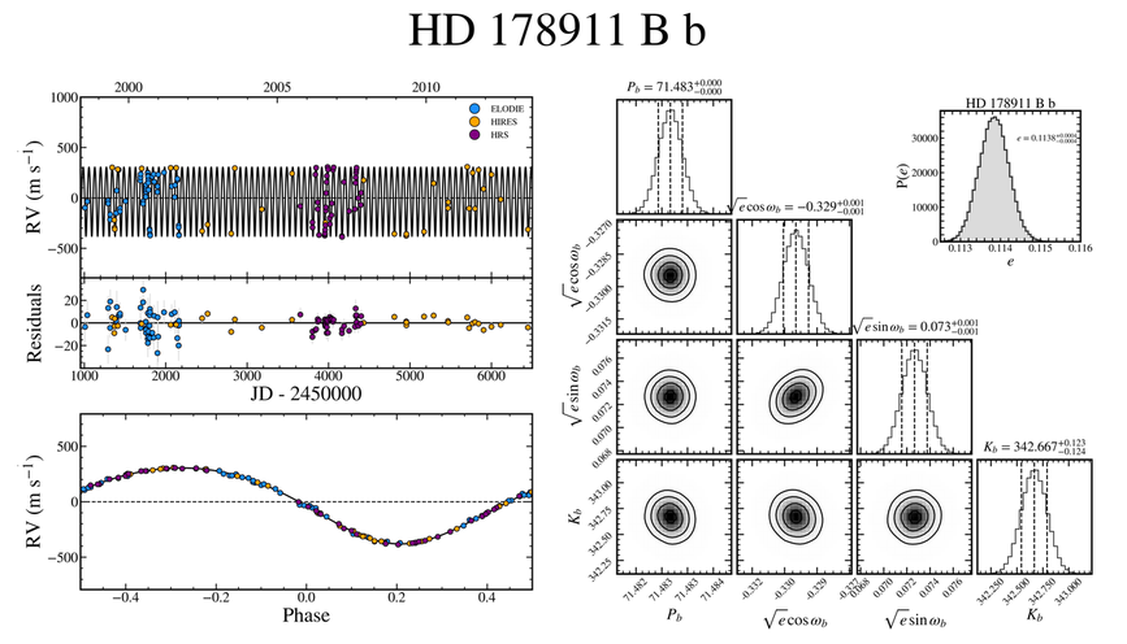}
 \end{minipage}
 \caption{Summary of results for the warm Jupiters HD 175541 b and HD 178911 B b.}
 \label{fig:Combined_Plots34}
\end{figure}

\begin{figure}
\hskip -0.8 in
 \centering
 \begin{minipage}{\textwidth}
   \centering
   \includegraphics[width=\linewidth]{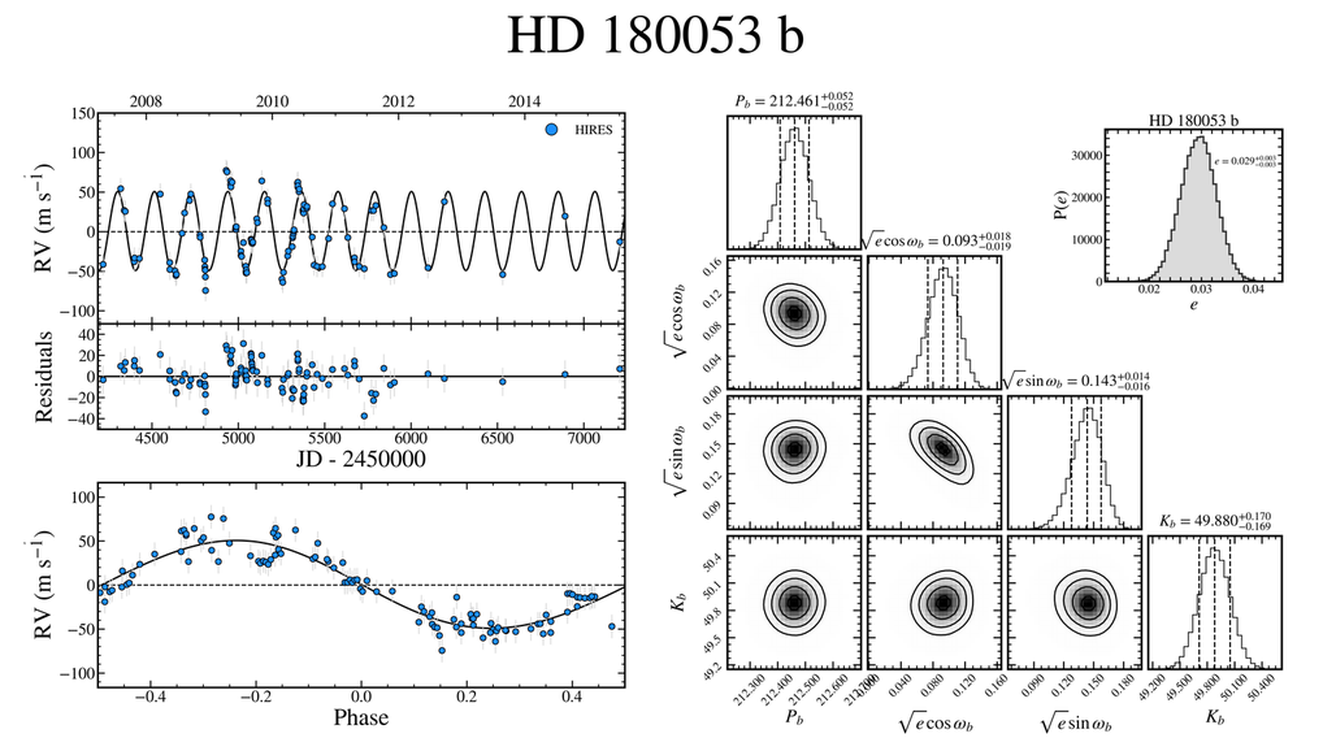}\\
   \vskip .3 in
   \includegraphics[width=\linewidth]{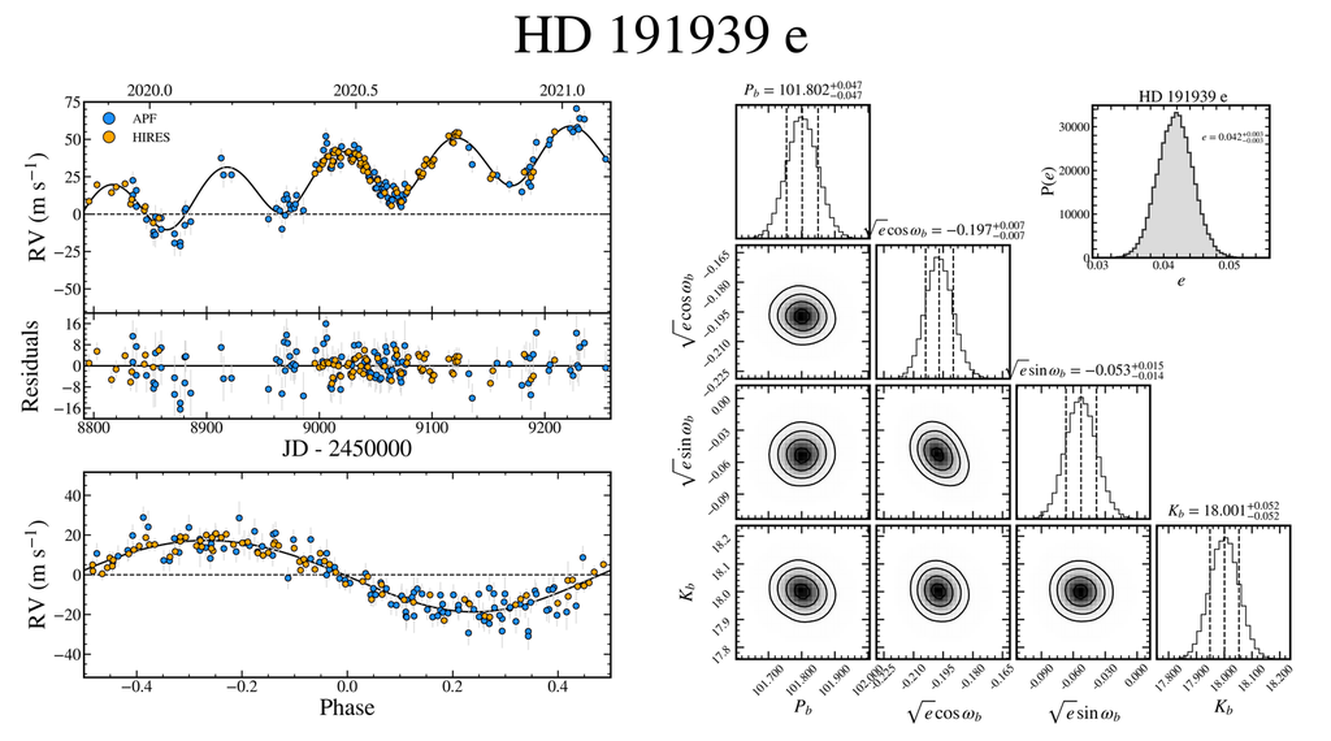}
 \end{minipage}
 \caption{Summary of results for the warm Jupiters HD 180053 b and HD 191939 e.}
 \label{fig:Combined_Plots35}
\end{figure}

\begin{figure}
\hskip -0.8 in
 \centering
 \begin{minipage}{\textwidth}
   \centering
   \includegraphics[width=\linewidth]{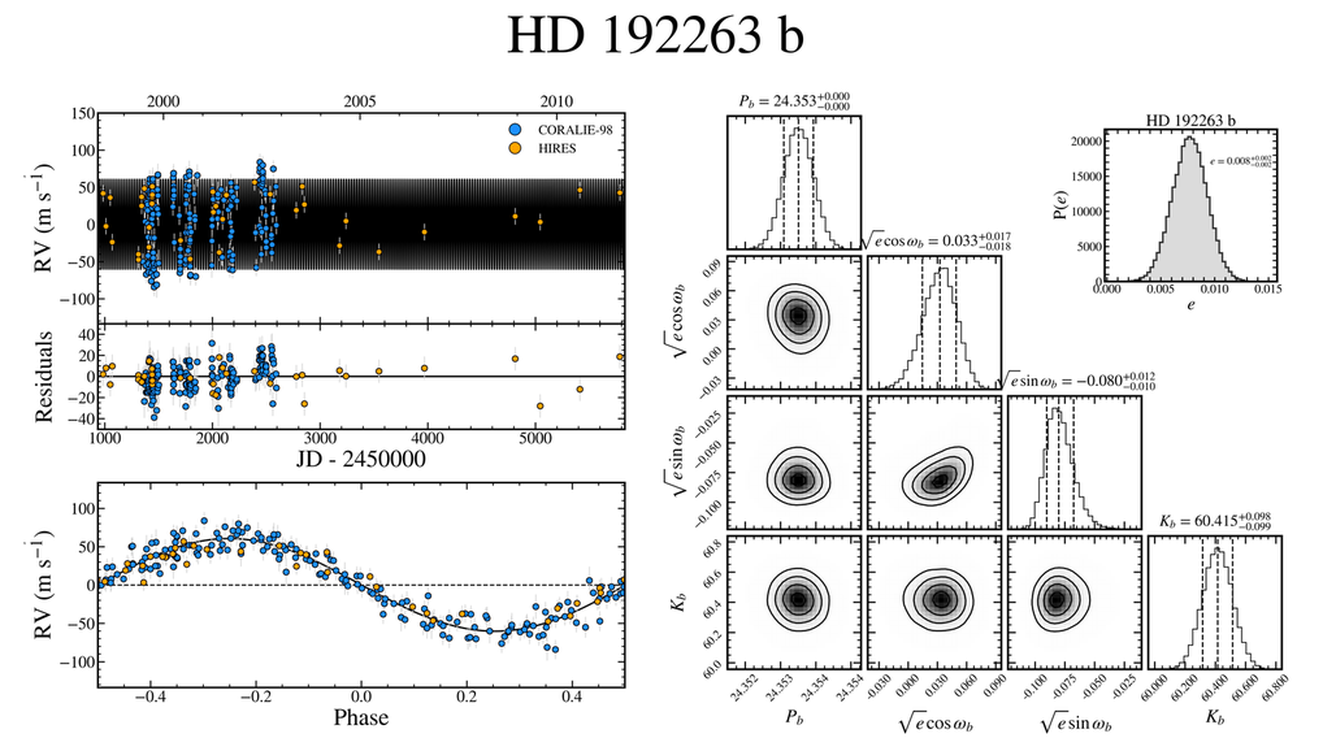}\\
   \vskip .3 in
   \includegraphics[width=\linewidth]{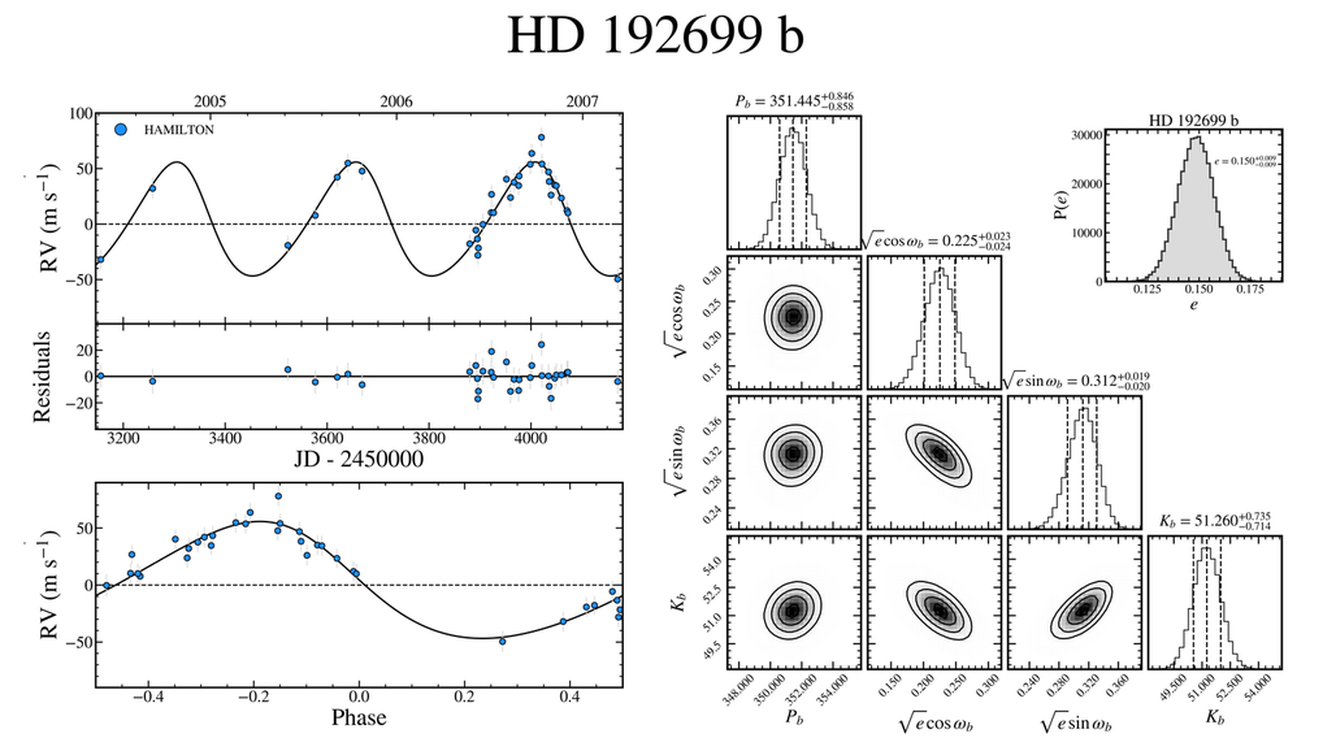}
 \end{minipage}
 \caption{Summary of results for the warm Jupiters HD 192263 b and HD 192699 b.}
 \label{fig:Combined_Plots36}
\end{figure}

\begin{figure}
\hskip -0.8 in
 \centering
 \begin{minipage}{\textwidth}
   \centering
   \includegraphics[width=\linewidth]{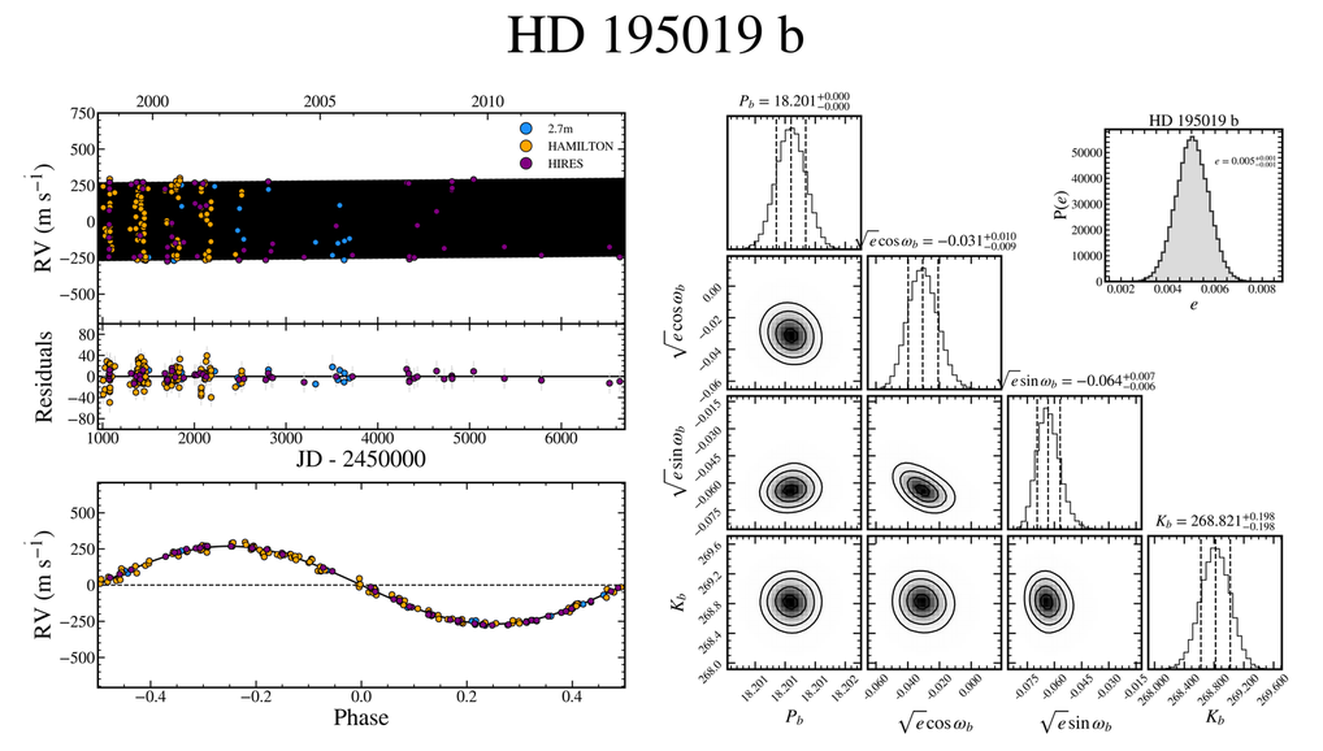}\\
   \vskip .3 in
   \includegraphics[width=\linewidth]{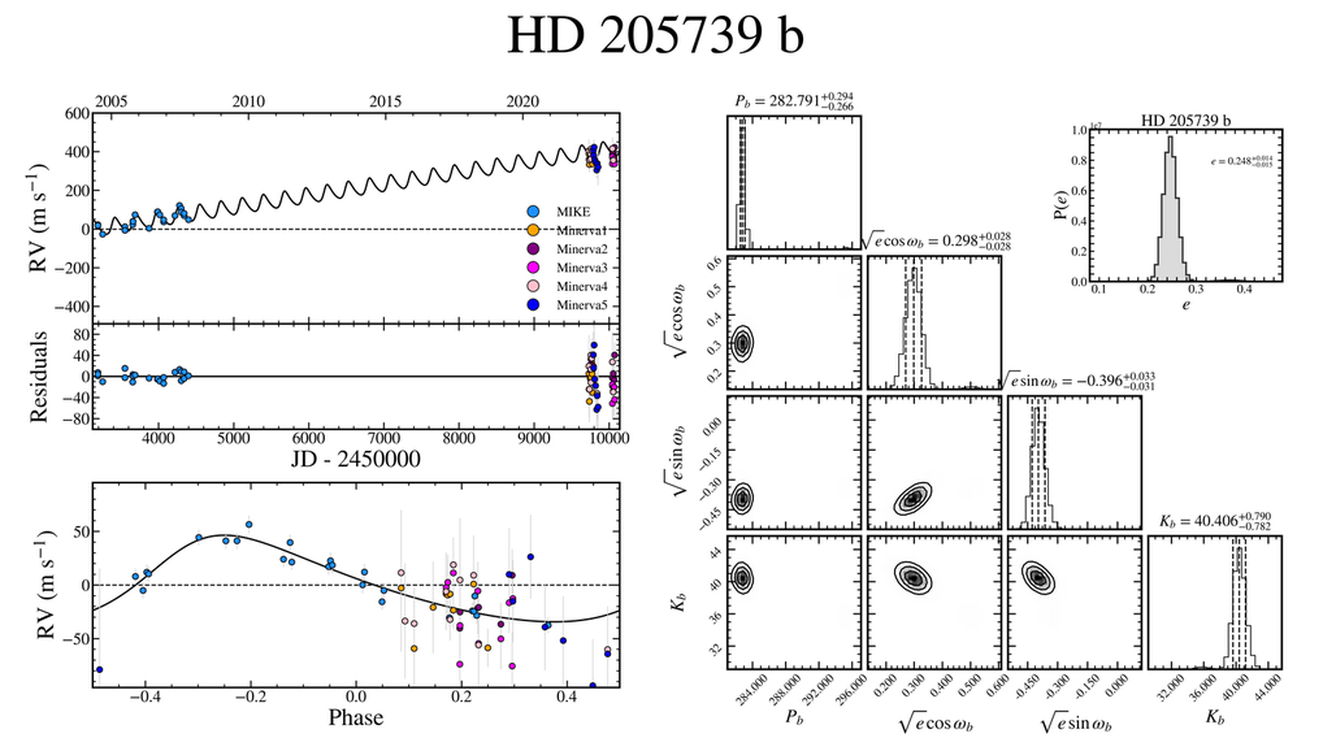}
 \end{minipage}
 \caption{Summary of results for the warm Jupiters HD 195019 b and HD 205739 b.}
 \label{fig:Combined_Plots37}
\end{figure}

\begin{figure}
\hskip -0.8 in
 \centering
 \begin{minipage}{\textwidth}
   \centering
   \includegraphics[width=\linewidth]{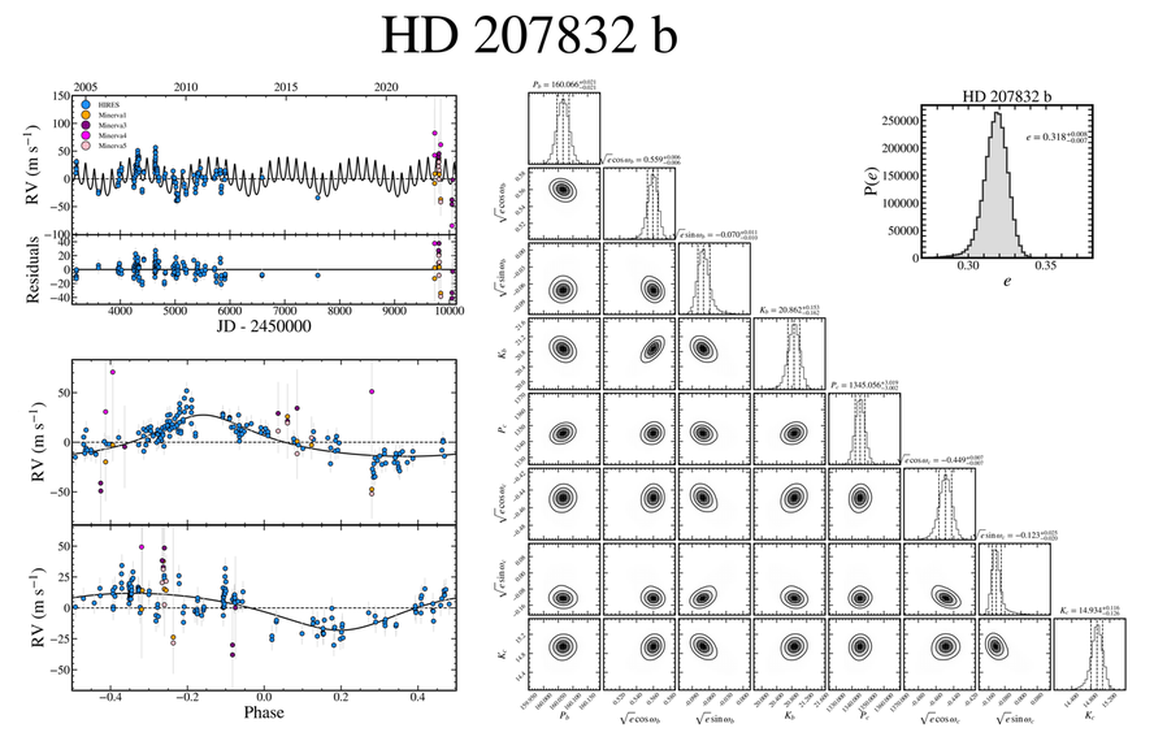}\\
   \vskip .3 in
   \includegraphics[width=\linewidth]{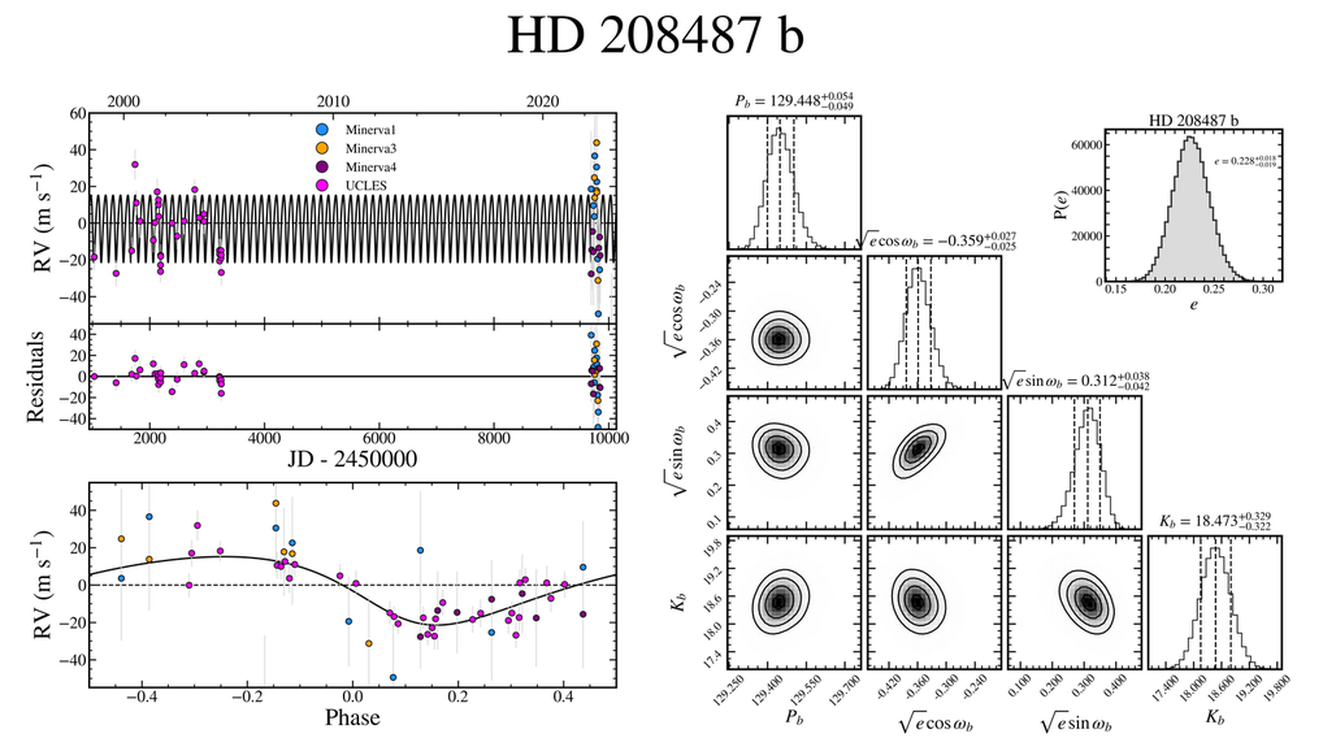}
 \end{minipage}
 \caption{Summary of results for the warm Jupiters HD 207832 b and HD 208487 b.}
 \label{fig:Combined_Plots38}
\end{figure}

\begin{figure}
\hskip -0.8 in
 \centering
 \begin{minipage}{\textwidth}
   \centering
   \includegraphics[width=\linewidth]{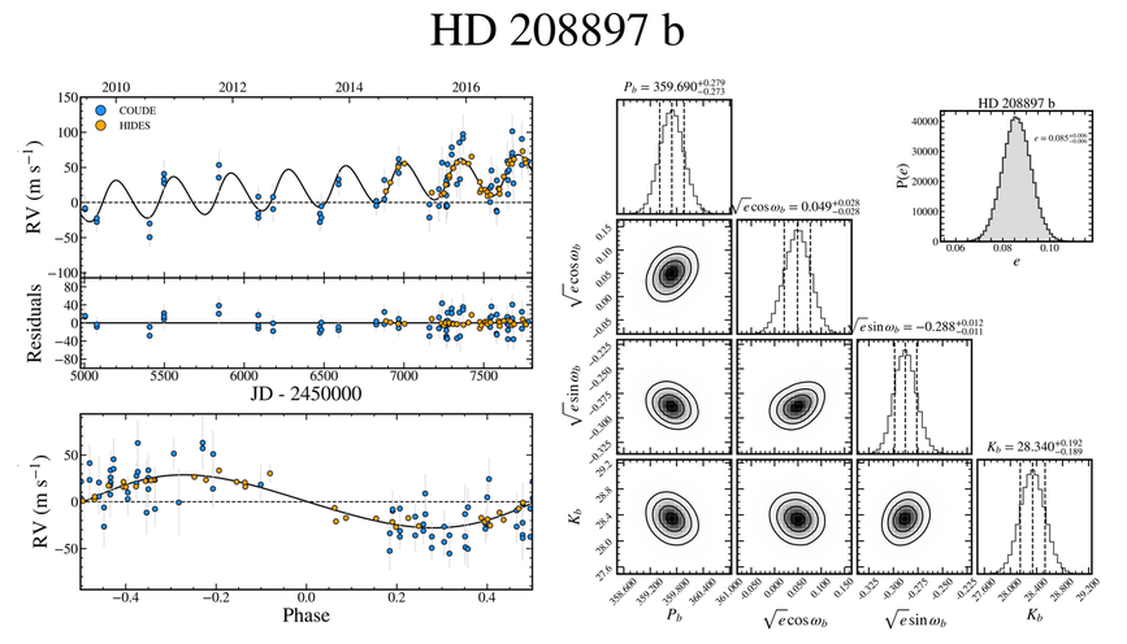}\\
   \vskip .3 in
   \includegraphics[width=\linewidth]{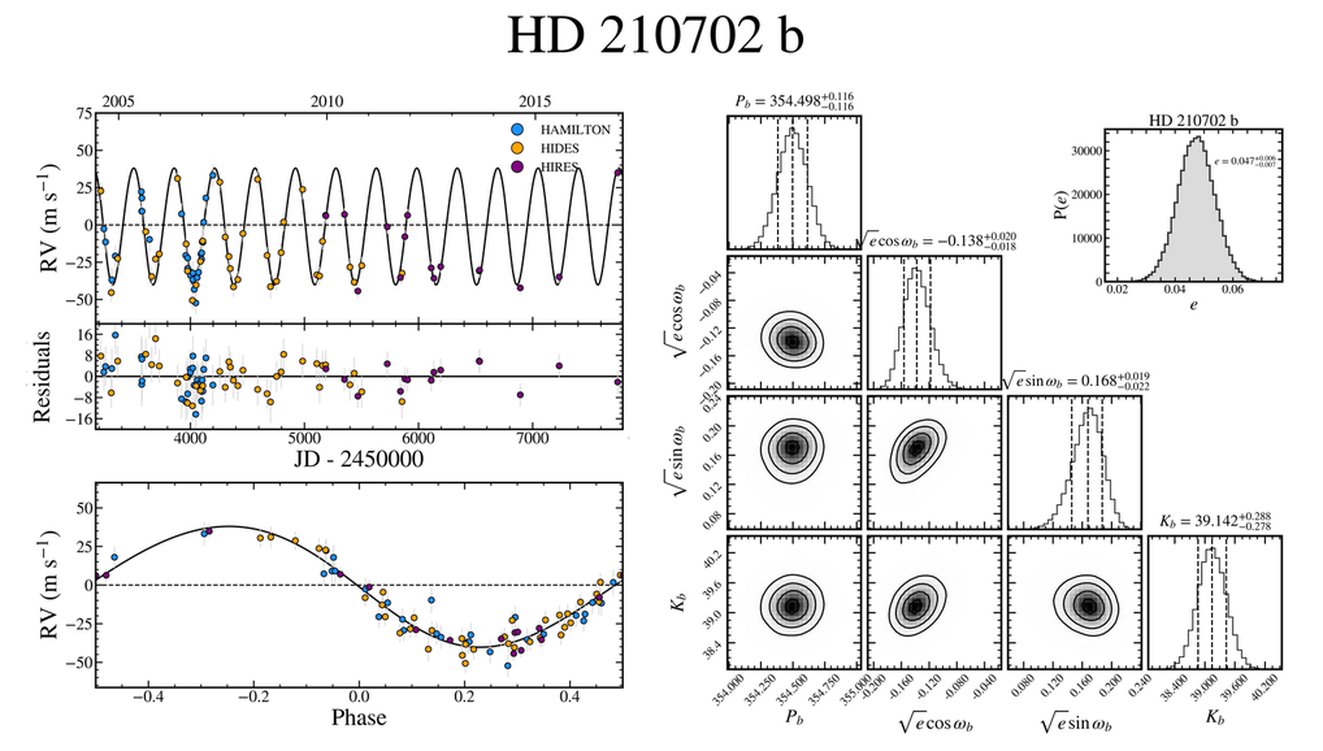}
 \end{minipage}
 \caption{Summary of results for the warm Jupiters HD 208897 b and HD 210702 b.}
 \label{fig:Combined_Plots39}
\end{figure}

\begin{figure}
\hskip -0.8 in
 \centering
 \begin{minipage}{\textwidth}
   \centering
   \includegraphics[width=\linewidth]{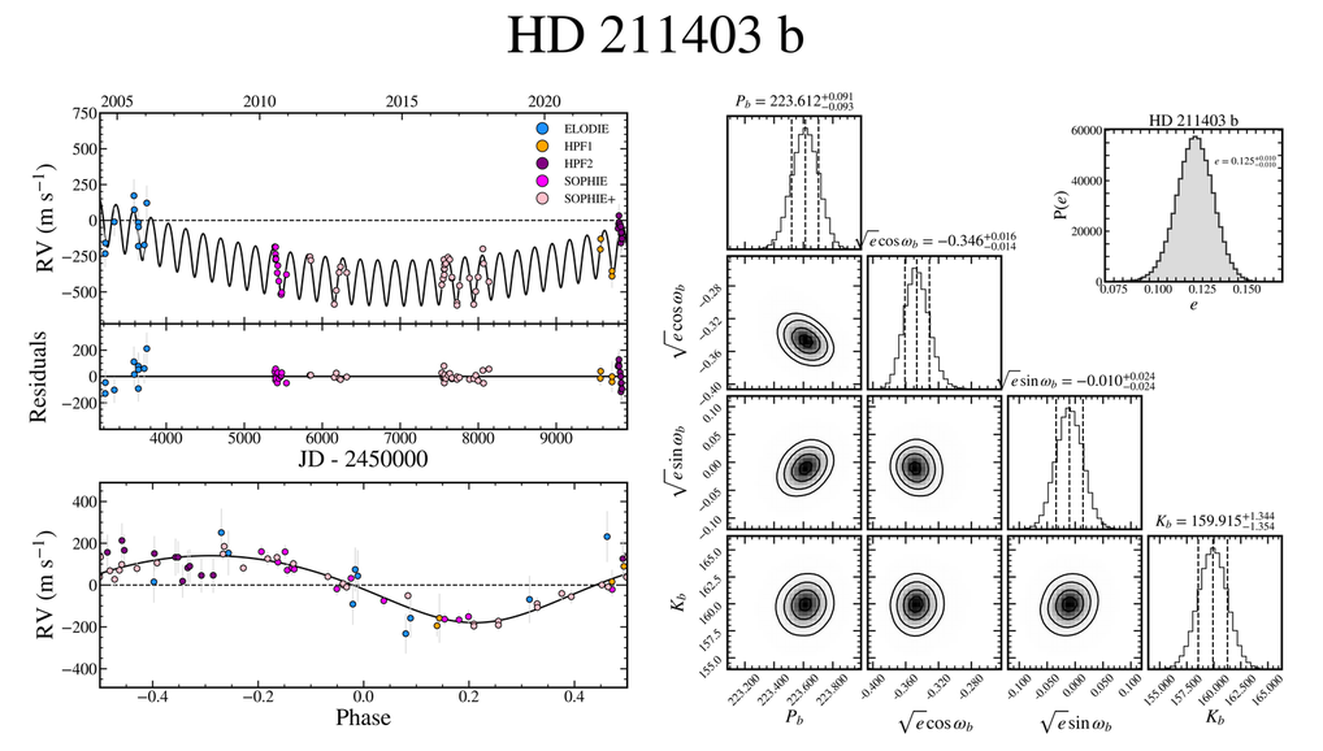}\\
   \vskip .3 in
   \includegraphics[width=\linewidth]{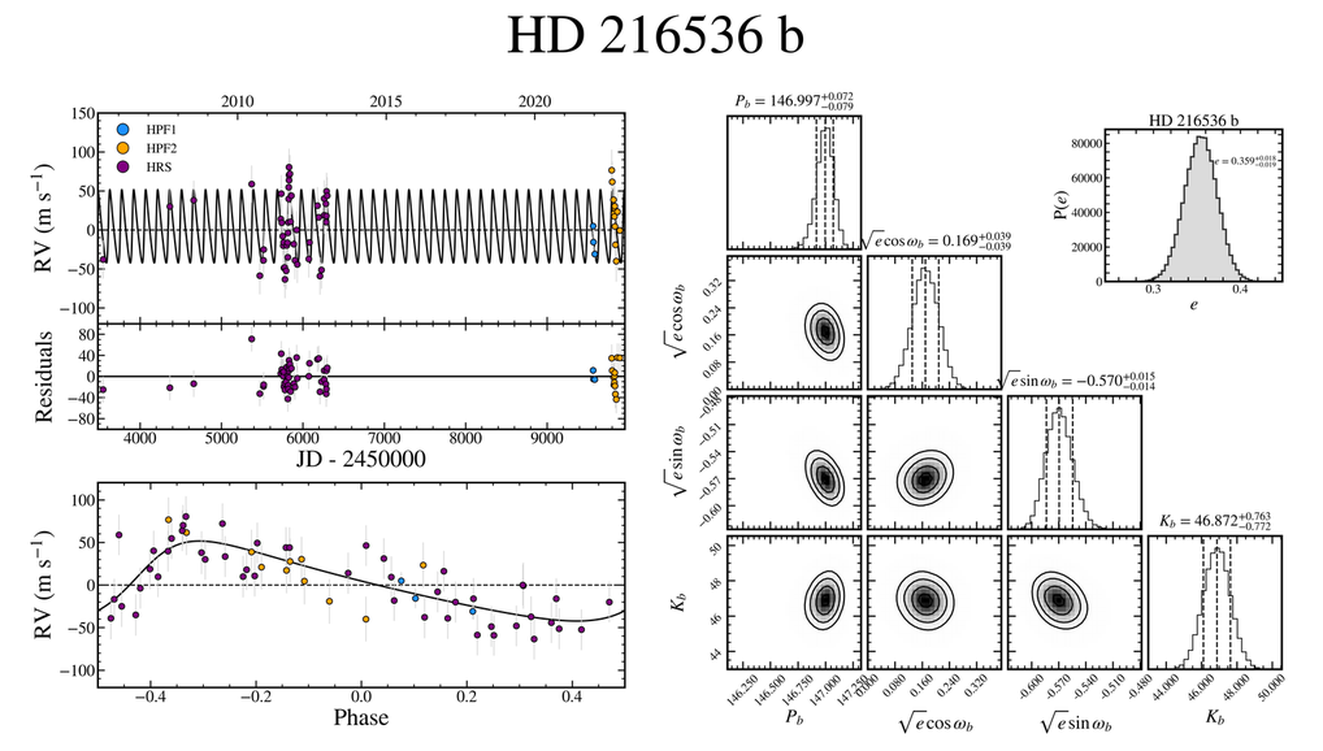}
 \end{minipage}
 \caption{Summary of results for the warm Jupiters HD 211403 b and HD 216536 b.}
 \label{fig:Combined_Plots40}
\end{figure}

\begin{figure}
\hskip -0.8 in
 \centering
 \begin{minipage}{\textwidth}
   \centering
   \includegraphics[width=\linewidth]{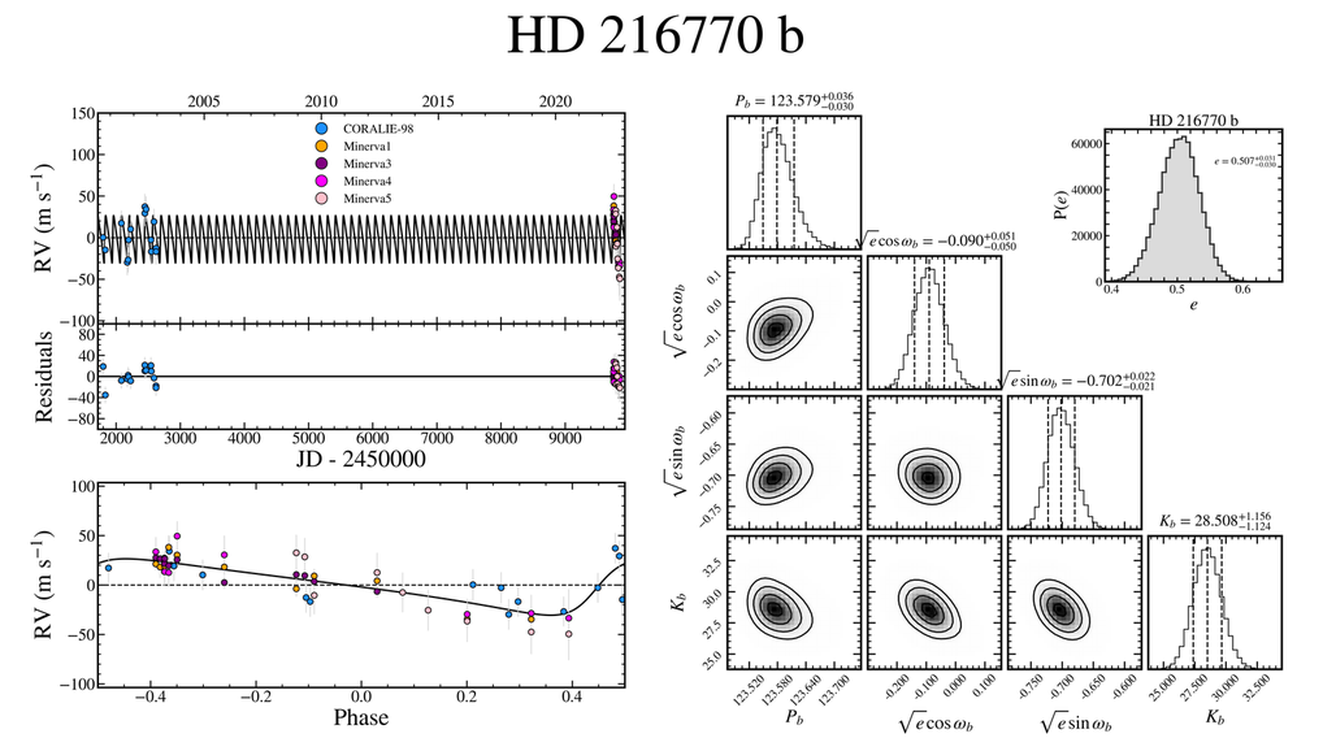}\\
   \vskip .3 in
   \includegraphics[width=\linewidth]{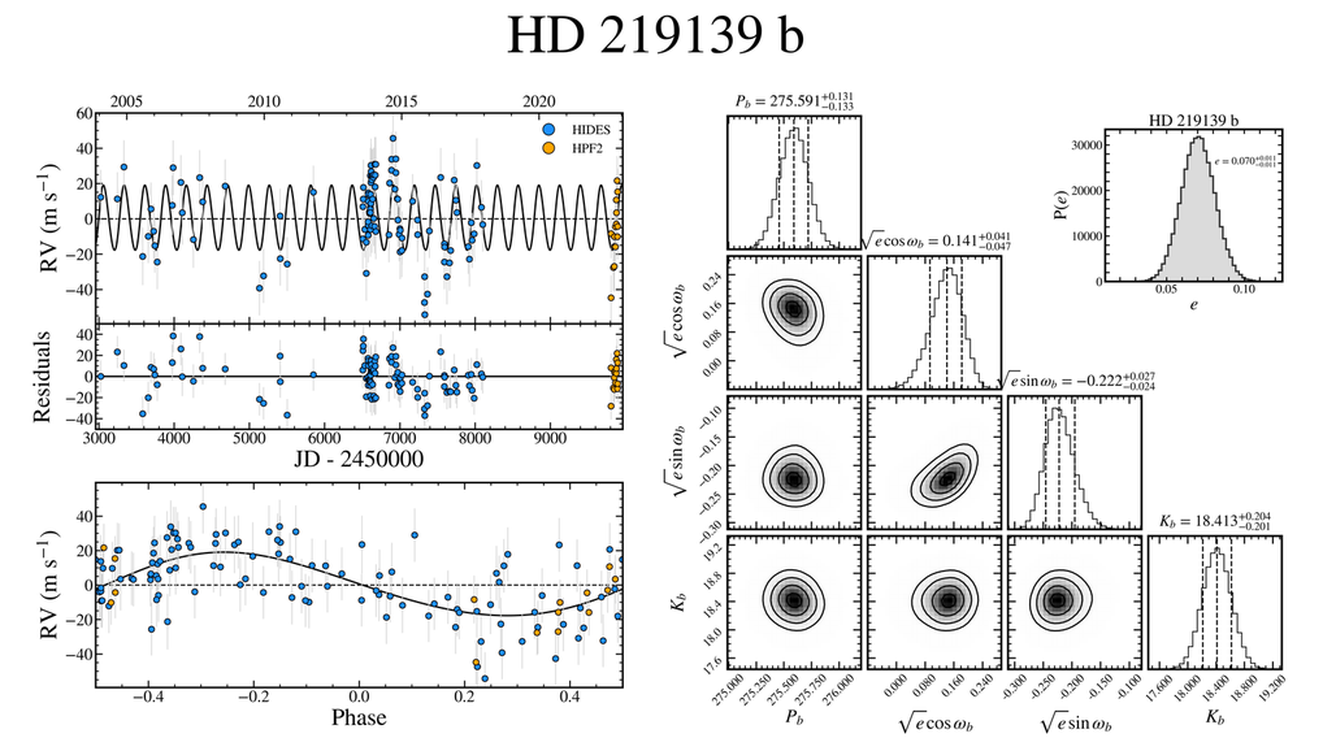}
 \end{minipage}
 \caption{Summary of results for the warm Jupiters HD 216770 b and HD 219139 b.}
 \label{fig:Combined_Plots41}
\end{figure}

\begin{figure}
\hskip -0.8 in
 \centering
 \begin{minipage}{\textwidth}
   \centering
   \includegraphics[width=\linewidth]{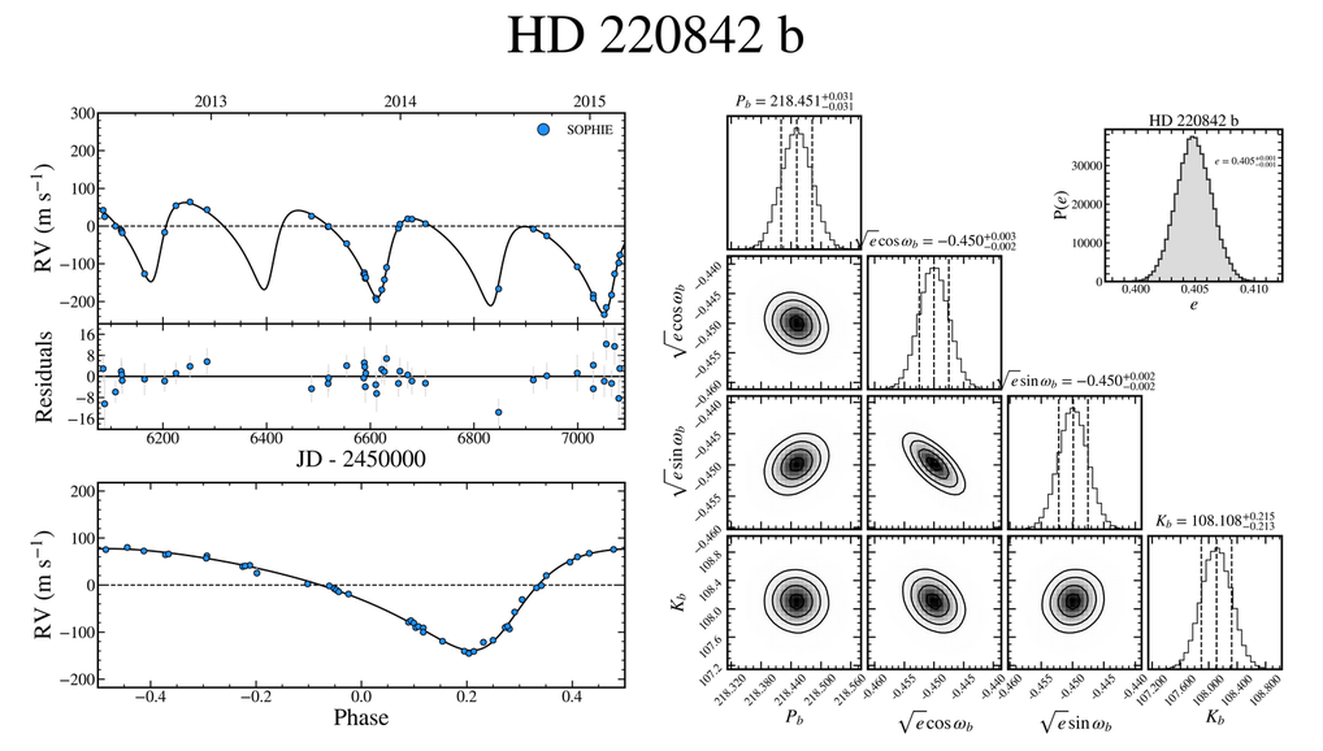}\\
   \vskip .3 in
   \includegraphics[width=\linewidth]{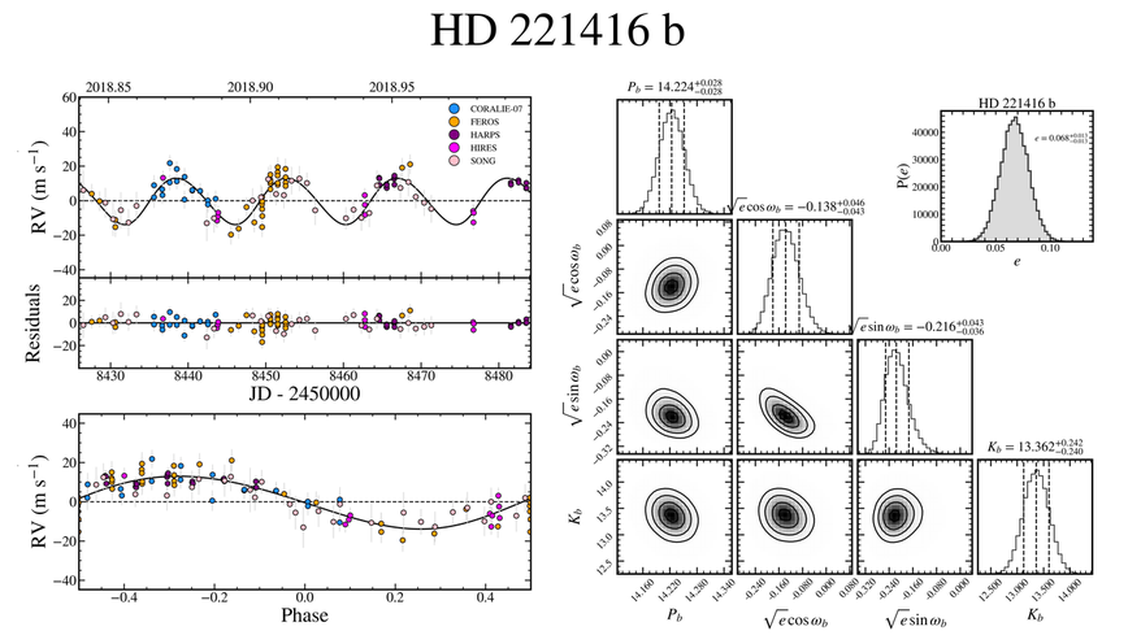}
 \end{minipage}
 \caption{Summary of results for the warm Jupiters HD 220842 b and HD 221416 b.}
 \label{fig:Combined_Plots42}
\end{figure}

\begin{figure}
\hskip -0.8 in
 \centering
 \begin{minipage}{\textwidth}
   \centering
   \includegraphics[width=\linewidth]{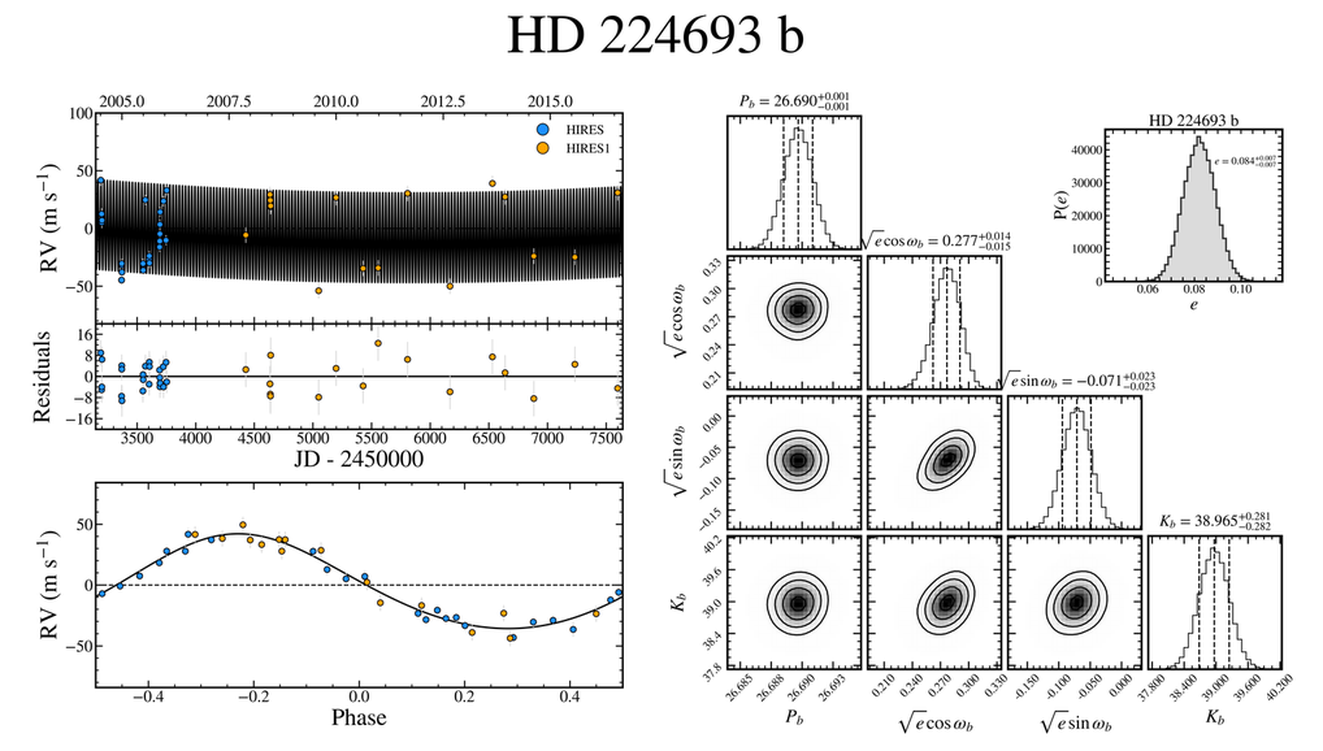}\\
   \vskip .3 in
   \includegraphics[width=\linewidth]{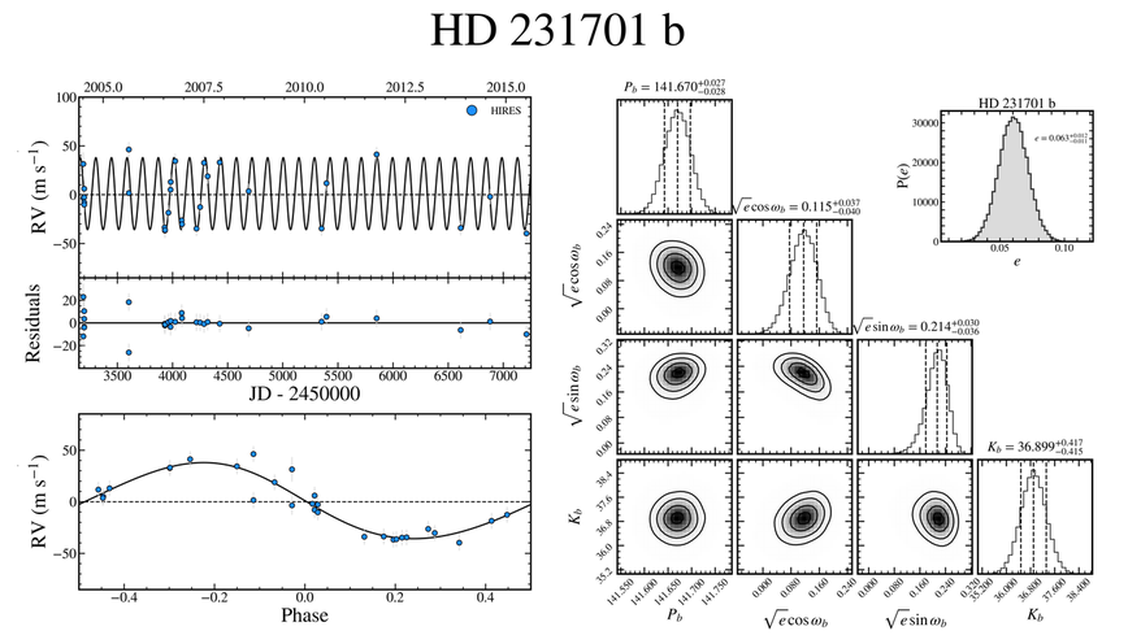}
 \end{minipage}
 \caption{Summary of results for the warm Jupiters HD 224693 b and HD 231701 b.}
 \label{fig:Combined_Plots43}
\end{figure}

\begin{figure}
\hskip -0.8 in
 \centering
 \begin{minipage}{\textwidth}
   \centering
   \includegraphics[width=\linewidth]{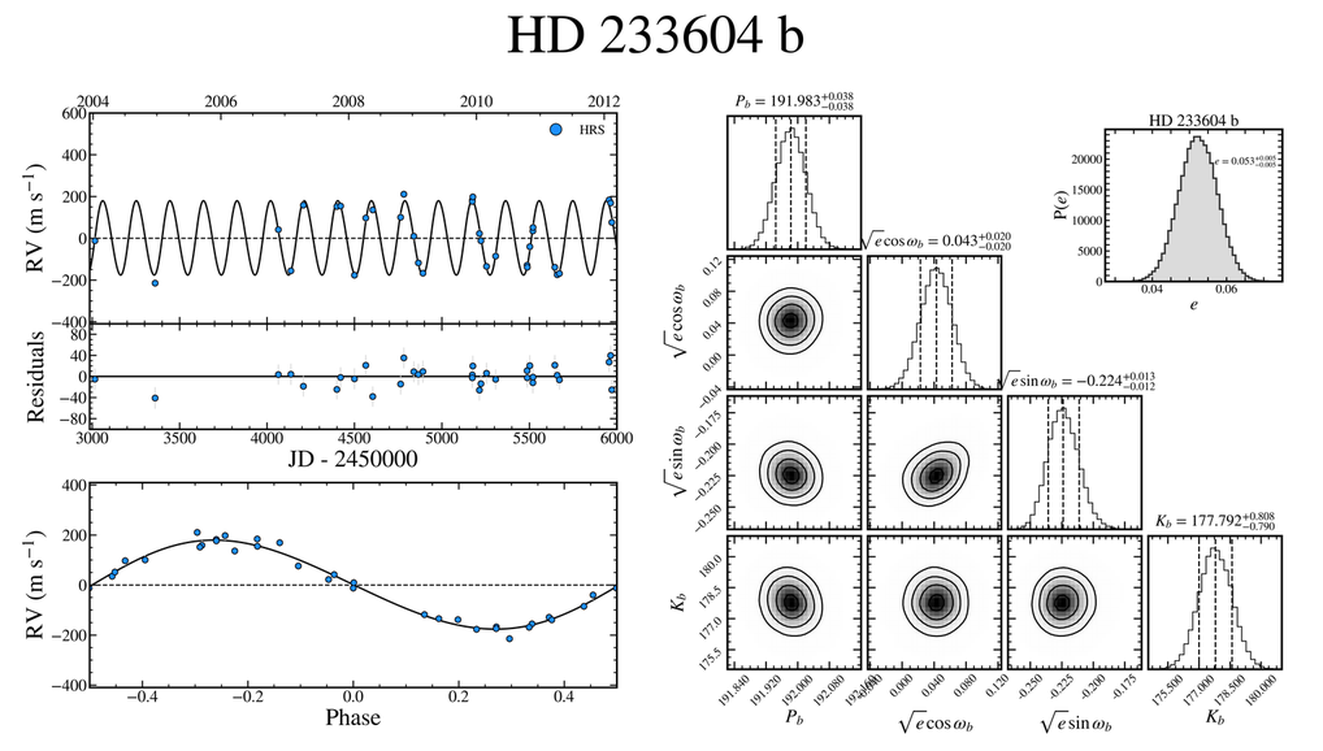}\\
   \vskip .3 in
   \includegraphics[width=\linewidth]{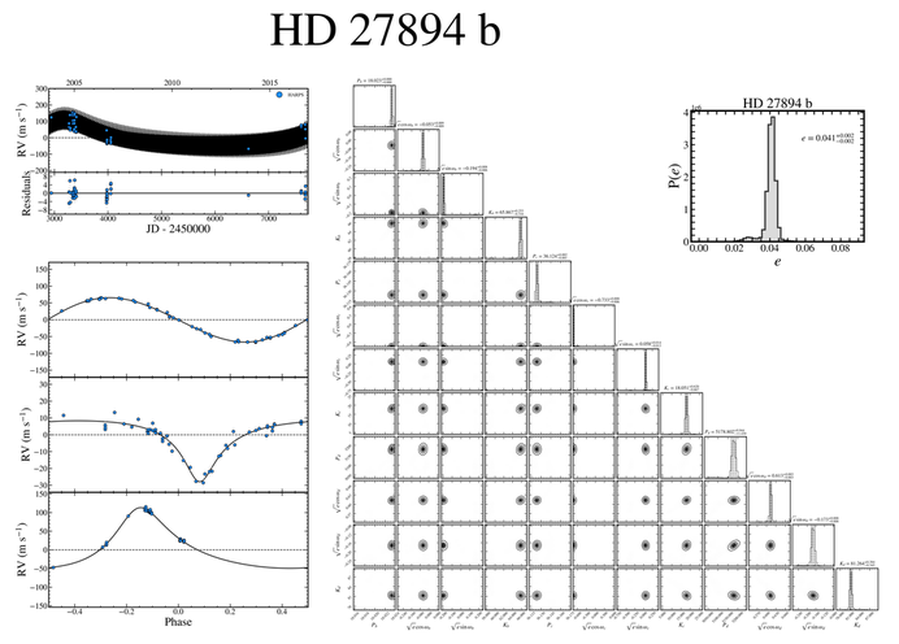}
 \end{minipage}
 \caption{Summary of results for the warm Jupiters HD 233604 b and HD 27894 b.}
 \label{fig:Combined_Plots44}
\end{figure}

\begin{figure}
\hskip -0.8 in
 \centering
 \begin{minipage}{\textwidth}
   \centering
   \includegraphics[width=\linewidth]{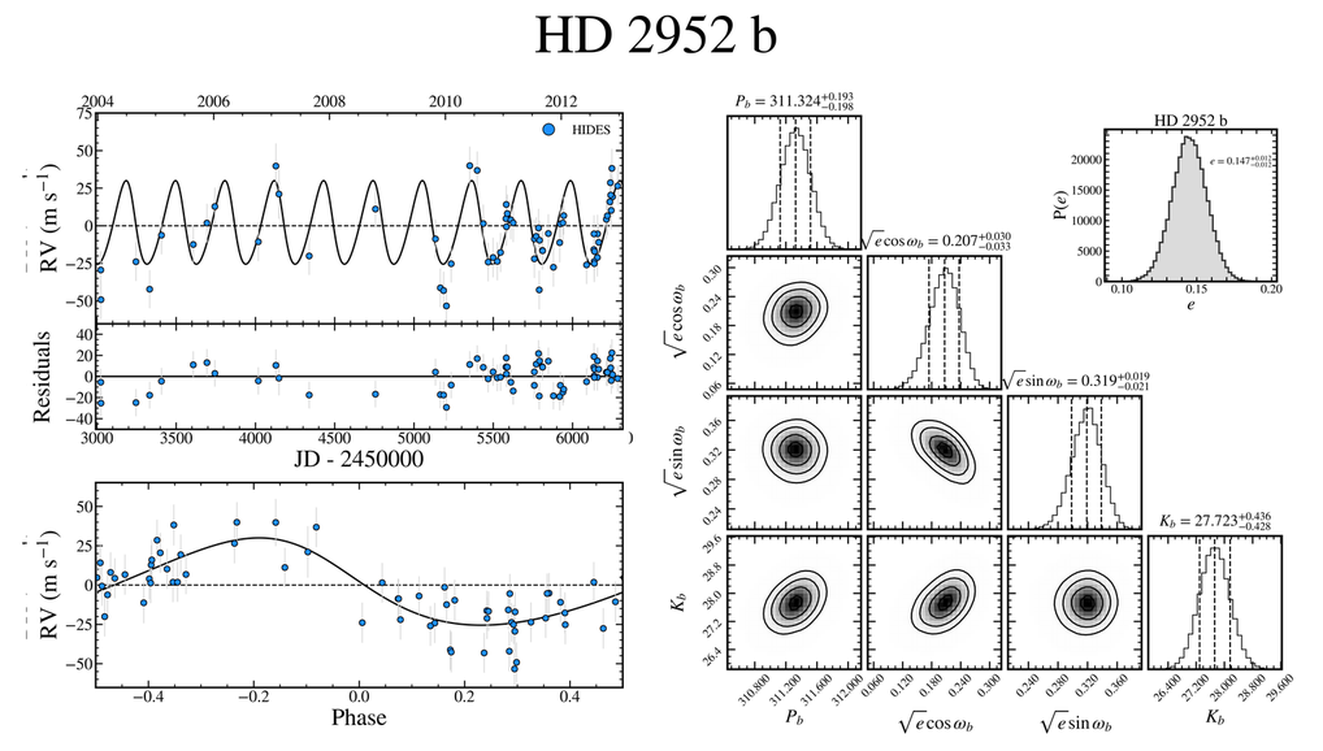}\\
   \vskip .3 in
   \includegraphics[width=\linewidth]{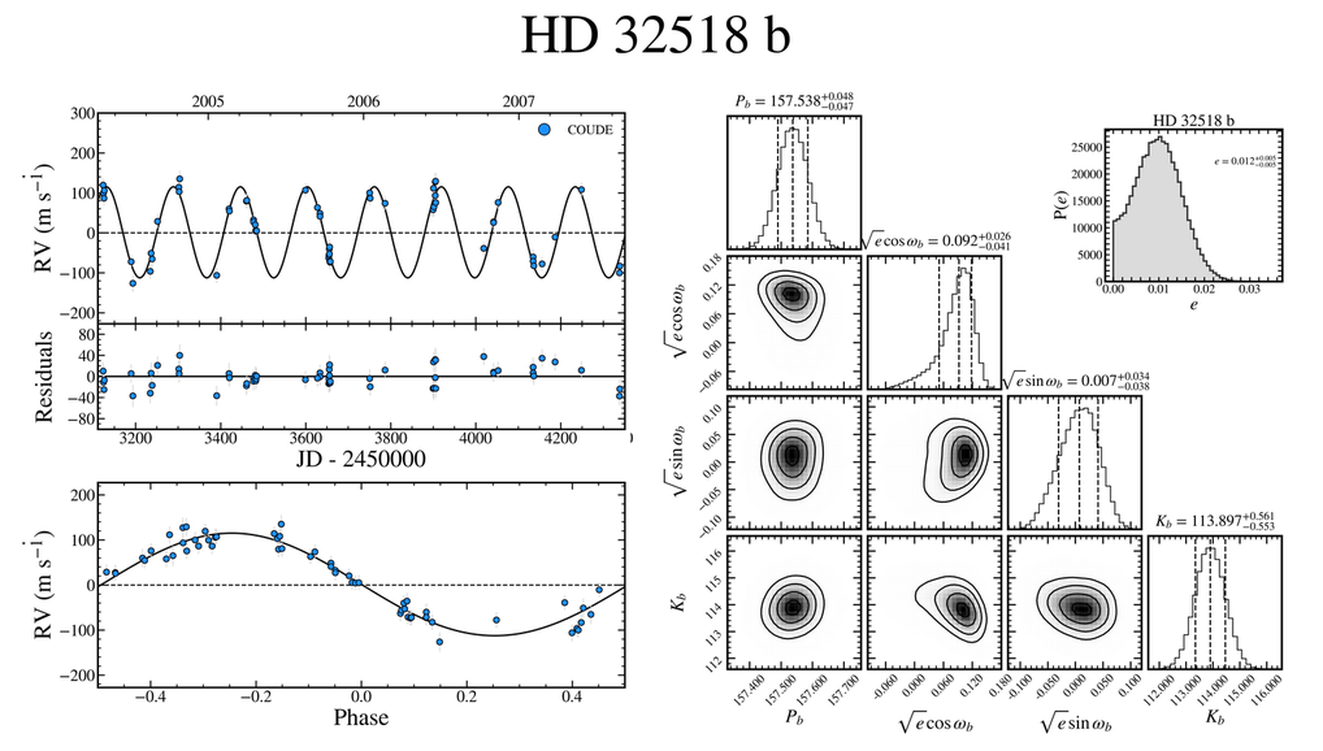}
 \end{minipage}
 \caption{Summary of results for the warm Jupiters HD 2952 b and HD 32518 b.}
 \label{fig:Combined_Plots45}
\end{figure}

\begin{figure}
\hskip -0.8 in
 \centering
 \begin{minipage}{\textwidth}
   \centering
   \includegraphics[width=\linewidth]{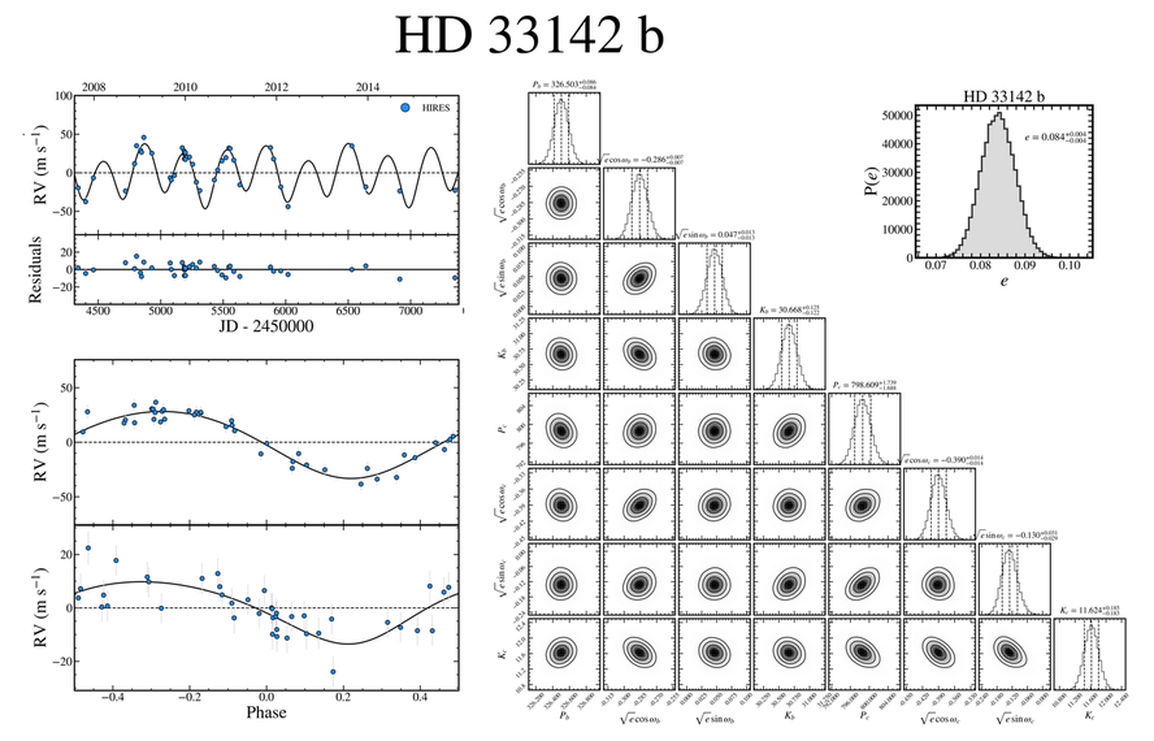}\\
   \vskip .3 in
   \includegraphics[width=\linewidth]{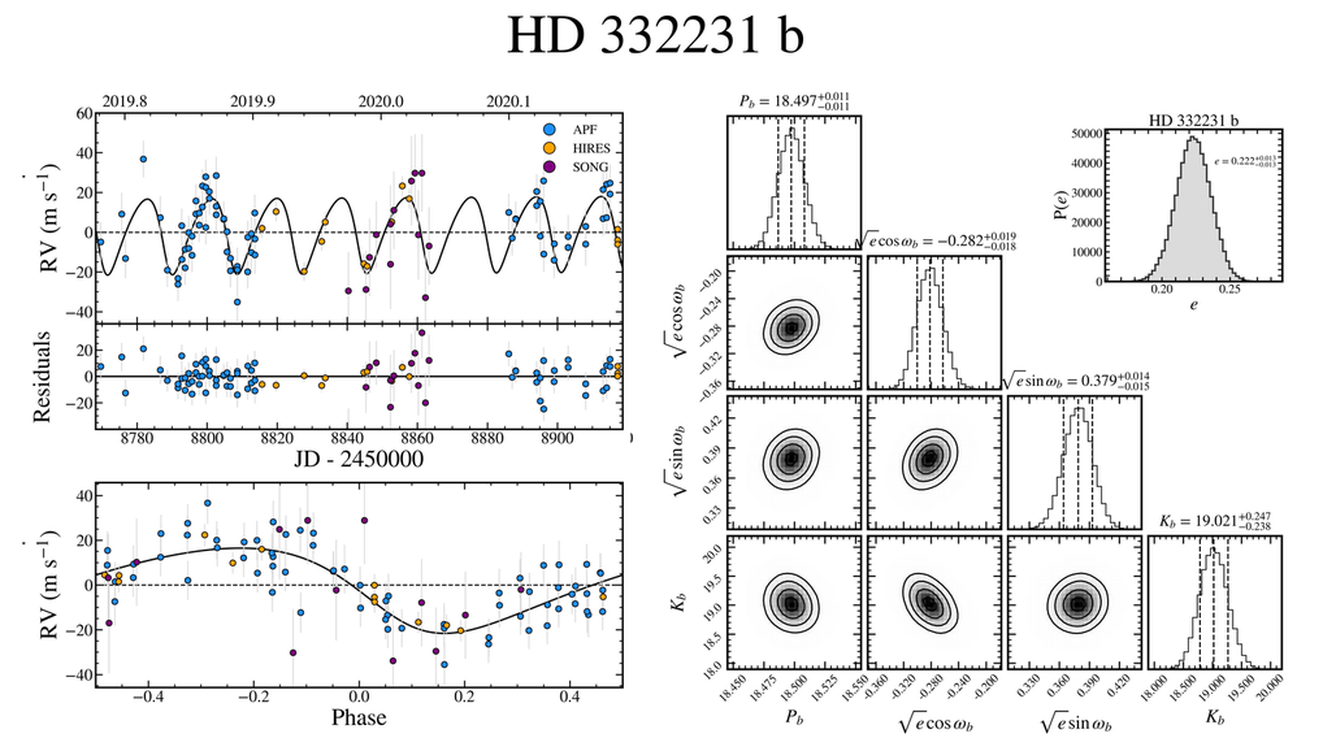}
 \end{minipage}
 \caption{Summary of results for the warm Jupiters HD 33142 b and HD 332231 b.}
 \label{fig:Combined_Plots46}
\end{figure}

\begin{figure}
\hskip -0.8 in
 \centering
 \begin{minipage}{\textwidth}
   \centering
   \includegraphics[width=\linewidth]{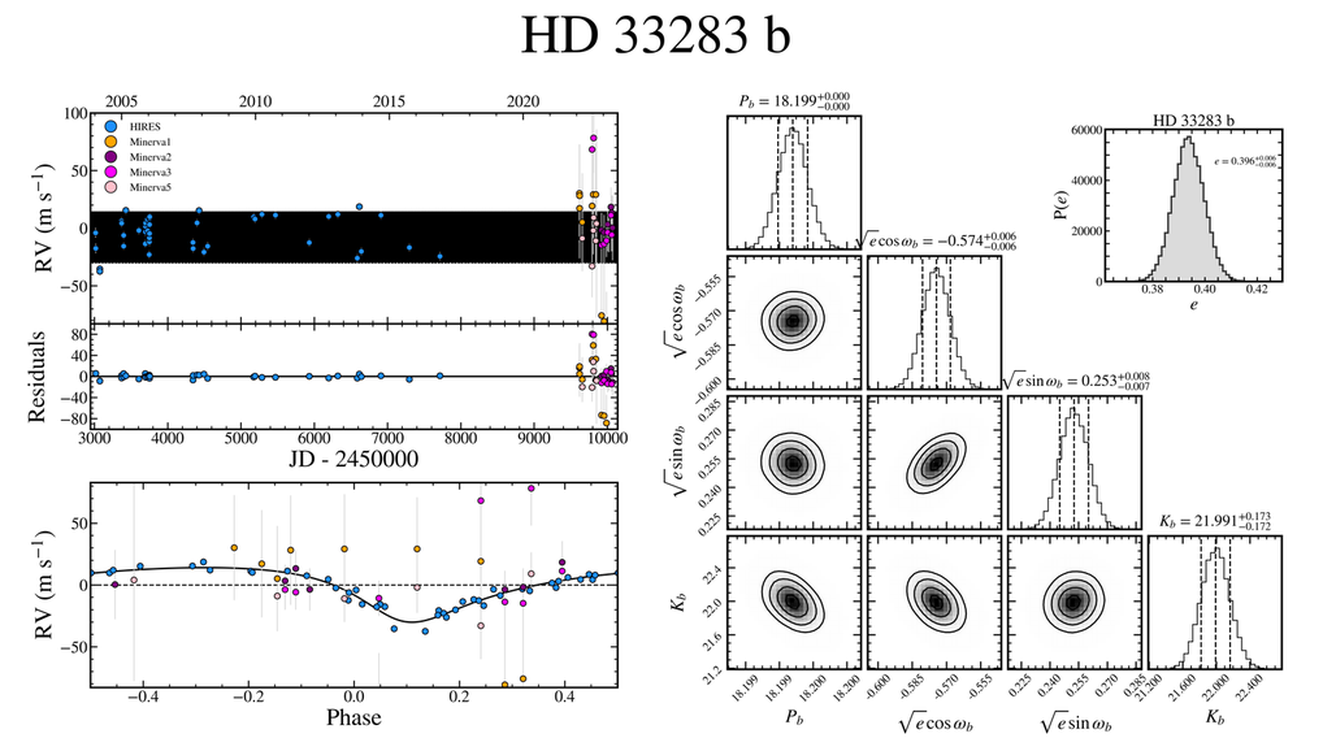}\\
   \vskip .3 in
   \includegraphics[width=\linewidth]{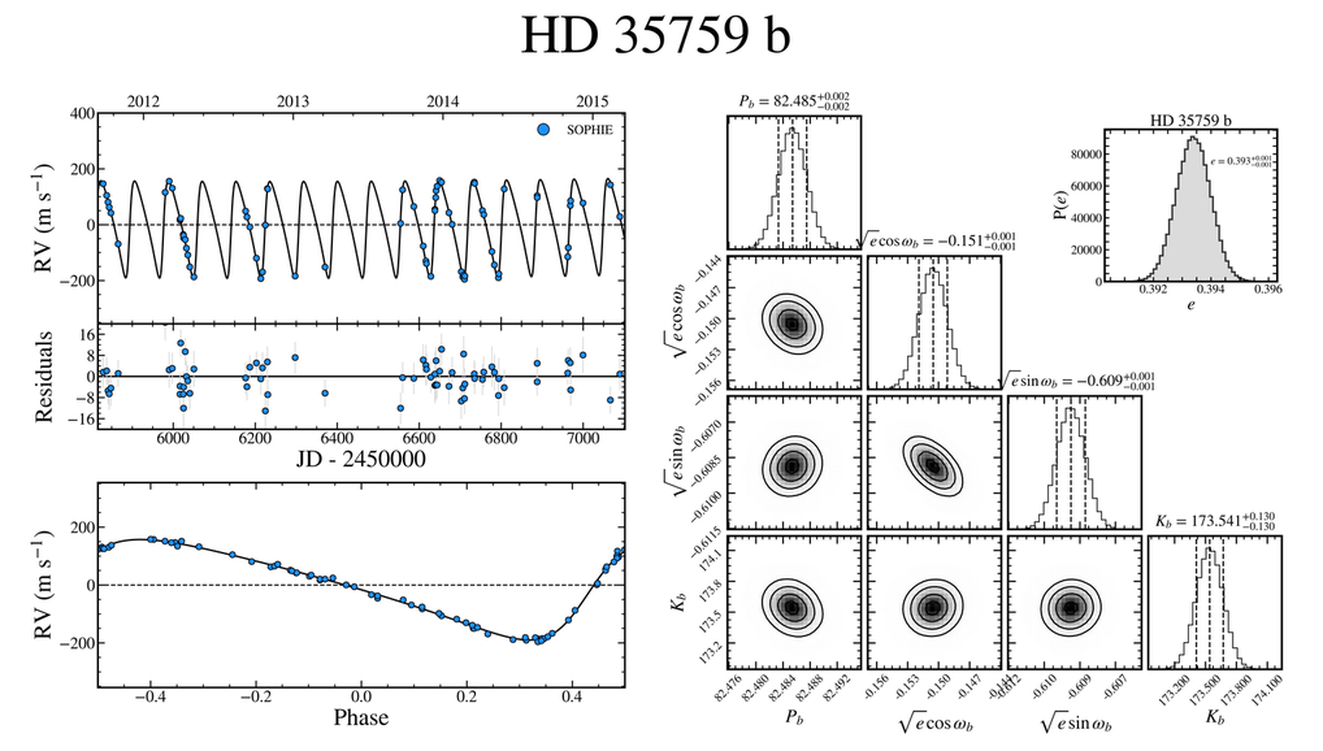}
 \end{minipage}
 \caption{Summary of results for the warm Jupiters HD 33283 b and HD 35759 b.}
 \label{fig:Combined_Plots47}
\end{figure}

\begin{figure}
\hskip -0.8 in
 \centering
 \begin{minipage}{\textwidth}
   \centering
   \includegraphics[width=\linewidth]{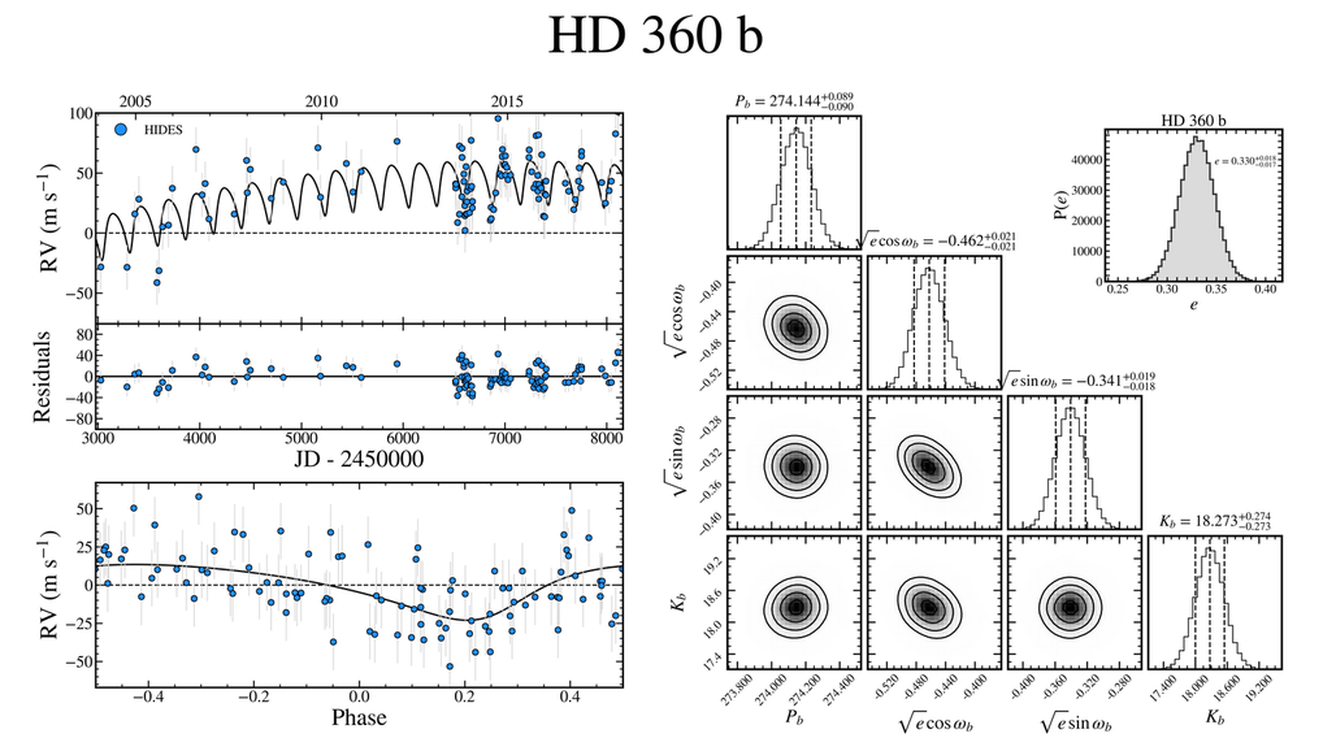}\\
   \vskip .3 in
   \includegraphics[width=\linewidth]{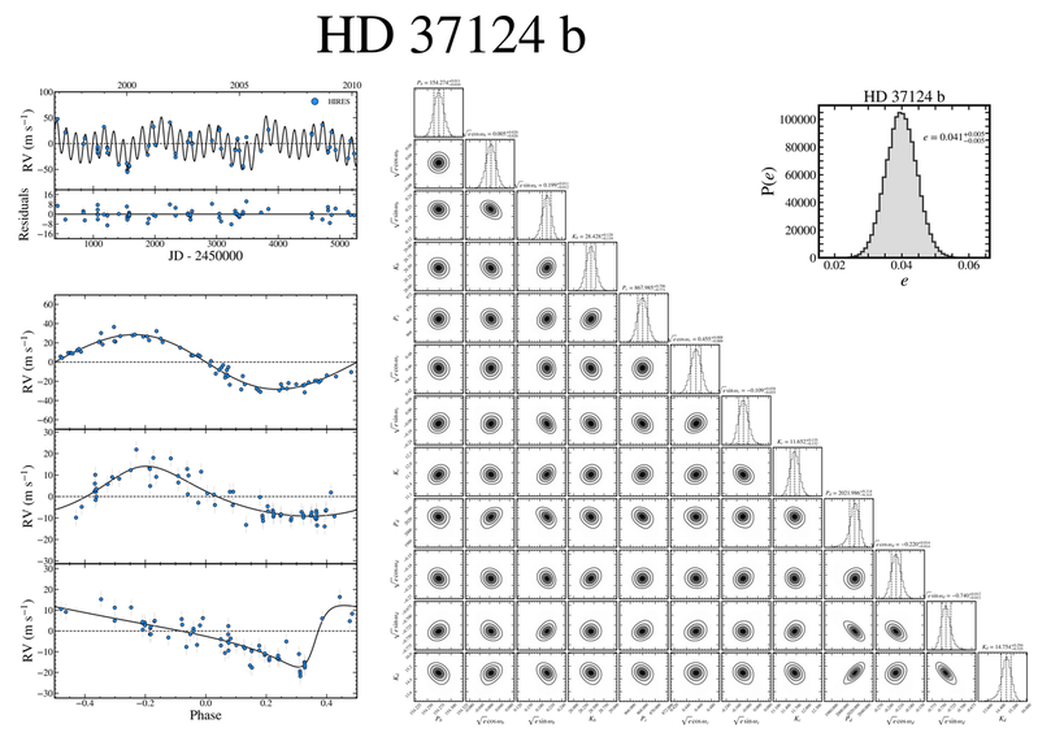}
 \end{minipage}
 \caption{Summary of results for the warm Jupiters HD 360 b and HD 37124 b.}
 \label{fig:Combined_Plots48}
\end{figure}

\begin{figure}
\hskip -0.8 in
 \centering
 \begin{minipage}{\textwidth}
   \centering
   \includegraphics[width=\linewidth]{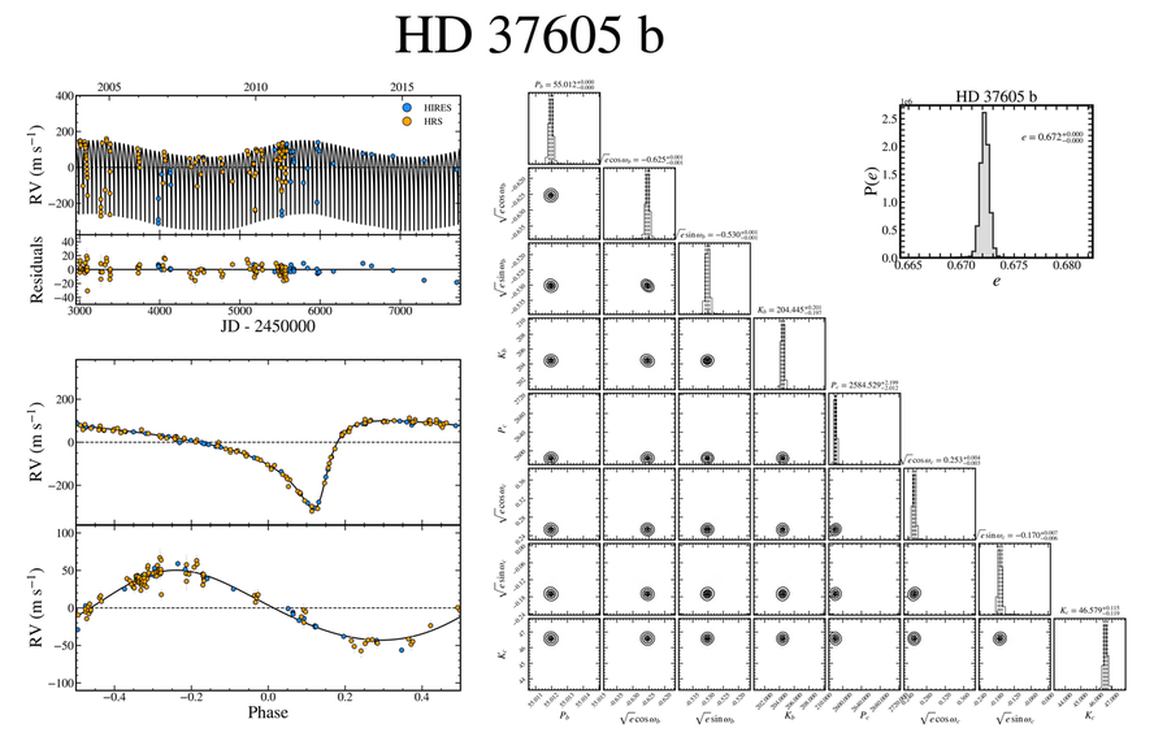}\\
   \vskip .3 in
   \includegraphics[width=\linewidth]{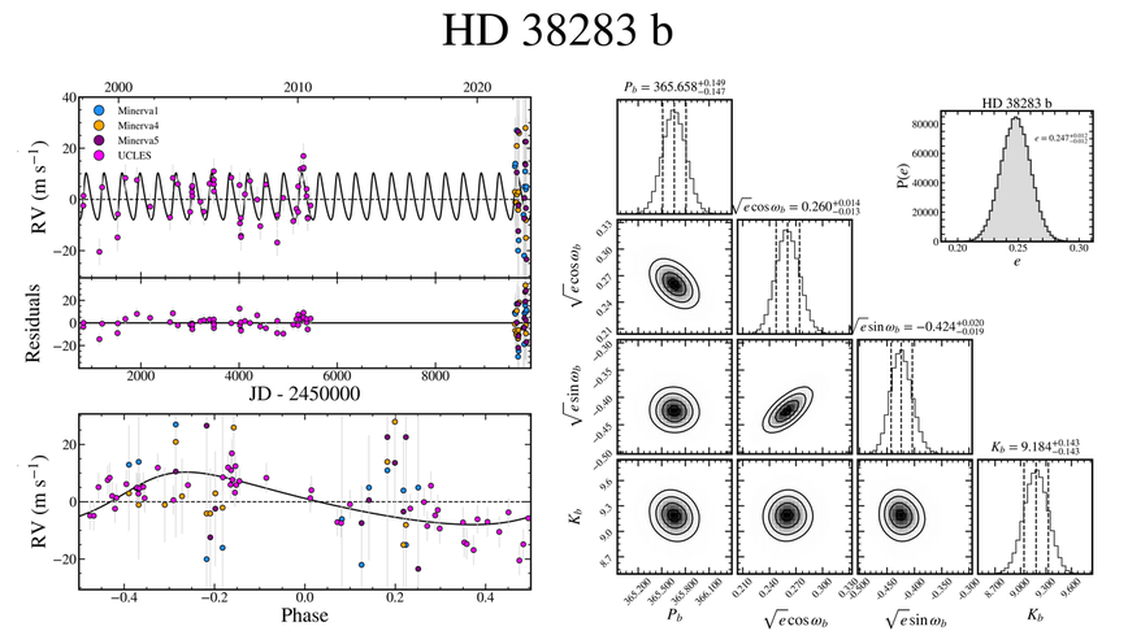}
 \end{minipage}
 \caption{Summary of results for the warm Jupiters HD 37605 b and HD 38283 b.}
 \label{fig:Combined_Plots49}
\end{figure}

\begin{figure}
\hskip -0.8 in
 \centering
 \begin{minipage}{\textwidth}
   \centering
   \includegraphics[width=\linewidth]{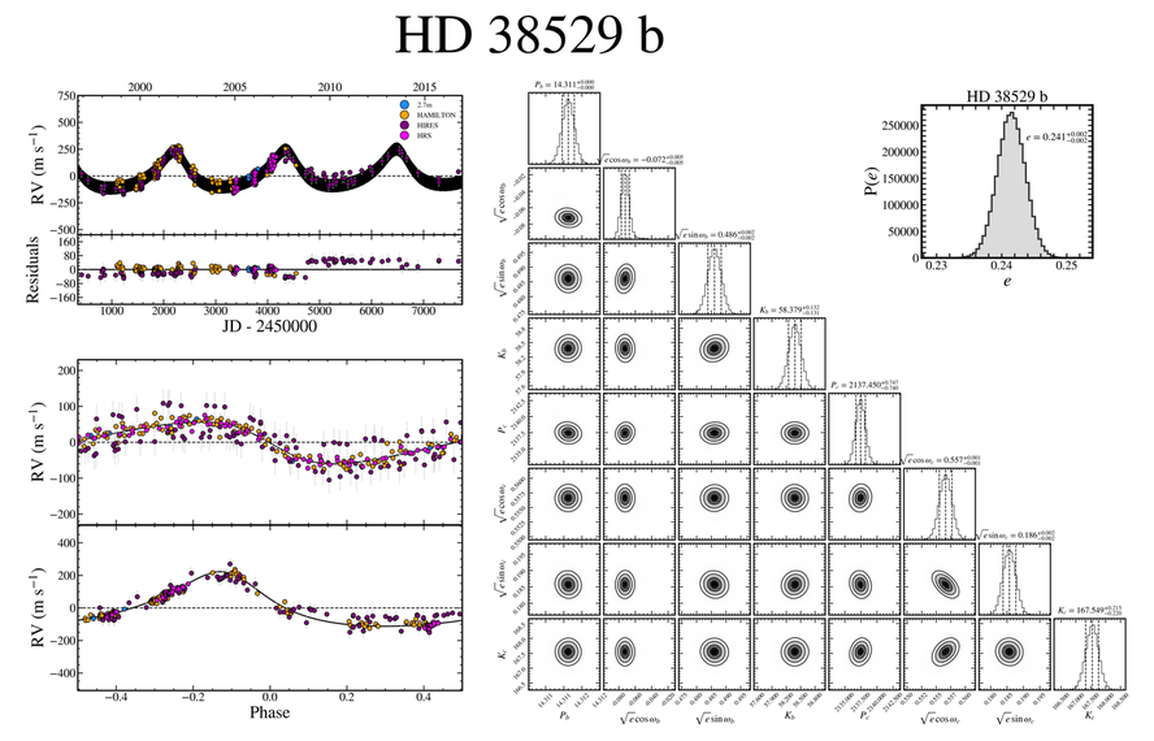}\\
   \vskip .3 in
   \includegraphics[width=\linewidth]{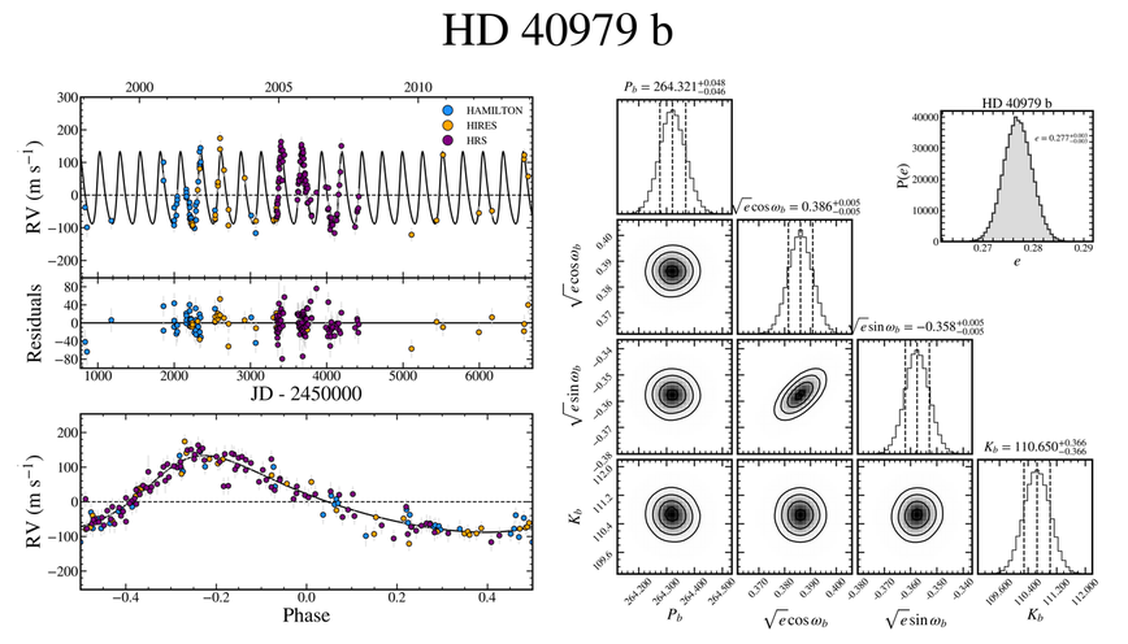}
 \end{minipage}
 \caption{Summary of results for the warm Jupiters HD 38529 b and HD 40979 b.}
 \label{fig:Combined_Plots50}
\end{figure}

\begin{figure}
\hskip -0.8 in
 \centering
 \begin{minipage}{\textwidth}
   \centering
   \includegraphics[width=\linewidth]{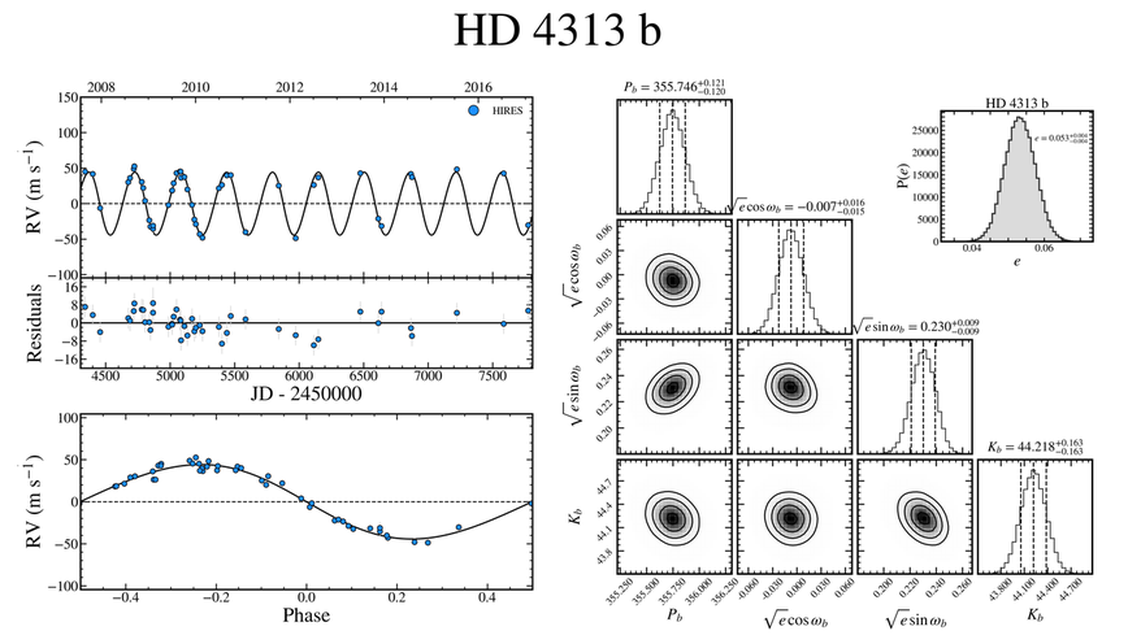}\\
   \vskip .3 in
   \includegraphics[width=\linewidth]{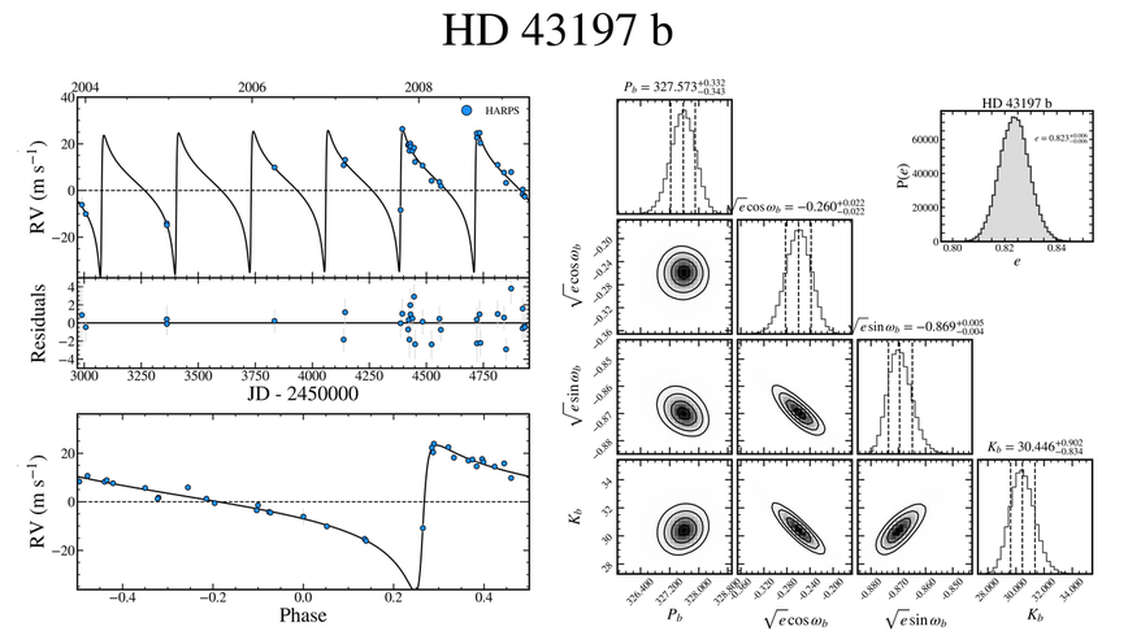}
 \end{minipage}
 \caption{Summary of results for the warm Jupiters HD 4313 b and HD 43197 b.}
 \label{fig:Combined_Plots51}
\end{figure}

\begin{figure}
\hskip -0.8 in
 \centering
 \begin{minipage}{\textwidth}
   \centering
   \includegraphics[width=\linewidth]{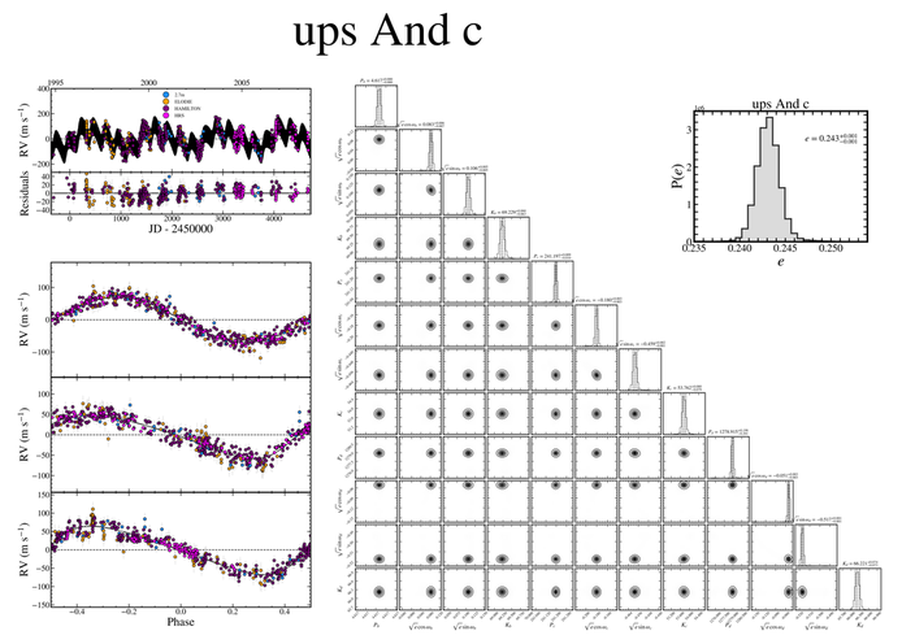}\\
   \vskip .3 in
   \includegraphics[width=\linewidth]{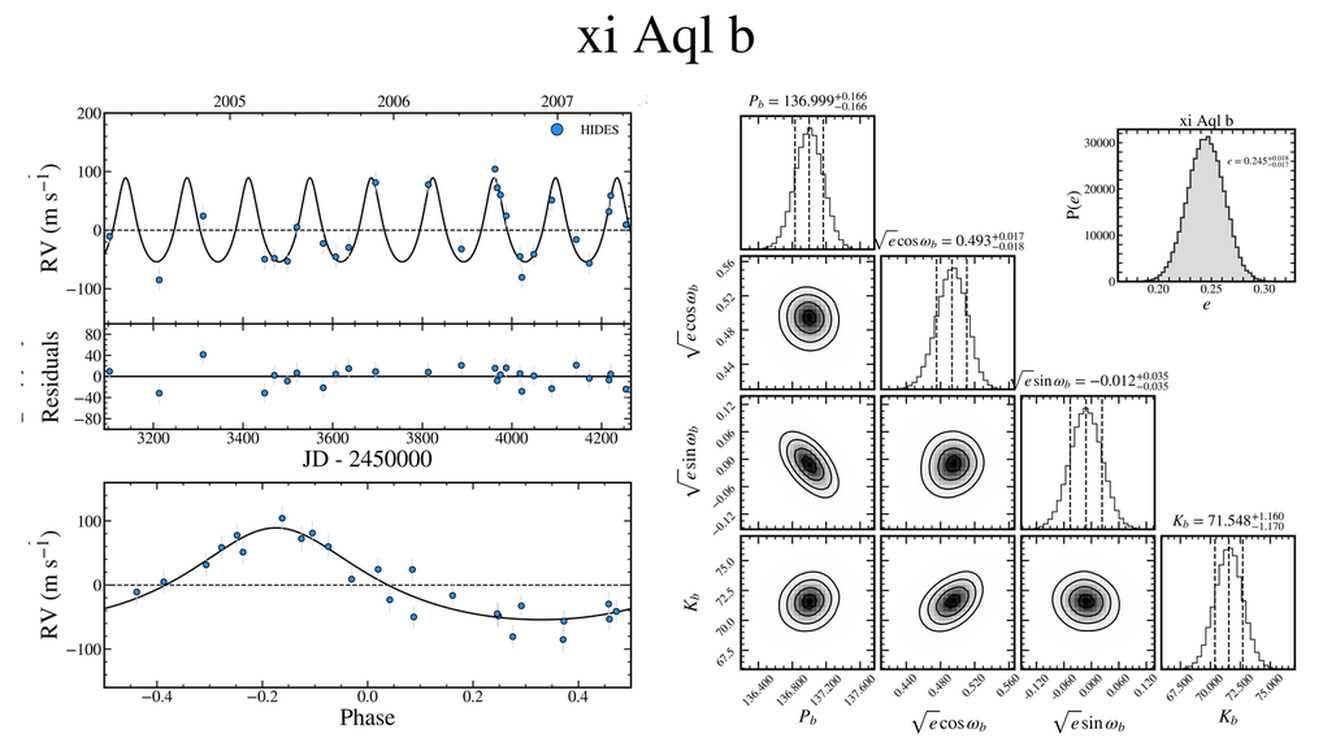}
 \end{minipage}
 \caption{Summary of results for the warm Jupiters ups And c and xi Aql b.}
 \label{fig:Combined_Plots100}
\end{figure}

Results of our uniform refitting for all 200 warm Jupiters.  When an RV trend or change in radial acceleration is preferred, this best-fit model is shown. Figures \ref{fig:Combined_Plots1}--\ref{fig:Combined_Plots100} show the best fit model compared to the time-series RVs and phase-folded light curve, along with corner plots for $P$, $\sqrt{e}$ $\cos \omega$, $\sqrt{e}$ $\sin \omega$, and $K$. A histogram of the eccentricity ``$e$" posterior distribution is also shown. We provide the posterior chains used in the Exploring Warm Jupiter Migration Pathways With Eccentricities series on Zenodo for public access (DOI: 10.5281/zenodo.17161693).


\clearpage
\clearpage
\clearpage
\clearpage
\clearpage
\clearpage
\clearpage
\clearpage
\clearpage
\clearpage
\clearpage
\clearpage
\clearpage
\clearpage
\clearpage
\clearpage
\clearpage
\clearpage
\clearpage
\clearpage
\clearpage
\clearpage
\clearpage
\clearpage
\clearpage
\clearpage
\clearpage
\clearpage
\clearpage
\clearpage
\clearpage
\clearpage
\clearpage
\clearpage
\clearpage
\clearpage
\clearpage
\clearpage
\clearpage
\clearpage
\clearpage
\clearpage
\clearpage
\clearpage
\clearpage
\clearpage
\clearpage
\clearpage
\clearpage
\clearpage
\clearpage
\begin{figure}
\hskip -0.8 in
 \centering
 \begin{minipage}{\textwidth}
   \centering
   \includegraphics[width=\linewidth]{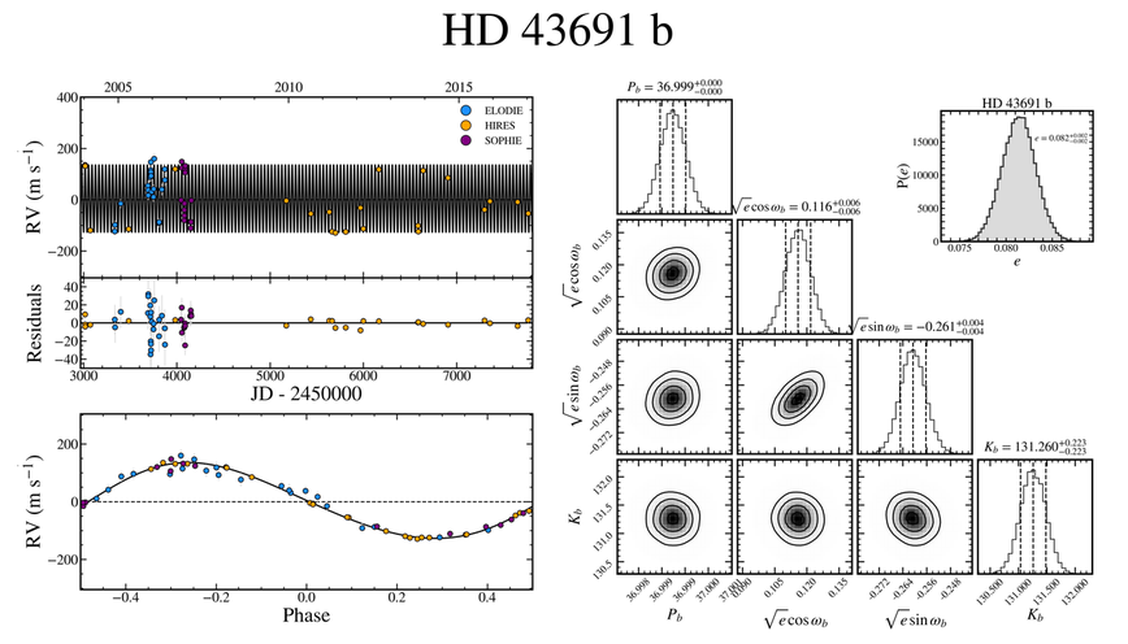}\\
   \vskip .3 in
   \includegraphics[width=\linewidth]{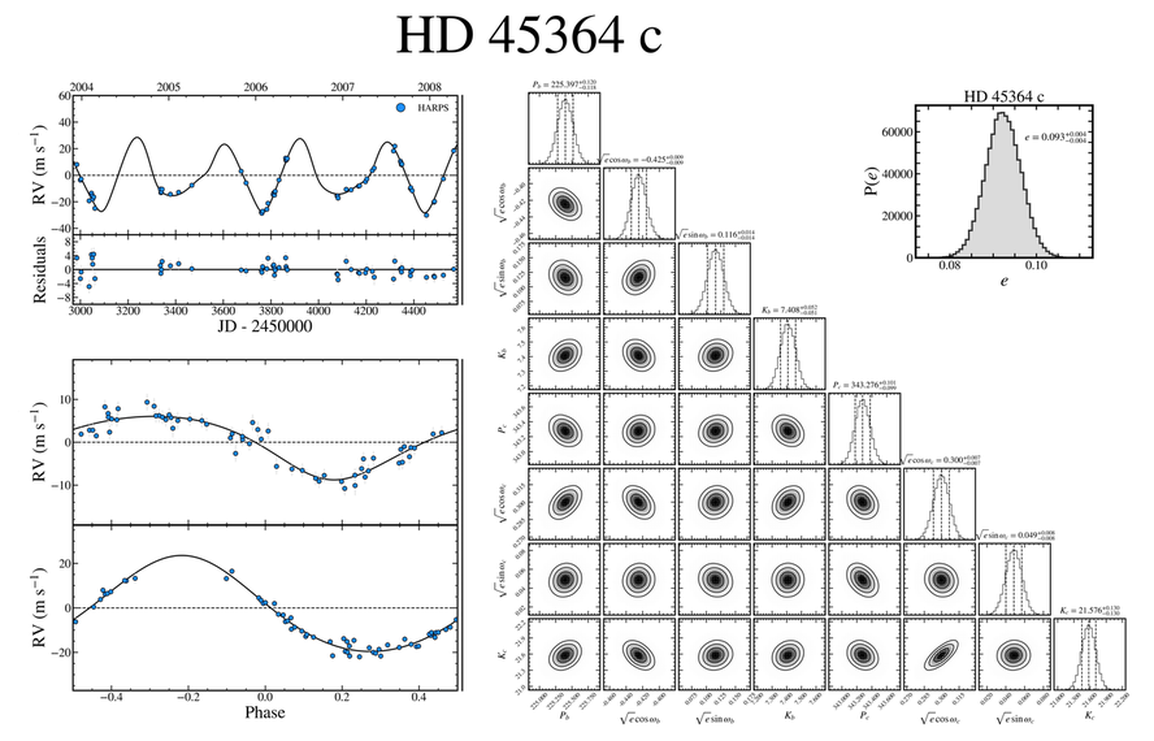}
 \end{minipage}
 \caption{Summary of results for the warm Jupiters HD 43691 b and HD 45364 c.}
 \label{fig:Combined_Plots52}
\end{figure}
\clearpage
\begin{figure}
\hskip -0.8 in
 \centering
 \begin{minipage}{\textwidth}
   \centering
   \includegraphics[width=\linewidth]{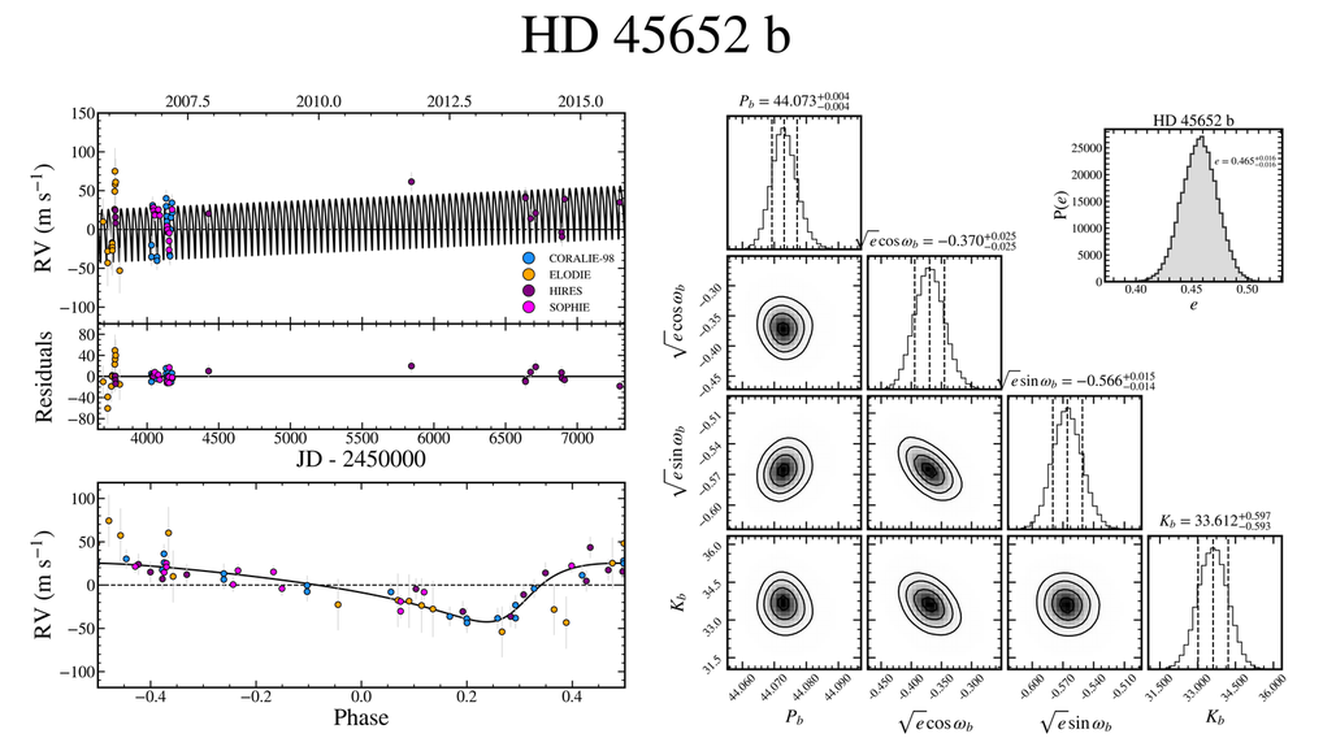}\\
   \vskip .3 in
   \includegraphics[width=\linewidth]{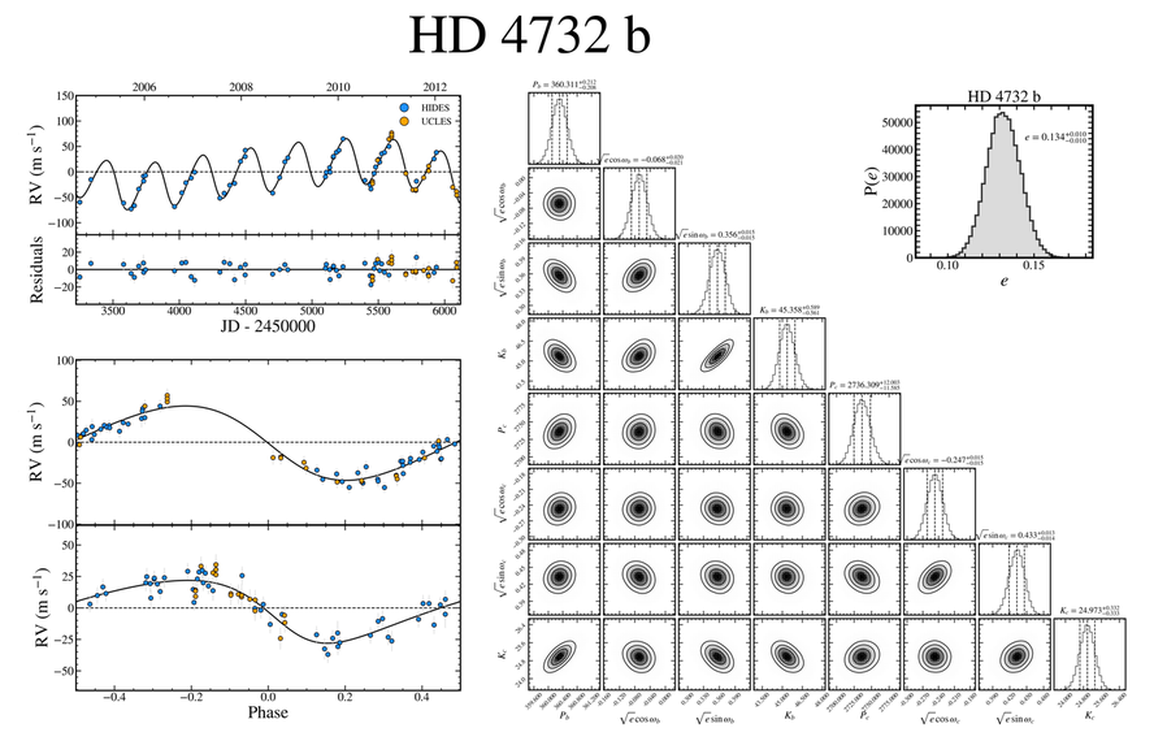}
 \end{minipage}
 \caption{Summary of results for the warm Jupiters HD 45652 b and HD 4732 b.}
 \label{fig:Combined_Plots53}
\end{figure}
\clearpage
\begin{figure}
\hskip -0.8 in
 \centering
 \begin{minipage}{\textwidth}
   \centering
   \includegraphics[width=\linewidth]{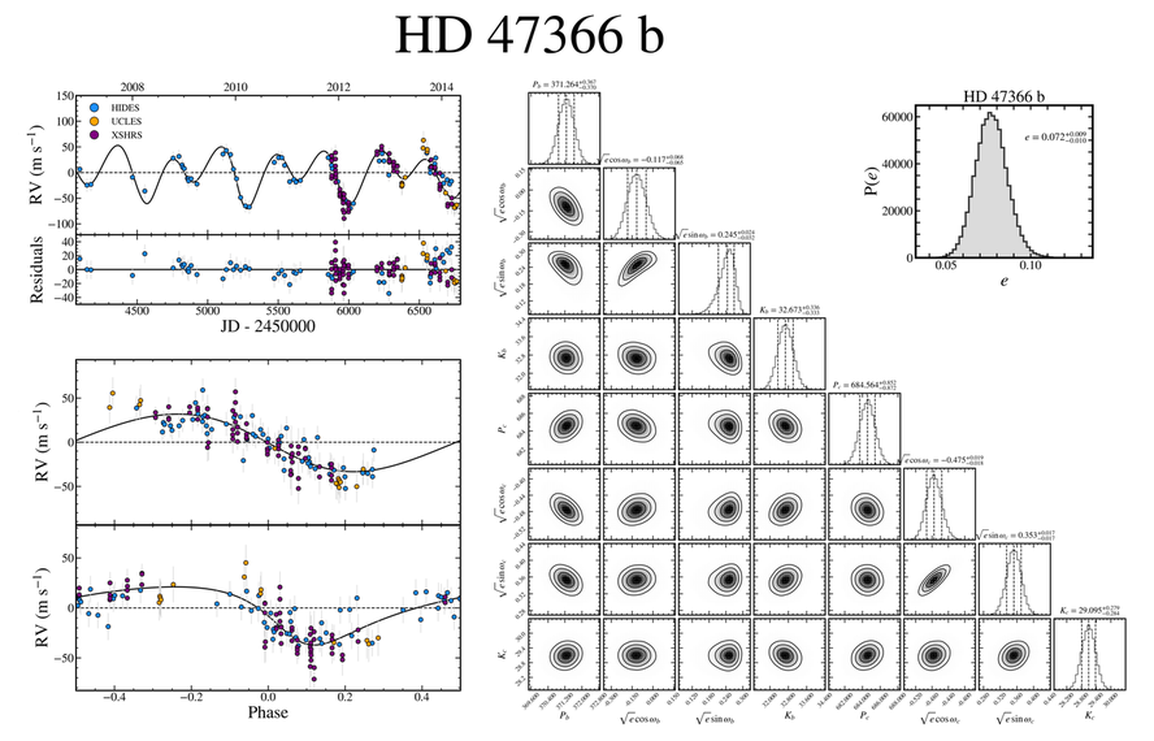}\\
   \vskip .3 in
   \includegraphics[width=\linewidth]{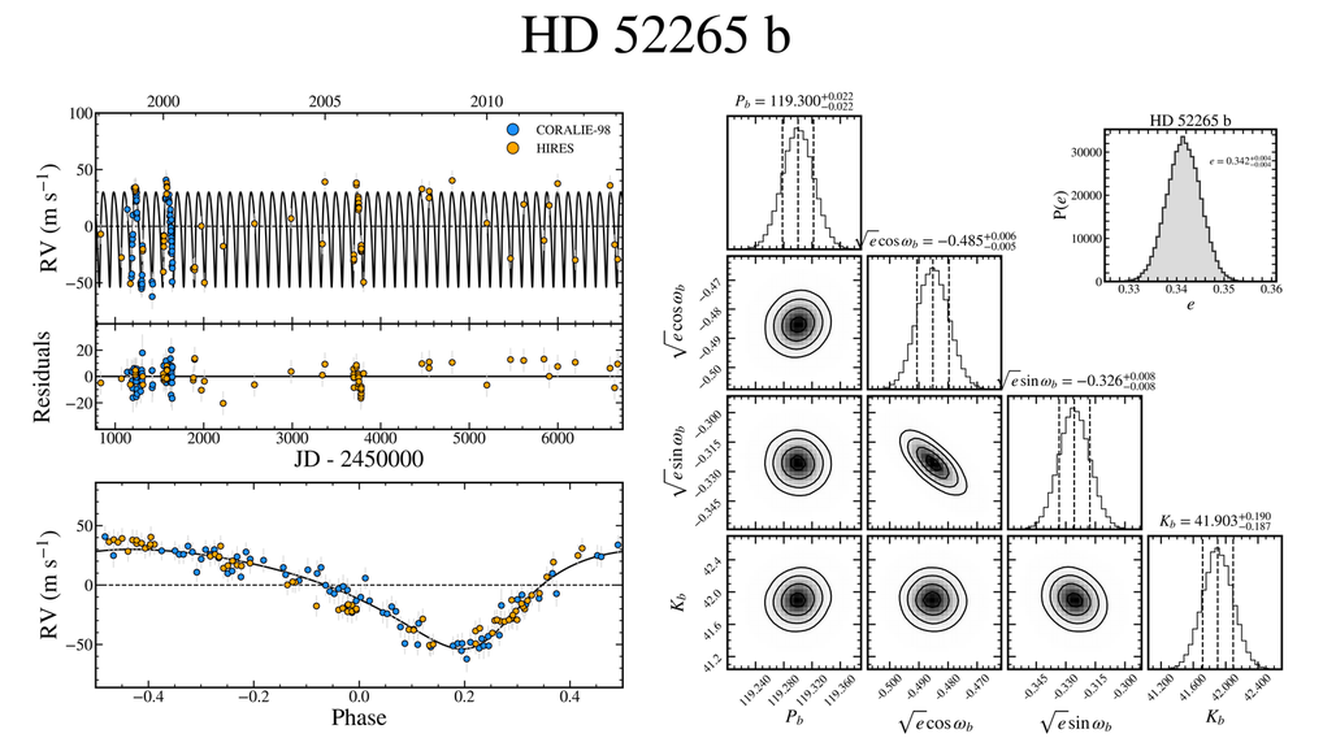}
 \end{minipage}
 \caption{Summary of results for the warm Jupiters HD 47366 b and HD 52265 b.}
 \label{fig:Combined_Plots54}
\end{figure}
\clearpage
\begin{figure}
\hskip -0.8 in
 \centering
 \begin{minipage}{\textwidth}
   \centering
   \includegraphics[width=\linewidth]{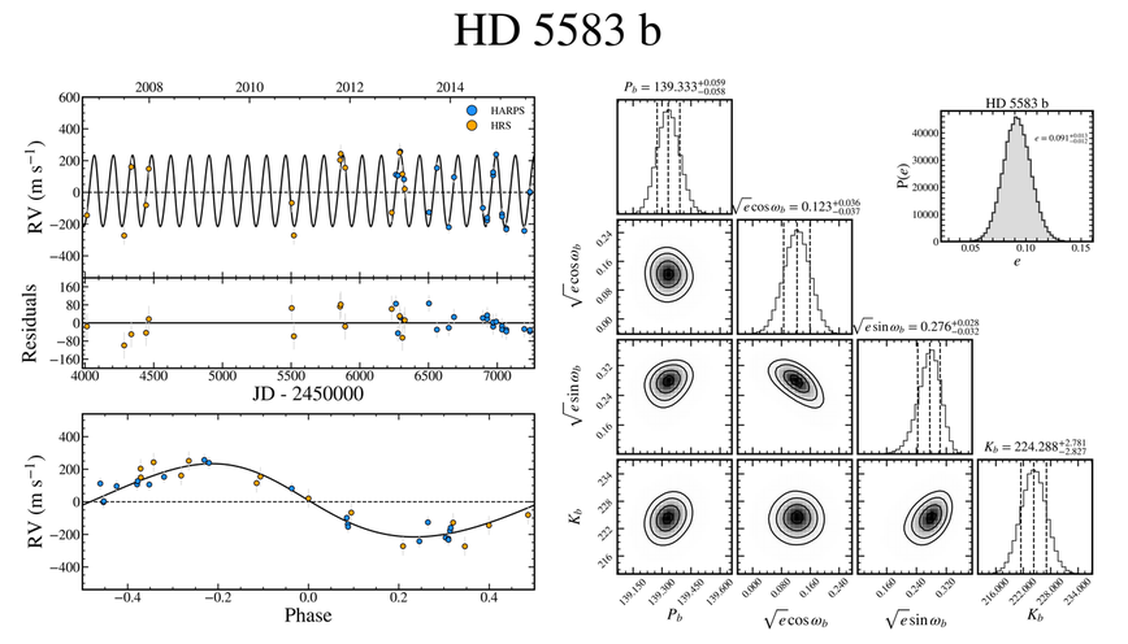}\\
   \vskip .3 in
   \includegraphics[width=\linewidth]{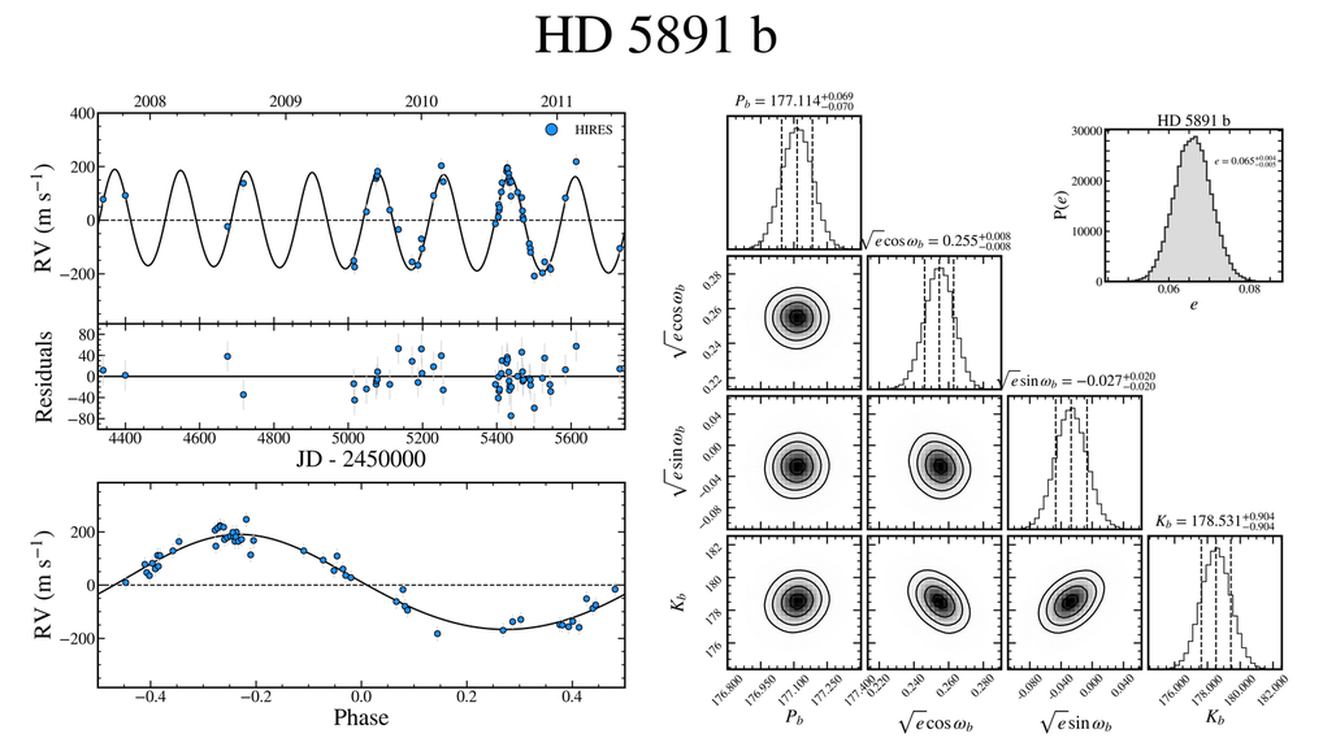}
 \end{minipage}
 \caption{Summary of results for the warm Jupiters HD 5583 b and HD 5891 b.}
 \label{fig:Combined_Plots55}
\end{figure}
\clearpage
\begin{figure}
\hskip -0.8 in
 \centering
 \begin{minipage}{\textwidth}
   \centering
   \includegraphics[width=\linewidth]{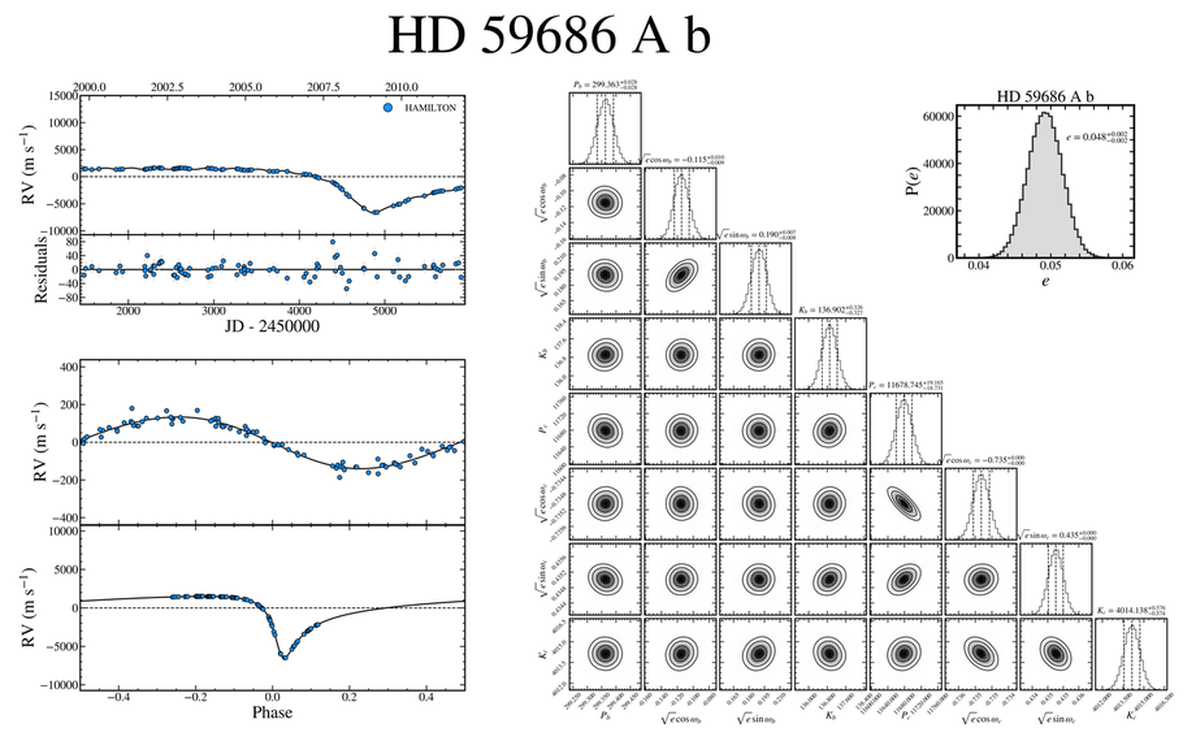}\\
   \vskip .3 in
   \includegraphics[width=\linewidth]{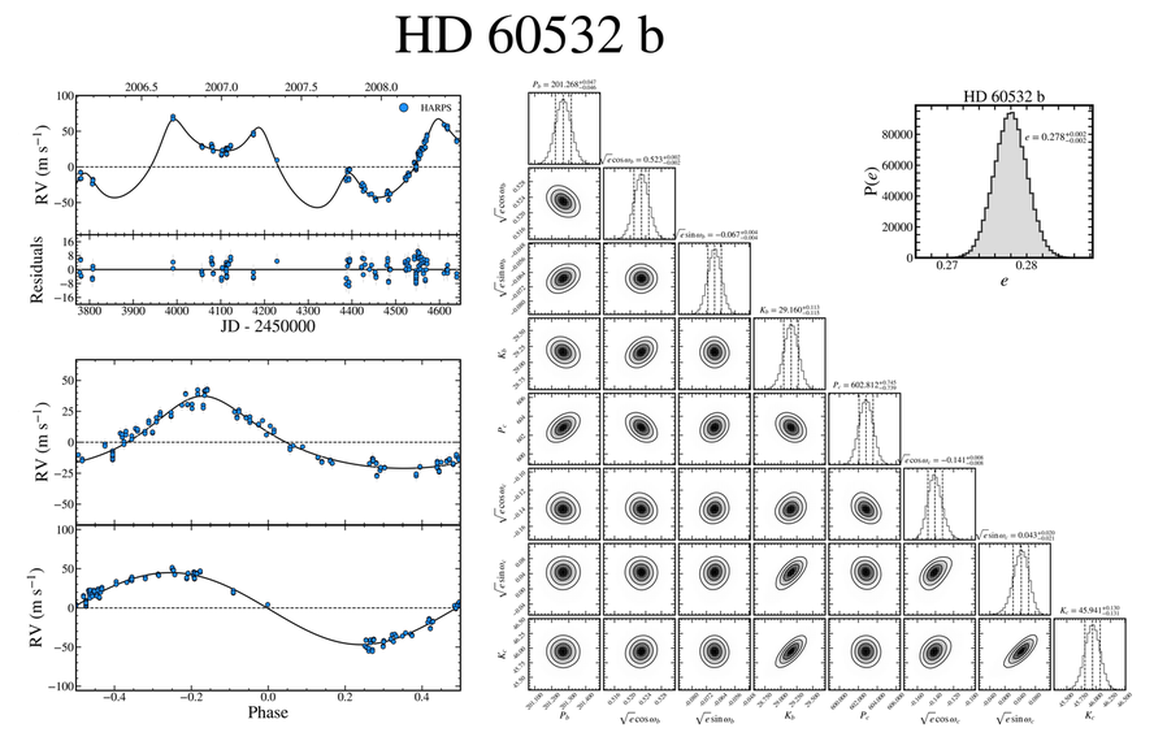}
 \end{minipage}
 \caption{Summary of results for the warm Jupiters HD 59686 A b and HD 60532 b.}
 \label{fig:Combined_Plots56}
\end{figure}
\clearpage
\begin{figure}
\hskip -0.8 in
 \centering
 \begin{minipage}{\textwidth}
   \centering
   \includegraphics[width=\linewidth]{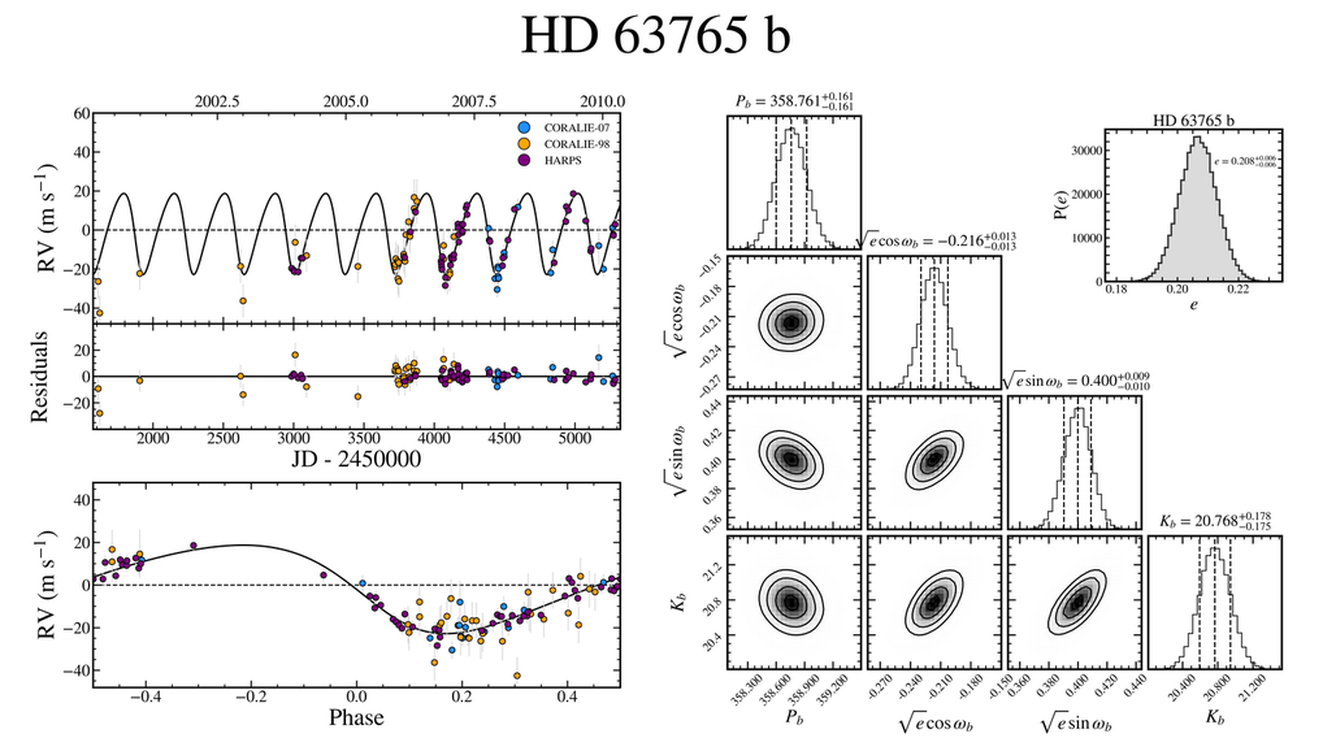}\\
   \vskip .3 in
   \includegraphics[width=\linewidth]{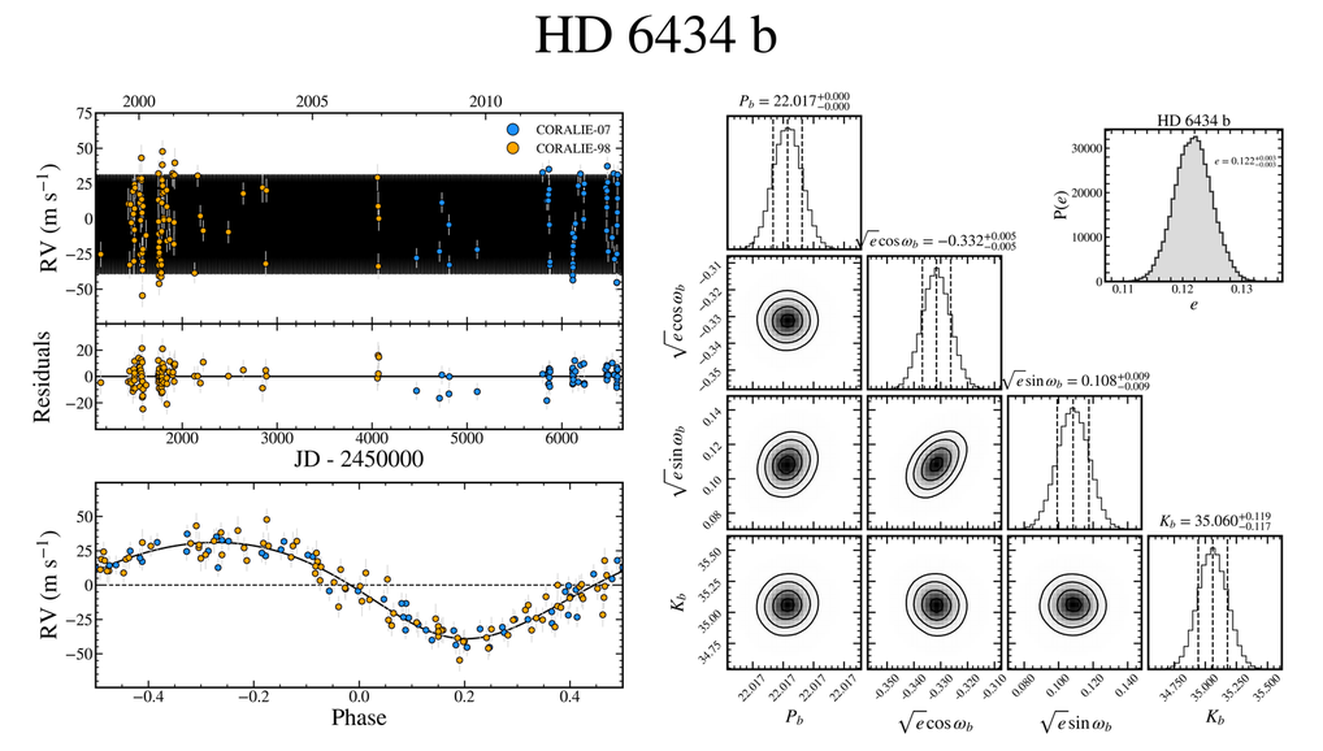}
 \end{minipage}
 \caption{Summary of results for the warm Jupiters HD 63765 b and HD 6434 b.}
 \label{fig:Combined_Plots57}
\end{figure}
\clearpage
\begin{figure}
\hskip -0.8 in
 \centering
 \begin{minipage}{\textwidth}
   \centering
   \includegraphics[width=\linewidth]{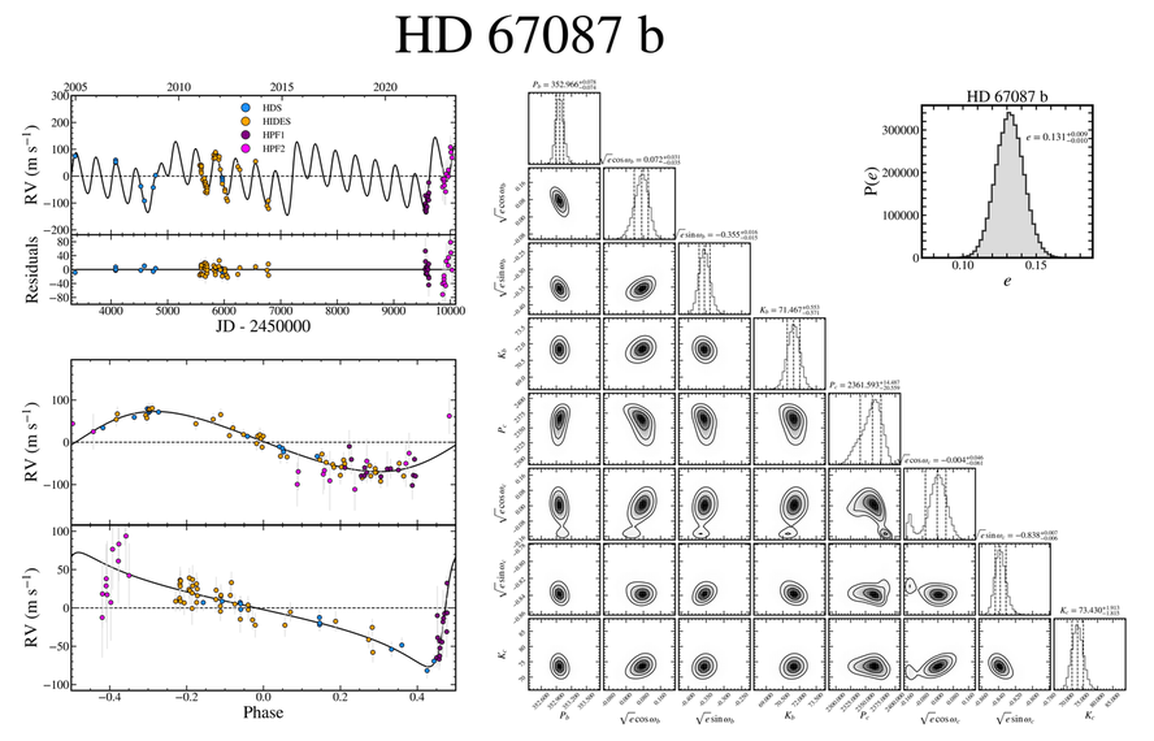}\\
   \vskip .3 in
   \includegraphics[width=\linewidth]{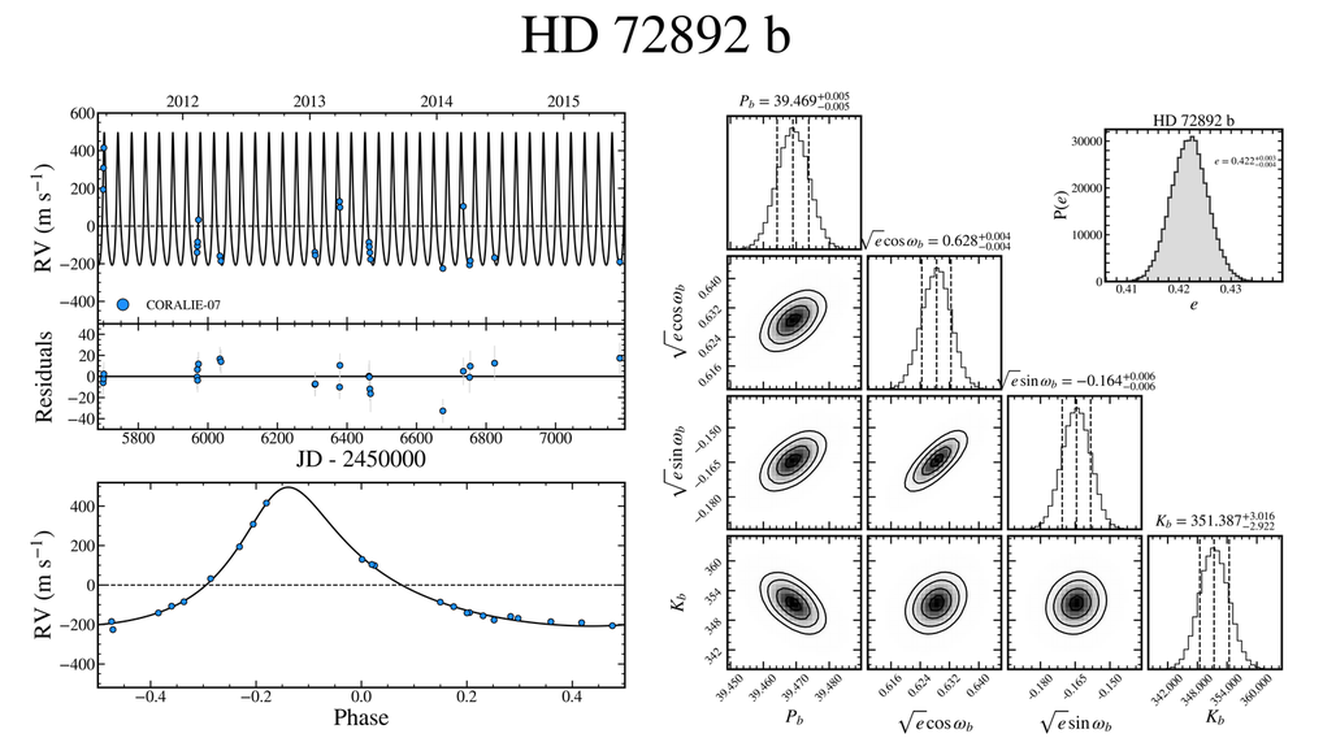}
 \end{minipage}
 \caption{Summary of results for the warm Jupiters HD 67087 b and HD 72892 b.}
 \label{fig:Combined_Plots58}
\end{figure}
\clearpage
\begin{figure}
\hskip -0.8 in
 \centering
 \begin{minipage}{\textwidth}
   \centering
   \includegraphics[width=\linewidth]{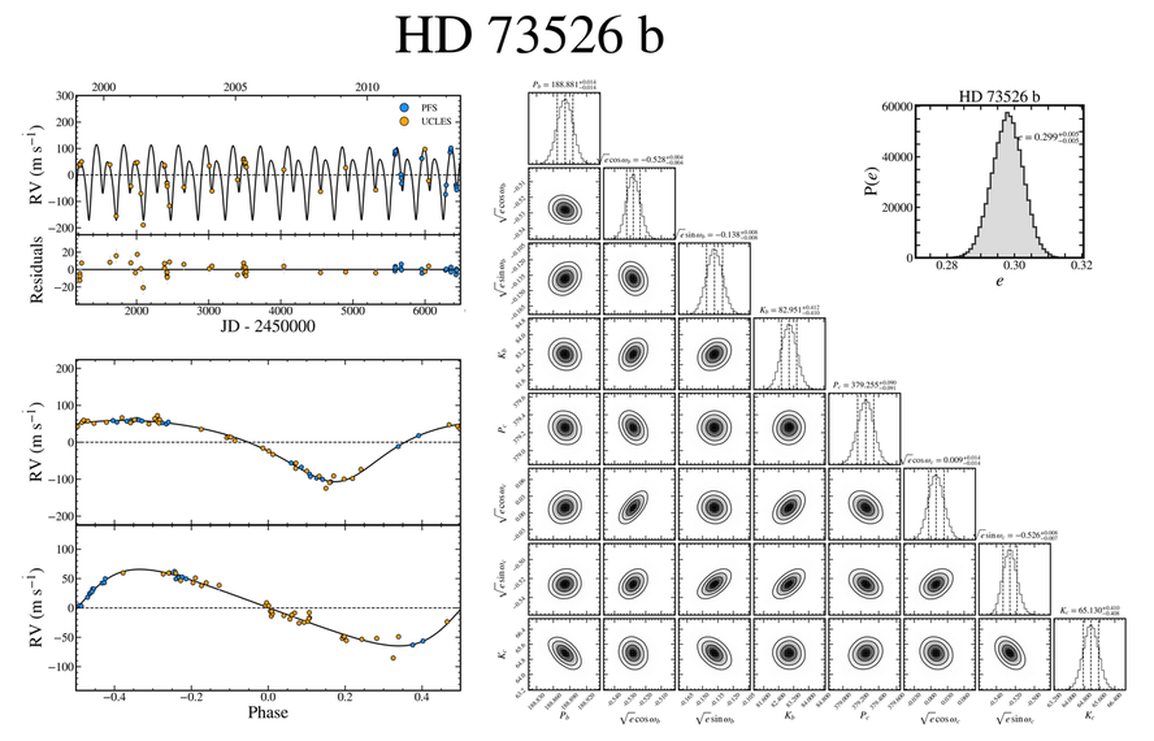}\\
   \vskip .3 in
   \includegraphics[width=\linewidth]{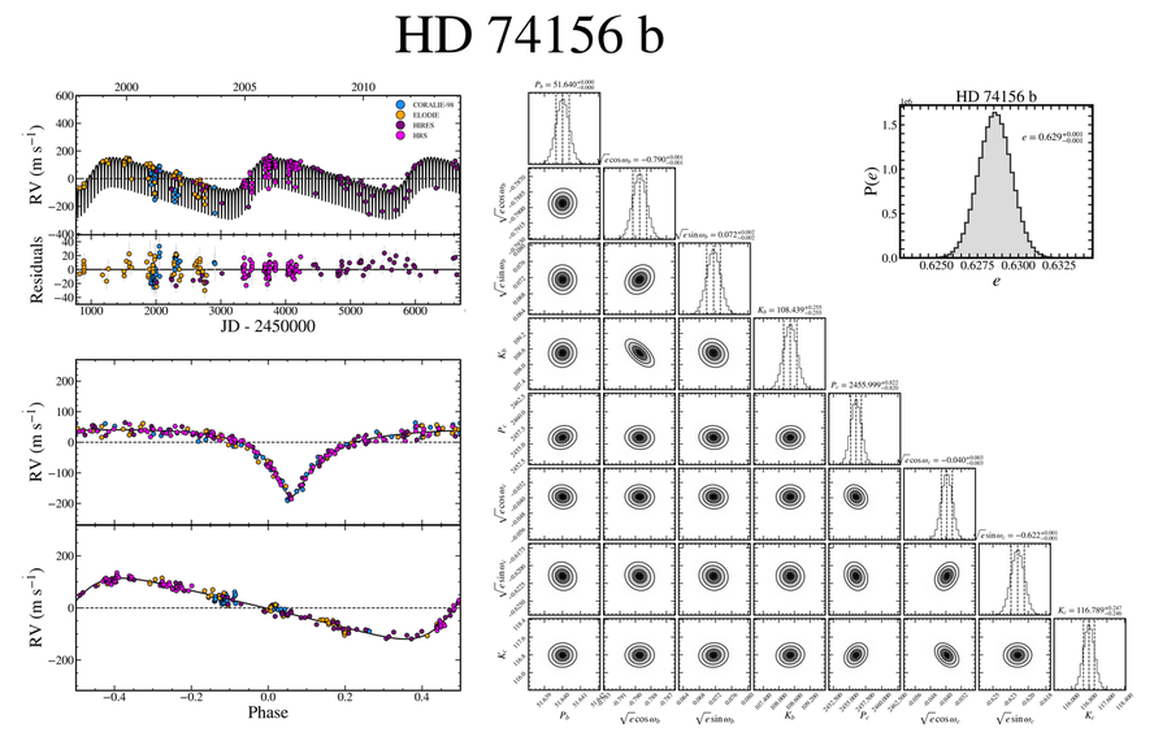}
 \end{minipage}
 \caption{Summary of results for the warm Jupiters HD 73526 b and HD 74156 b.}
 \label{fig:Combined_Plots59}
\end{figure}
\clearpage
\begin{figure}
\hskip -0.8 in
 \centering
 \begin{minipage}{\textwidth}
   \centering
   \includegraphics[width=\linewidth]{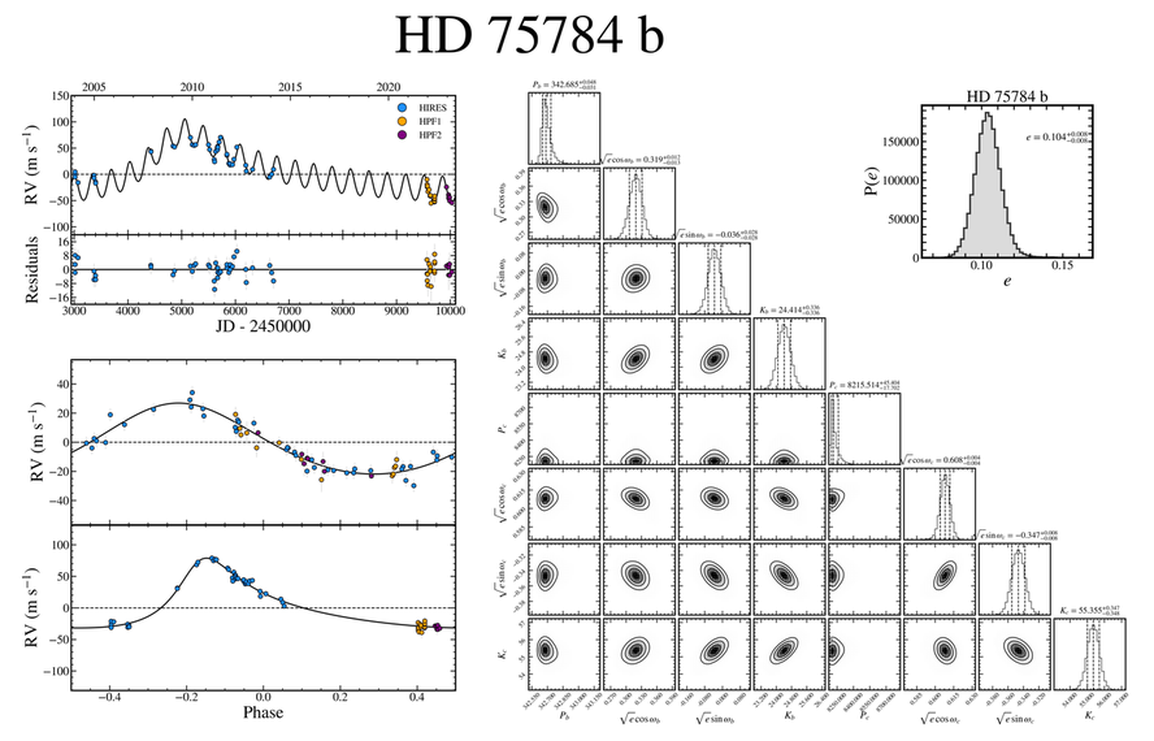}\\
   \vskip .3 in
   \includegraphics[width=\linewidth]{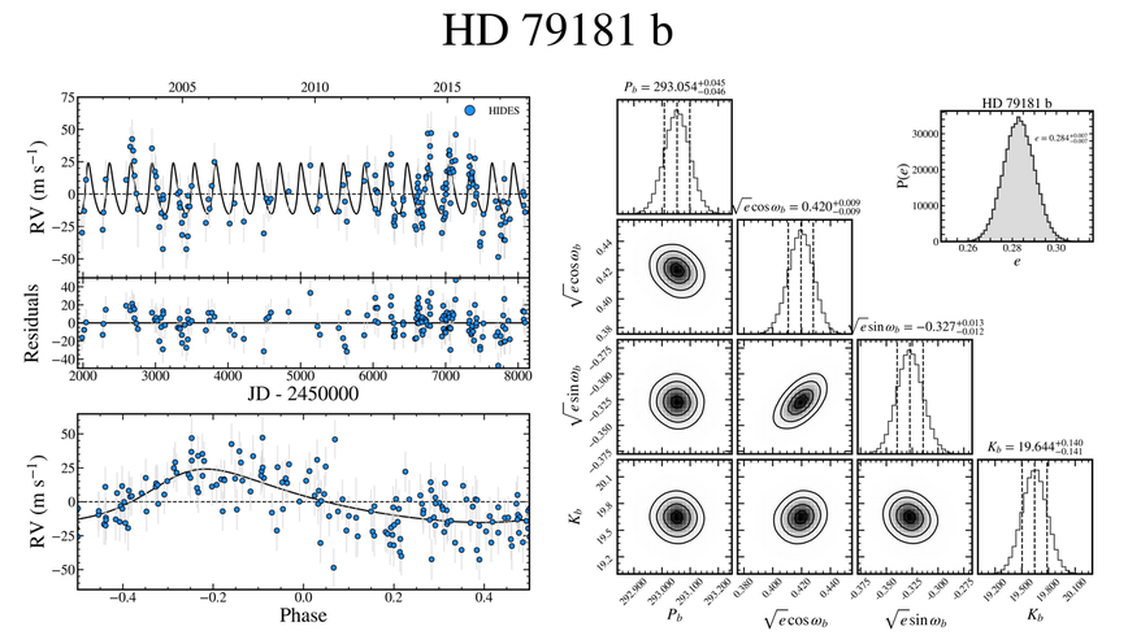}
 \end{minipage}
 \caption{Summary of results for the warm Jupiters HD 75784 b and HD 79181 b.}
 \label{fig:Combined_Plots60}
\end{figure}
\clearpage
\begin{figure}
\hskip -0.8 in
 \centering
 \begin{minipage}{\textwidth}
   \centering
   \includegraphics[width=\linewidth]{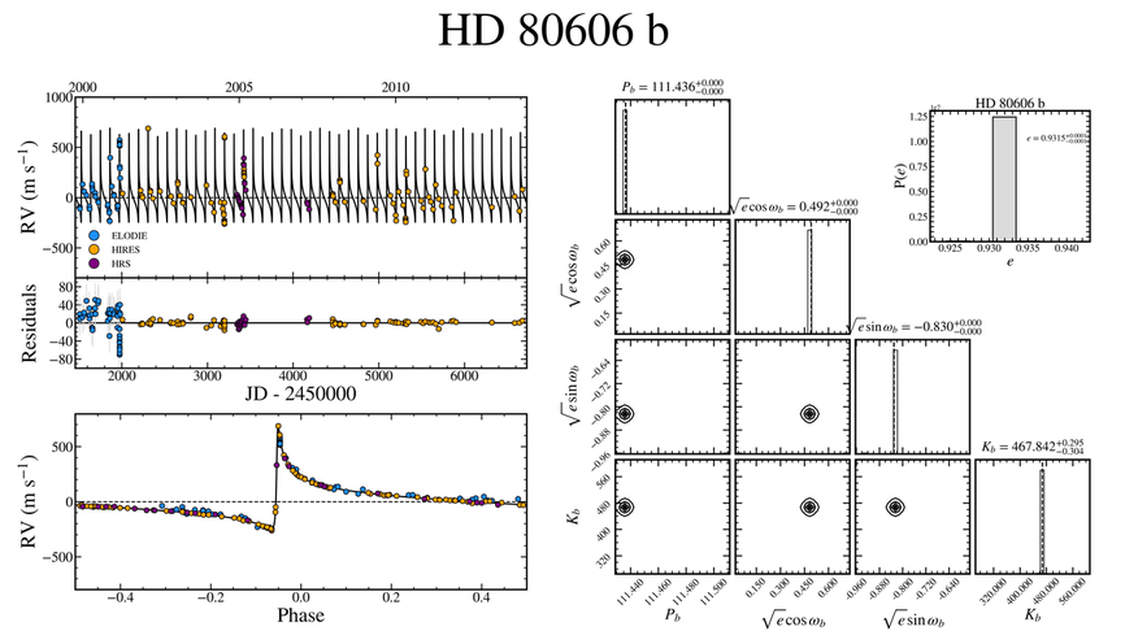}\\
   \vskip .3 in
   \includegraphics[width=\linewidth]{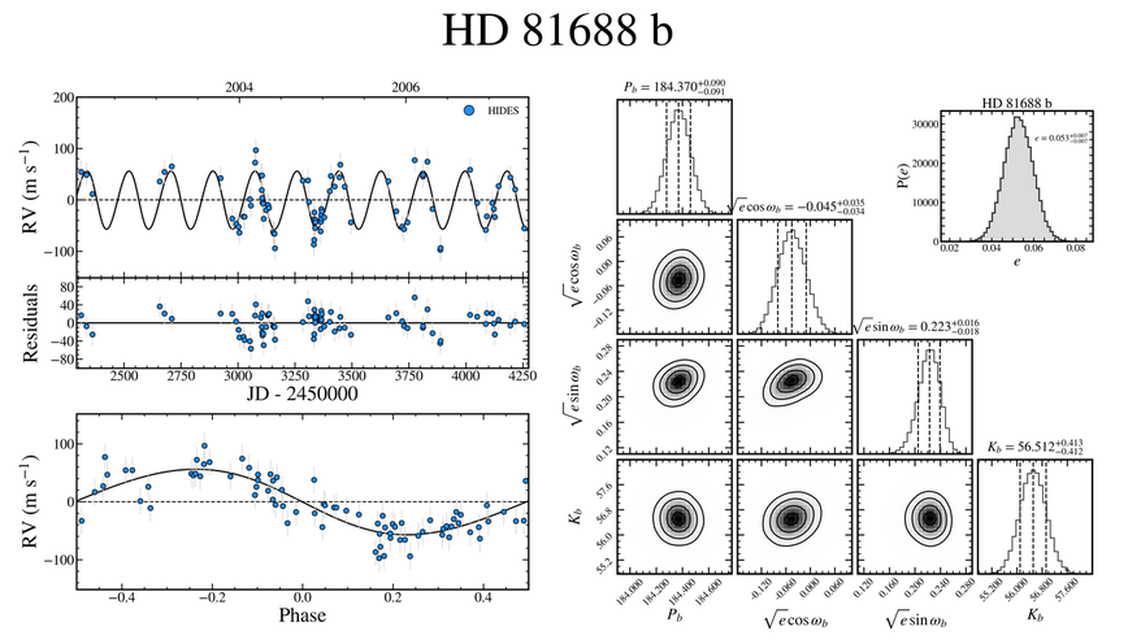}
 \end{minipage}
 \caption{Summary of results for the warm Jupiters HD 80606 b and HD 81688 b.}
 \label{fig:Combined_Plots61}
\end{figure}
\clearpage
\begin{figure}
\hskip -0.8 in
 \centering
 \begin{minipage}{\textwidth}
   \centering
   \includegraphics[width=\linewidth]{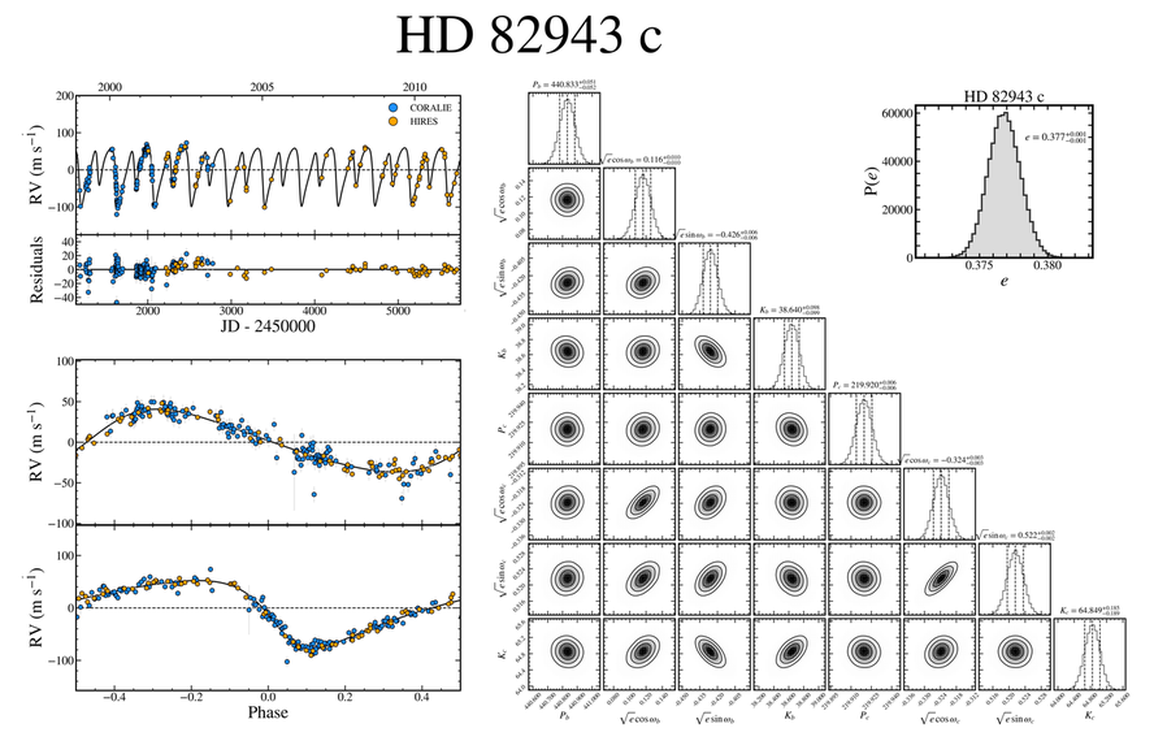}\\
   \vskip .3 in
   \includegraphics[width=\linewidth]{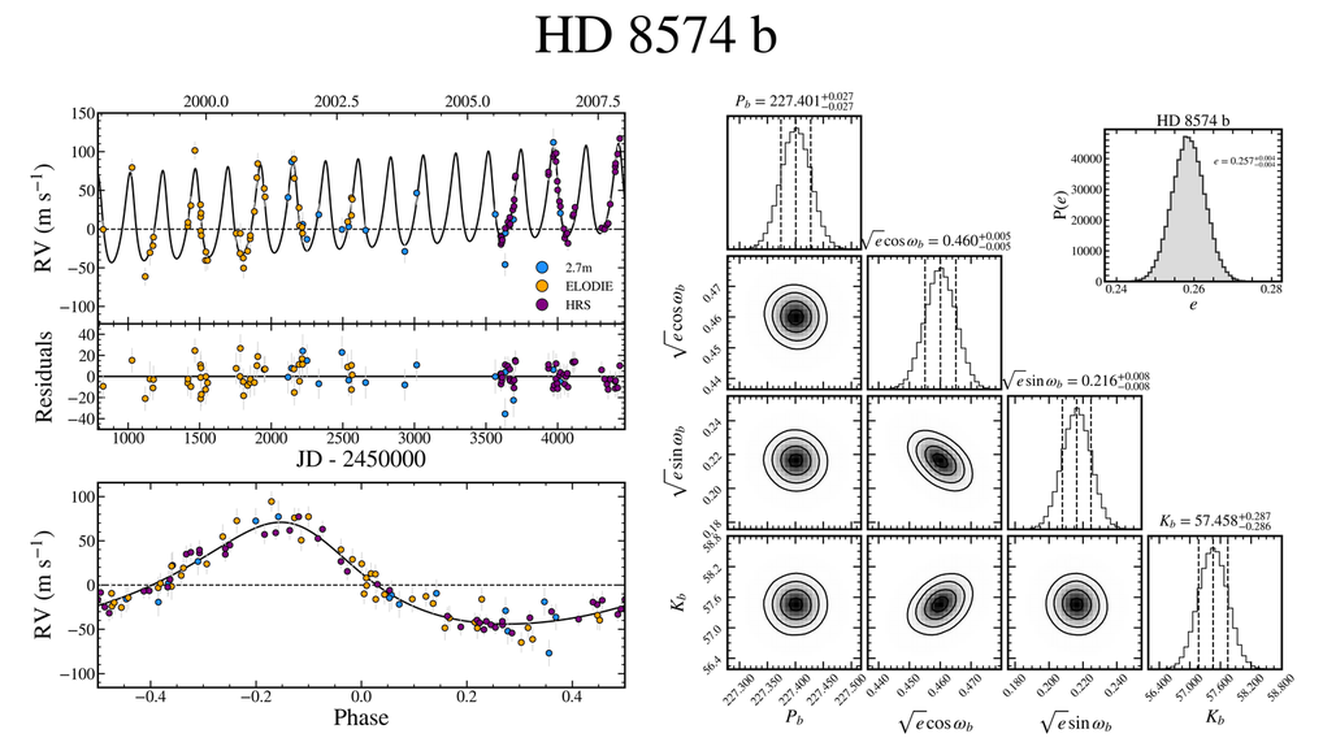}
 \end{minipage}
 \caption{Summary of results for the warm Jupiters HD 82943 c and HD 8574 b.}
 \label{fig:Combined_Plots62}
\end{figure}
\clearpage
\begin{figure}
\hskip -0.8 in
 \centering
 \begin{minipage}{\textwidth}
   \centering
   \includegraphics[width=\linewidth]{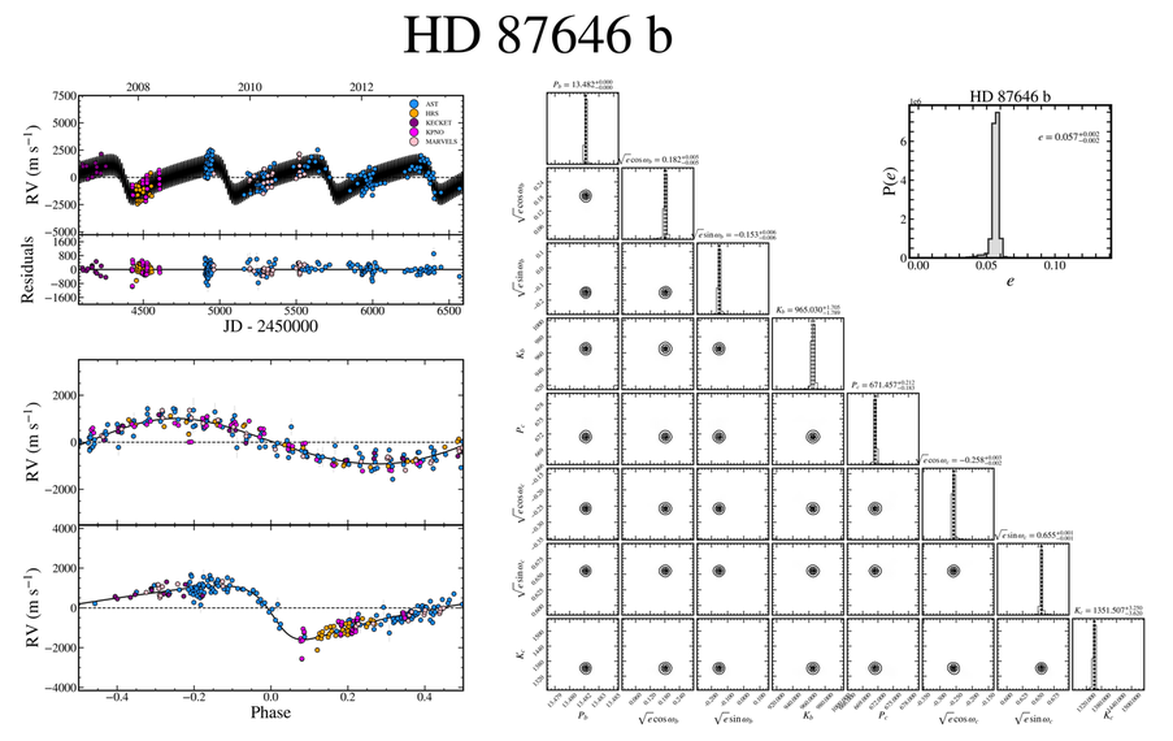}\\
   \vskip .3 in
   \includegraphics[width=\linewidth]{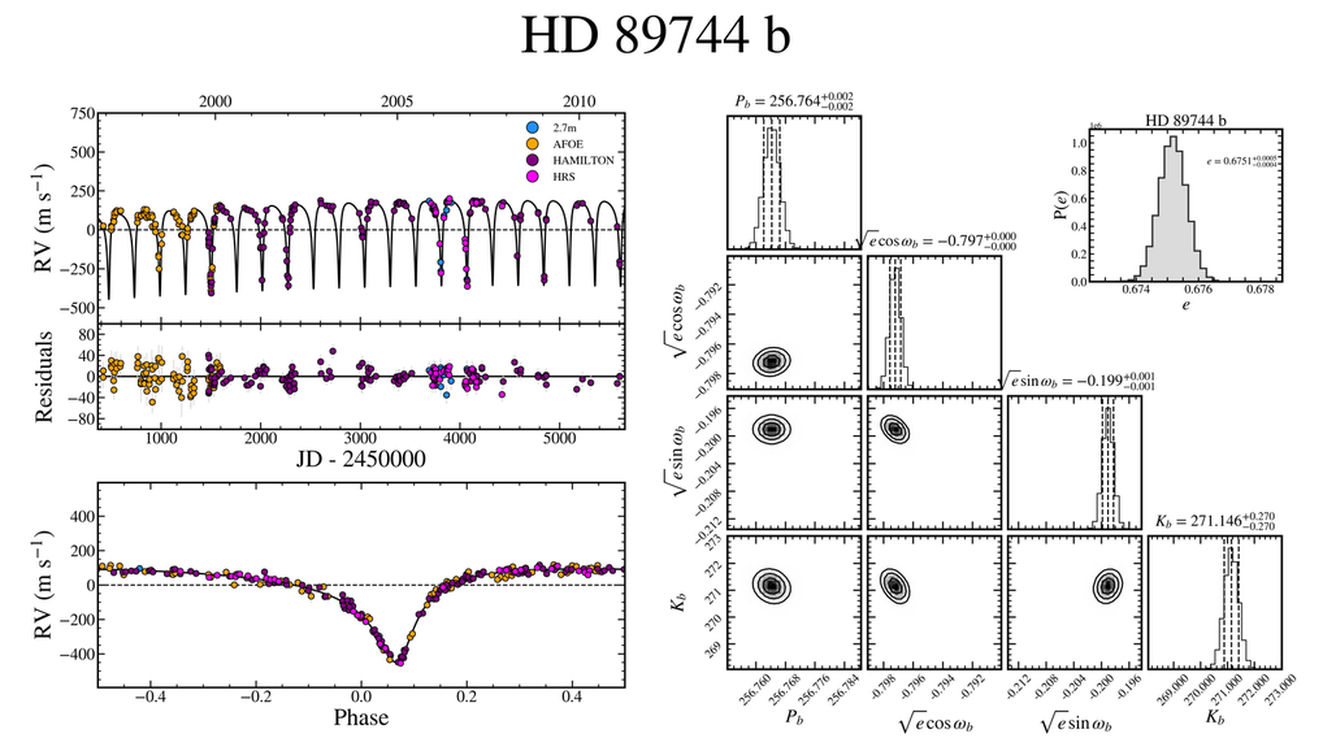}
 \end{minipage}
 \caption{Summary of results for the warm Jupiters HD 87646 b and HD 89744 b.}
 \label{fig:Combined_Plots63}
\end{figure}
\clearpage
\begin{figure}
\hskip -0.8 in
 \centering
 \begin{minipage}{\textwidth}
   \centering
   \includegraphics[width=\linewidth]{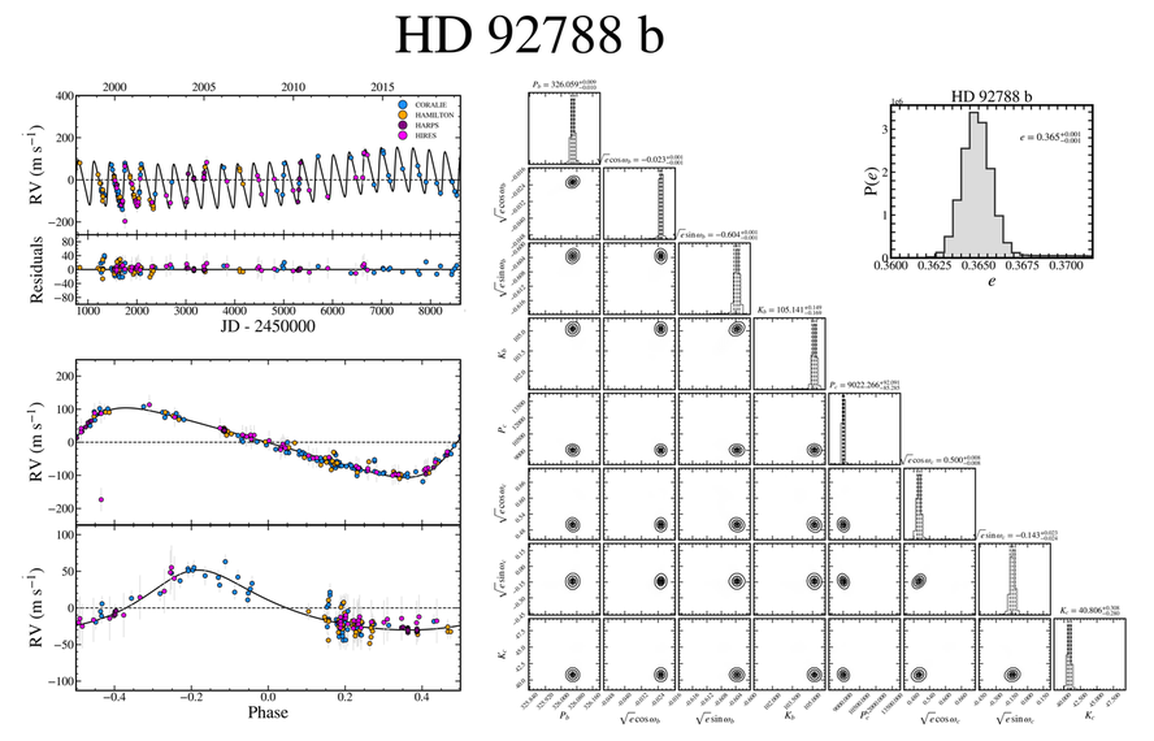}\\
   \vskip .3 in
   \includegraphics[width=\linewidth]{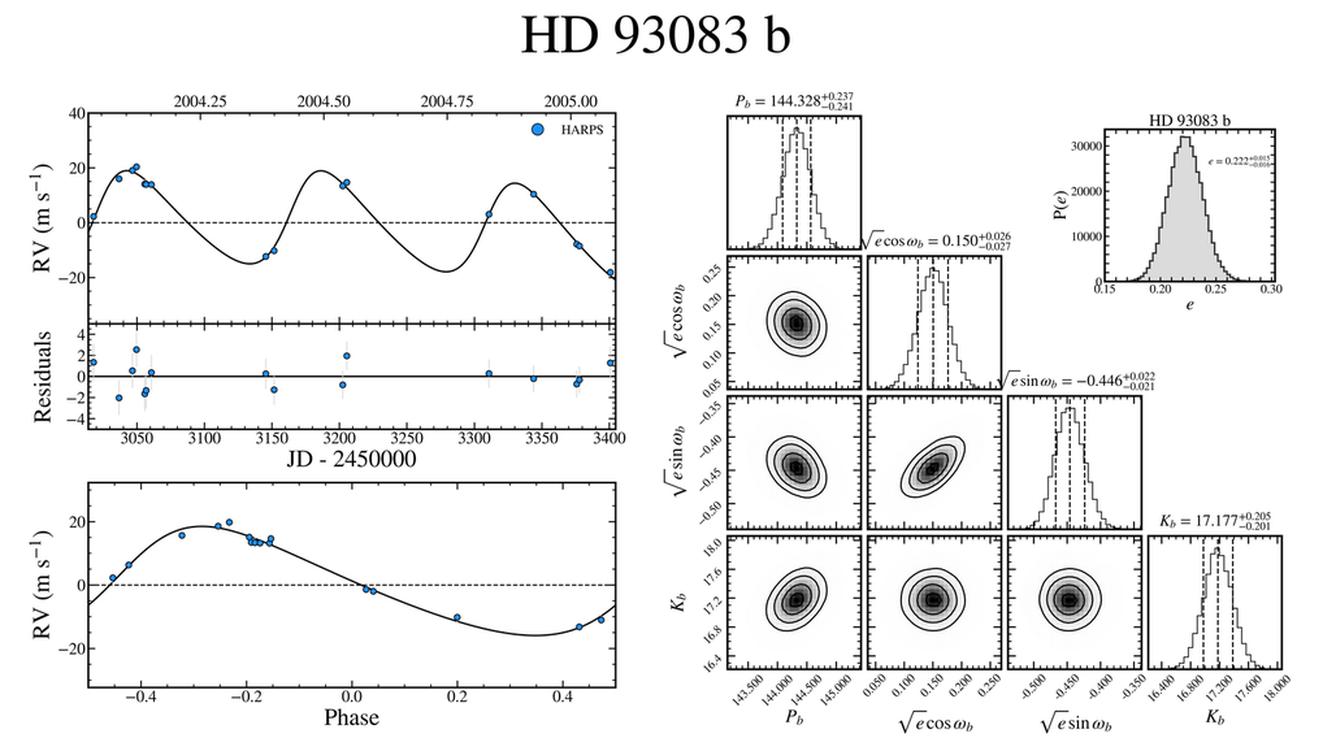}
 \end{minipage}
 \caption{Summary of results for the warm Jupiters HD 92788 b and HD 93083 b.}
 \label{fig:Combined_Plots64}
\end{figure}
\clearpage
\begin{figure}
\hskip -0.8 in
 \centering
 \begin{minipage}{\textwidth}
   \centering
   \includegraphics[width=\linewidth]{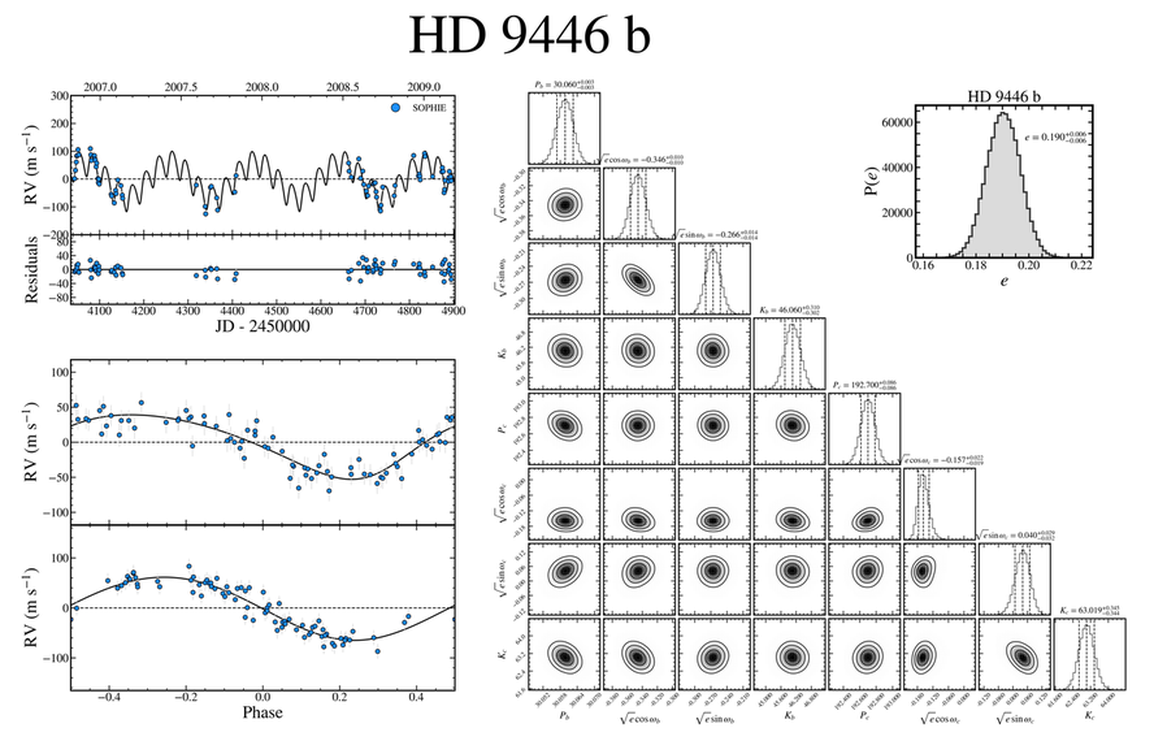}\\
   \vskip .3 in
   \includegraphics[width=\linewidth]{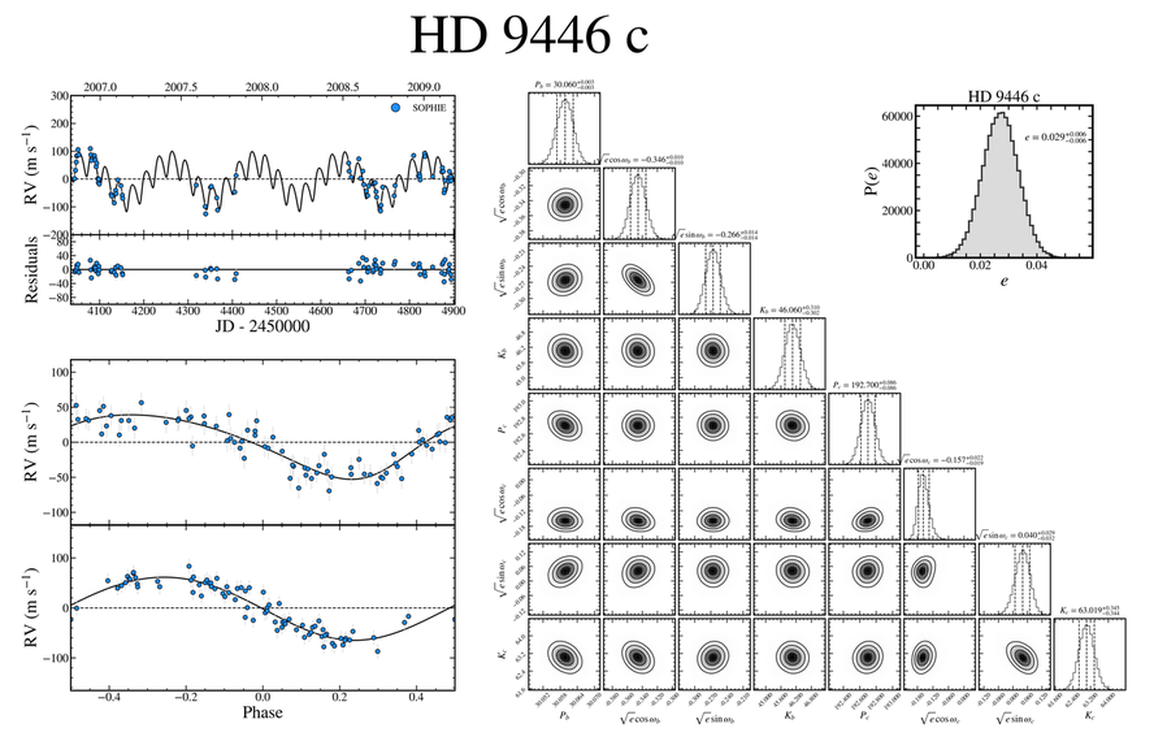}
 \end{minipage}
 \caption{Summary of results for the warm Jupiters HD 9446 b and HD 9446 c.}
 \label{fig:Combined_Plots65}
\end{figure}
\clearpage
\begin{figure}
\hskip -0.8 in
 \centering
 \begin{minipage}{\textwidth}
   \centering
   \includegraphics[width=\linewidth]{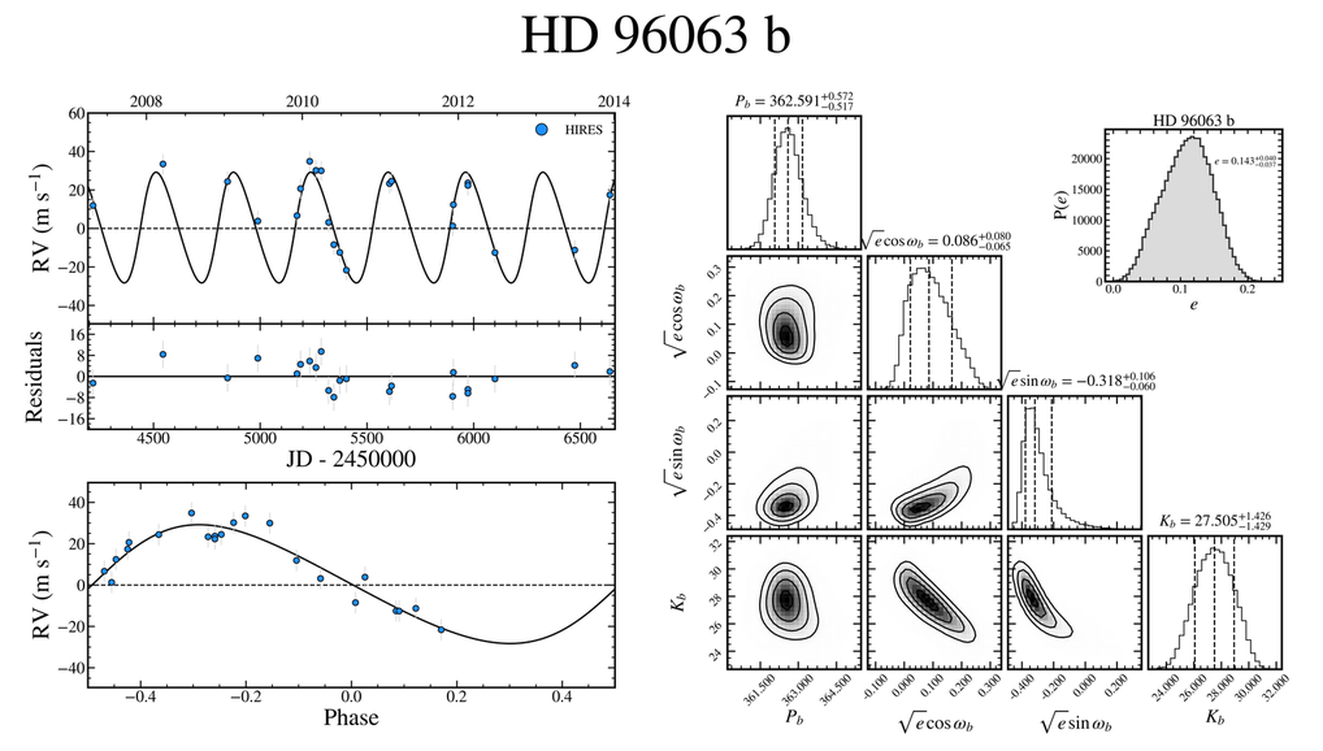}\\
   \vskip .3 in
   \includegraphics[width=\linewidth]{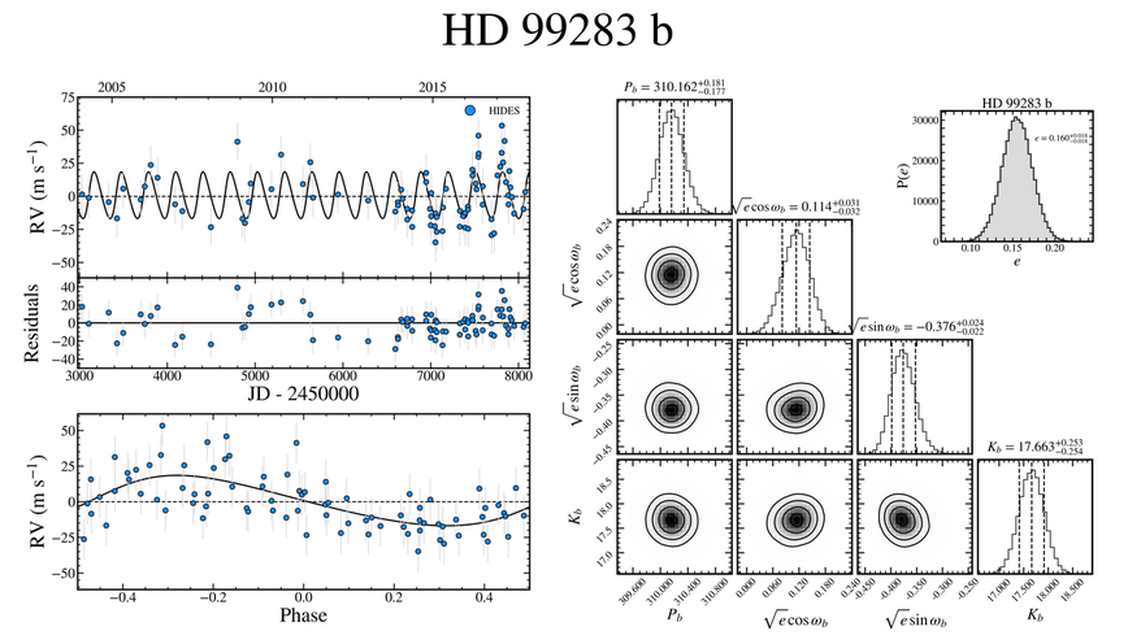}
 \end{minipage}
 \caption{Summary of results for the warm Jupiters HD 96063 b and HD 99283 b.}
 \label{fig:Combined_Plots66}
\end{figure}
\clearpage
\begin{figure}
\hskip -0.8 in
 \centering
 \begin{minipage}{\textwidth}
   \centering
   \includegraphics[width=\linewidth]{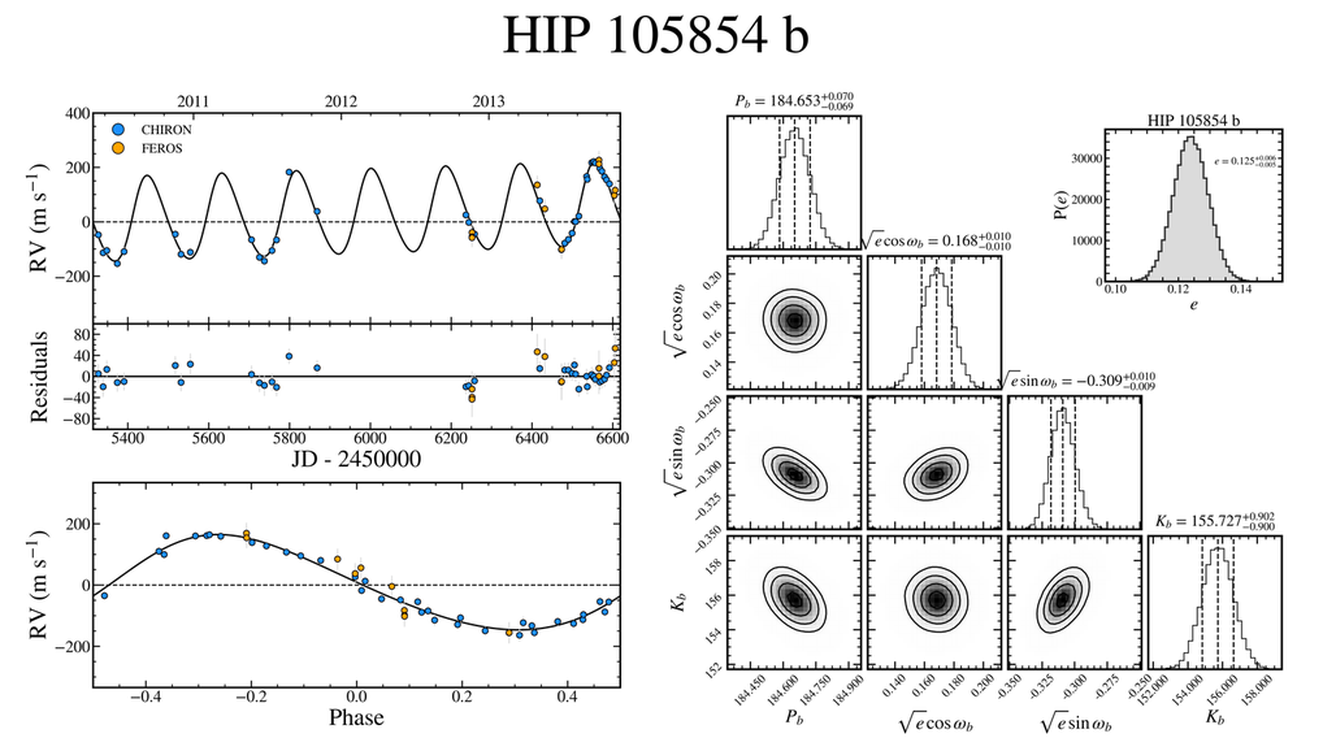}\\
   \vskip .3 in
   \includegraphics[width=\linewidth]{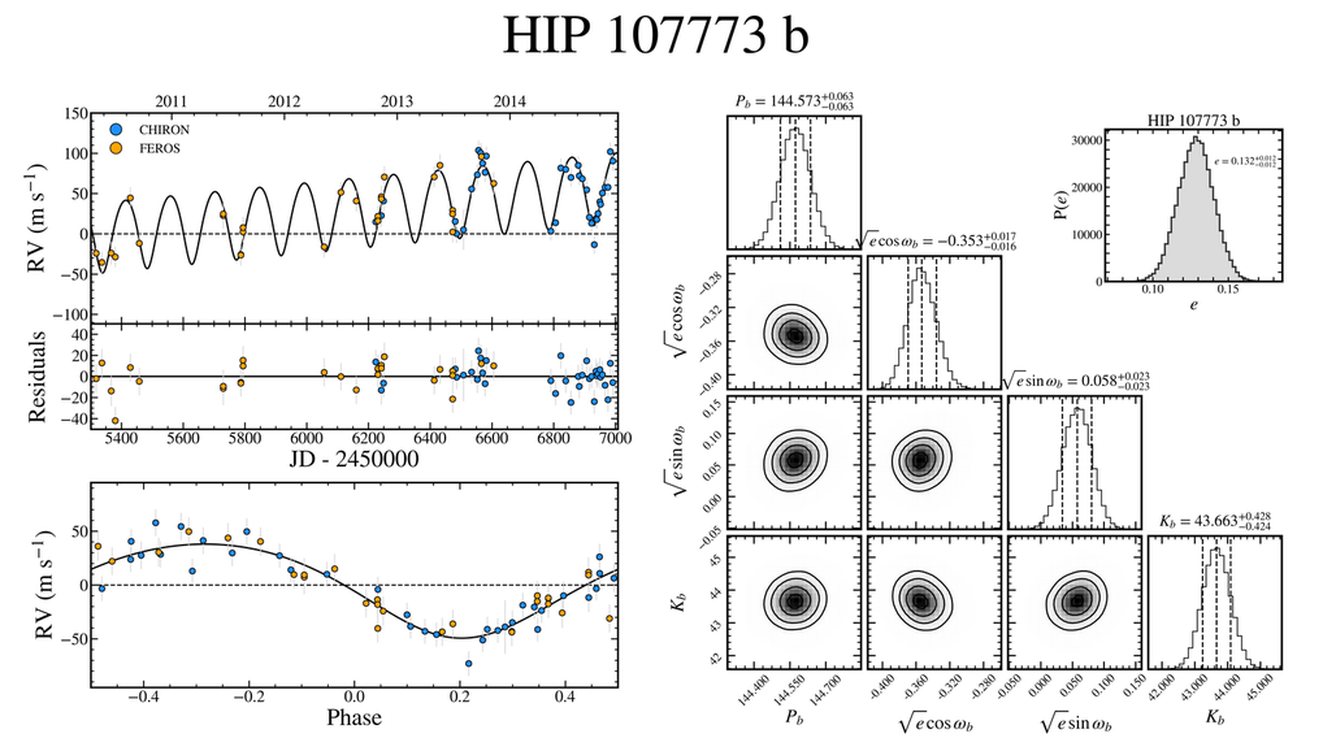}
 \end{minipage}
 \caption{Summary of results for the warm Jupiters HIP 105854 b and HIP 107773 b.}
 \label{fig:Combined_Plots67}
\end{figure}
\clearpage
\begin{figure}
\hskip -0.8 in
 \centering
 \begin{minipage}{\textwidth}
   \centering
   \includegraphics[width=\linewidth]{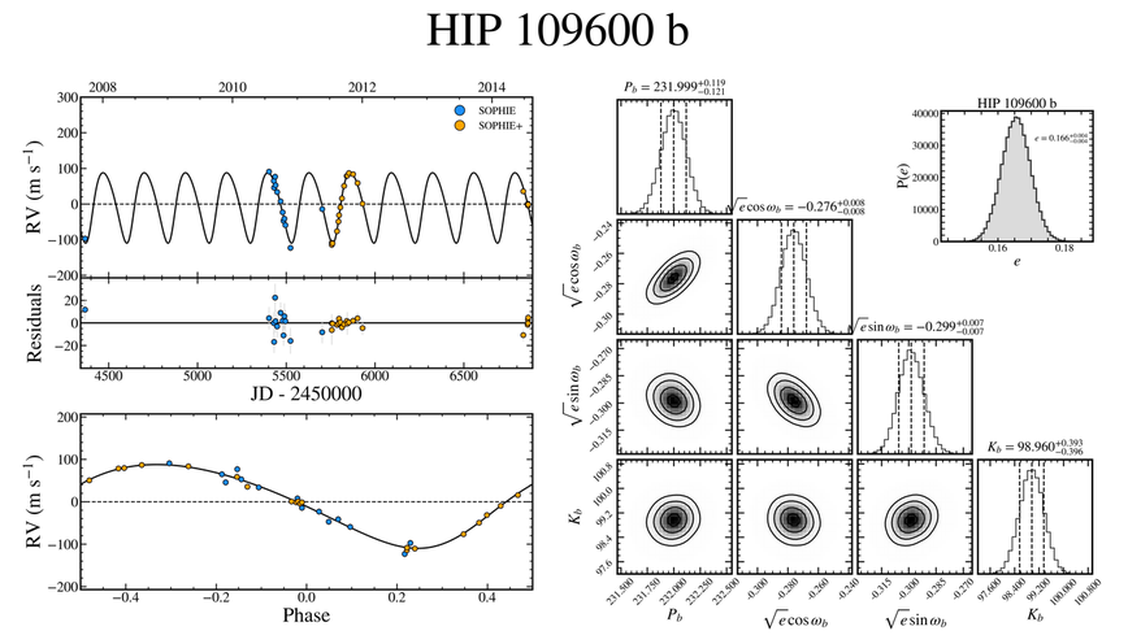}\\
   \vskip .3 in
   \includegraphics[width=\linewidth]{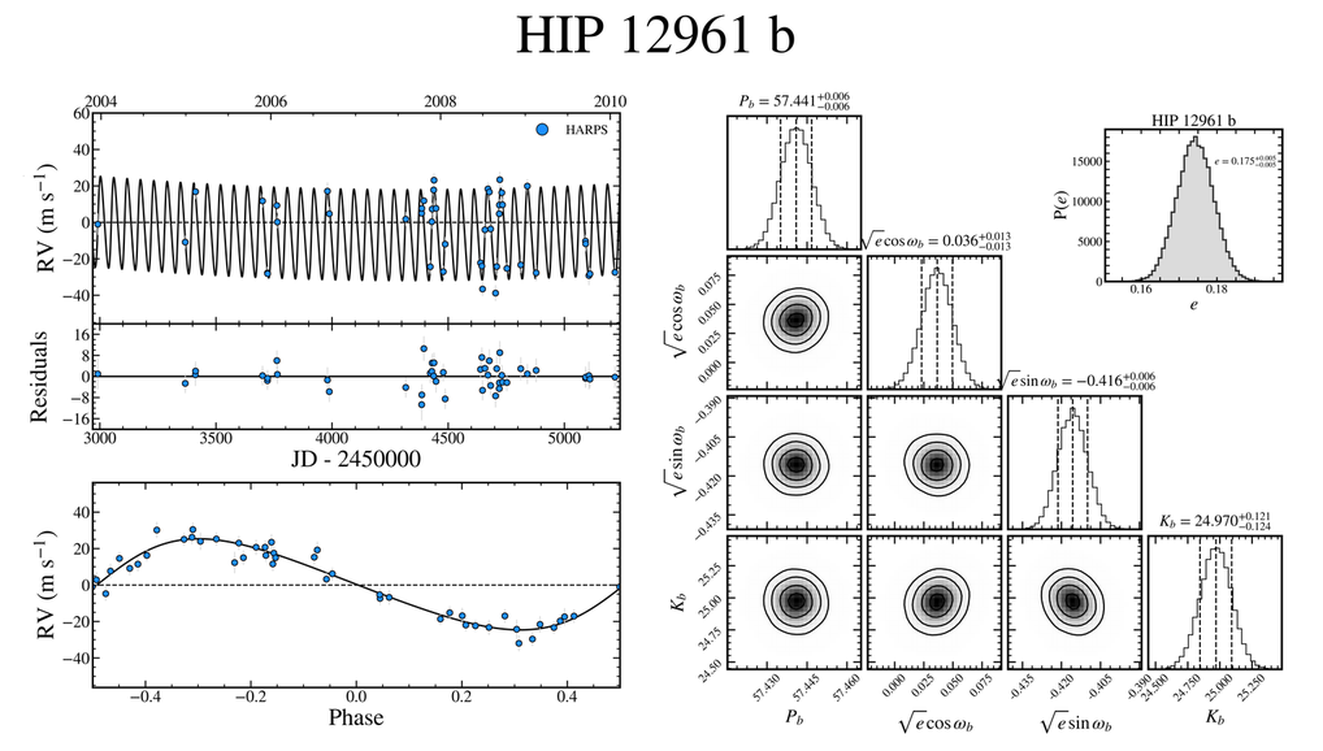}
 \end{minipage}
 \caption{Summary of results for the warm Jupiters HIP 109600 b and HIP 12961 b.}
 \label{fig:Combined_Plots68}
\end{figure}
\clearpage
\begin{figure}
\hskip -0.8 in
 \centering
 \begin{minipage}{\textwidth}
   \centering
   \includegraphics[width=\linewidth]{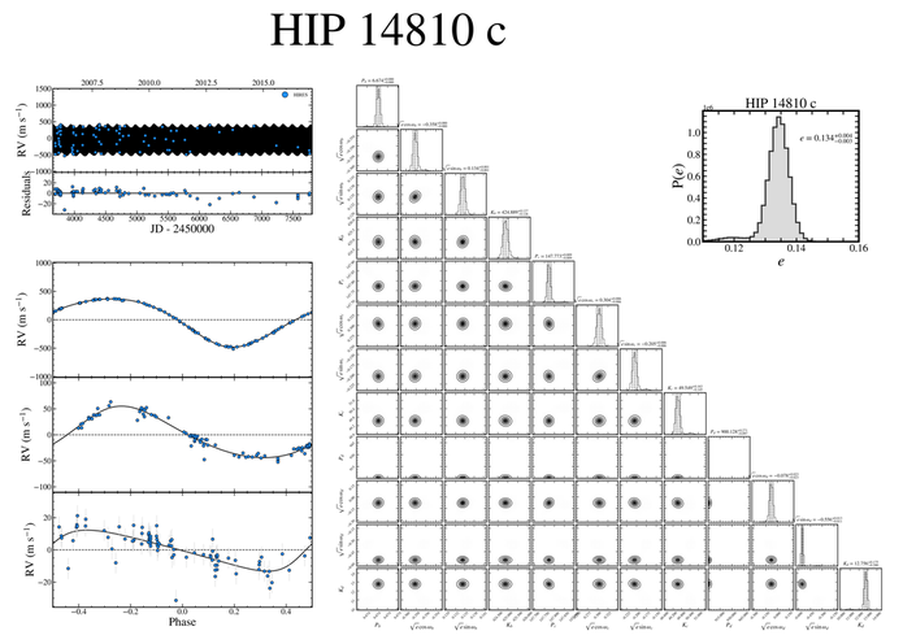}\\
   \vskip .3 in
   \includegraphics[width=\linewidth]{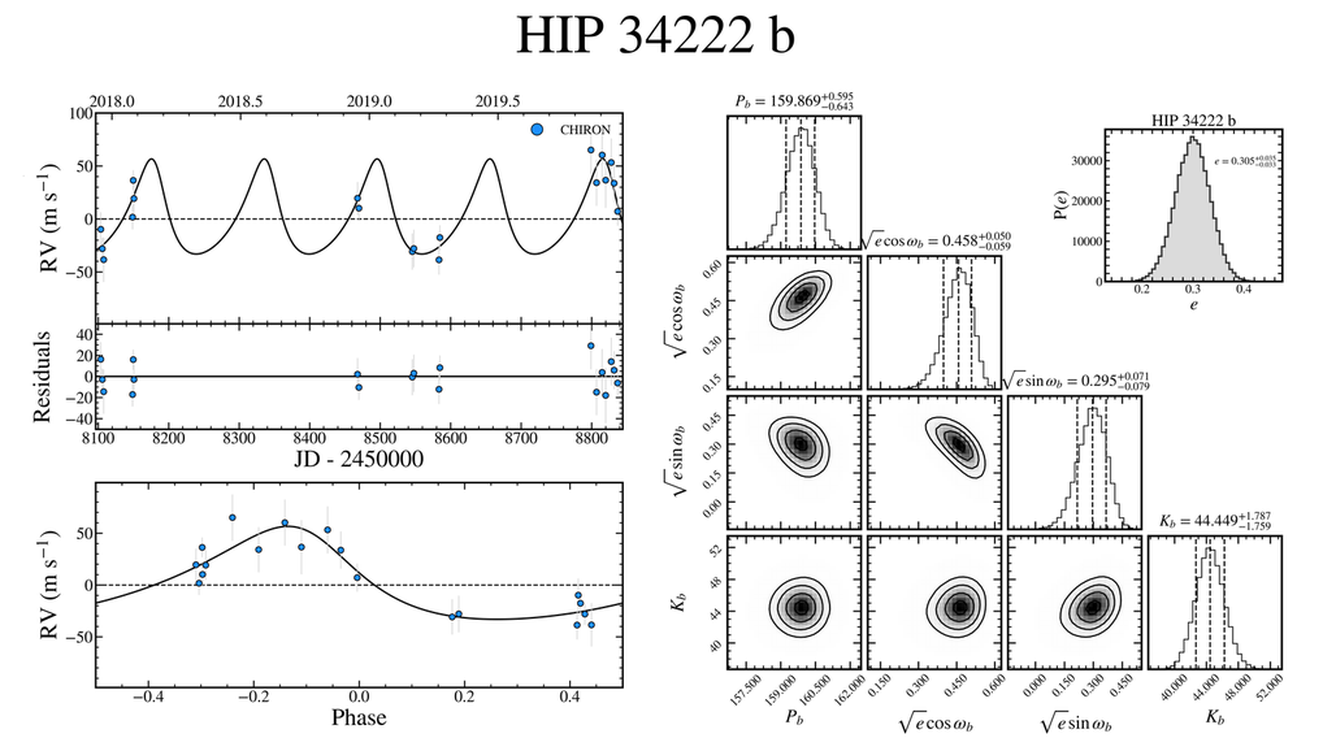}
 \end{minipage}
 \caption{Summary of results for the warm Jupiters HIP 14810 c and HIP 34222 b.}
 \label{fig:Combined_Plots69}
\end{figure}
\clearpage
\begin{figure}
\hskip -0.8 in
 \centering
 \begin{minipage}{\textwidth}
   \centering
   \includegraphics[width=\linewidth]{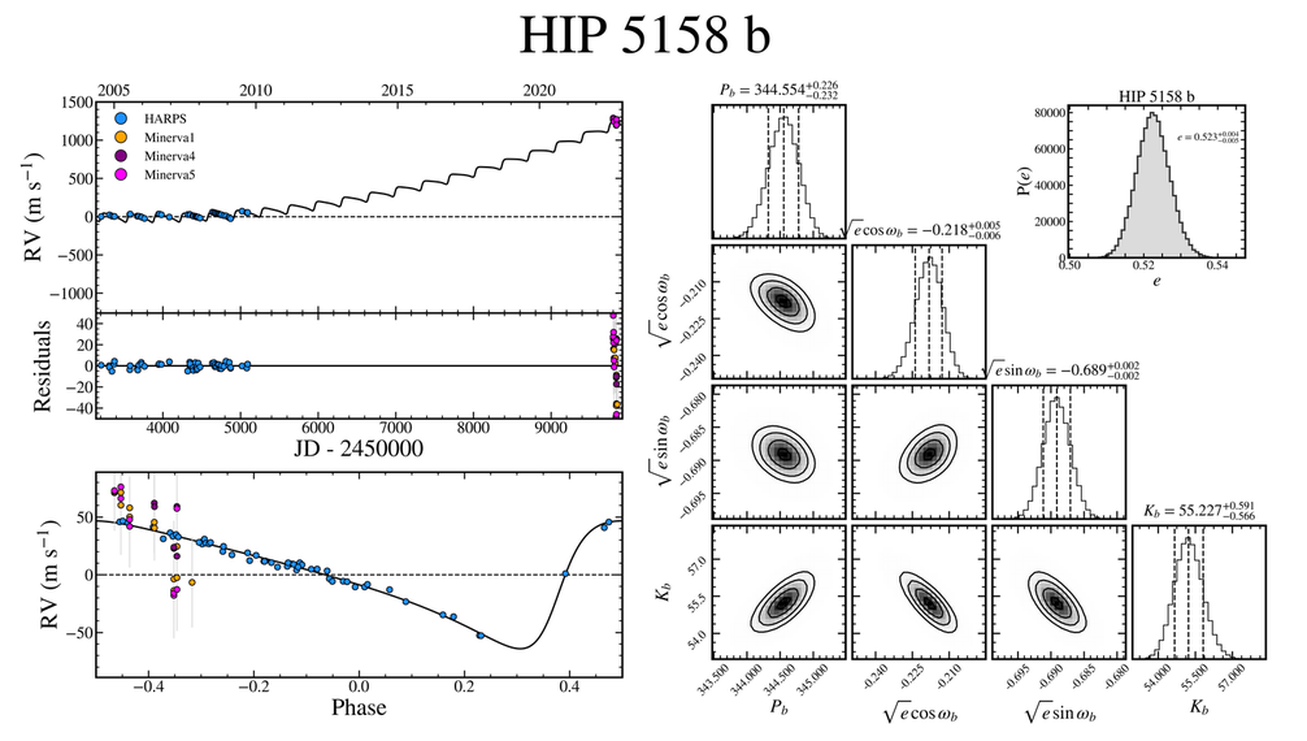}\\
   \vskip .3 in
   \includegraphics[width=\linewidth]{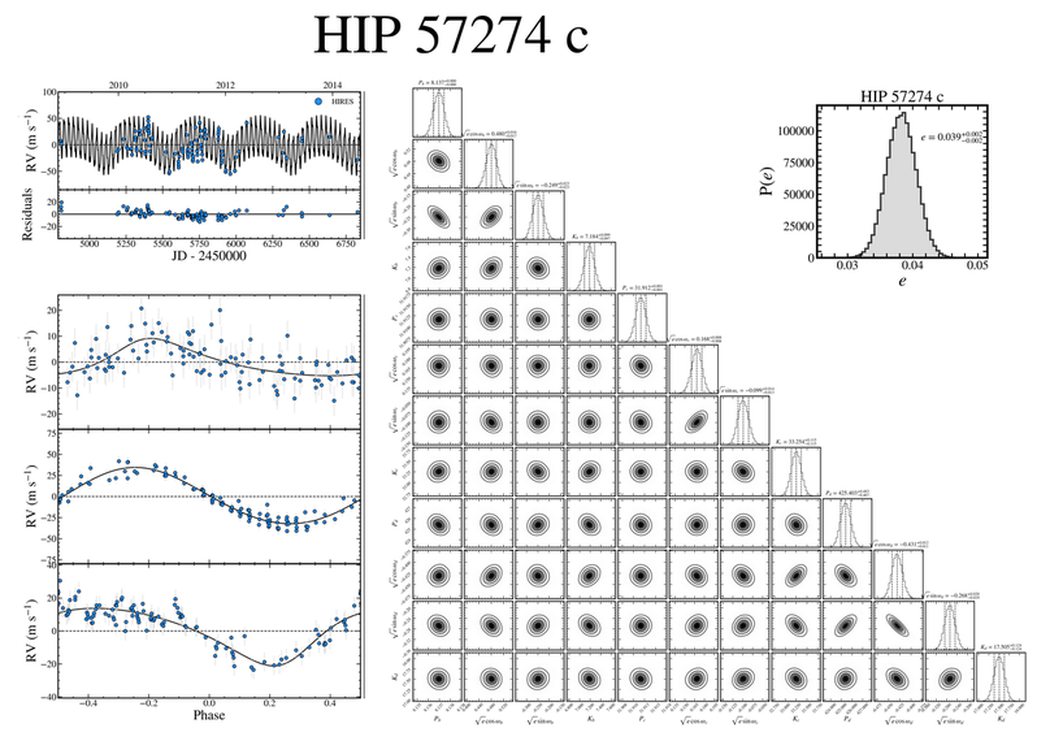}
 \end{minipage}
 \caption{Summary of results for the warm Jupiters HIP 5158 b and HIP 57274 c.}
 \label{fig:Combined_Plots70}
\end{figure}
\clearpage
\begin{figure}
\hskip -0.8 in
 \centering
 \begin{minipage}{\textwidth}
   \centering
   \includegraphics[width=\linewidth]{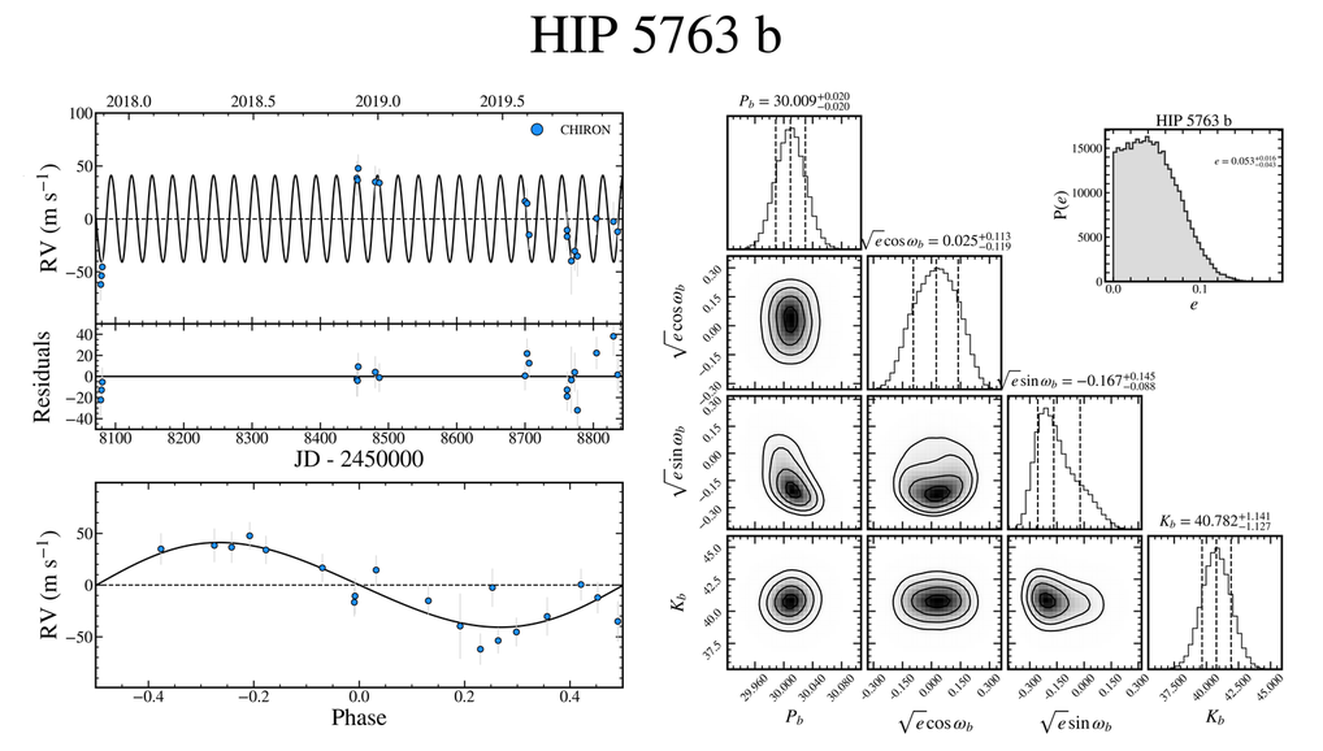}\\
   \vskip .3 in
   \includegraphics[width=\linewidth]{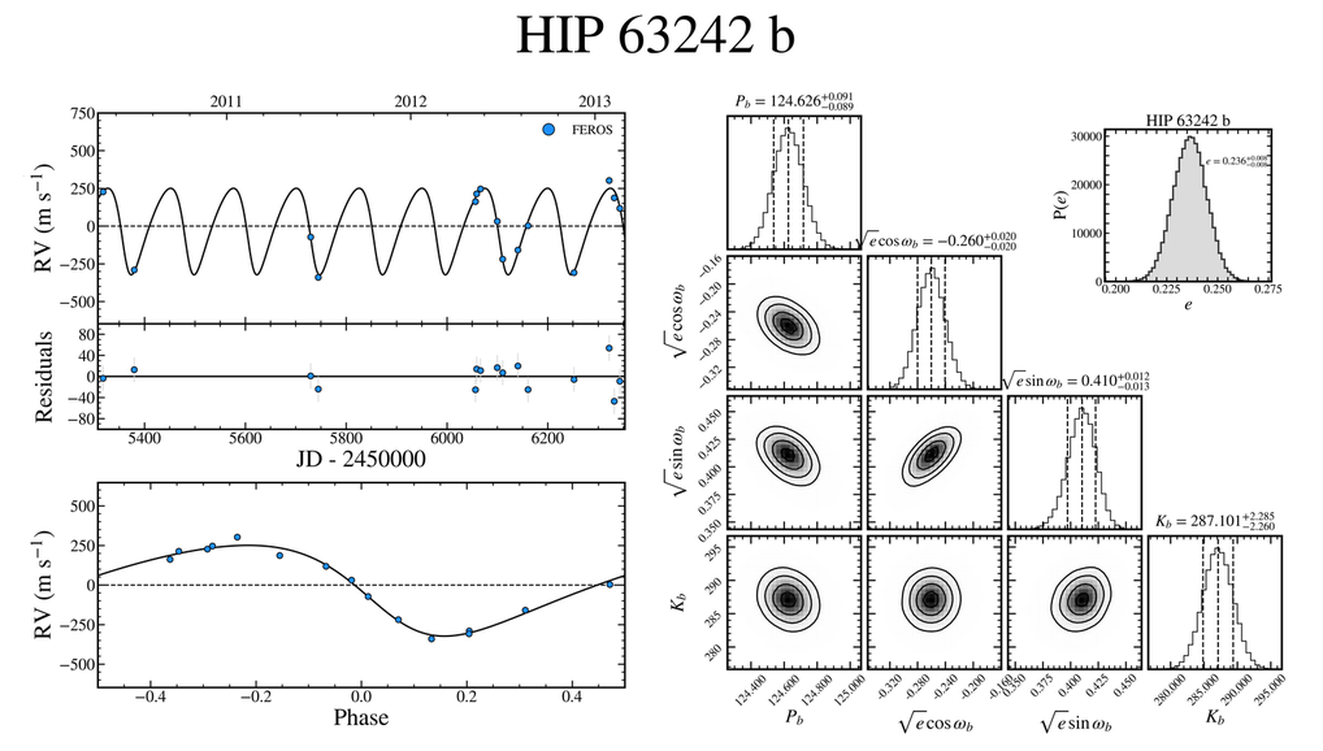}
 \end{minipage}
 \caption{Summary of results for the warm Jupiters HIP 5763 b and HIP 63242 b.}
 \label{fig:Combined_Plots71}
\end{figure}
\clearpage
\begin{figure}
\hskip -0.8 in
 \centering
 \begin{minipage}{\textwidth}
   \centering
   \includegraphics[width=\linewidth]{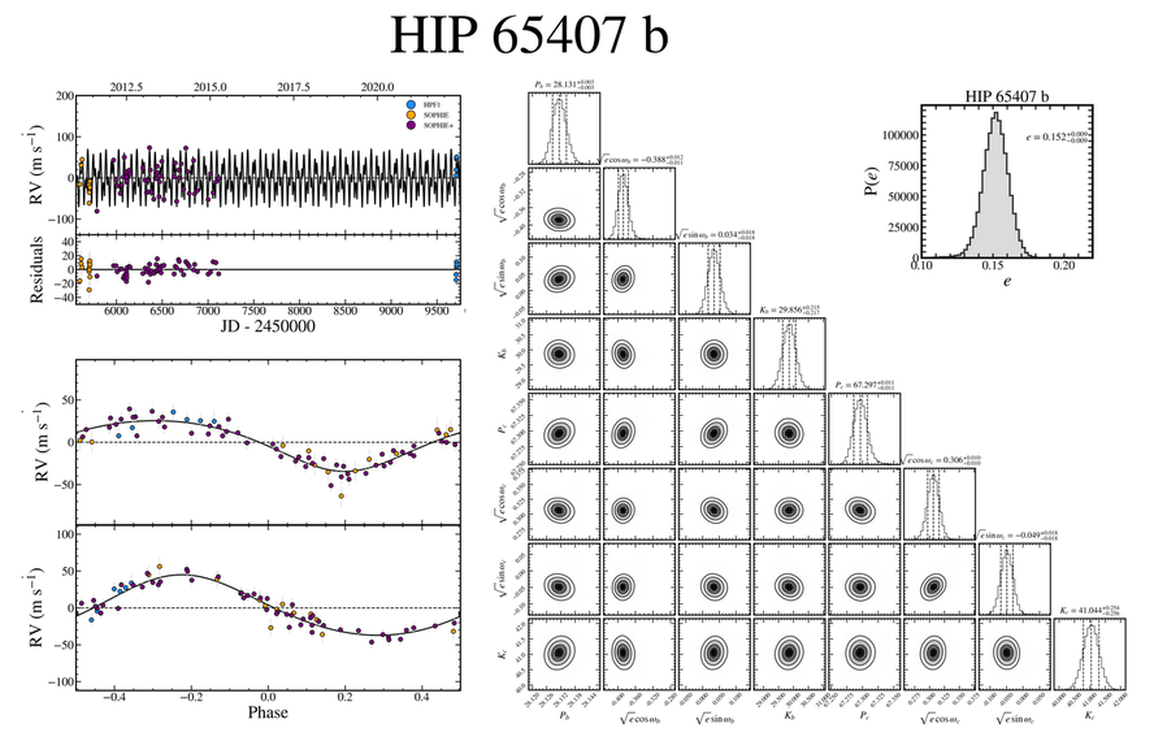}\\
   \vskip .3 in
   \includegraphics[width=\linewidth]{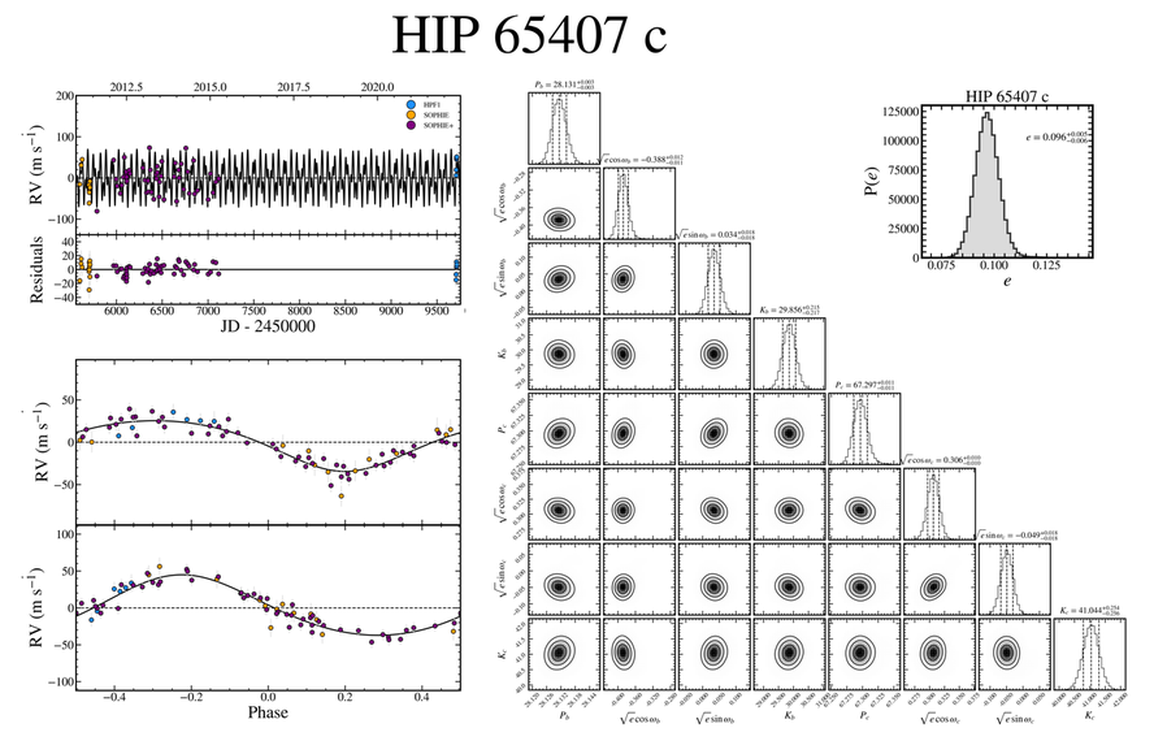}
 \end{minipage}
 \caption{Summary of results for the warm Jupiters HIP 65407 b and HIP 65407 c.}
 \label{fig:Combined_Plots72}
\end{figure}
\clearpage
\begin{figure}
\hskip -0.8 in
 \centering
 \begin{minipage}{\textwidth}
   \centering
   \includegraphics[width=\linewidth]{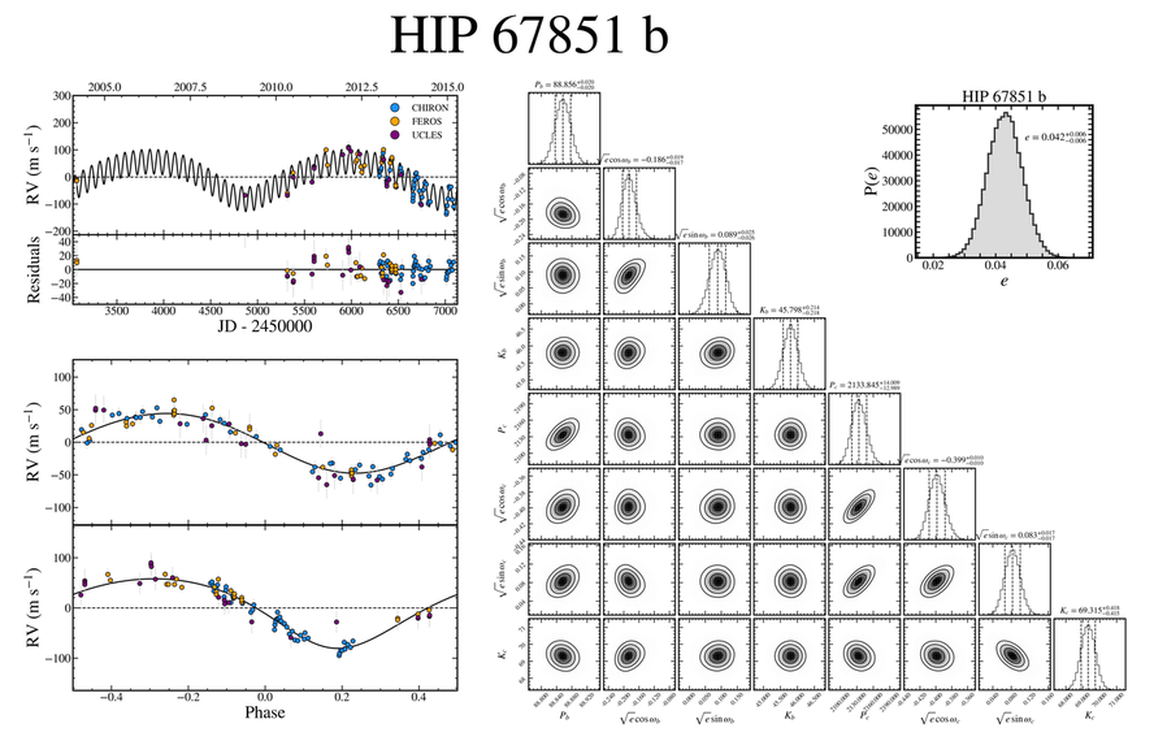}\\
   \vskip .3 in
   \includegraphics[width=\linewidth]{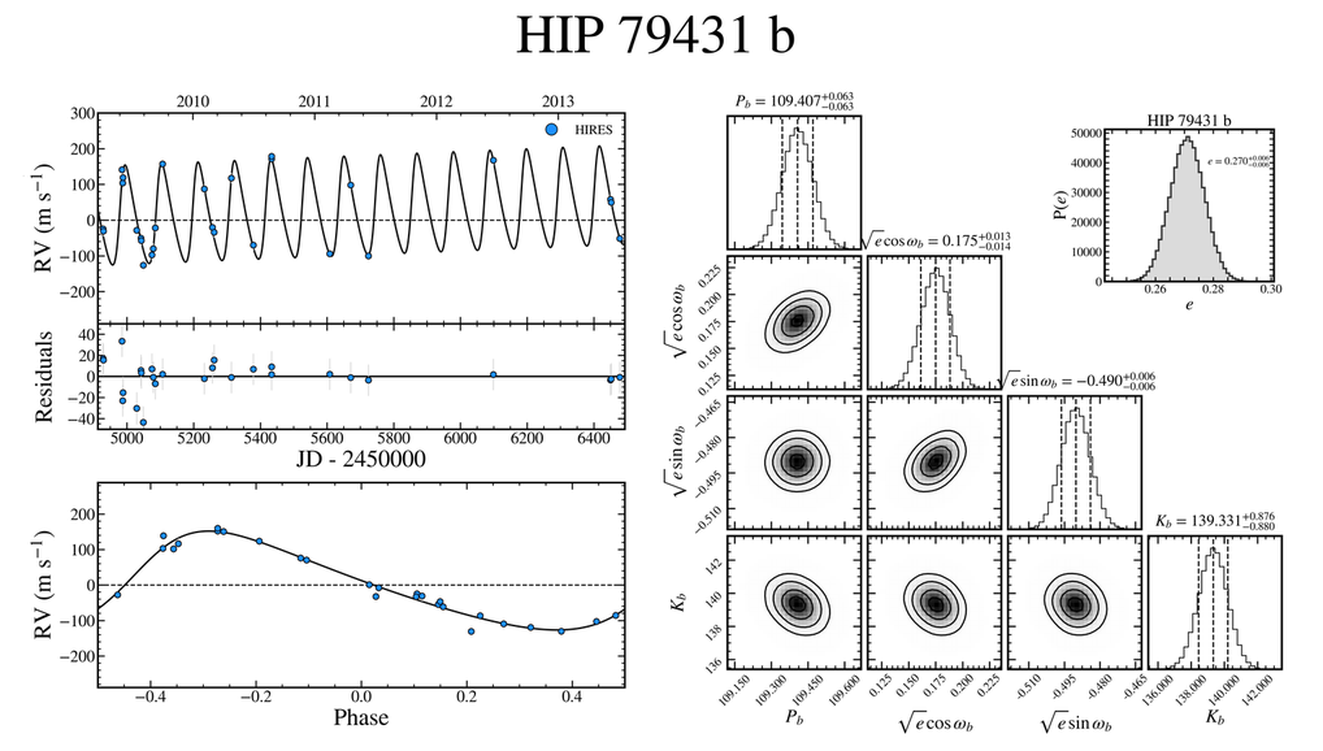}
 \end{minipage}
 \caption{Summary of results for the warm Jupiters HIP 67851 b and HIP 79431 b.}
 \label{fig:Combined_Plots73}
\end{figure}
\clearpage
\begin{figure}
\hskip -0.8 in
 \centering
 \begin{minipage}{\textwidth}
   \centering
   \includegraphics[width=\linewidth]{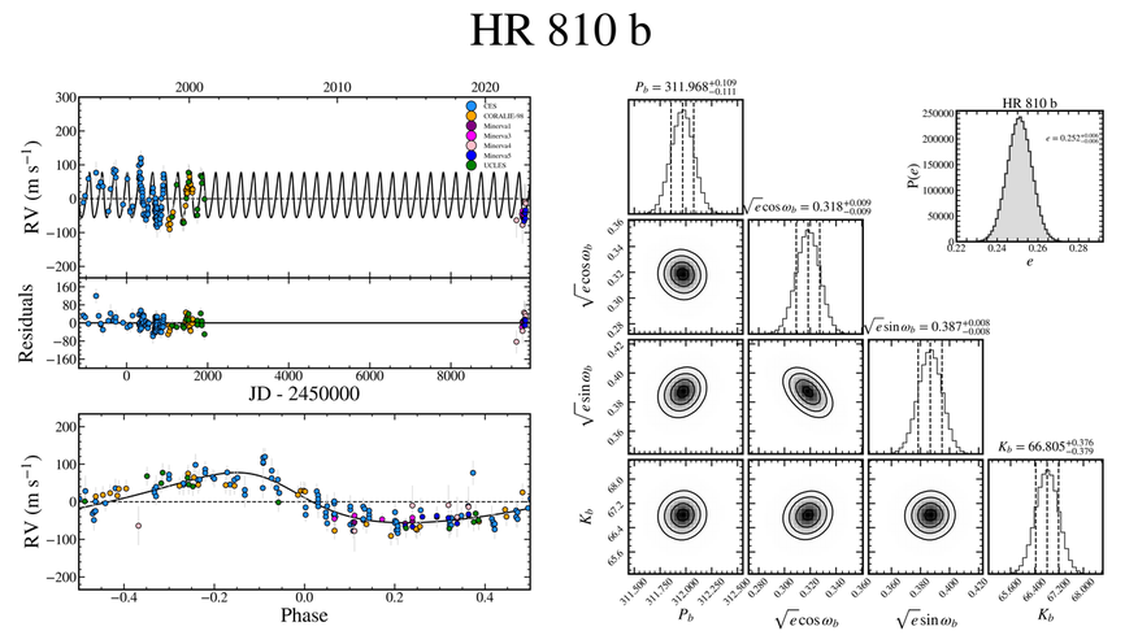}\\
   \vskip .3 in
   \includegraphics[width=\linewidth]{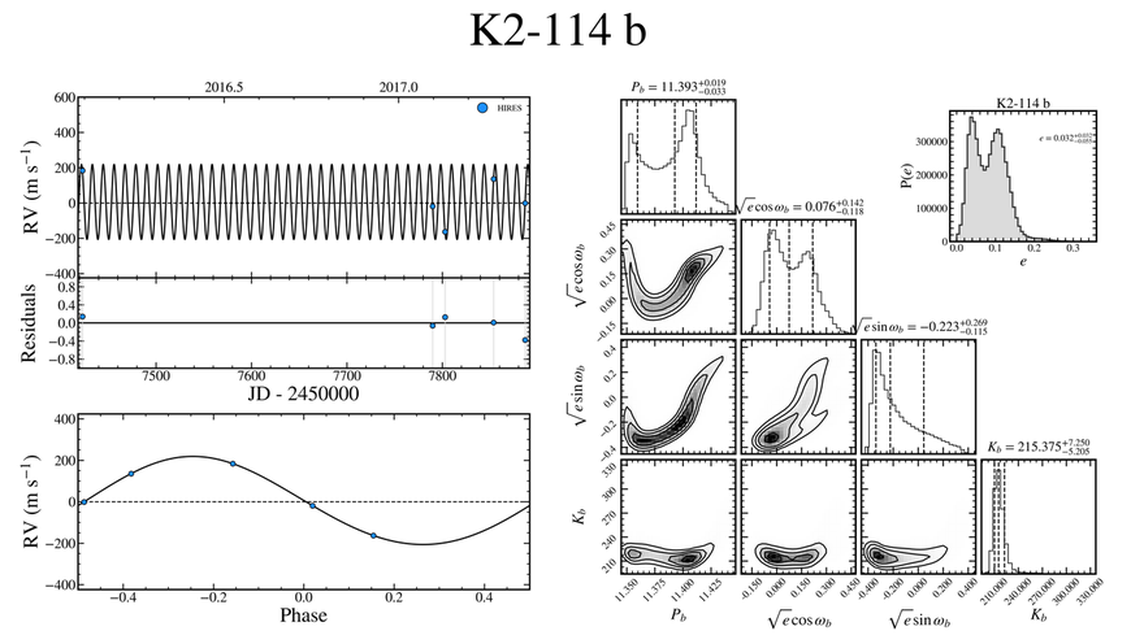}
 \end{minipage}
 \caption{Summary of results for the warm Jupiters HR 810 b and K2-114 b.}
 \label{fig:Combined_Plots74}
\end{figure}
\clearpage
\begin{figure}
\hskip -0.8 in
 \centering
 \begin{minipage}{\textwidth}
   \centering
   \includegraphics[width=\linewidth]{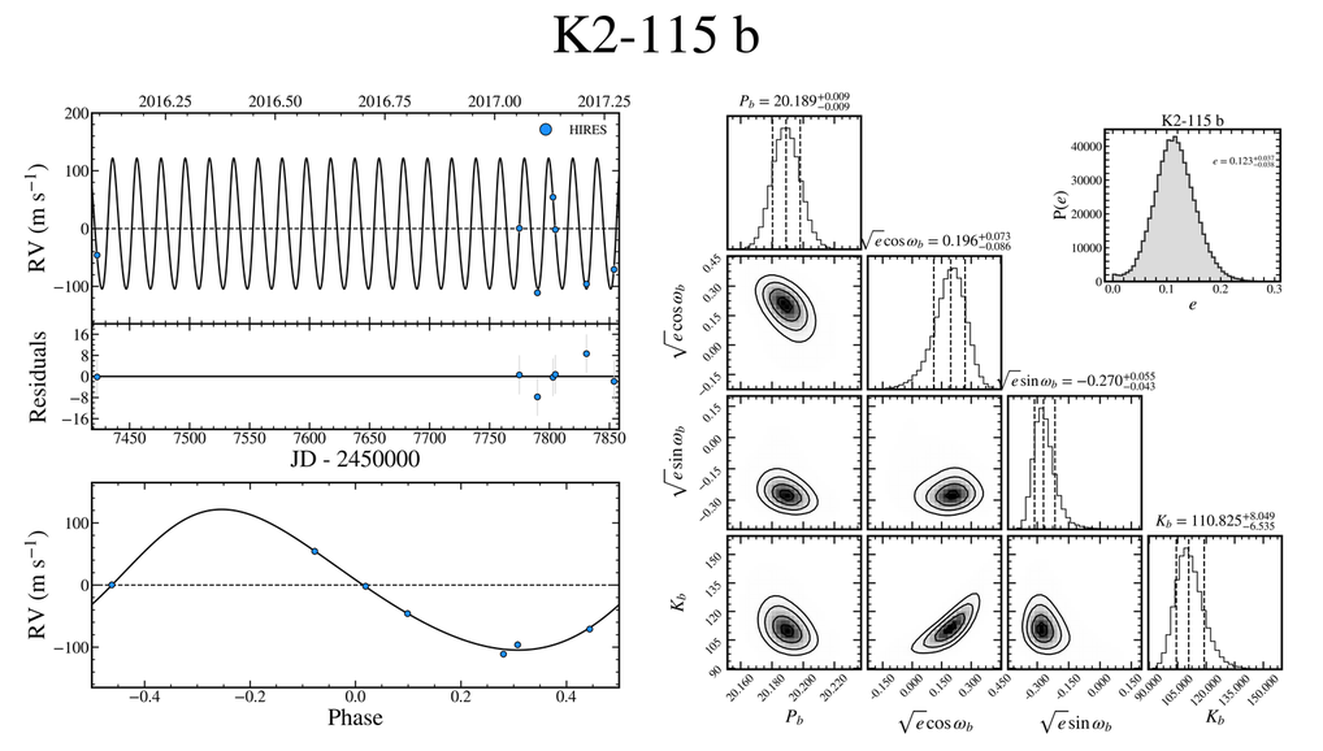}\\
   \vskip .3 in
   \includegraphics[width=\linewidth]{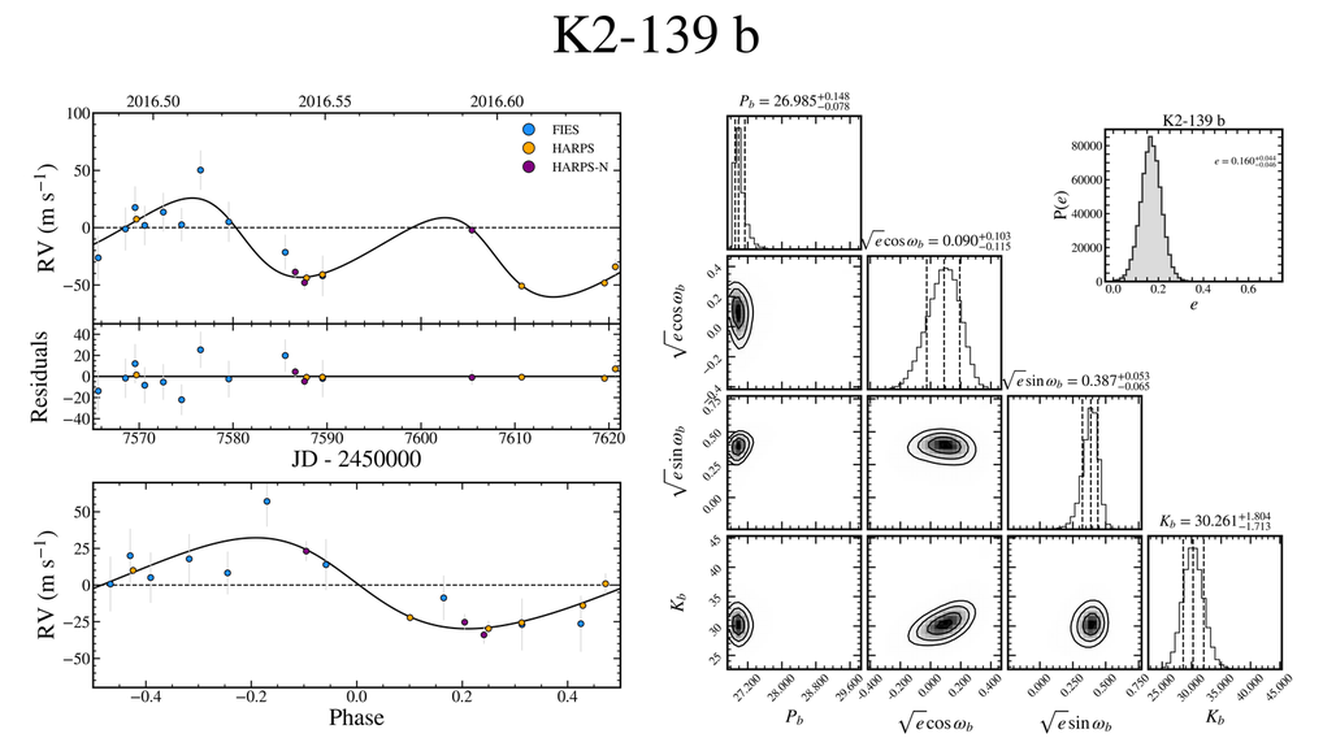}
 \end{minipage}
 \caption{Summary of results for the warm Jupiters K2-115 b and K2-139 b.}
 \label{fig:Combined_Plots75}
\end{figure}
\clearpage
\begin{figure}
\hskip -0.8 in
 \centering
 \begin{minipage}{\textwidth}
   \centering
   \includegraphics[width=\linewidth]{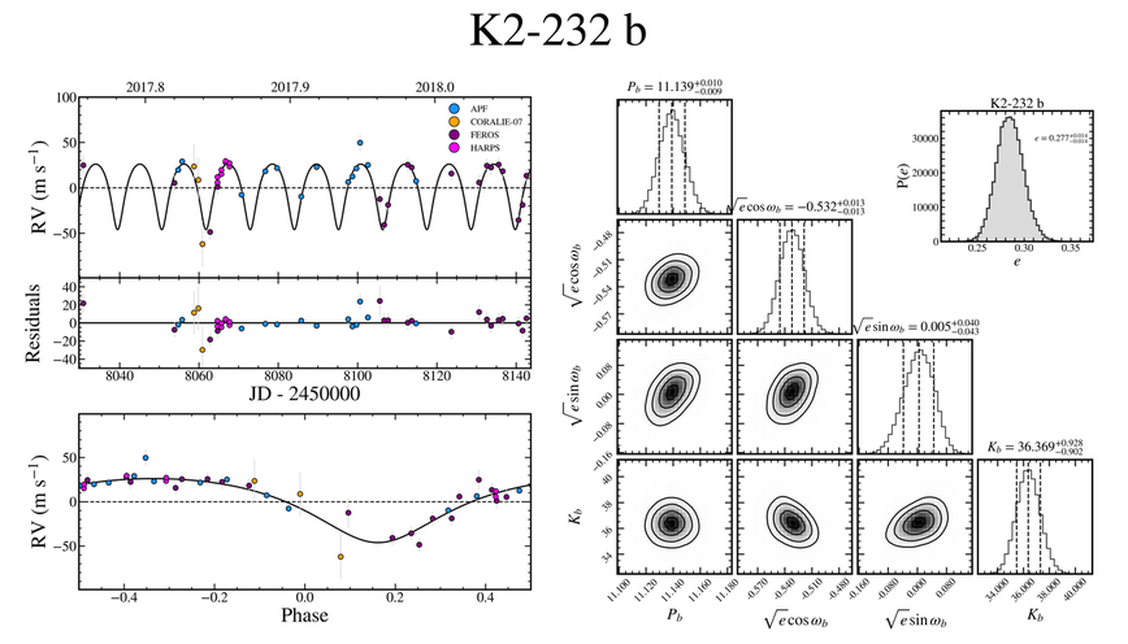}\\
   \vskip .3 in
   \includegraphics[width=\linewidth]{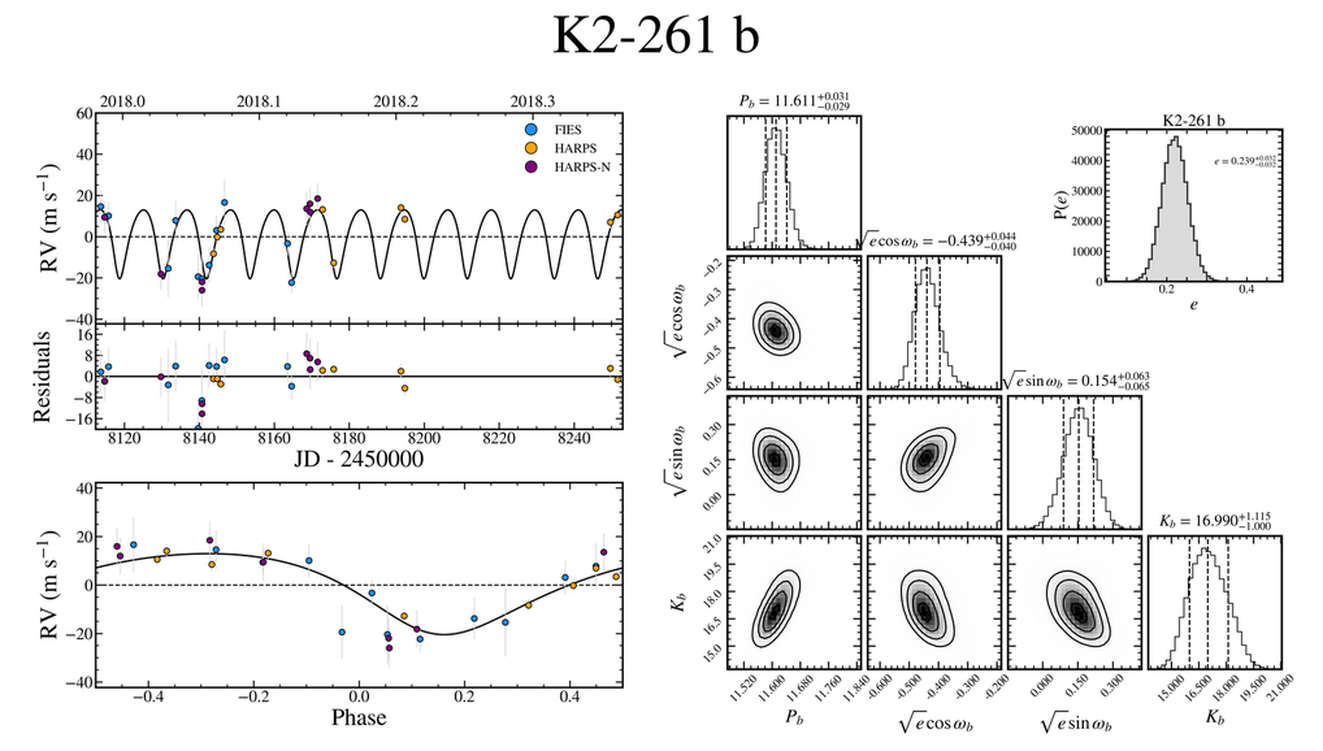}
 \end{minipage}
 \caption{Summary of results for the warm Jupiters K2-232 b and K2-261 b.}
 \label{fig:Combined_Plots76}
\end{figure}
\clearpage
\begin{figure}
\hskip -0.8 in
 \centering
 \begin{minipage}{\textwidth}
   \centering
   \includegraphics[width=\linewidth]{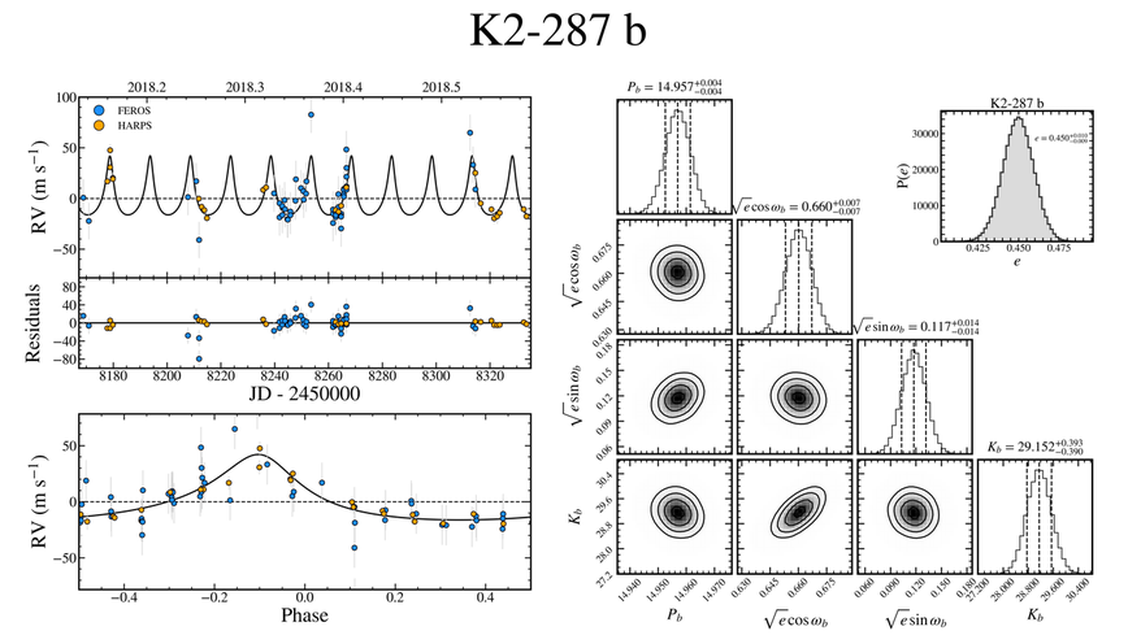}\\
   \vskip .3 in
   \includegraphics[width=\linewidth]{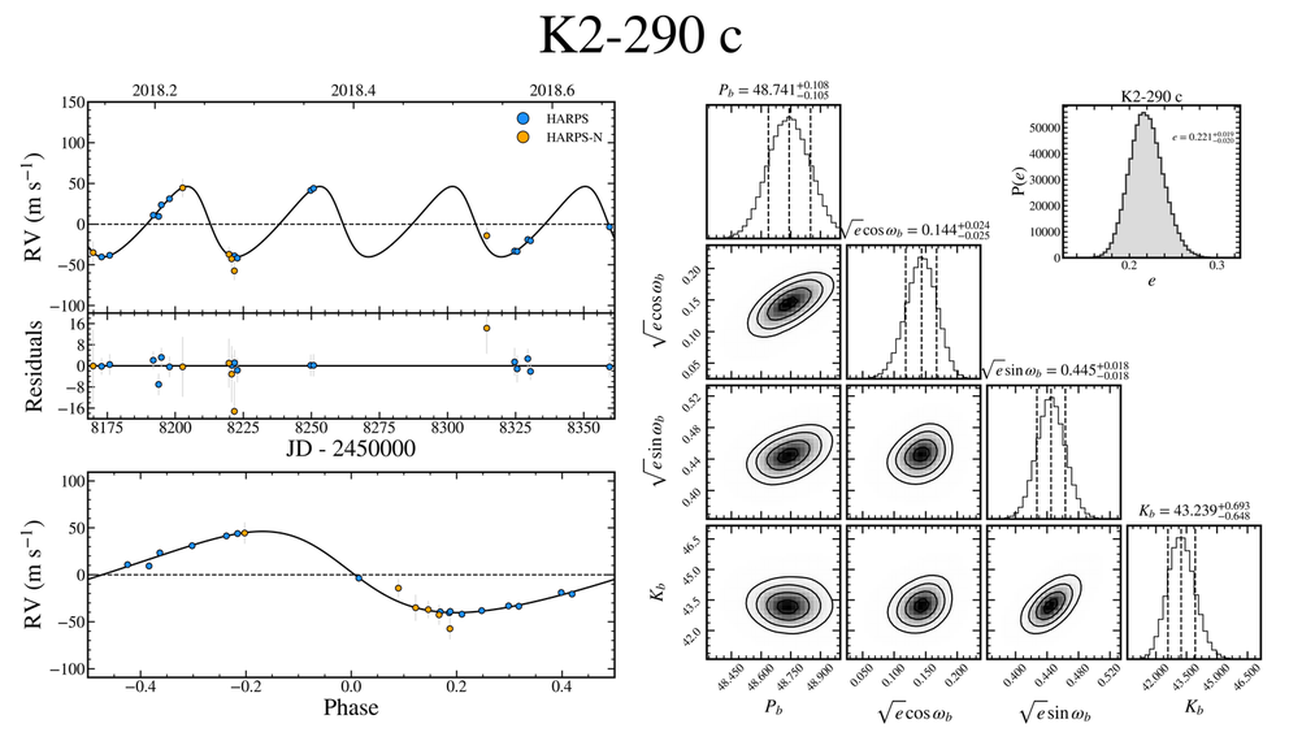}
 \end{minipage}
 \caption{Summary of results for the warm Jupiters K2-287 b and K2-290 c.}
 \label{fig:Combined_Plots77}
\end{figure}
\clearpage
\begin{figure}
\hskip -0.8 in
 \centering
 \begin{minipage}{\textwidth}
   \centering
   \includegraphics[width=\linewidth]{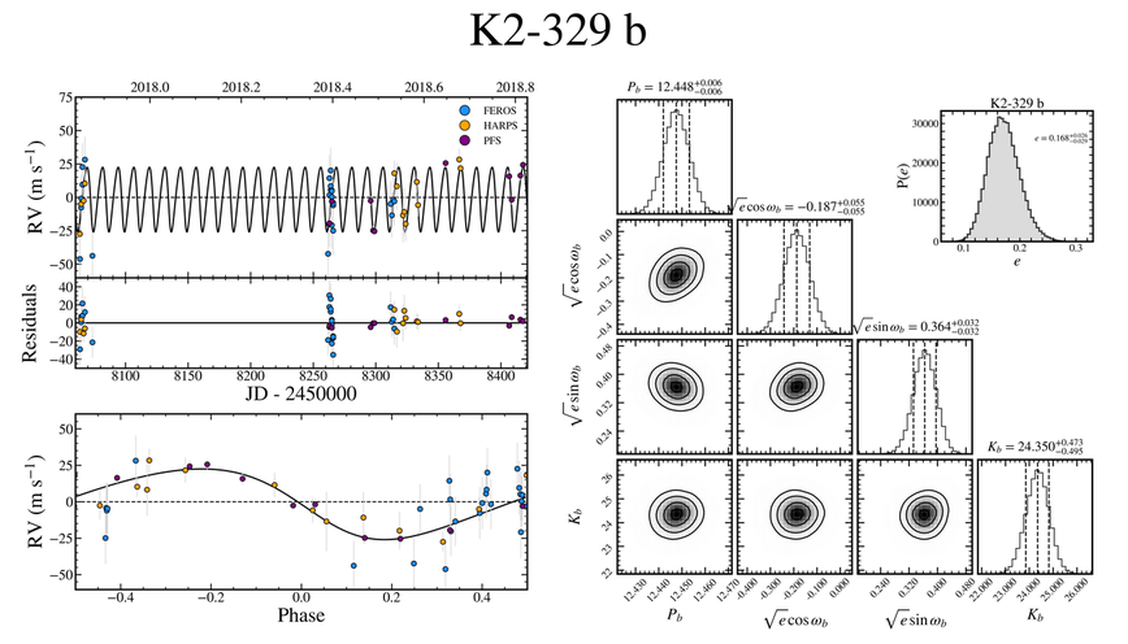}\\
   \vskip .3 in
   \includegraphics[width=\linewidth]{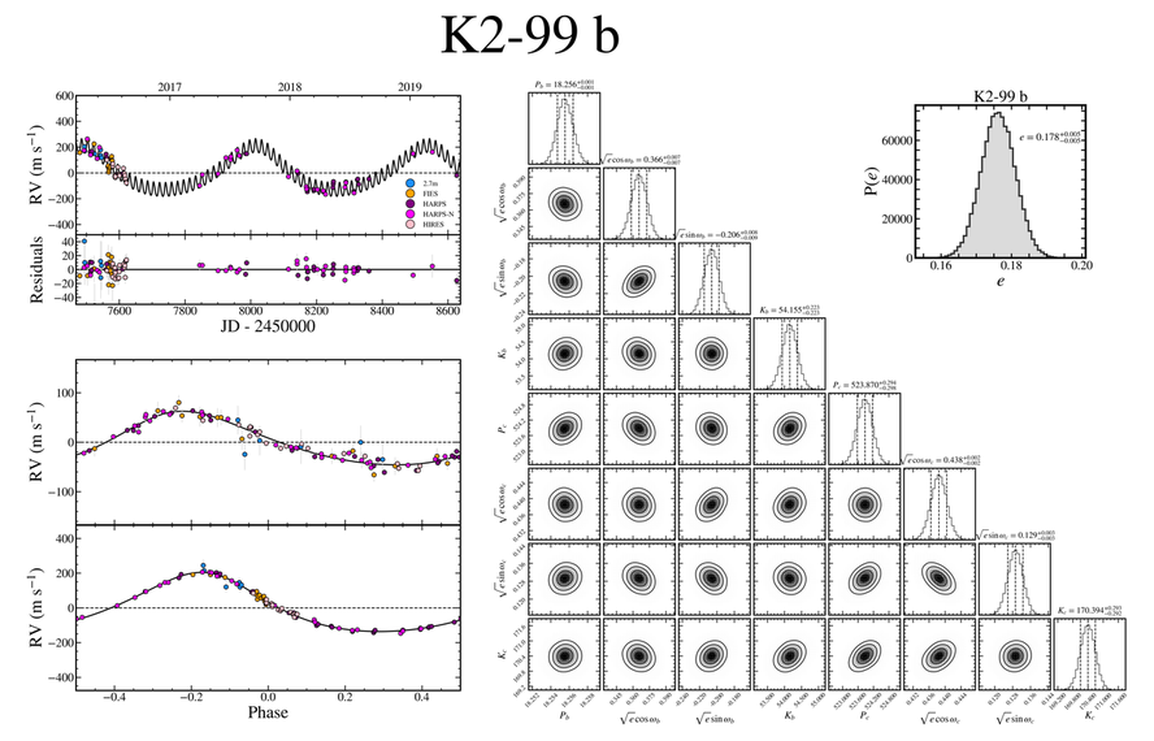}
 \end{minipage}
 \caption{Summary of results for the warm Jupiters K2-329 b and K2-99 b.}
 \label{fig:Combined_Plots78}
\end{figure}
\clearpage
\begin{figure}
\hskip -0.8 in
 \centering
 \begin{minipage}{\textwidth}
   \centering
   \includegraphics[width=\linewidth]{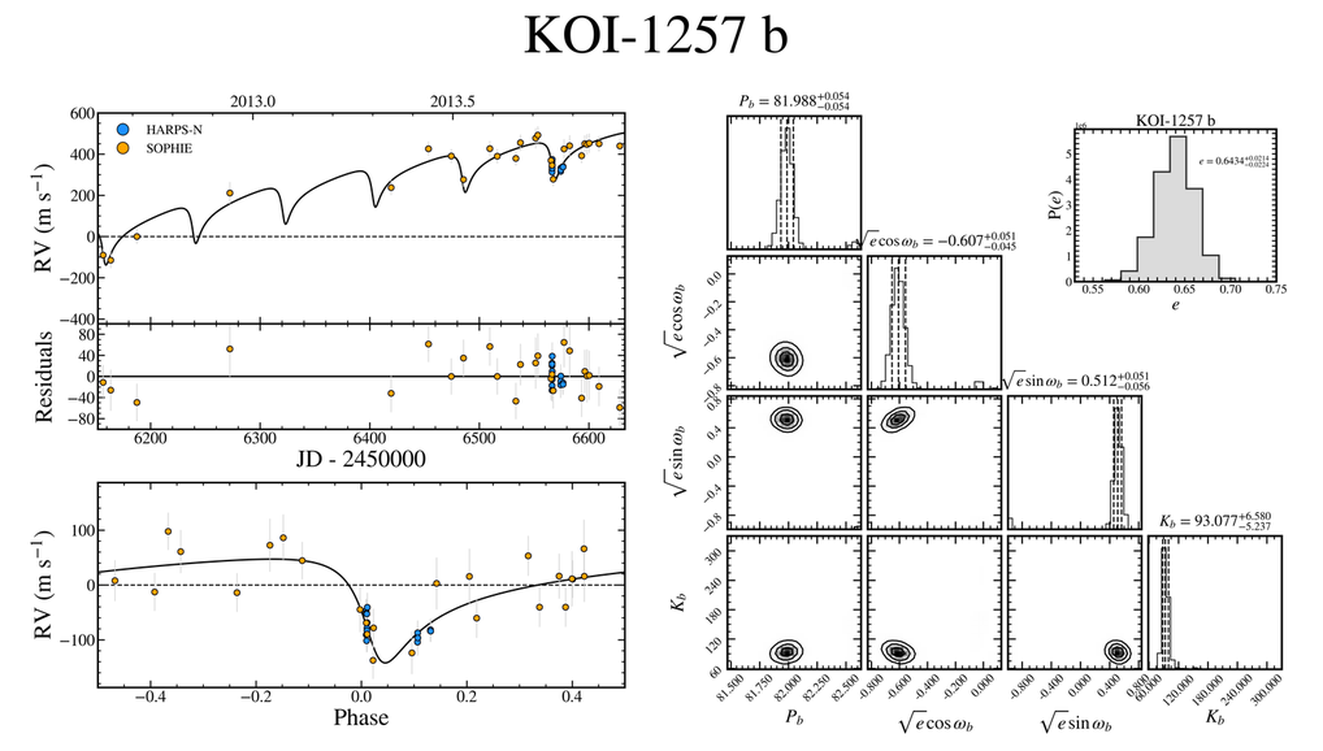}\\
   \vskip .3 in
   \includegraphics[width=\linewidth]{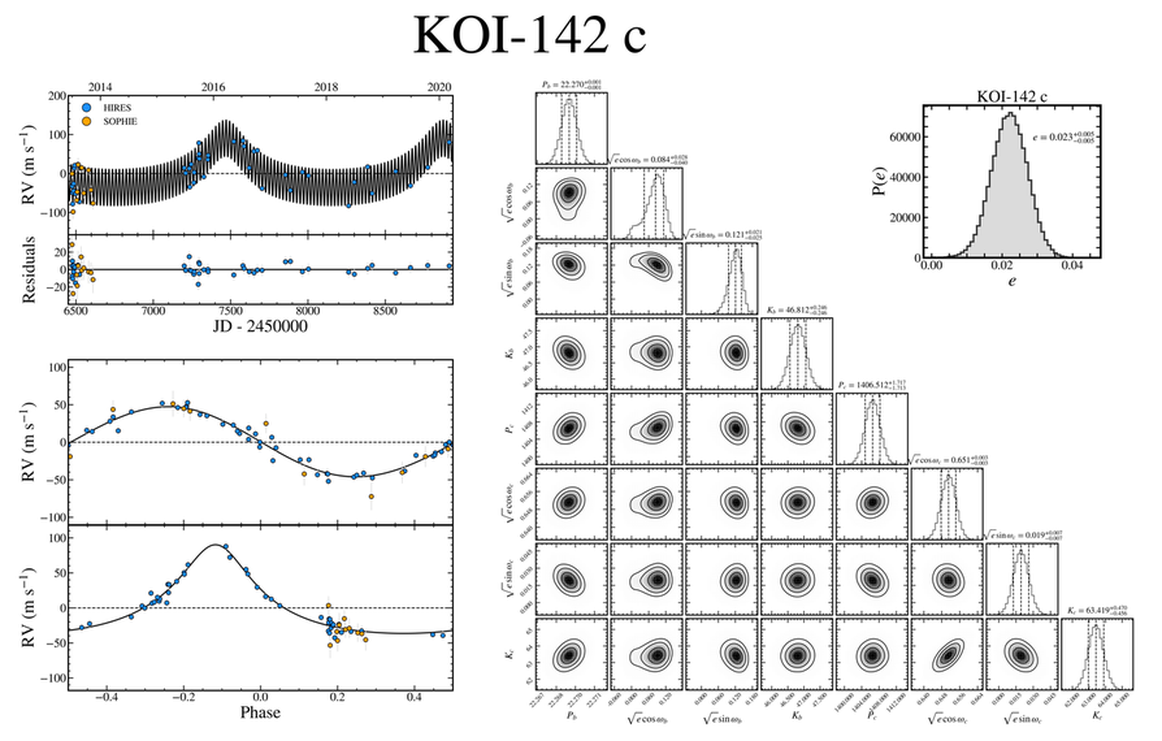}
 \end{minipage}
 \caption{Summary of results for the warm Jupiters KOI-1257 b and KOI-142 c.}
 \label{fig:Combined_Plots79}
\end{figure}
\clearpage
\begin{figure}
\hskip -0.8 in
 \centering
 \begin{minipage}{\textwidth}
   \centering
   \includegraphics[width=\linewidth]{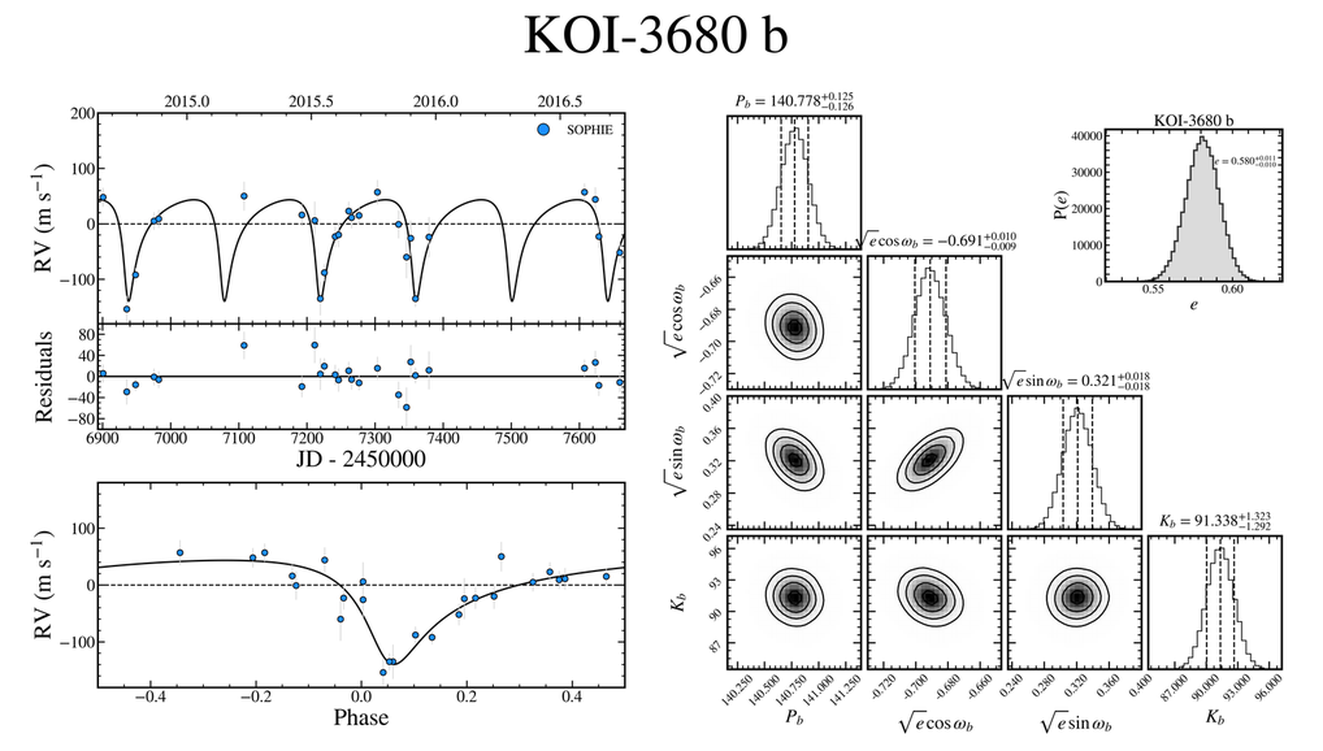}\\
   \vskip .3 in
   \includegraphics[width=\linewidth]{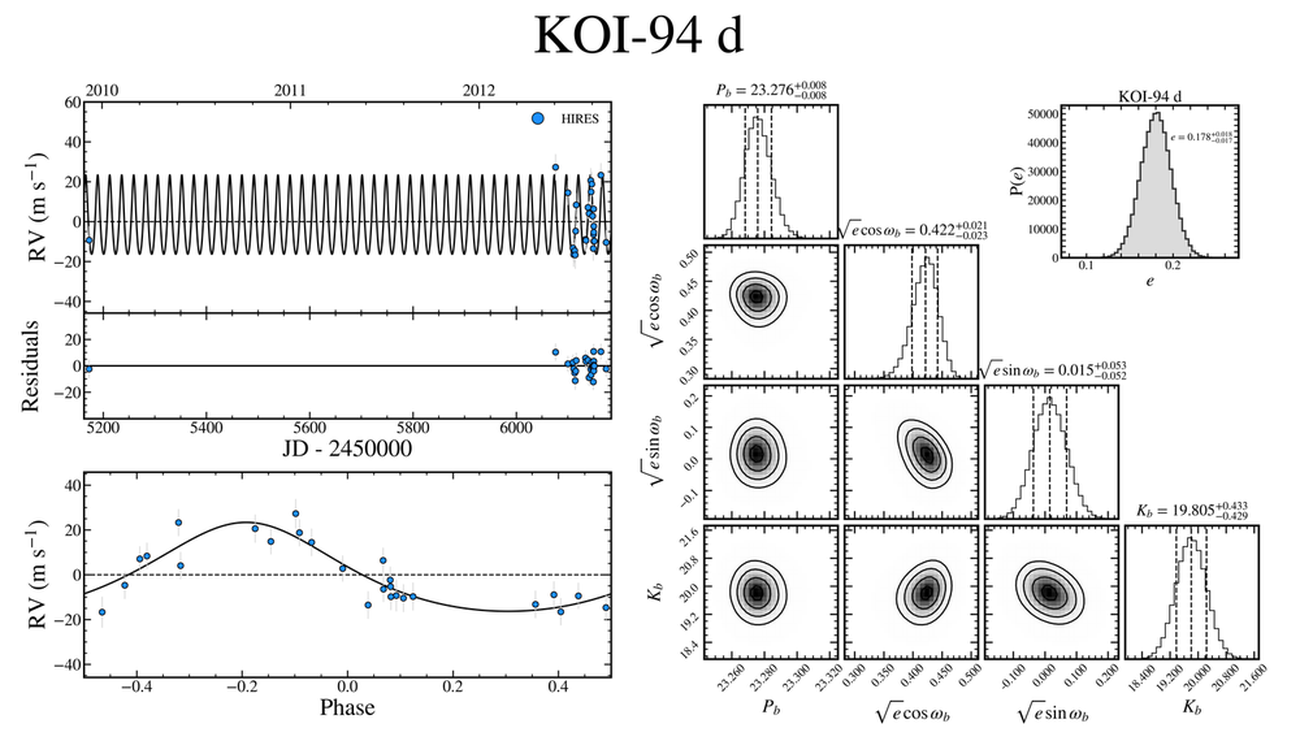}
 \end{minipage}
 \caption{Summary of results for the warm Jupiters KOI-3680 b and KOI-94 d.}
 \label{fig:Combined_Plots80}
\end{figure}
\clearpage
\begin{figure}
\hskip -0.8 in
 \centering
 \begin{minipage}{\textwidth}
   \centering
   \includegraphics[width=\linewidth]{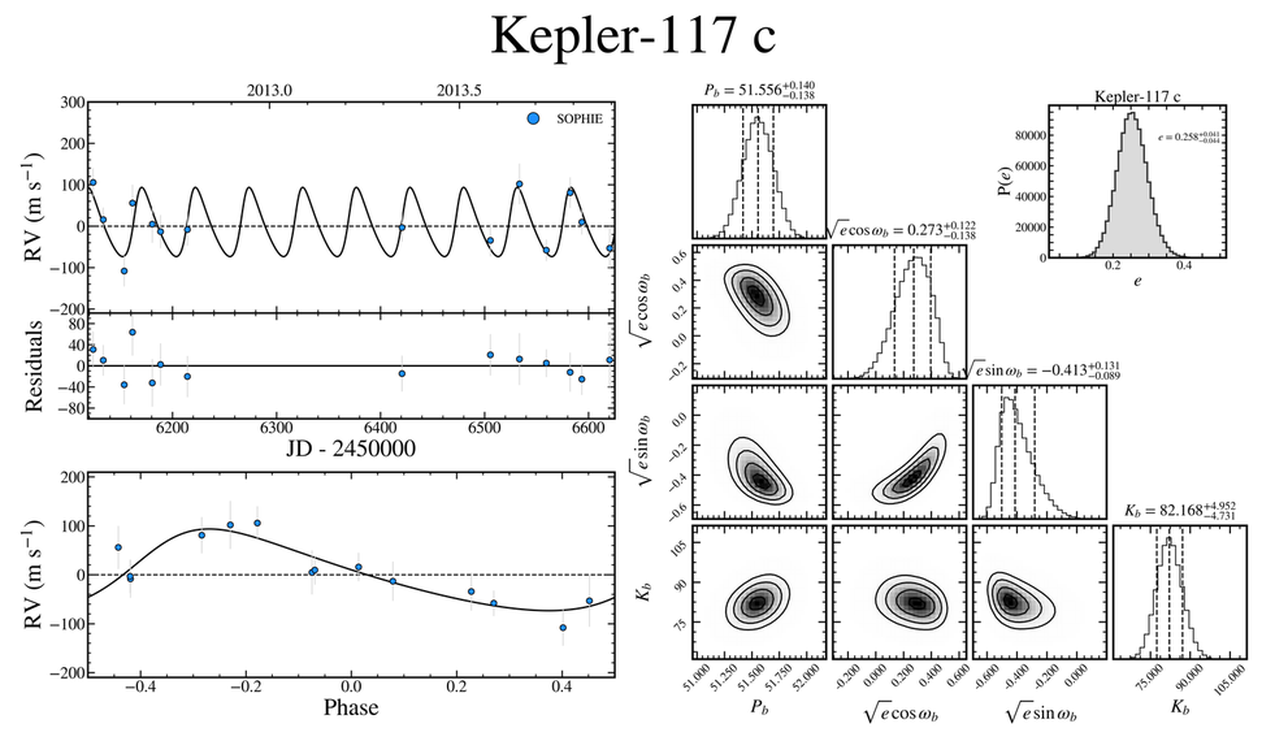}\\
   \vskip .3 in
   \includegraphics[width=\linewidth]{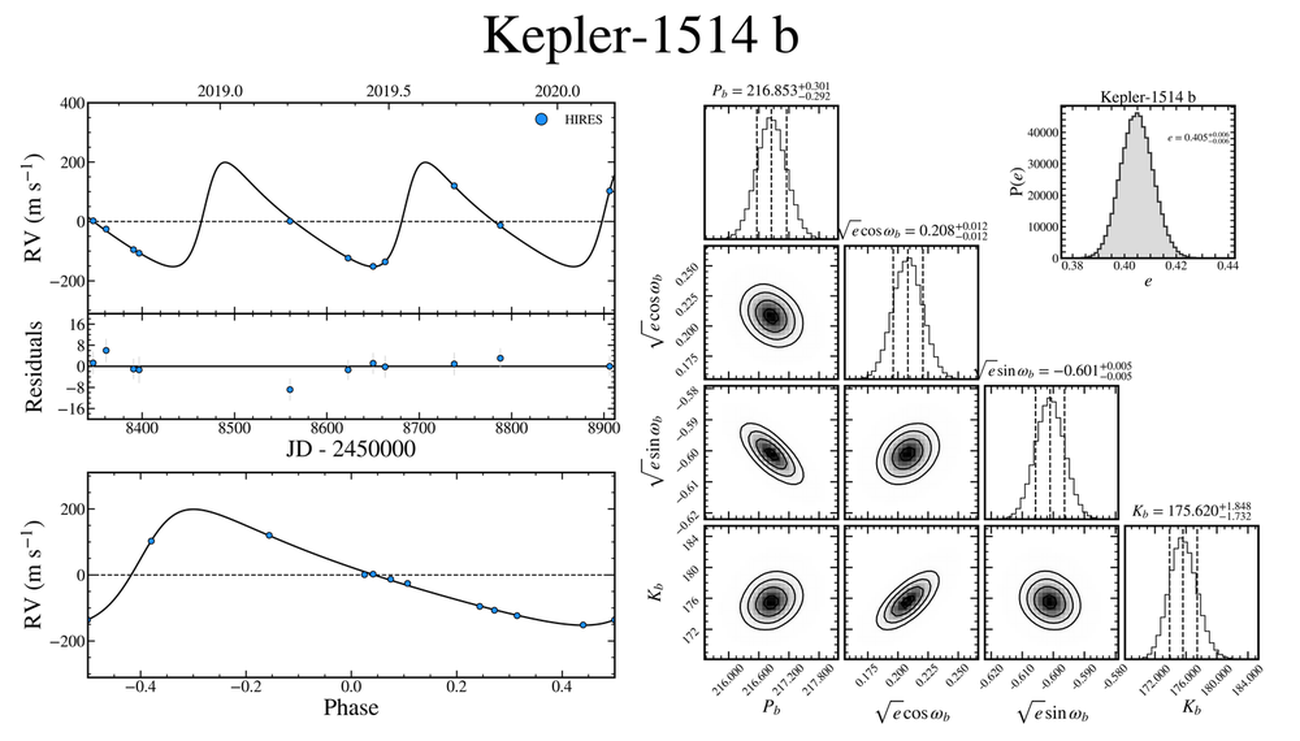}
 \end{minipage}
 \caption{Summary of results for the warm Jupiters Kepler-117 c and Kepler-1514 b.}
 \label{fig:Combined_Plots81}
\end{figure}
\clearpage
\begin{figure}
\hskip -0.8 in
 \centering
 \begin{minipage}{\textwidth}
   \centering
   \includegraphics[width=\linewidth]{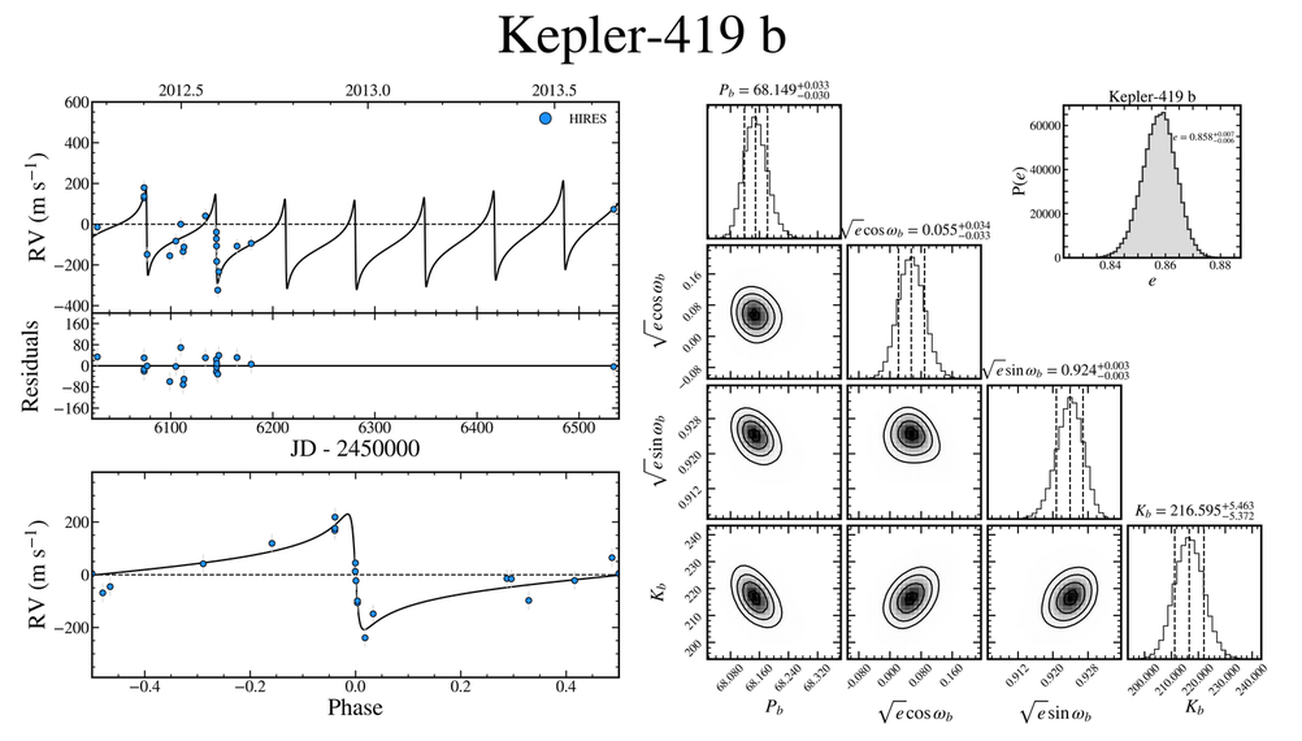}\\
   \vskip .3 in
   \includegraphics[width=\linewidth]{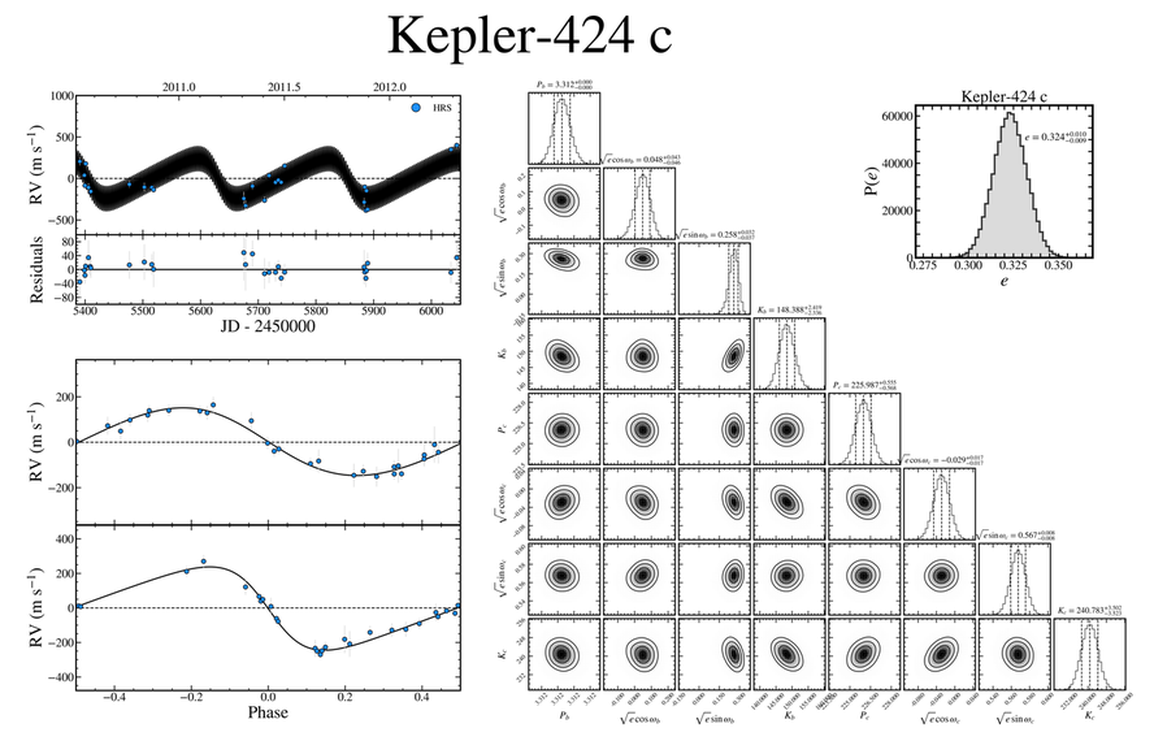}
 \end{minipage}
 \caption{Summary of results for the warm Jupiters Kepler-419 b and Kepler-424 c.}
 \label{fig:Combined_Plots82}
\end{figure}
\clearpage
\begin{figure}
\hskip -0.8 in
 \centering
 \begin{minipage}{\textwidth}
   \centering
   \includegraphics[width=\linewidth]{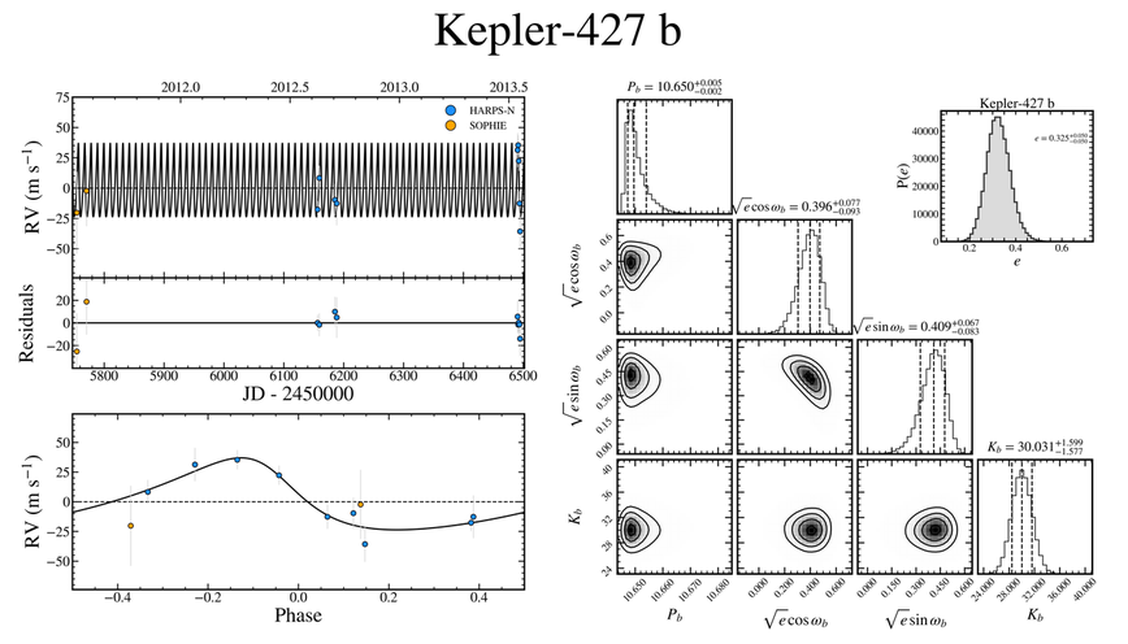}\\
   \vskip .3 in
   \includegraphics[width=\linewidth]{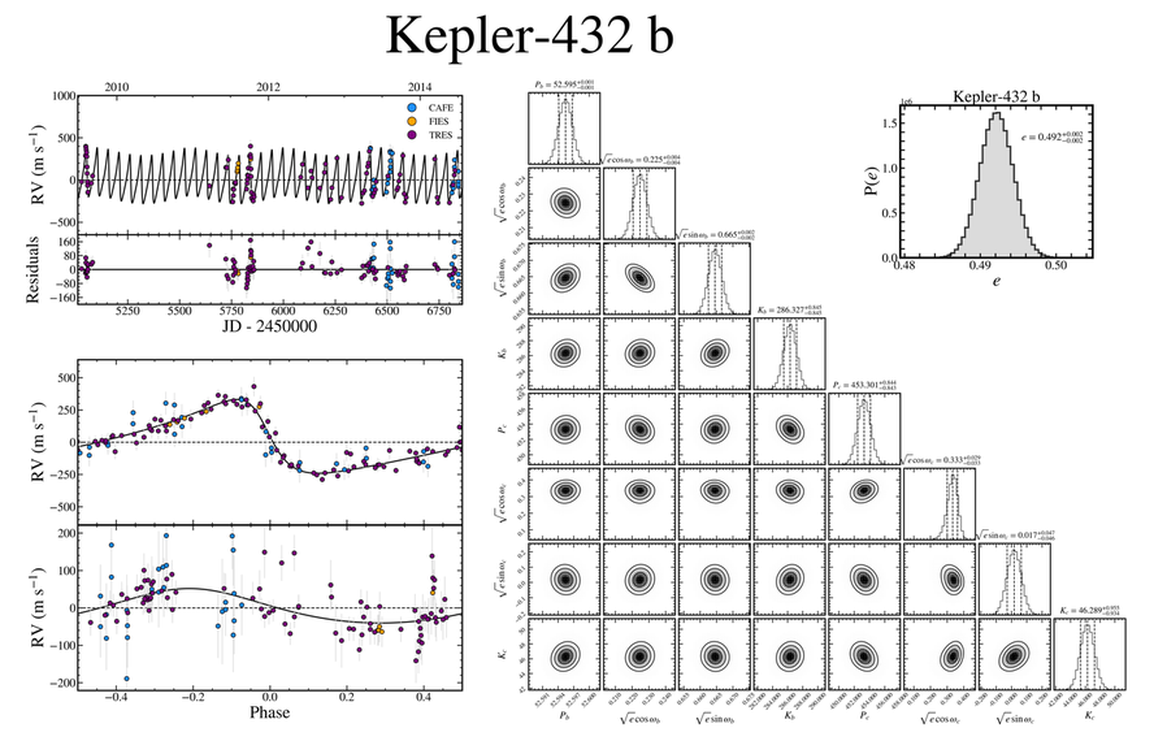}
 \end{minipage}
 \caption{Summary of results for the warm Jupiters Kepler-427 b and Kepler-432 b.}
 \label{fig:Combined_Plots83}
\end{figure}
\clearpage
\begin{figure}
\hskip -0.8 in
 \centering
 \begin{minipage}{\textwidth}
   \centering
   \includegraphics[width=\linewidth]{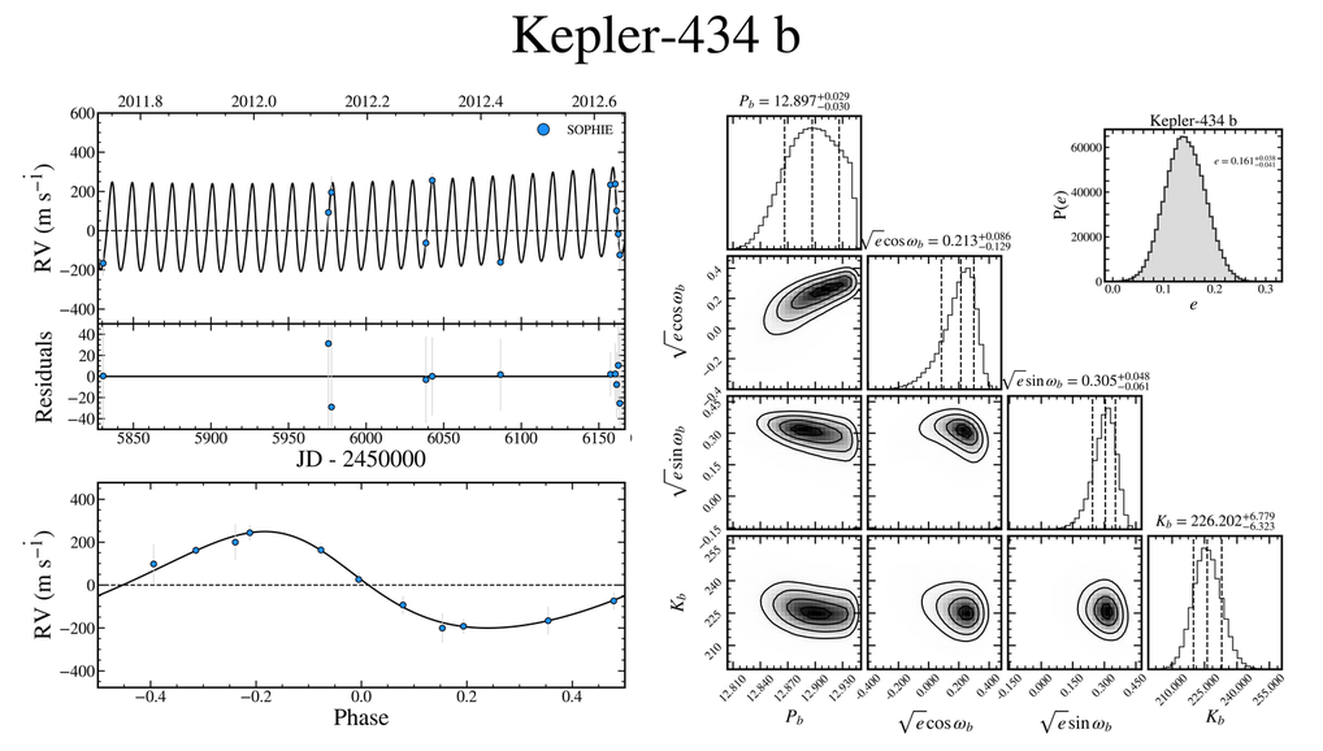}\\
   \vskip .3 in
   \includegraphics[width=\linewidth]{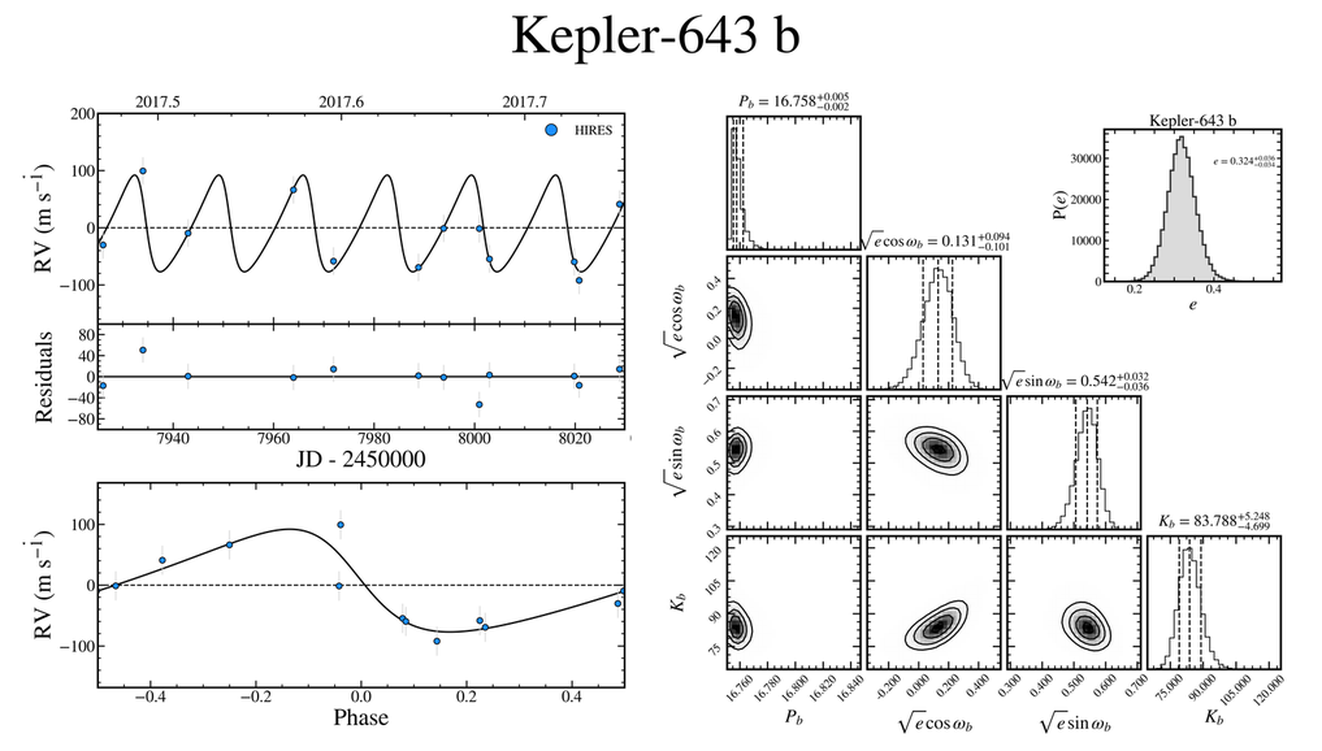}
 \end{minipage}
 \caption{Summary of results for the warm Jupiters Kepler-434 b and Kepler-643 b.}
 \label{fig:Combined_Plots84}
\end{figure}
\clearpage
\begin{figure}
\hskip -0.8 in
 \centering
 \begin{minipage}{\textwidth}
   \centering
   \includegraphics[width=\linewidth]{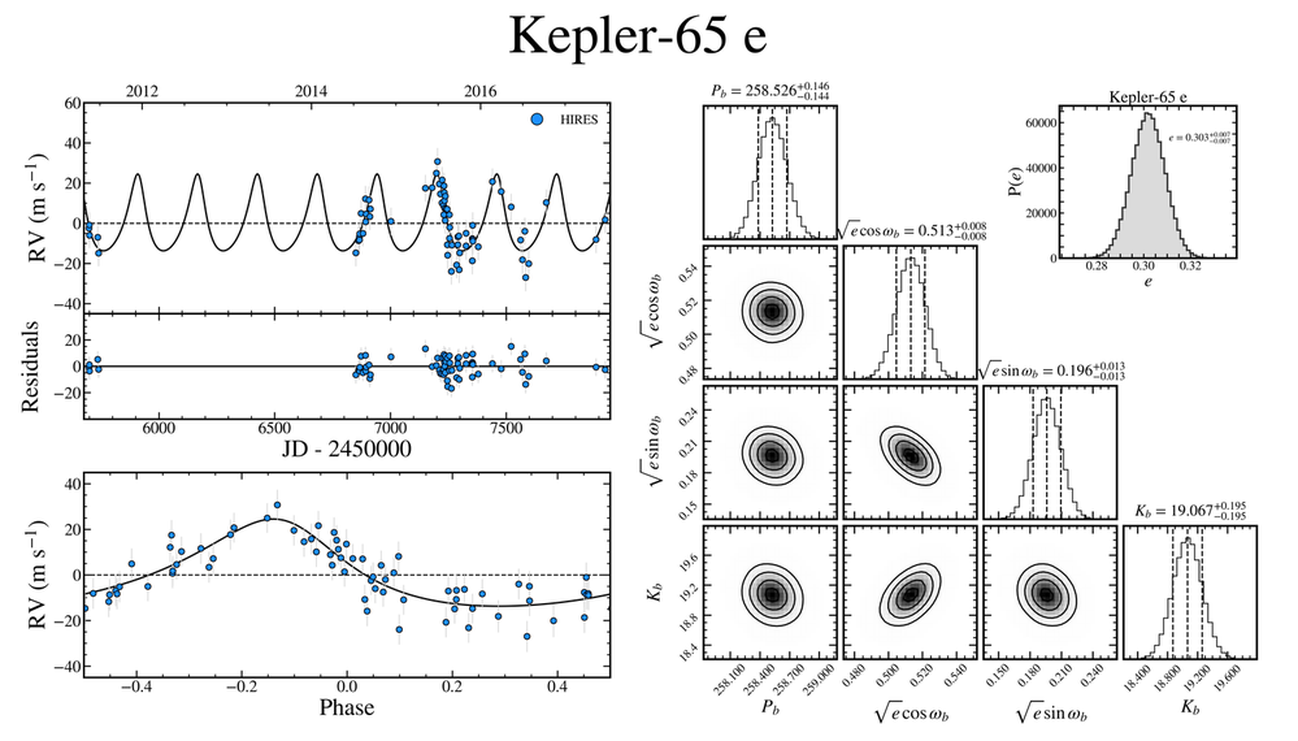}\\
   \vskip .3 in
   \includegraphics[width=\linewidth]{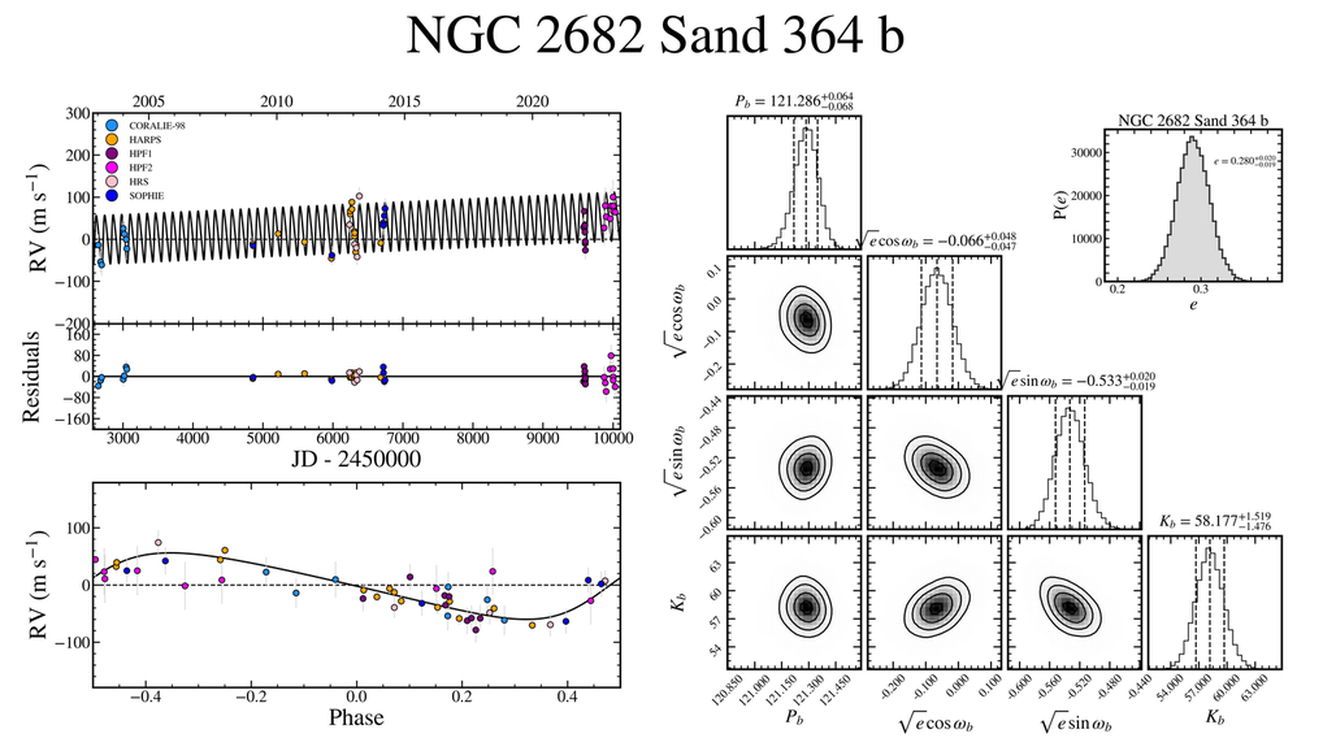}
 \end{minipage}
 \caption{Summary of results for the warm Jupiters Kepler-65 e and NGC 2682 Sand 364 b.}
 \label{fig:Combined_Plots85}
\end{figure}
\clearpage
\begin{figure}
\hskip -0.8 in
 \centering
 \begin{minipage}{\textwidth}
   \centering
   \includegraphics[width=\linewidth]{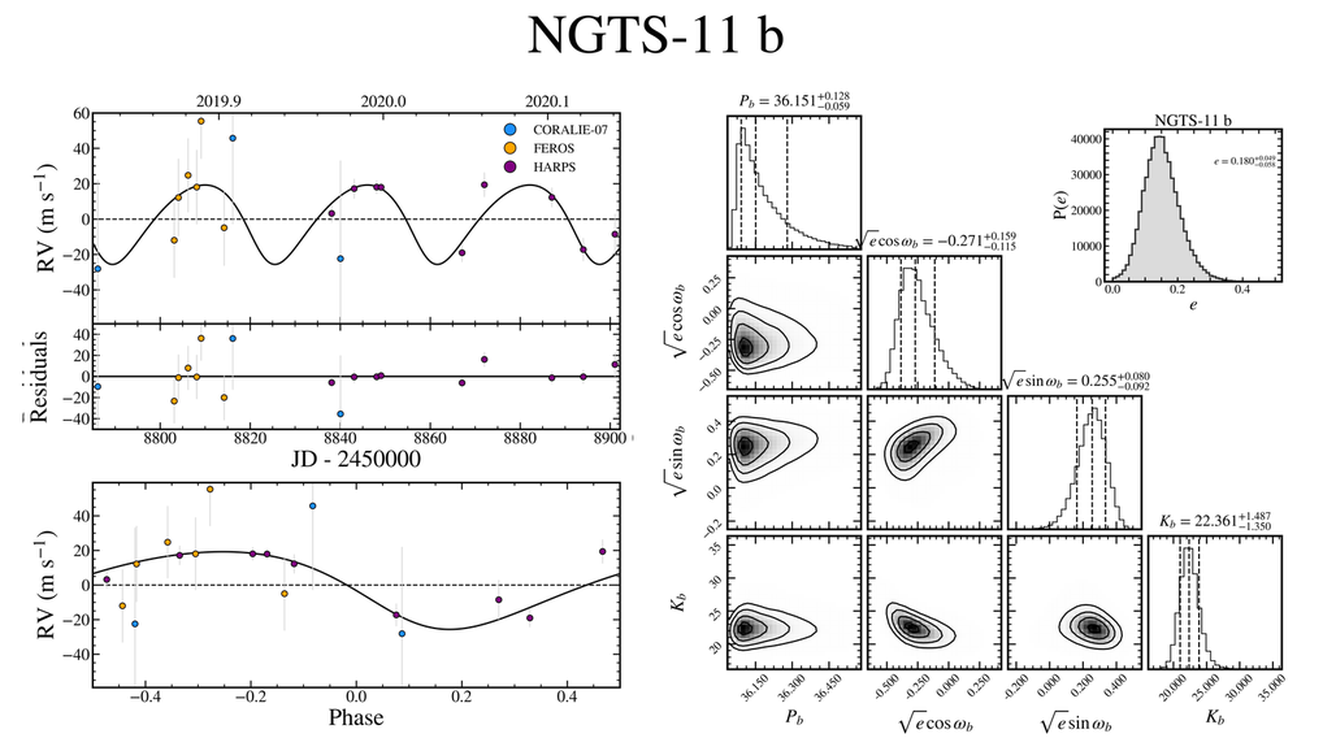}\\
   \vskip .3 in
   \includegraphics[width=\linewidth]{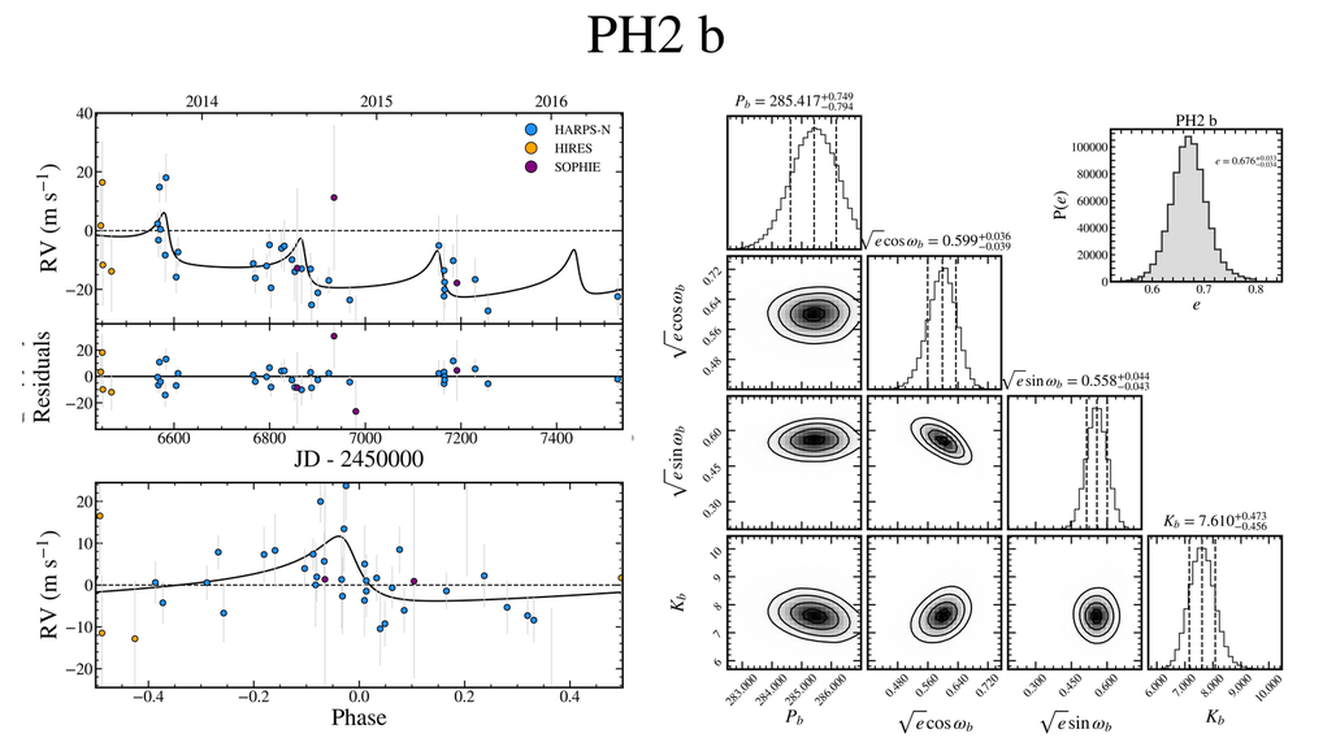}
 \end{minipage}
 \caption{Summary of results for the warm Jupiters NGTS-11 b and PH2 b.}
 \label{fig:Combined_Plots86}
\end{figure}
\clearpage
\begin{figure}
\hskip -0.8 in
 \centering
 \begin{minipage}{\textwidth}
   \centering
   \includegraphics[width=\linewidth]{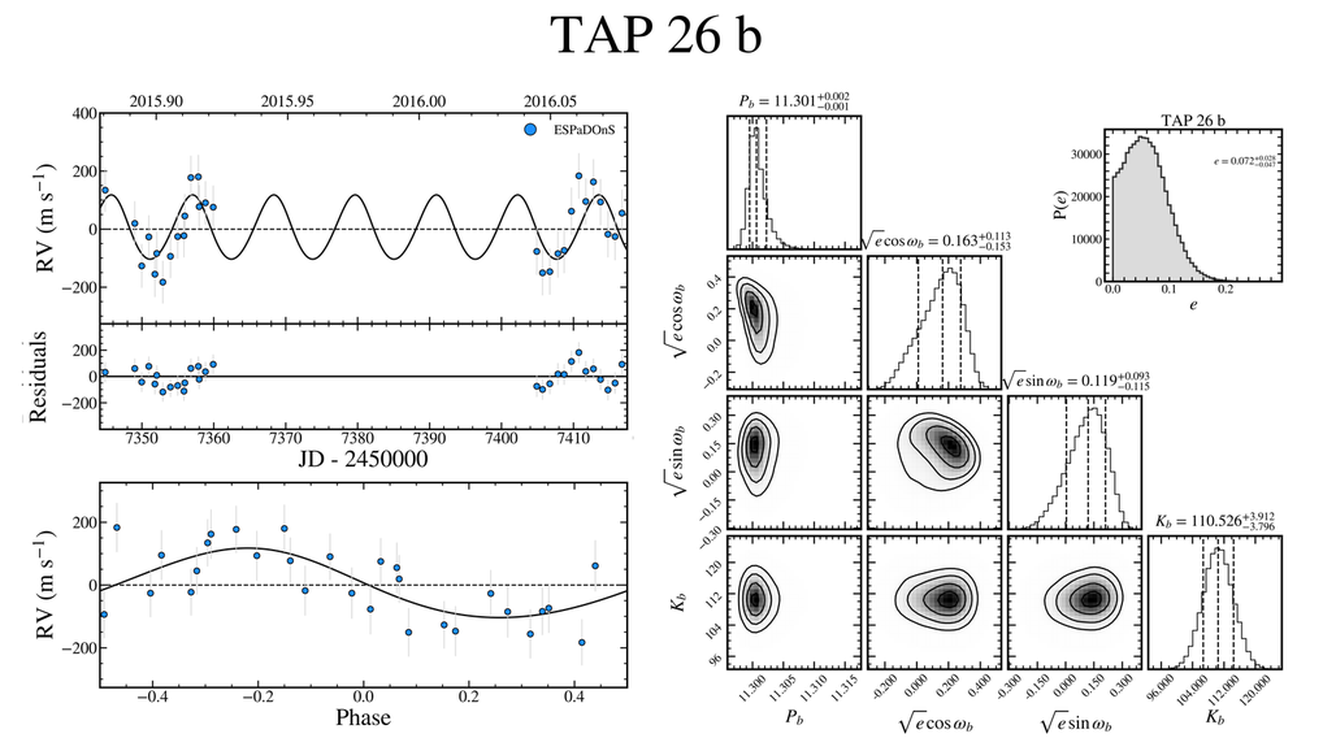}\\
   \vskip .3 in
   \includegraphics[width=\linewidth]{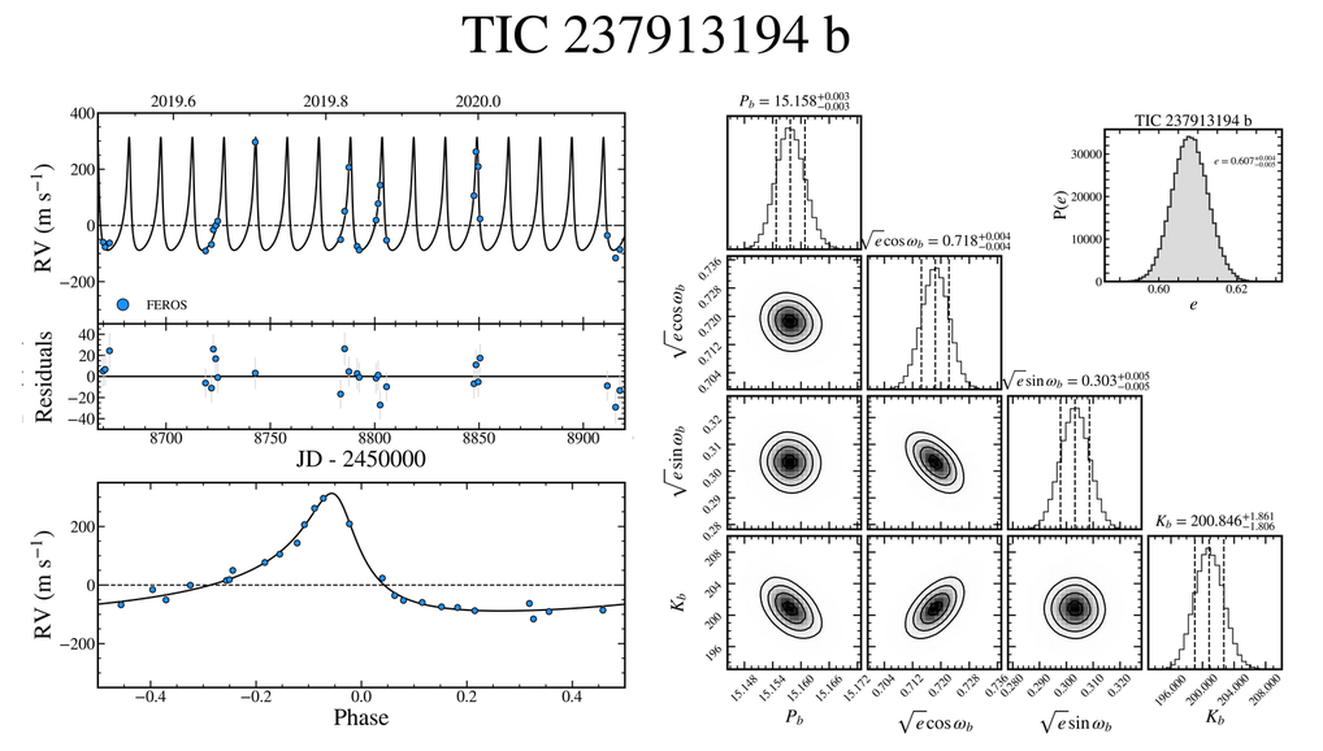}
 \end{minipage}
 \caption{Summary of results for the warm Jupiters TAP 26 b and TIC 237913194 b.}
 \label{fig:Combined_Plots87}
\end{figure}
\clearpage
\begin{figure}
\hskip -0.8 in
 \centering
 \begin{minipage}{\textwidth}
   \centering
   \includegraphics[width=\linewidth]{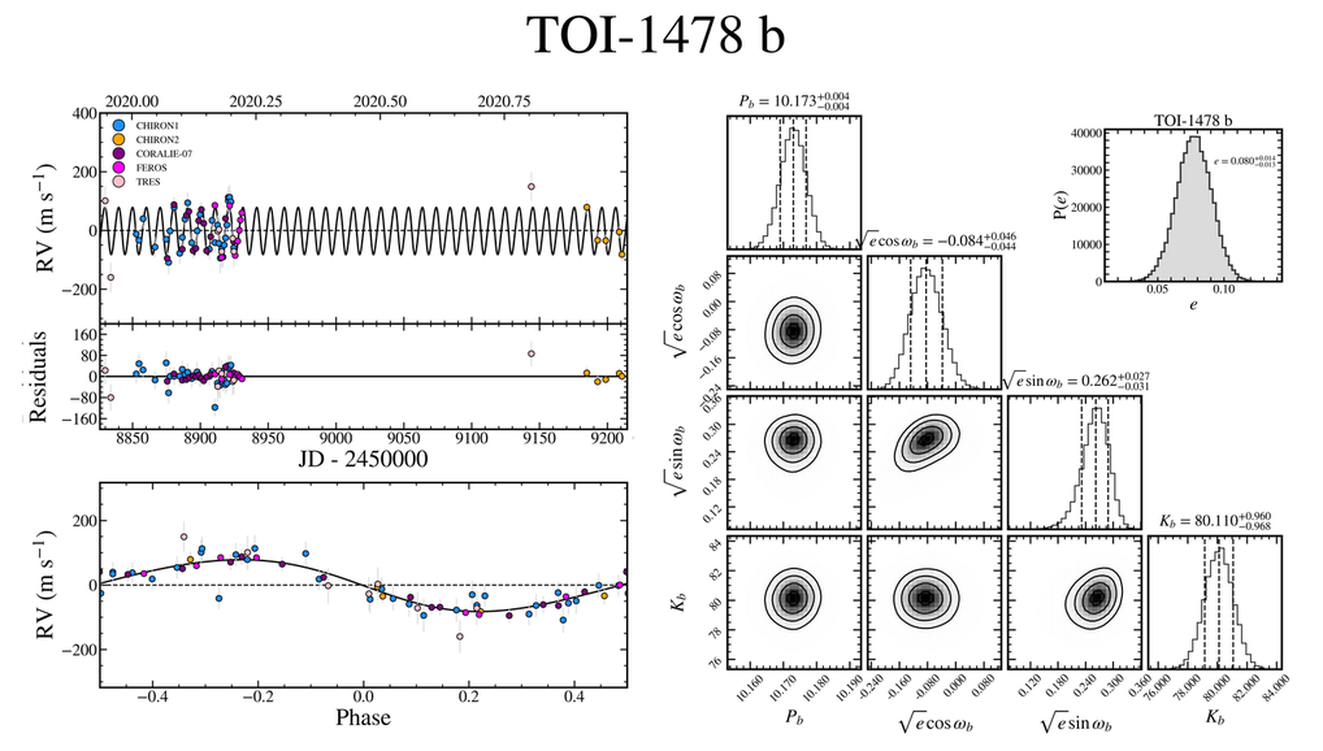}\\
   \vskip .3 in
   \includegraphics[width=\linewidth]{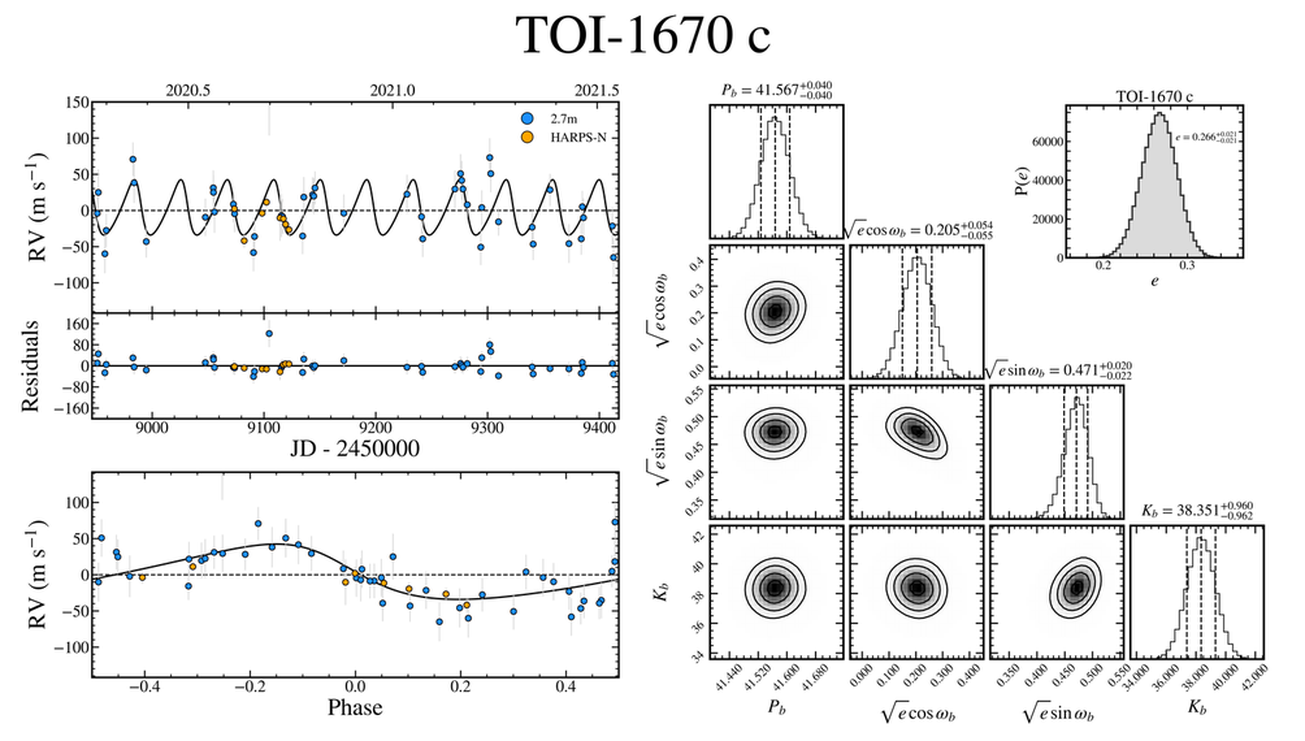}
 \end{minipage}
 \caption{Summary of results for the warm Jupiters TOI-1478 b and TOI-1670 c.}
 \label{fig:Combined_Plots88}
\end{figure}
\clearpage
\begin{figure}
\hskip -0.8 in
 \centering
 \begin{minipage}{\textwidth}
   \centering
   \includegraphics[width=\linewidth]{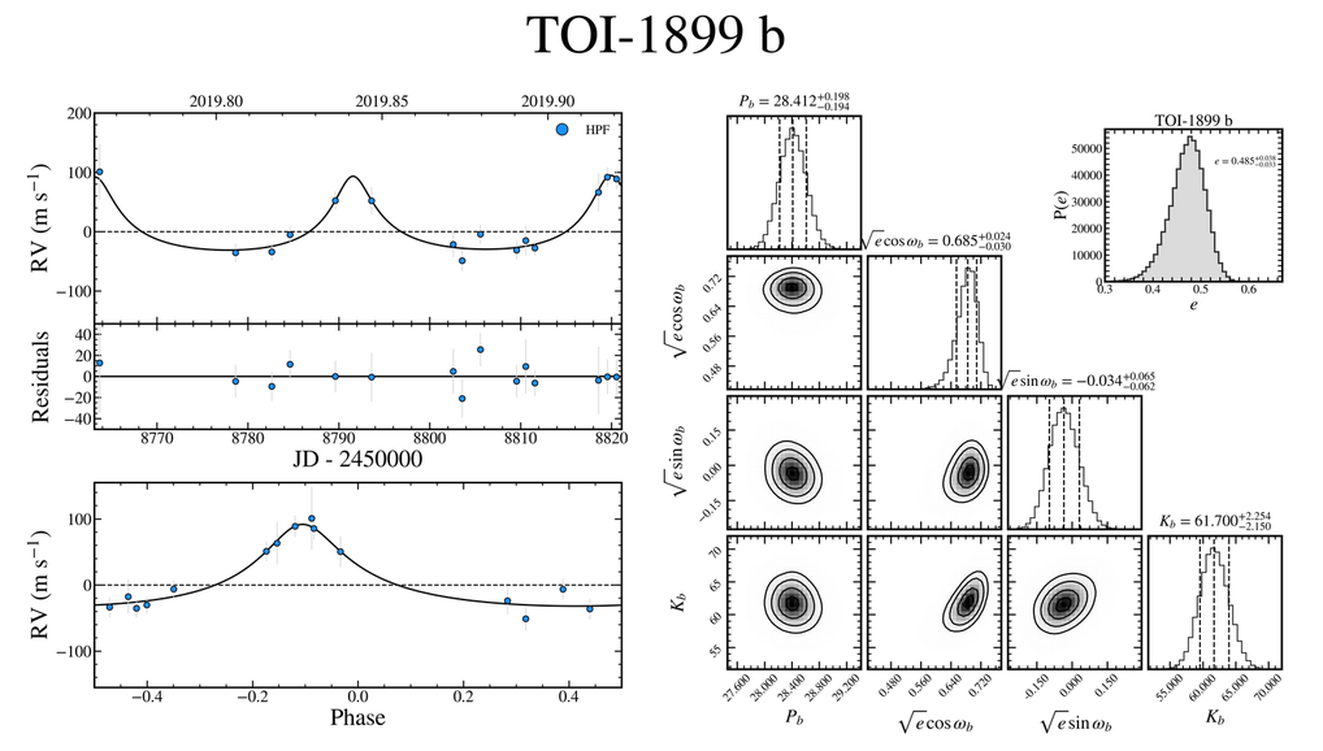}\\
   \vskip .3 in
   \includegraphics[width=\linewidth]{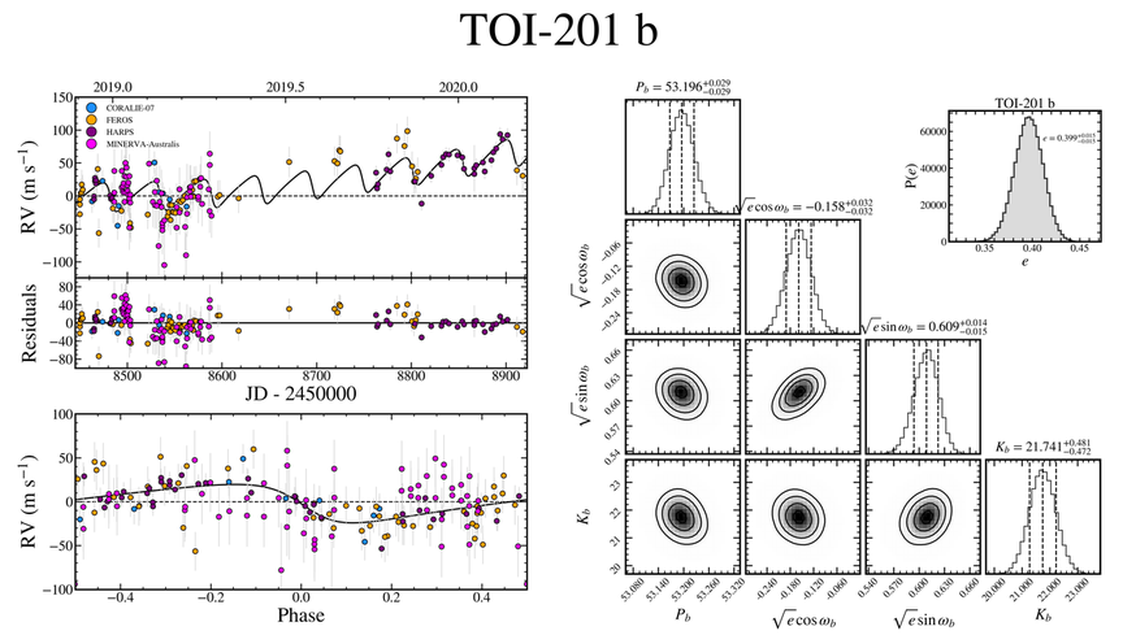}
 \end{minipage}
 \caption{Summary of results for the warm Jupiters TOI-1899 b and TOI-201 b.}
 \label{fig:Combined_Plots89}
\end{figure}
\clearpage
\begin{figure}
\hskip -0.8 in
 \centering
 \begin{minipage}{\textwidth}
   \centering
   \includegraphics[width=\linewidth]{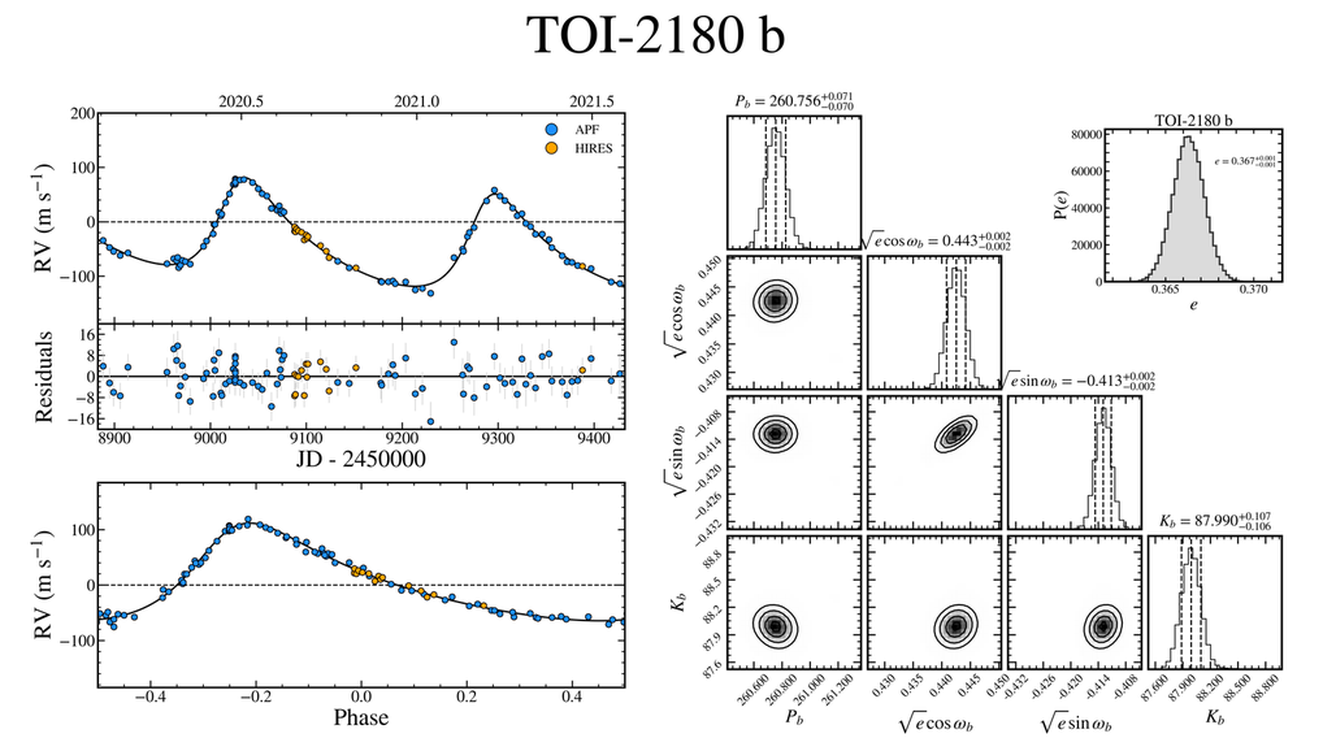}\\
   \vskip .3 in
   \includegraphics[width=\linewidth]{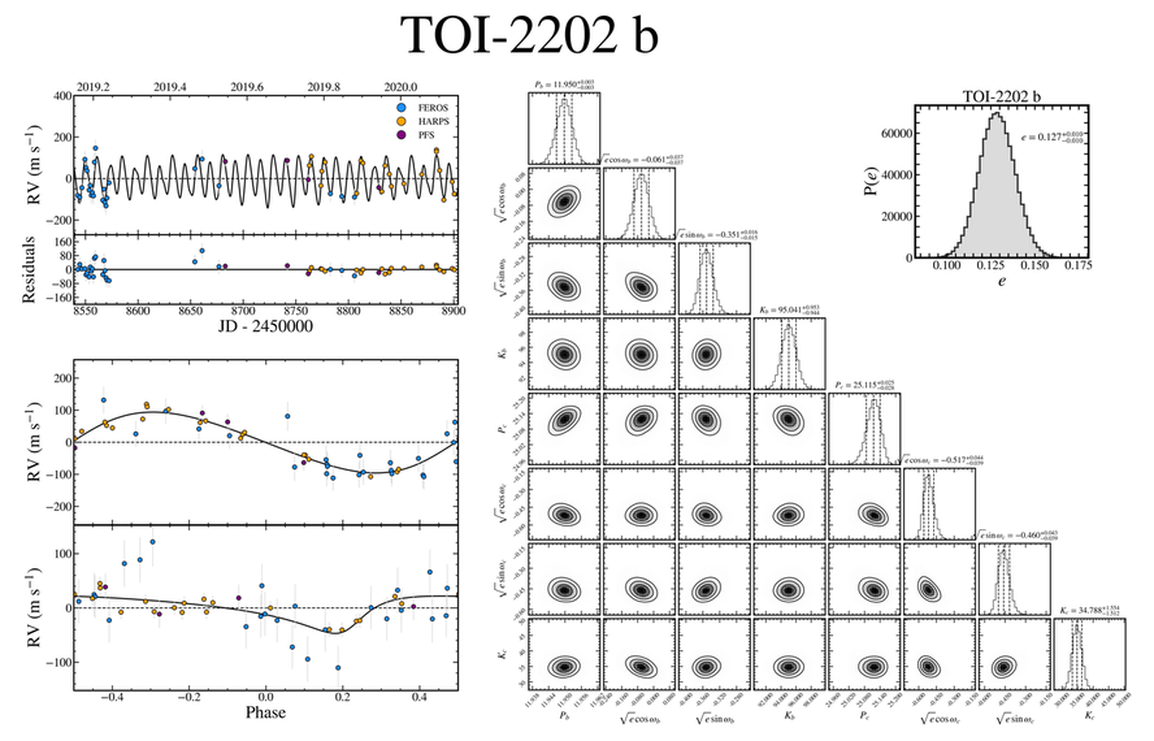}
 \end{minipage}
 \caption{Summary of results for the warm Jupiters TOI-2180 b and TOI-2202 b.}
 \label{fig:Combined_Plots90}
\end{figure}
\clearpage
\begin{figure}
\hskip -0.8 in
 \centering
 \begin{minipage}{\textwidth}
   \centering
   \includegraphics[width=\linewidth]{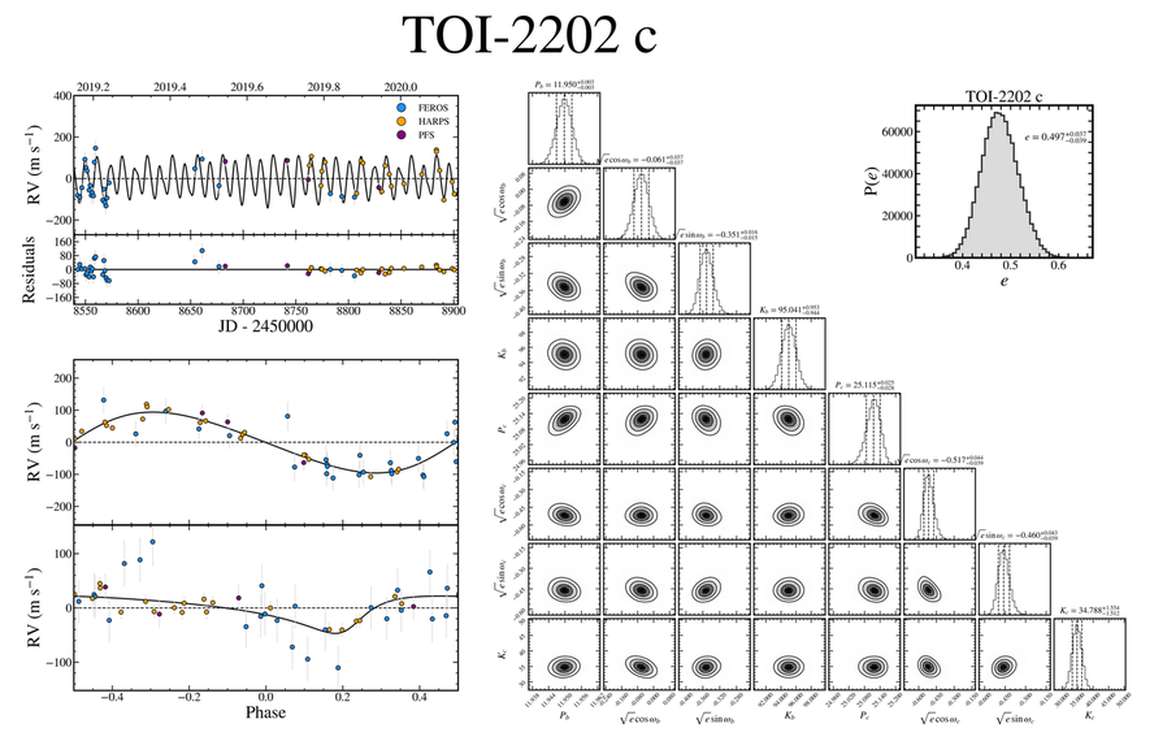}\\
   \vskip .3 in
   \includegraphics[width=\linewidth]{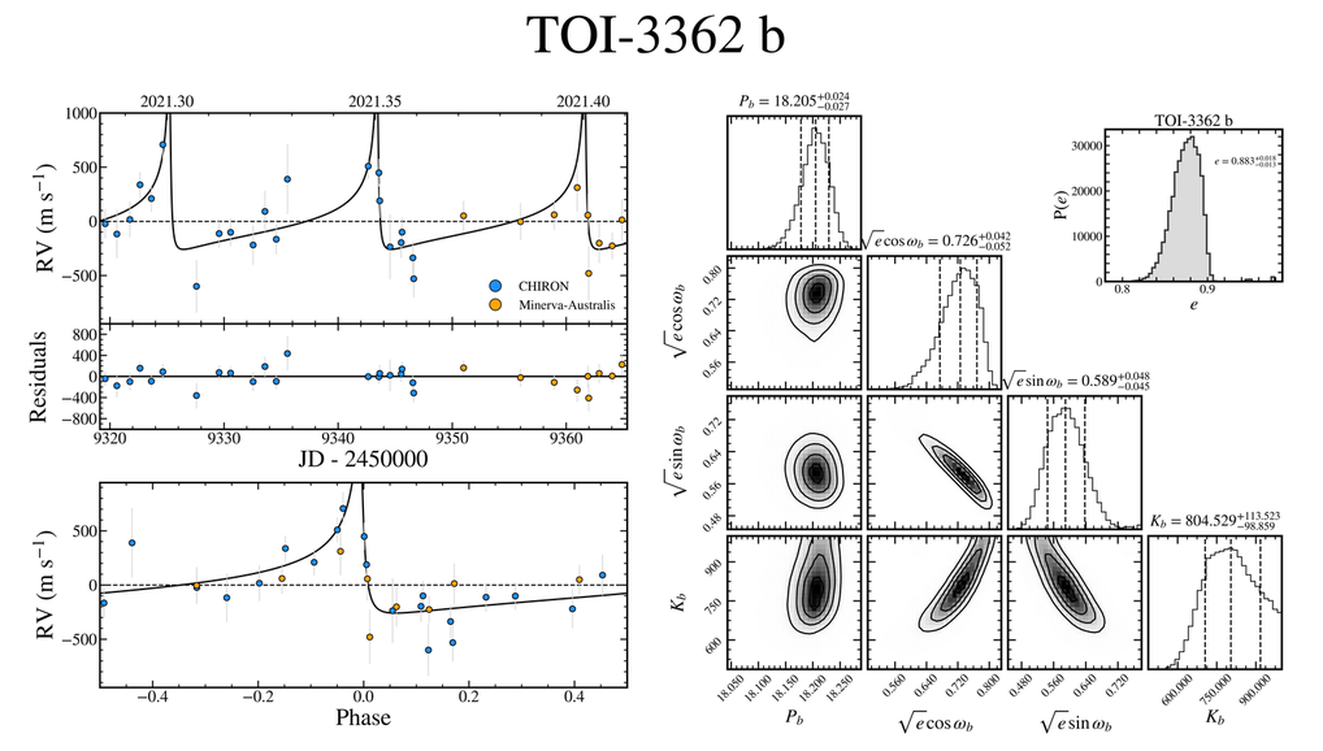}
 \end{minipage}
 \caption{Summary of results for the warm Jupiters TOI-2202 c and TOI-3362 b.}
 \label{fig:Combined_Plots91}
\end{figure}
\clearpage
\begin{figure}
\hskip -0.8 in
 \centering
 \begin{minipage}{\textwidth}
   \centering
   \includegraphics[width=\linewidth]{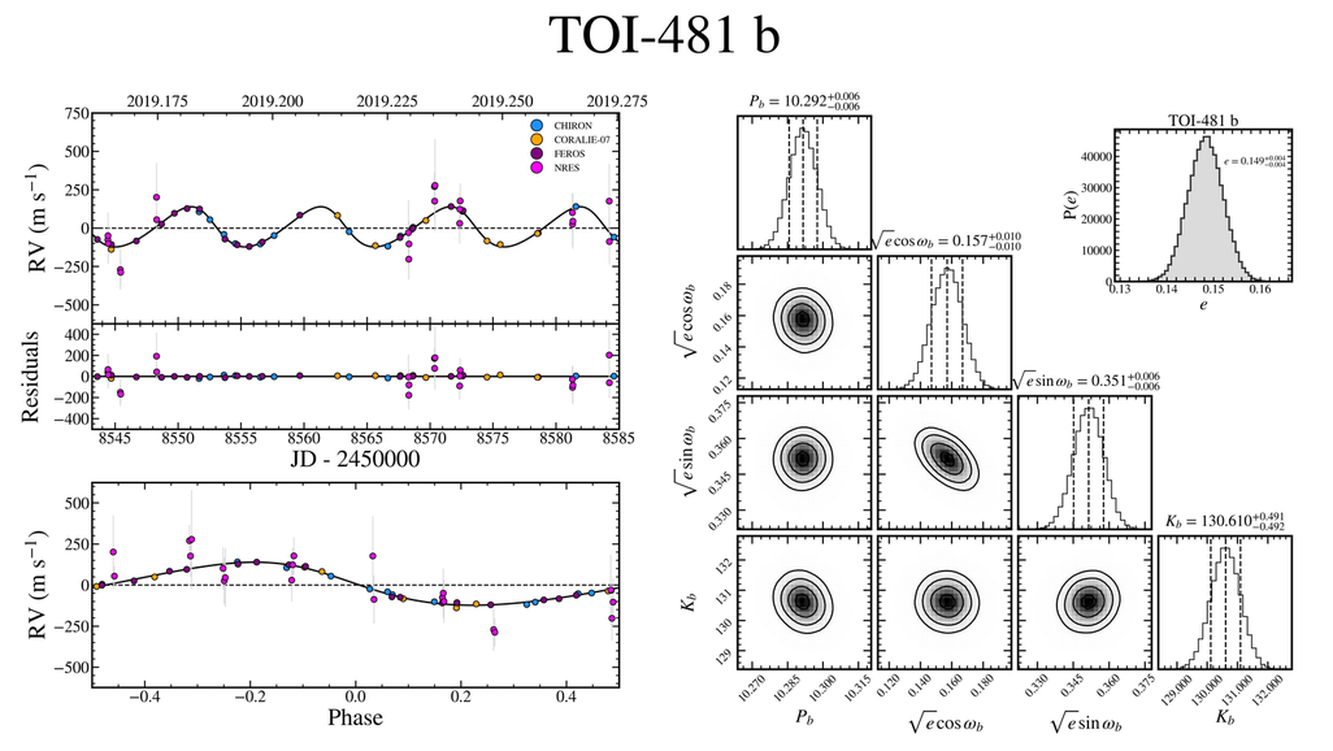}\\
   \vskip .3 in
   \includegraphics[width=\linewidth]{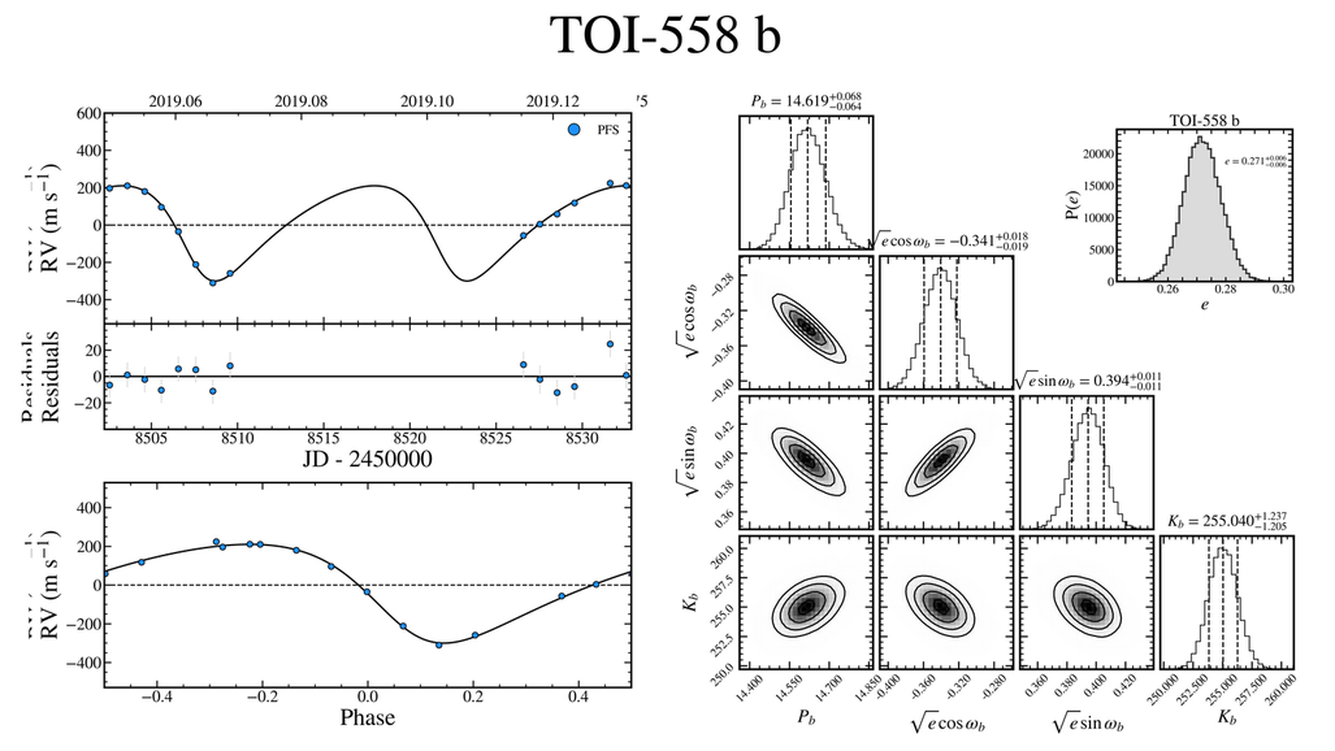}
 \end{minipage}
 \caption{Summary of results for the warm Jupiters TOI-481 b and TOI-558 b.}
 \label{fig:Combined_Plots92}
\end{figure}
\clearpage
\begin{figure}
\hskip -0.8 in
 \centering
 \begin{minipage}{\textwidth}
   \centering
   \includegraphics[width=\linewidth]{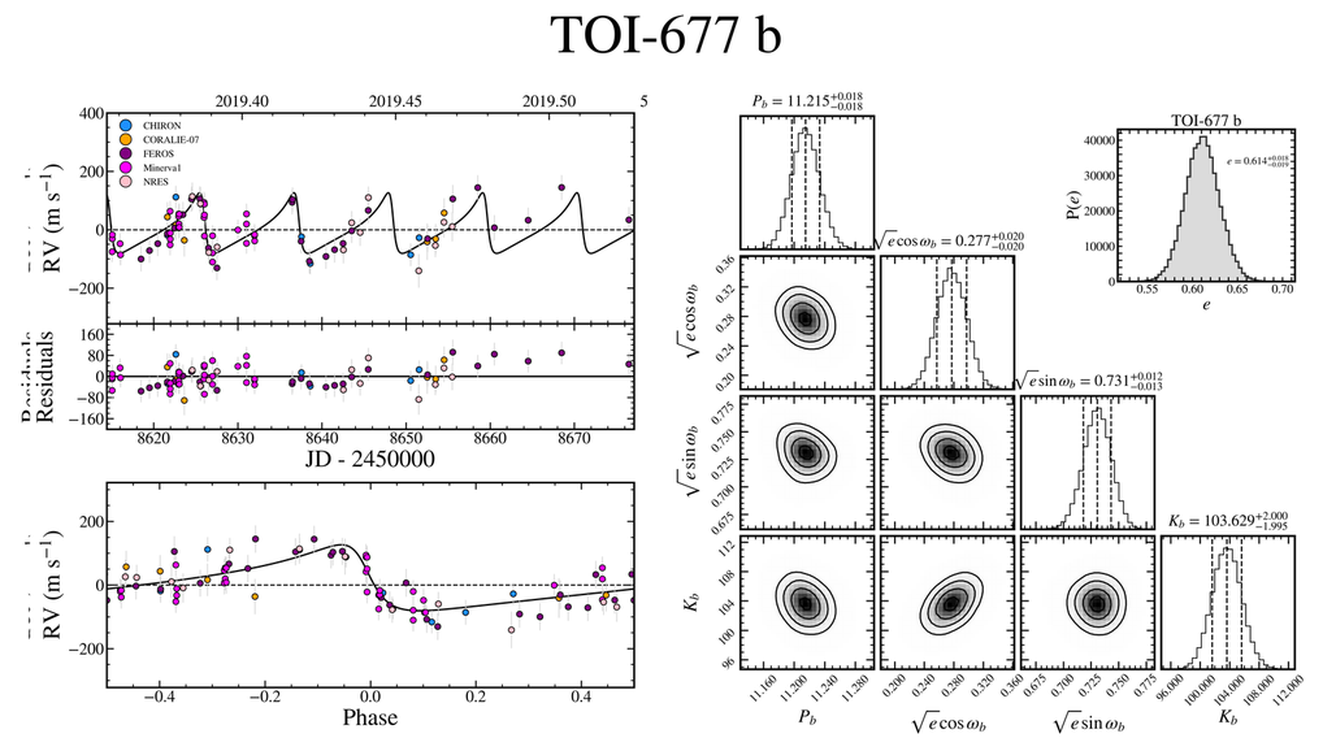}\\
   \vskip .3 in
   \includegraphics[width=\linewidth]{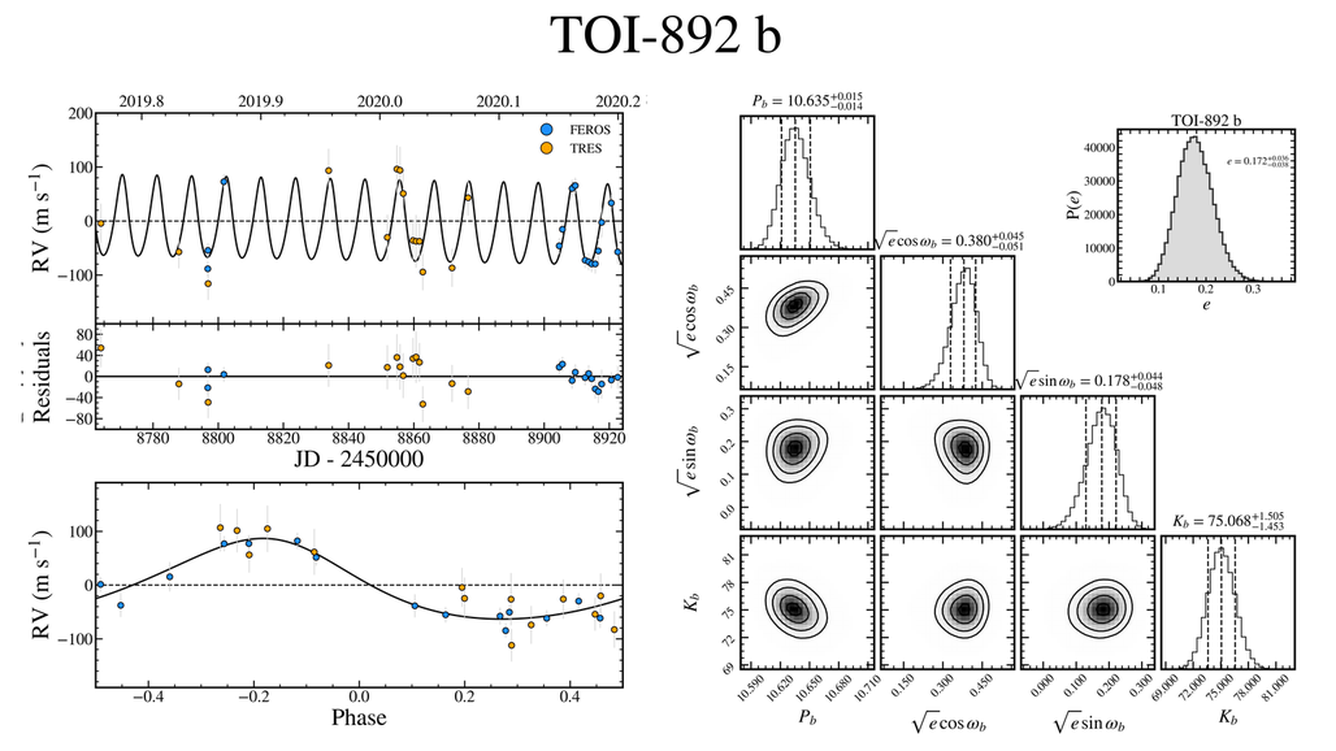}
 \end{minipage}
 \caption{Summary of results for the warm Jupiters TOI-677 b and TOI-892 b.}
 \label{fig:Combined_Plots93}
\end{figure}
\clearpage
\begin{figure}
\hskip -0.8 in
 \centering
 \begin{minipage}{\textwidth}
   \centering
   \includegraphics[width=\linewidth]{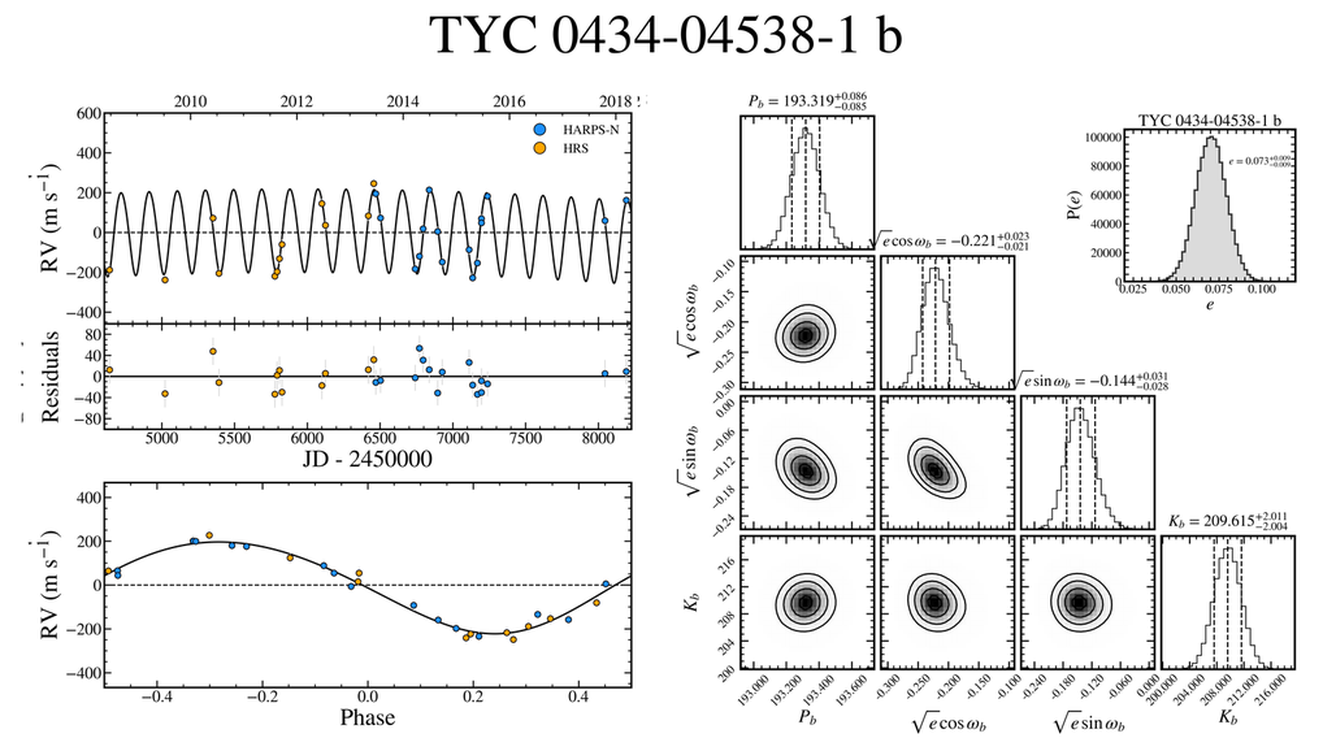}\\
   \vskip .3 in
   \includegraphics[width=\linewidth]{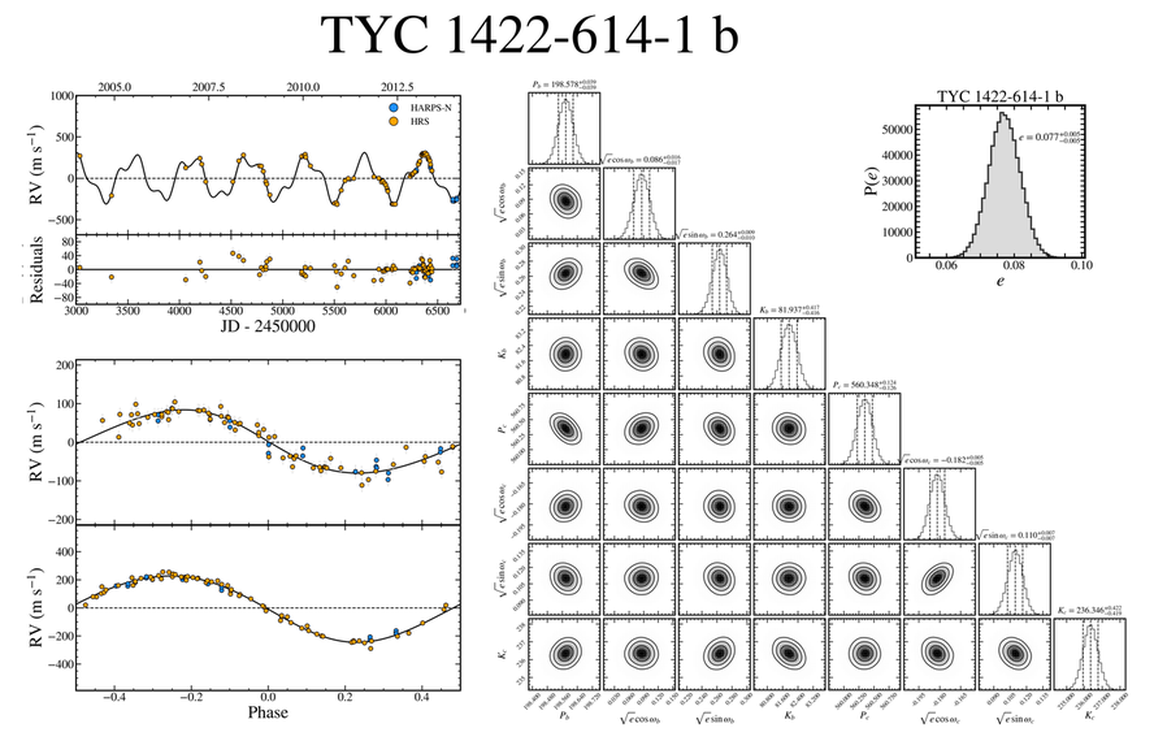}
 \end{minipage}
 \caption{Summary of results for the warm Jupiters TYC 0434-04538-1 b and TYC 1422-614-1 b.}
 \label{fig:Combined_Plots94}
\end{figure}
\clearpage
\begin{figure}
\hskip -0.8 in
 \centering
 \begin{minipage}{\textwidth}
   \centering
   \includegraphics[width=\linewidth]{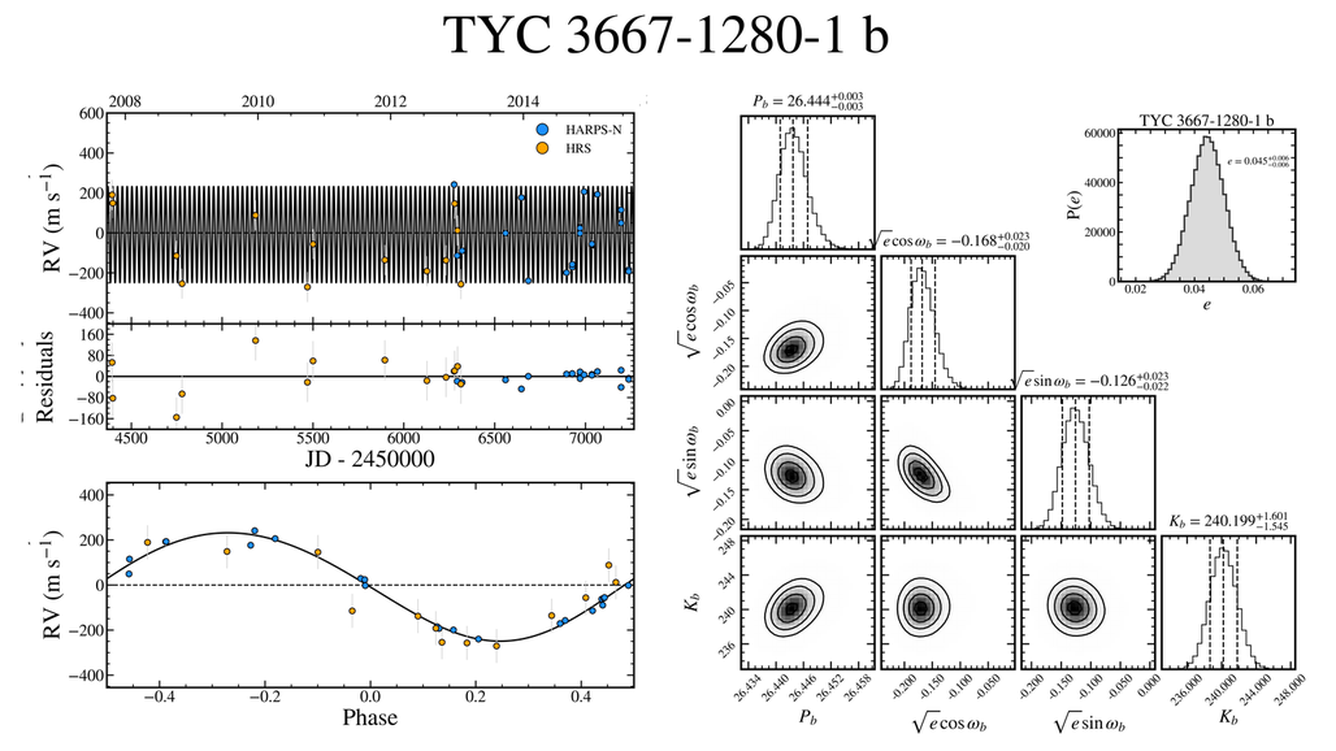}\\
   \vskip .3 in
   \includegraphics[width=\linewidth]{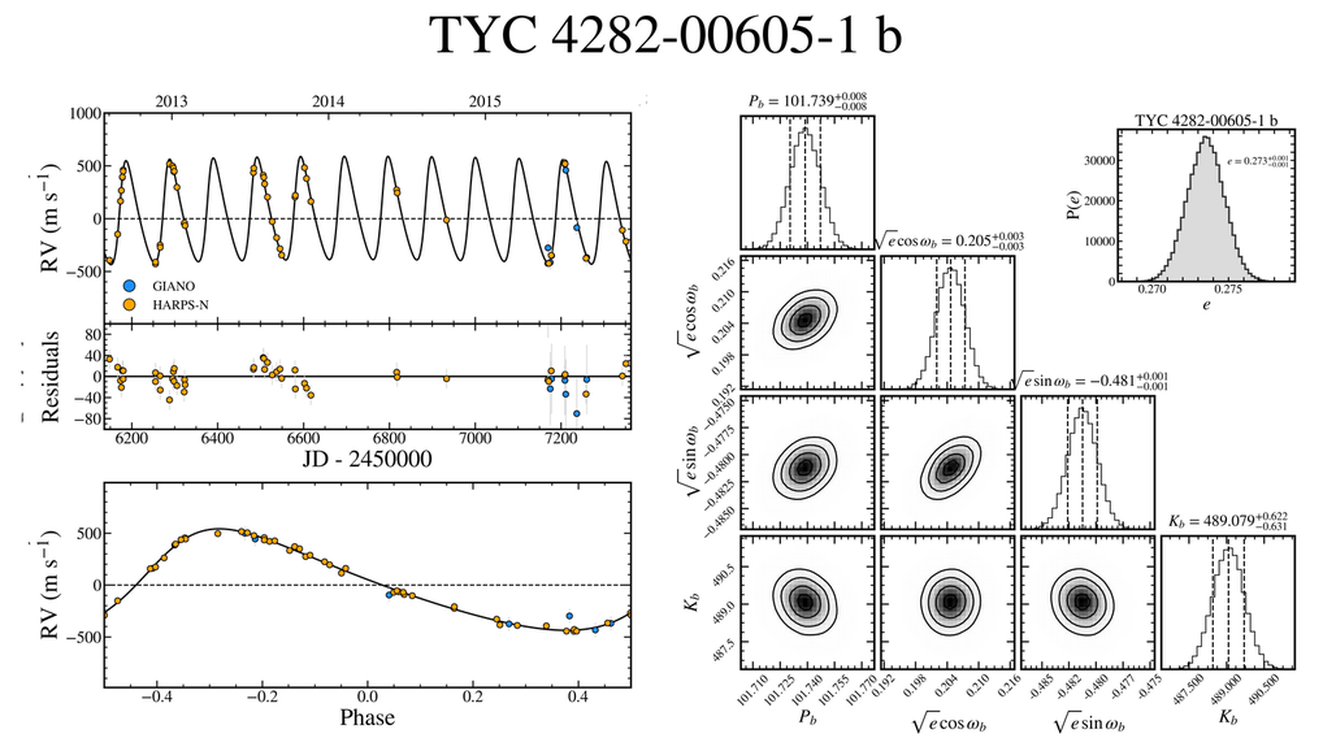}
 \end{minipage}
 \caption{Summary of results for the warm Jupiters TYC 3667-1280-1 b and TYC 4282-00605-1 b.}
 \label{fig:Combined_Plots95}
\end{figure}
\clearpage
\begin{figure}
\hskip -0.8 in
 \centering
 \begin{minipage}{\textwidth}
   \centering
   \includegraphics[width=\linewidth]{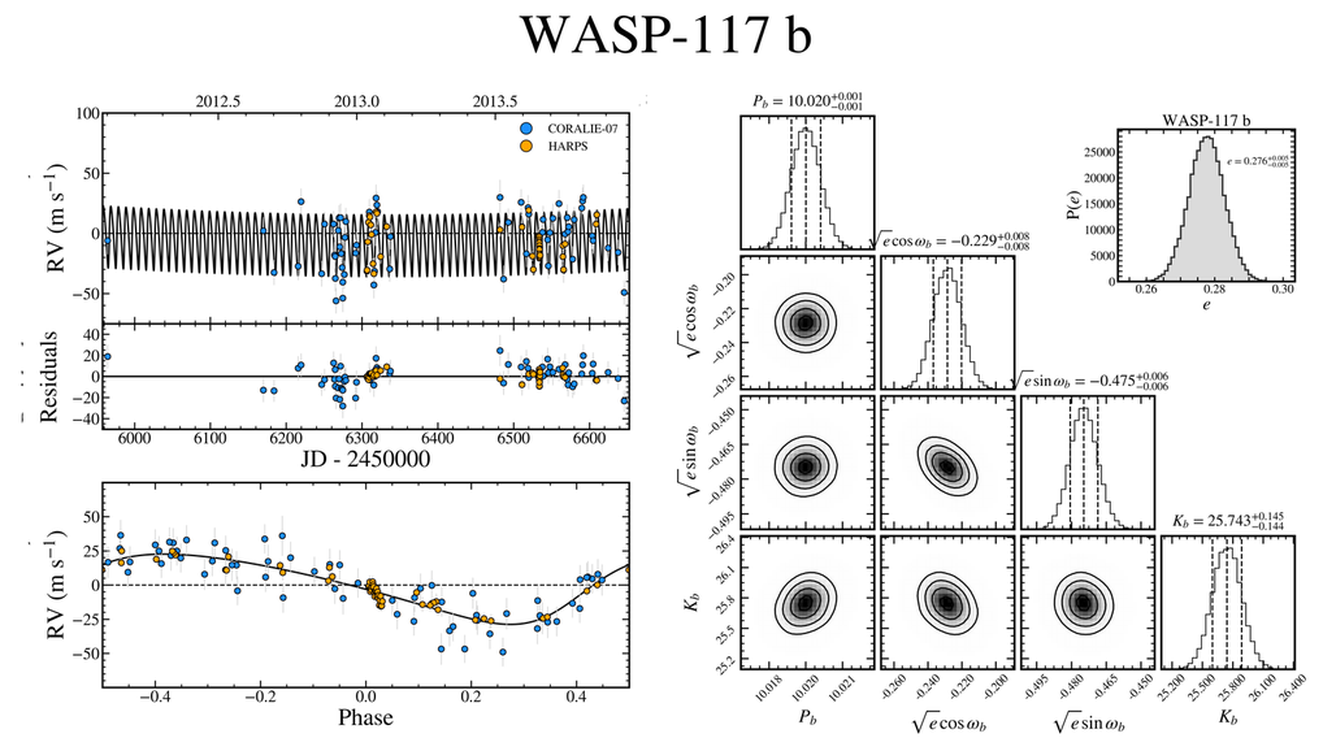}\\
   \vskip .3 in
   \includegraphics[width=\linewidth]{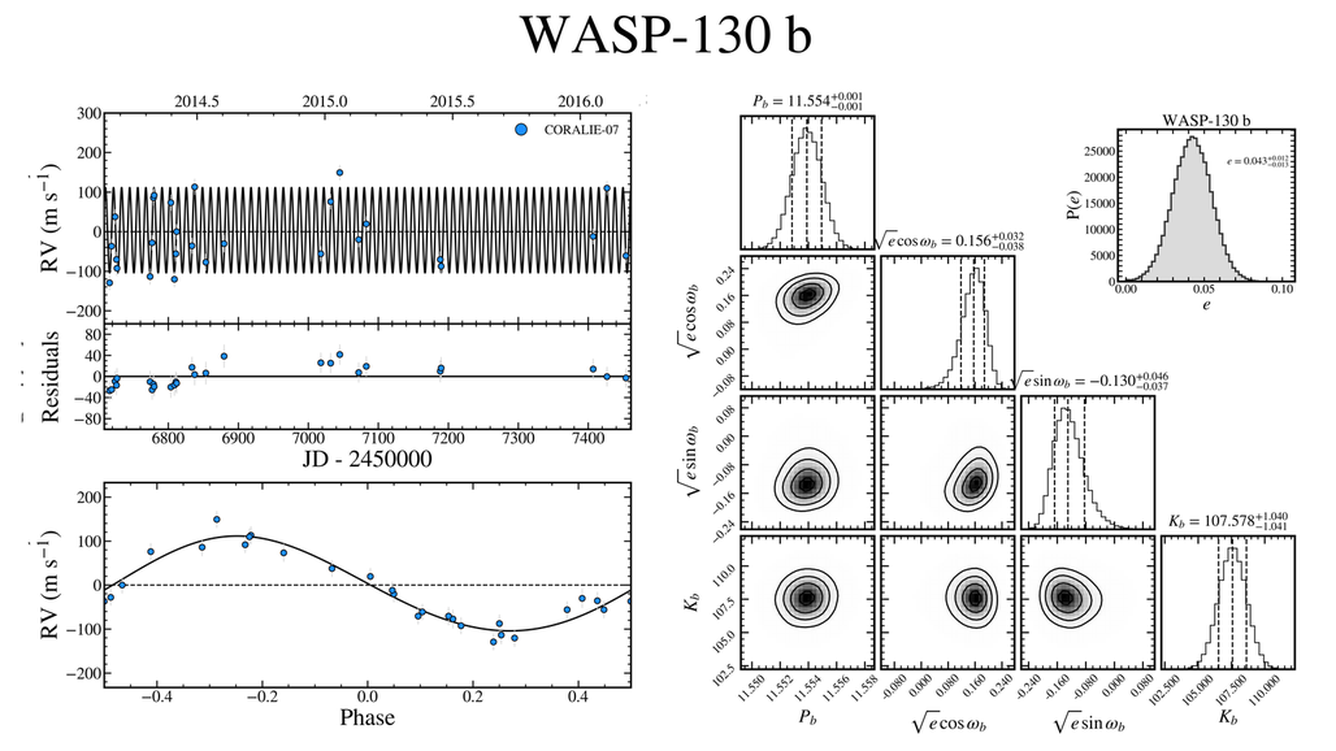}
 \end{minipage}
 \caption{Summary of results for the warm Jupiters WASP-117 b and WASP-130 b.}
 \label{fig:Combined_Plots96}
\end{figure}
\clearpage
\begin{figure}
\hskip -0.8 in
 \centering
 \begin{minipage}{\textwidth}
   \centering
   \includegraphics[width=\linewidth]{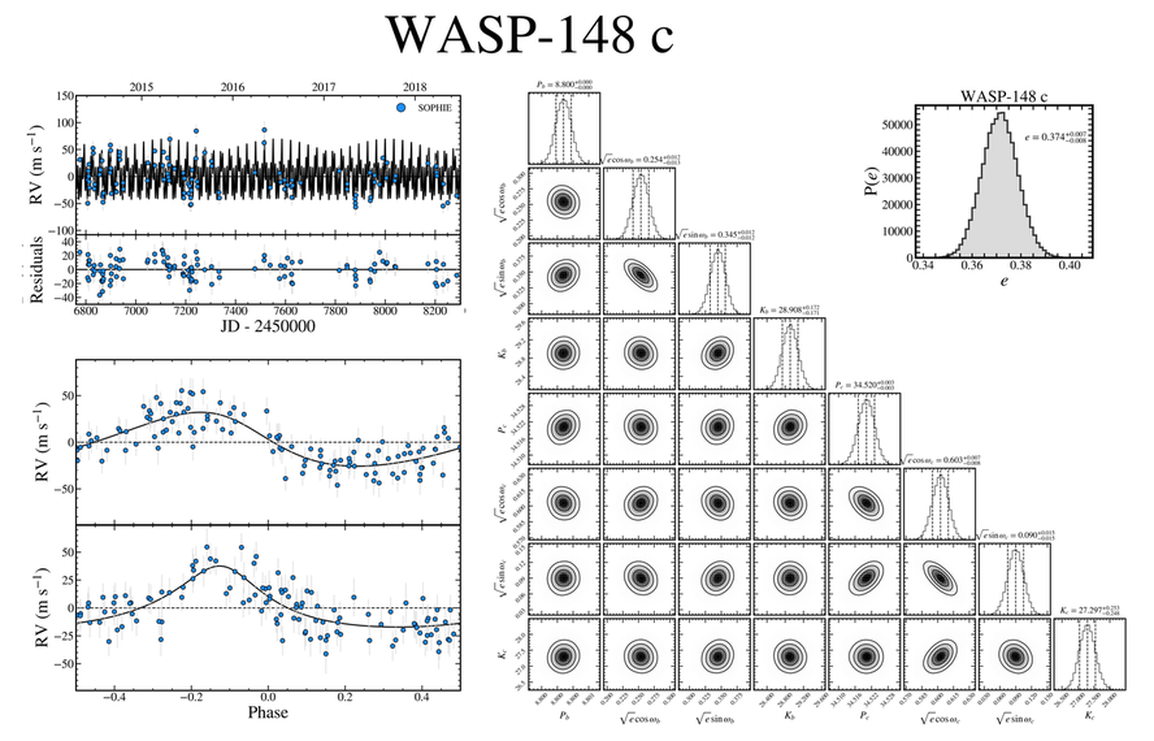}\\
   \vskip .3 in
   \includegraphics[width=\linewidth]{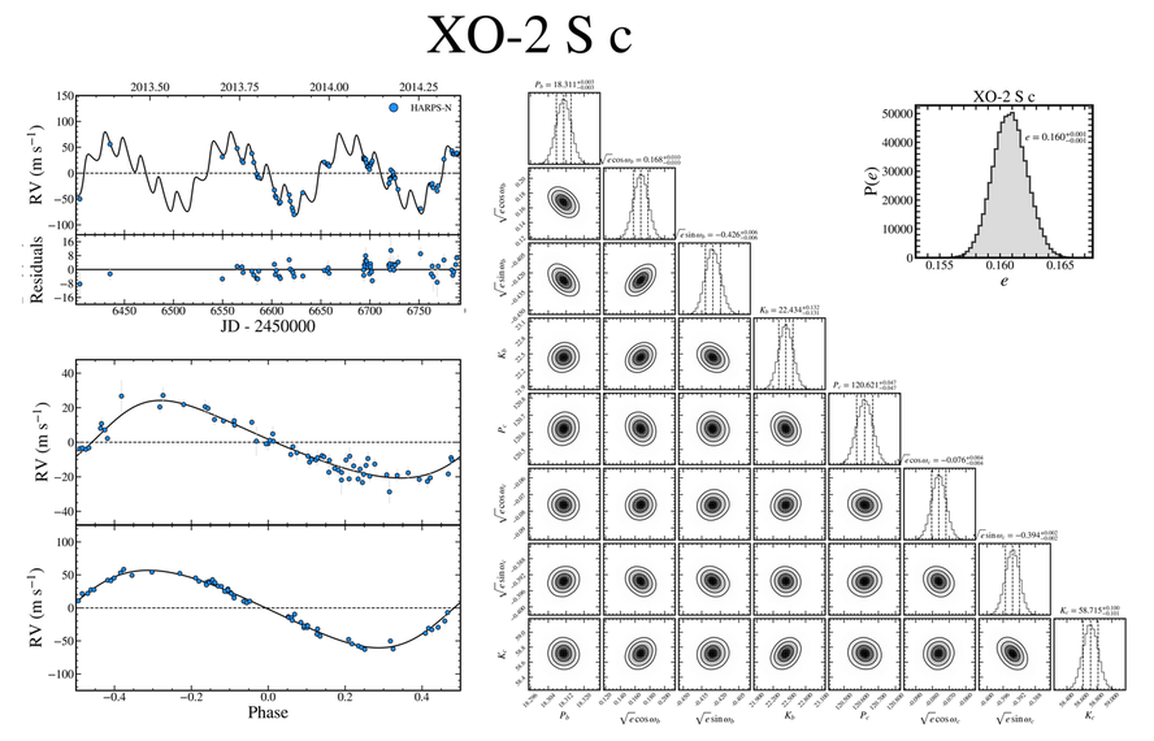}
 \end{minipage}
 \caption{Summary of results for the warm Jupiters WASP-148 c and XO-2 S c.}
 \label{fig:Combined_Plots97}
\end{figure}
\clearpage
\begin{figure}
\hskip -0.8 in
 \centering
 \begin{minipage}{\textwidth}
   \centering
   \includegraphics[width=\linewidth]{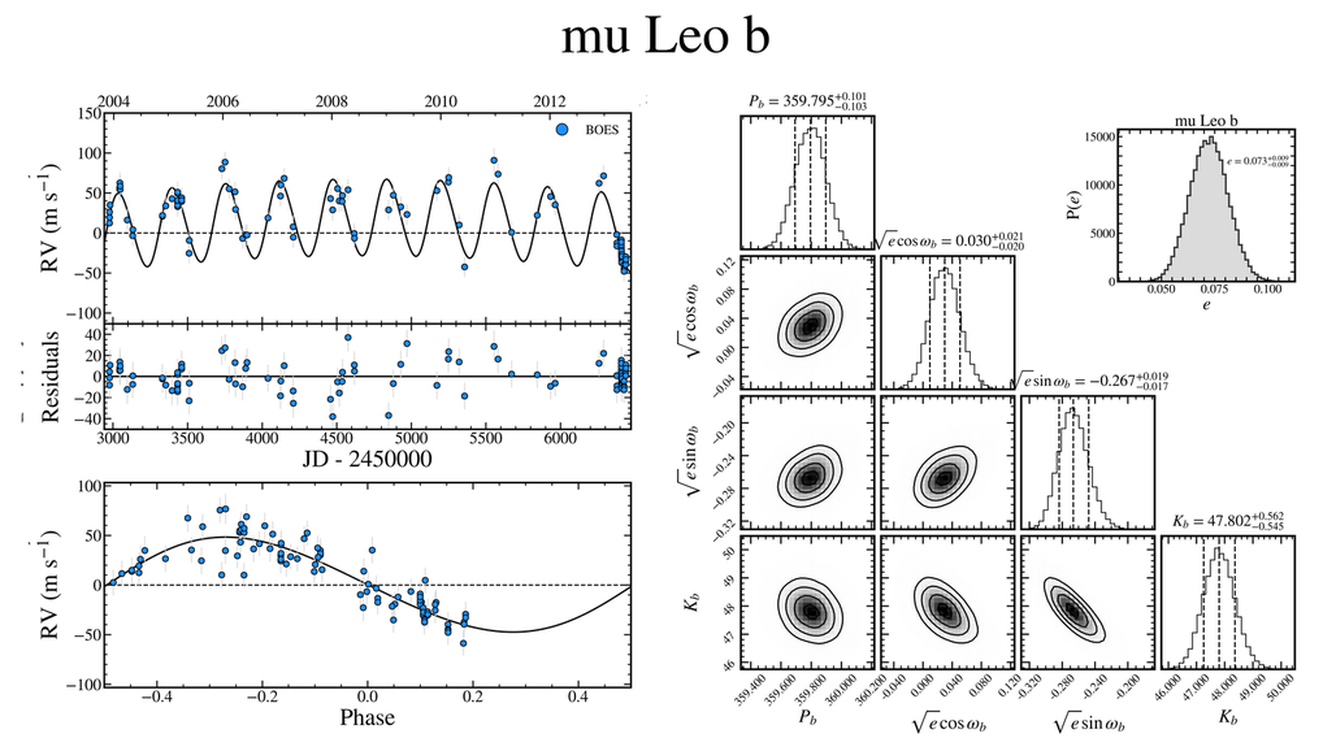}\\
   \vskip .3 in
   \includegraphics[width=\linewidth]{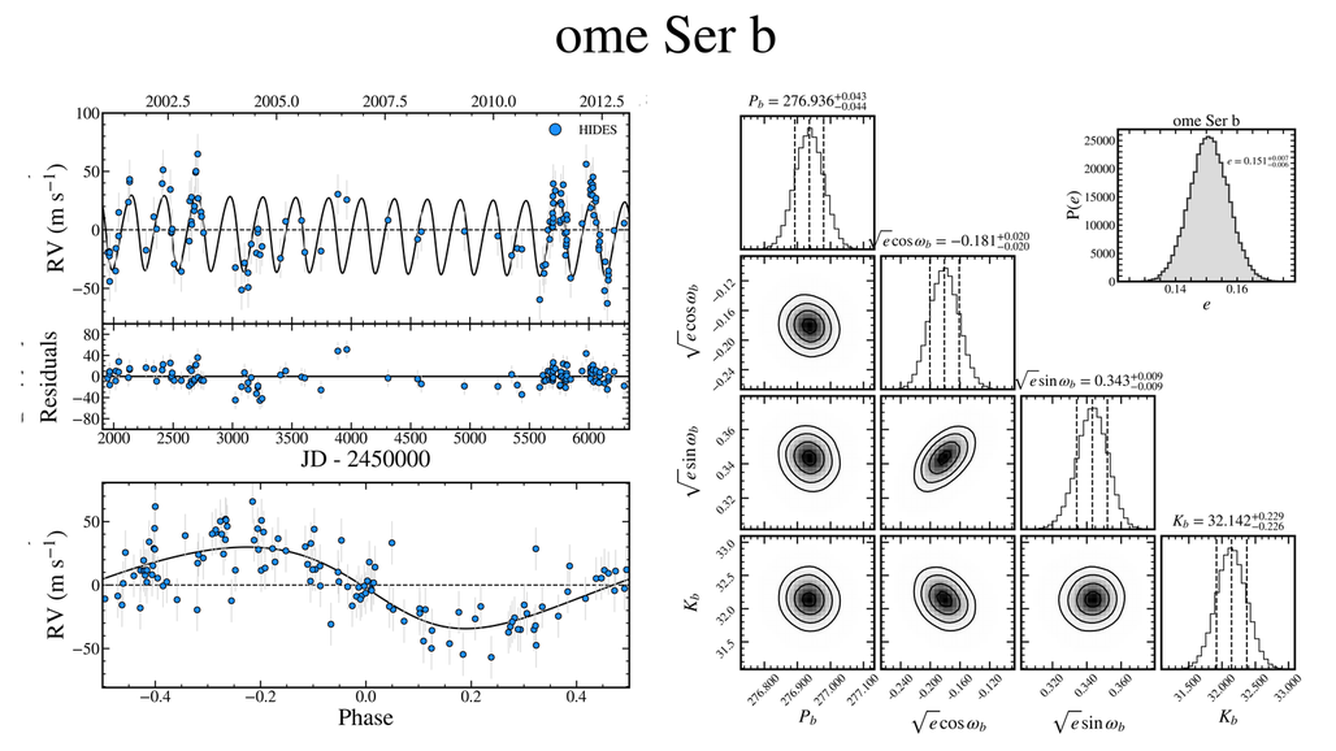}
 \end{minipage}
 \caption{Summary of results for the warm Jupiters mu Leo b and ome Ser b.}
 \label{fig:Combined_Plots98}
\end{figure}
\clearpage
\begin{figure}
\hskip -0.8 in
 \centering
 \begin{minipage}{\textwidth}
   \centering
   \includegraphics[width=\linewidth]{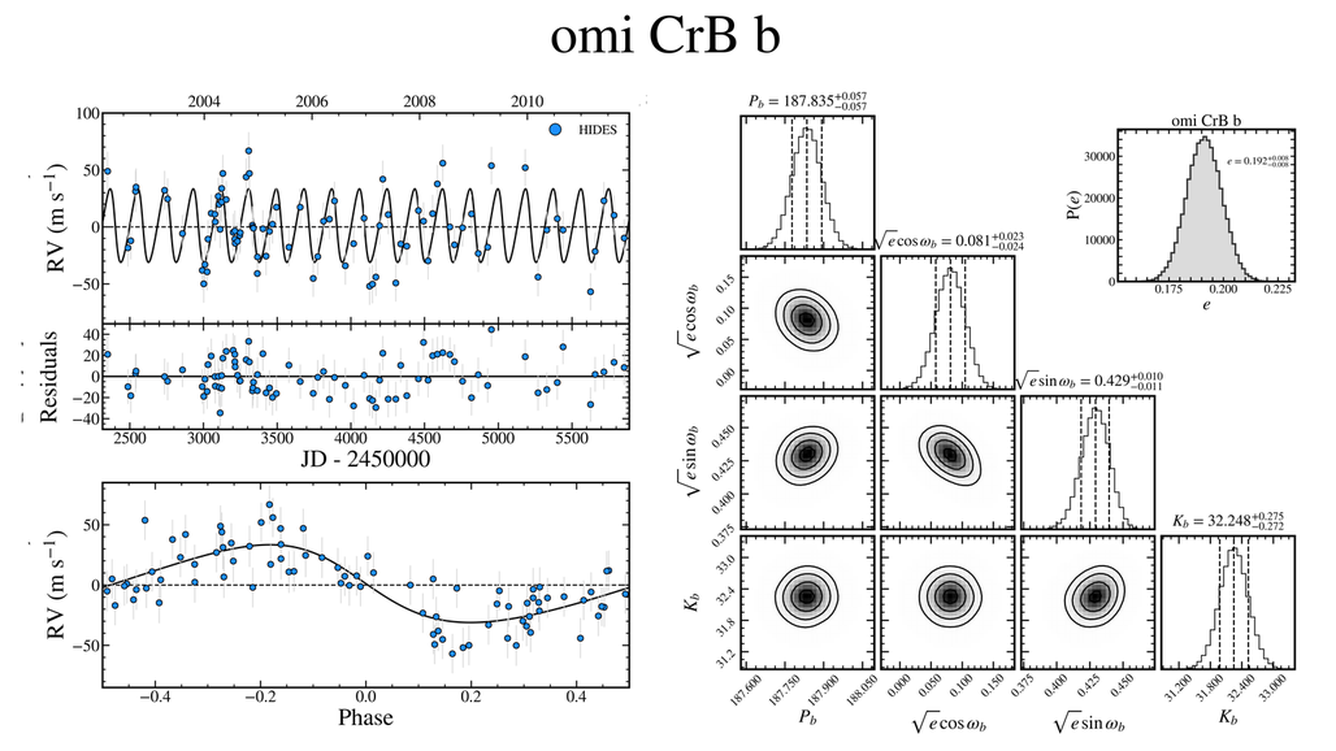}\\
   \vskip .3 in
   \includegraphics[width=\linewidth]{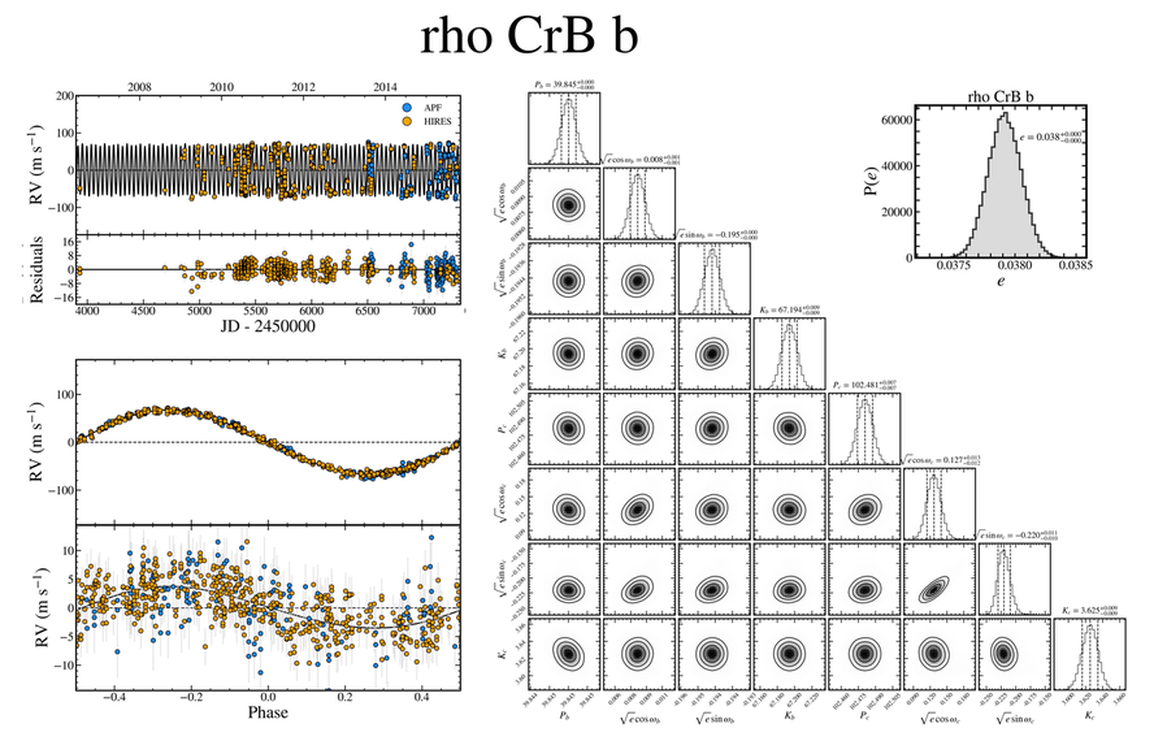}
 \end{minipage}
 \caption{Summary of results for the warm Jupiters omi CrB b and rho CrB b.}
 \label{fig:Combined_Plots99}
\end{figure}
\clearpage
\clearpage

\bibliography{sample63}{}
\bibliographystyle{aasjournal}

\end{document}